%% file: thesis-version-2-3.tex
\documentclass[a4paper,12pt,oldfontcommands]{memoir}
\pdfoutput=1 

\usepackage{thesisstyle} 
\usepackage{matdefs} 
\begin{document}

\frontmatter


\cleardoublepage
\thispagestyle{empty}
{
\centering

{\noindent\large\scshape Université de Genève \hfill Faculté des Sciences}\\[0.25\baselineskip]
{\noindent Section de Mathématiques \hfill Professeur Marcos Mari\~no Beiras}\\[\baselineskip]

\hrule
\vspace*{\baselineskip}
\parbox{0.95\textwidth}{\centering\Huge\bfseries\sffamily On the resurgence of renormalons\\ in integrable theories}
\vspace*{\baselineskip}
\hrule

\vspace*{2\baselineskip}
{\includegraphics[scale=0.625]{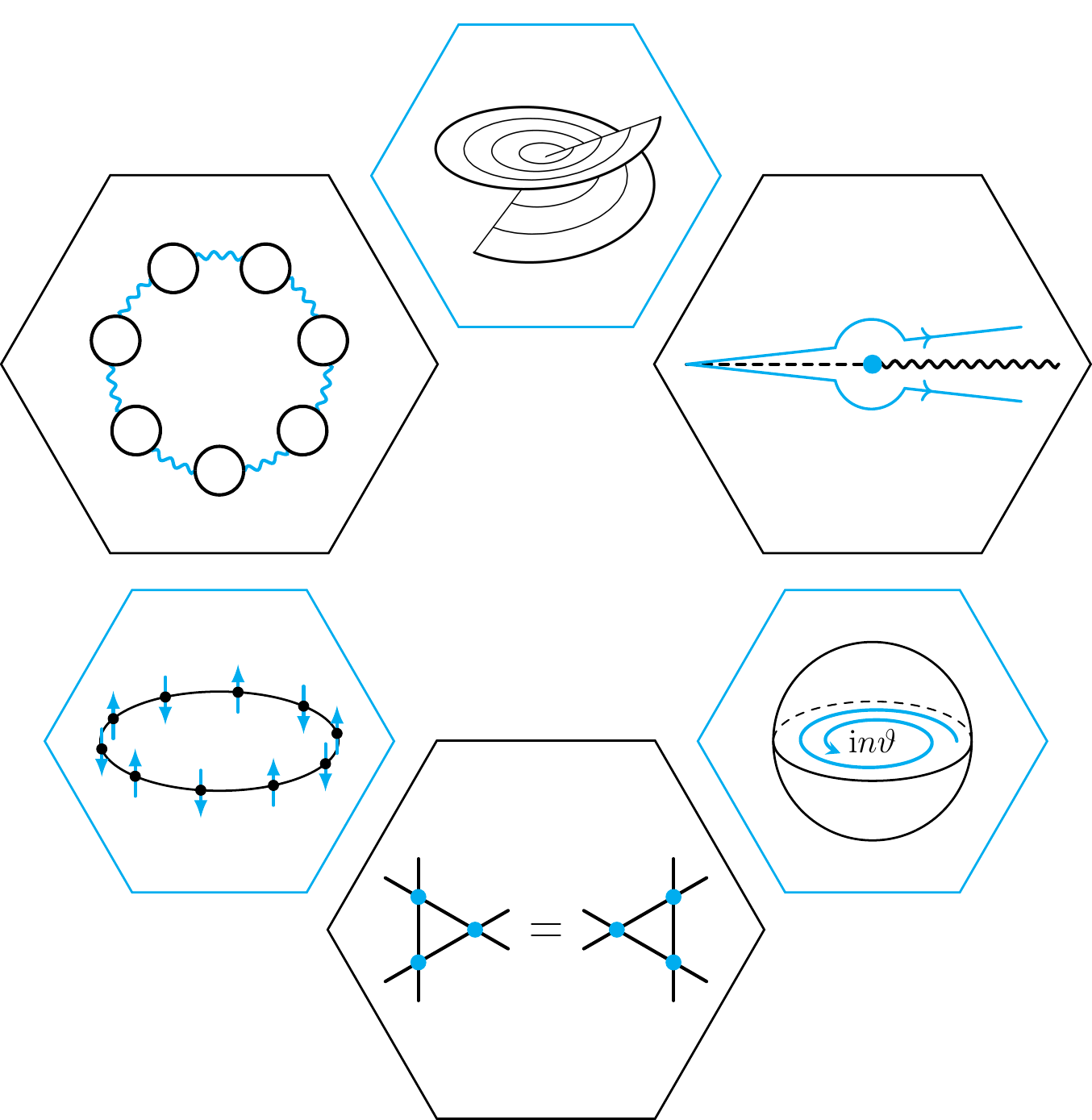}}
\vspace*{2\baselineskip}

\begin{center}
{\parbox{0.825\textwidth}{\centering
{\huge\scshape Tomás Reis}
}}
\end{center}

\vfill

\begin{center}
{\parbox{0.825\textwidth}{
\centering
{\Large Thèse N$^{\circ}$ $5676$}\\
\centering\Large\scshape 2022}}
\end{center}

}
\clearpage


\introsec{Members of the jury}
\raggedbottom 
{
\vspace{\baselineskip}
{\noindent\bf\sffamily Professor Marcos Mariño Beiras}\\
{\noindent\bf\sffamily Directeur de thèse}\\
Départment de Physique Théorique et Section de Mathématiques,\\
Université de Genève\\
Genève, Switzerland\\
\\

{\noindent\bf\sffamily Professor Alba Grassi}\\
Section de Mathématiques, Université de Genève\\
CERN, Theoretical Physics Department\\
Genève, Switzerland\\
\\

{\noindent\bf\sffamily Professor János Balog}\\
Institute for Particle and Nuclear Physics, \\
Wigner Research Centre for Physics\\
Budapest, Hungary

}

\pagebreak

\introsec{Abstract}
\raggedbottom 
\input{abstract}

\pagebreak

\introsec{Resumé en Français}
\input{resume-fr}

\pagebreak

\introsec{Acknowledgements}
\input{thanks}

\pagebreak

\tableofcontents

\pagebreak

\input{conventions}

\flushbottom

\mainmatter
\setcounter{secnumdepth}{3}

\input{thesis-text}

\linespread{1}
\printbibliography[keyword=own,
resetnumbers=true,
title = {Publications}]

\linespread{1}
\printbibliography[notkeyword=own,
resetnumbers=true,
title={Bibliography}]

\end{document}

%% file: abstract.tex
In this thesis we explore the physics of renormalons in integrable models under the framework of resurgence. In the first part, we review some background on resurgence, integrability and renormalons, including a discussion of renormalon at large $N$ and ring diagrams.
In the second part, we start from the Bethe ansatz integral equations, provided by the exact description of the ground state of integrable relativistic field theories, and obtain exact trans-series for the free energy in multiple models. These trans-series include non-perturbative effects which correspond to renormalons at unexpected positions in the Borel plane.
We validate the trans-series by inspecting their large $N$ limit, comparing them to the large order behaviour of perturbation theory and testing them against numerical solutions of the integral equations. We also study what happens to these trans-series under the effect of a $\vartheta$ topological angle. In the third part, we use the results from integrability to apply the techniques of resurgence in non-relativistic field theories. We find a relation between the energy gap of the system and the positions of singularities in the Borel plane. By studying the asymptotic behaviour of ring diagrams, we identify this relation with renormalons.
This thesis is based on the publications
\cite{mr-ll,mr-long,mr-ren,mr-2dren,mr-hubbard,mr-three,mmr,mmr-antrans,mmr-theta}
.

%% file: resume-fr.tex
La théorie quantique des champs décrit les champs fondamentaux qui composent l'univers. L'assomption que les champs interagissent faiblement, la théorie des perturbations, permet de faire des calculs, parfois confirmés par les expériences avec une précision époustouflante. Néanmoins, la théorie des perturbations ne parvient pas à inclure certains phénomènes, appelés non-perturbatifs. De plus, son produit est une série naïvement divergente.
Grace à la théorie de résurgence, les deux problèmes sont résolus d’un seul coup. La resommation de Borel transforme les trans-séries, qui incluent soit les termes perturbatifs soit les termes non-perturbatifs, en des fonctions bien définies. Cette théorie rend possible l’exploration de la physique non-perturbative en supposant des conditions de consistance sur la théorie des perturbations.

Il est naturel de demander quelle est l’origine des phénomènes non-perturbatifs dans la théorie quantique. Il y a une classe de tels effets, les instantons, qui sont dus à des solutions des equations Euclidiennes du mouvement. Dans le cas de la mécanique quantique, ils sont bien compris. Cependant, il y a une autre classe, les renormalons, sur lesquelles peu est connu. Au niveau de la théorie des perturbations, ils se manifestent comme des séquences divergentes de diagrammes de Feynman, e.g. les diagrammes en anneaux. Malgré leurs propriétés insaisissables, il est connu que les renormalons sont frequentes dans les théories de champs, incluant la chromodynamique quantique, qui explique la physique des quarks et protons dans notre univers.

Comme les renormalons sont difficiles à comprendre et les théories des champs sont des objets complexes, il est préférable de les étudier dans un cas de complexité réduite, mais pas trop non plus. Nous avons choisi d’analyser les théories des champs intégrables comme cas d’étude. Il s’agit des théories dans un espace-temps 1+1 dimensionnel qui satisfont la factorisation de la matrice-S. Ça veut dire que tous les processus de collision dans une théorie intégrable peuvent être partitionnés en les collision de 2 à 2. Ces théories sont résolues par l’ansatz de Bethe, qui décrit l’état fondamental comme une equation intégrale.

En adaptant les méthodes de Wiener--Hopf, nous avons extrait la trans-series analytique pour l’énergie libre à partir des équations intégrales données par l’ansatz de Bethe. Dans ces trans-series, on a identifié soit des effets semblables à des instantons, soit des effets semblables à des renormalons. Étonnamment, les positions dans l’espace de Borel associés aux renormalons ne sont pas en accord avec les prédictions par Parise et 't Hooft. Nous avons testés ces ``nouvelles renormalons'' avec les relations de résurgence, en utilisant la méthode de Volin pour obtenir un grand nombre de termes dans la théorie des perturbations.

Nous avons aussi analysé les systèmes non-relativistes, provenant de la théorie de la matière condensée. Dans ces systèmes, nous avons identifié les renormalons soit avec des termes exacts dans la théorie des perturbations, soit avec les diagrammes en anneaux. De plus, nous avons lié la présence de renormalons à l’existence d’un écart d’énergie dans la théorie. Un example d’un tel écart est le écart BCS qui caractérise les systèmes supraconducteurs.


%% file: thanks.tex
First and foremost, I would like to thank my advisor Marcos Mari\~no. For his insightful guidance and collaboration in the scientific projects and beyond, for helping me develop as an researcher, and for his constant camaraderie. I must also thank Ramon Miravitllas Mas as a collaborator and a friend, with whom it was always easy to work. 

I thank Jie Gu for helping me with the many perils of the Borel plane, and Szabolcs Zakany for helping me get started in research. 
I also thank Yoan Emery and Ramon for helping me with the redaction of the thesis, and Janos Balog for the comments on the final version.
I thank as well the HPC cluster Baobab, both its team and its emerging consciousness, without which many calculations would have been impossible.
Lastly, I would like to thank José Mour\~ao for the academic mentoring over the years. 

As a PhD student I had the privilege of taking my first steps in the scientific community. I would then like to thank all the groups and conferences that invited me to give talks or participate in scientific events, providing great opportunities for discussions.
I would like to thank in particular the organization and lecturers of the Solvay School of 2018, as well as my colleagues and friends of the ``Solvay class''.

My stay in Geneva was marked by the great colleagues and friends with whom I shared a significant share of my time at the University and outside of it.  In addition to those already mentioned, I would thus like to thank the good company of Adrian, Alex, Arunabha, Ben, Bharat, Claudia, Francesco, Igor, João, Julian, Manuel, Manus, Marco, Massi, Paulo, Pietro, Pranjal and Rahel, among others. I commend those who shared the office with me for their tolerance to my endless stream of pointless remarks, and for the wit of their retorts.

Also in Geneva was the iron-bound clan who I thank for warmly welcoming me into their Saturday dinners. Beyond the  alpine confines, I thank my friends all over the world who kept in touch, through all means physical and virtual. I also thank Anorien, Eudrastos, Igor, Jorgmund, Wuqrin and Owyn, wherever they might be.

A special thanks must be reserved to my family. Not only for having fostered the curiosity and scholarship on which I relied,  but also for being constant examples of humanity and perseverance.

Since my PhD coincided with the SARS-CoV-2 pandemic, I would finally thank those who kept the world running or who worked in the development of the vaccines.

%% file: conventions.tex
\introsec{Notation}
\addcontentsline{toc}{section}{Notation}
\begin{itemize}
\item $\IN$ are the natural numbers $1,2,3,\dots$.
\item $\re$ is the Euler-Napier constant. $e$ is used for variables such as the $e$nergy.
\item $\ri$ is the complex root $\sqrt{-1}$. $i$ is used for indices.
\item $\gamma_E$ is the Euler-Mascheroni constant.
\item $\Theta(x)$ is the Heaviside theta function.
\item $\psi^{(n)}(z)$ is the polygamma function.
\item We use the following convention for Fourier transforms
\begin{equation*}
\ba
\tilde f(\omega) &= \int_\IR \re^{\ri\omega\theta} f(\theta) \rd\theta,\\
f(\theta) &= \frac{1}{2\pi} \int_\IR \re^{-\ri\omega\theta} \tilde f(\omega) \rd\omega.
\ea
\end{equation*}
\item $\mathbb{H}_\pm$ is upper (lower) complex half plane.
\item Terms such as $\omega +\ri 0$ should be read as a limit with a small positive real $\epsilon$,
\begin{equation*}
\omega +\ri 0 = \lim_{\epsilon\rightarrow 0^+} (\omega+\ri\epsilon).
\end{equation*}
\item $\mathcal{H}$ is the Hankel contour, drawn below in the complex plane.
\item We use $(\approx)$ for asymptotically identical. $(\sim)$ is used more broadly, usually for leading contributions and approximations. We use $(\propto)$ instead of $(\sim)$ when we ignore overall constants. 
When we include the error in big $\CO$ notation we use both $(=)$ or $(\sim)$, depending on context. 
\end{itemize}
\begin{figure*}[h]
\centering
\begin{tikzpicture}[scale=0.75,
decoration = {markings,
mark = at position 0.25 with {\arrow{>}},
mark = at position 0.75 with {\arrow{>}}
},
]
\draw[help lines,->] (-1,0) -- (5,0) coordinate (xaxis);
\draw[help lines,->] (0,-2) -- (0,2) coordinate (yaxis);
\begin{scope}[very thick]
\path[draw,line width=0.8pt,postaction=decorate,color=cyan] (5,0.3) -- (0,0.3) arc (90:270:0.3) -- (5,-0.3);
\end{scope}
\node at (2,0.8) {$\mathcal{H}$};
\end{tikzpicture}
\end{figure*}
\pagebreak

\introsec{Acronyms}
\addcontentsline{toc}{section}{Acronyms}

\begin{multicols}{2}
\begin{description}
\item[BAE] Bethe Ansatz Equations
\item[BCS] Bardeen--Cooper--Schrieffer
\item[FKW] Fateev--Kazakov--Wiegmann 
\item[GN] Gross--Neveu model 
\item[GY] Gaudin--Yang model
\item[IR] Infra-Red
\item[lhs] left hand side
\item[LL] Lieb--Liniger model
\item[NLO] Next to Leading Order
\item[NLSM] Non-Linear Sigma Model
\item[ODE] Ordinary Differential Equation
\item[OPE] Operator Product Expansion
\item[PCF] Principal Chiral Field 
\item[QED] Quantum ElectroDynamics
\item[QCD] Quantum ChromoDynamics
\item[QFT] Quantum Field Theory
\item[RG] Renormalization Group
\item[rhs] right hand side
\item[RPA] Random Phase Approximation
\item[SUSY] SUperSYmmetry
\item[SYM] Supersymmetric Yang--Mills theory
\item[TBA] Thermodynamic Bethe Ansatz
\item[TQFT] Topological QFT
\item[UV] Ultra-Violet
\item[vev] vacuum expectation value
\item[WH] Wiener--Hopf
\item[YBE] Yang Baxter Equations
\end{description}
\end{multicols}

%% file: thesis-text.tex
\part{Resurgence, Integrability and Renormalons}
\label{part-intro}

\chapter{Introduction}
\label{cha_intro}
\setlength{\epigraphwidth}{.7\textwidth}
\epigraph{\raggedleft
Our imagination is struck only by what is great; but the lover of natural philosophy should reflect equally on little things.}
{Alexander von Humboldt,\\
Personal Narrative of Travels to the Equinoctial Regions of America, During the Years 1799-1804 -- Vol. 2 (1820)}
\setlength{\epigraphwidth}{.6\textwidth}

Millennia before the concept of science had come to be, philosophers
wondered: What is everything made of? How do the fundamental elements behave? 
Aristotelian physics had four elements, whose movement followed the tendency towards their natural sphere of being \cite{on-heavens,on-generation}.
And the idea of the indivisible constituent, the atom, goes back at least to Leucippus and Democritus \cite{sep-atomism-ancient} or to late Vedic literature
\cite{sep-naturalism-india}. 
To our best current knowledge, the answer to the first question is \textit{quantum fields}. The known universe seems to consist of about a dozen fermion fields, a handful of gauge fields and a Brout--Englert--Higgs field to wrap up the Standard Model \cite{higgs,brout-englert,higgs-3,weinberg-leptons,glashow-lepton,salam-weak,qcd}. The quantum theory of these fields, Quantum Field Theory (QFT), is among humanity's greatest scientific achievements, being both theoretically fundamental and experimentally accurate \cite{pdg2020}. A celebrated example, the electron magnetic moment, matches the Standard Model prediction to 3 parts in $10^{13}$ \cite{electron-g}, which is roughly the same as knowing the length of a cobra to the precision of a hydrogen atom. And while there are many mysteries left when it comes to the first question, from the eldritch depths of quantum gravity to the tallying of dark matter, our understanding of the second is surprisingly lacking. Despite almost a century of study since their first formulation \cite{born-jordan1,born-jordan2}, quantum fields are still befuddling objects.

Our mastery of quantum field theory is particularly focused on the perturbative aspects.
In the early days, even perturbation theory itself was controversial, giving QFT the fame of a dark art filled with faustian bargains to dispel incomprehensible infinities. The understanding of renormalization in the latter part of the XX century clarified most of this confusion \cite{wilson-rg,wilson-rg2,polchinski-rg,thooft-veltman,asym-freedom,asym-freedom2} and perturbation theory is now a mathematically rigorous endeavor \cite{costello-book}.
However, non-perturbative aspects are crucial and remain a challenge. For an example of their importance,  we simply point to confinement in quantum chromodynamics (QCD), which is responsible for the formation of protons and thus the stability of all quotidian matter. Fortunately, the span of non-perturbative tools has been growing over the years, now ranging from lattice methods and variational techniques to the bootstrap programs and intricate dualities, among many others. In this thesis, we combine two particular tools to study non-perturbative QFT, resurgence and integrability, and we use them to study one of the most elusive conundrums in QFT, renormalons.

With resurgence, we can study non-perturbative physics by starting from perturbative information.
The problem, but also the opportunity, starts with the output of perturbative quantum field theory, which are asymptotic series. These formal power series are divergent and at best provide approximations to a function and, prima facie, not well defined numbers.
For the practicing experimentalist, this is certainly sufficient to test perturbative QFT.
But conceptually and mathematically, it is worrisome that no matter how many perturbative corrections we calculate, they can never go beyond limited approximations, particularly if our goal is to delve into non-perturbative effects.

The mathematical framework of resurgence, based on the groundbreaking work of Écalle \cite{ecalle}, allows us to turn the divergent formal power series of perturbative QFT into proper functions. The object of study is extended from asymptotic series to trans-series, which also include non-analytic terms. In doing so, it permits and requires the incorporation of non-perturbative effects. The resulting consistency requirements for non-perturbative sectors coming from the perturbative expansion mean that we can learn about the former by studying the asymptotic behavior of the latter. Non-perturbative physics ``resurges'' in the high orders of the perturbative expansion, as was first explored in quantum mechanics \cite{bw1,bw2}. Resurgence is then both a mathematical foundation of quantum field theory and a bridge the known world of perturbation theory with the unknown expanse of non-perturbative phenomena.

For decades, it was thought that non-perturbative effects came from saddle points of the action. These solutions to the Euclidean equations of motion, instantons, would provide a basis on which to do further perturbation and build a semi-classical expansion.  Renormalon effects, discovered by late the 1970's \cite{gross-neveu,thooft,parisi1,parisi2}, curtailed this perspective. Renormalons are a feature of many quantum field theories, among which QCD, which are both non-perturbative and not saddle points of the action, as far as we currently know. 
Even half a century past their discovery, we still cannot say much about what renormalons actually are, much less how to calculate their contribution in a general quantum field theory. 
And, as we explore in this thesis, much of the little existing ``common lore'' turns out not to be general.
Thus, a key piece in making sense of perturbative and non-perturbative QFT, united through resurgence, remains shrouded in confusion.

Renormalons are a hard to grasp aspect of an already complicated object, a quantum field theory. Although the goal might be to one day understand renormalons in four dimensional gauge theories, it is better to start with more tractable cases. Integrable models are such a class of models. These are models heavily constrained by an infinite sequence of symmetries: all multi-particle collisions reduce to sequences of 2-to-2 collisions. Nonetheless, these are intricate interacting theories which share many features with realistic theories, particularly being asymptotically free and having a ``number of colours'' parameter $N$ \cite{thooft-largen}. This makes them ideal toy models, evading the Scylla of  oversimplification and the Charybdis of unwieldy complications.

We study integrable models because the best way to solve a problem in physics is to know the answer first. In integrable models, the answer is the Bethe ansatz \cite{bethe}. This ansatz provides exact wave-functions, from which in principle any observable can be obtained. In practice, this is a difficult procedure. The Bethe ansatz description of the ground state of these theories is given by integral equations whose weakly coupled limit is a problem with a long history in mathematical physics \cite{loveeq}. Only thanks to recent developments by Volin \cite{volin,volin-thesis}, did it become tractable to calculate the perturbative series to very high order, which welcomed the use of the resurgence techniques.
As we review in the second part of this thesis, even the non-perturbative effect can be cleanly extracted from the Bethe ansatz integral equations. 
One of the main results explored is the calculation of non-perturbative contributions to the free energy of the ground state, including renormalon effects, in the form of an analytic trans-series.

Furthermore, quantum field theory is not just a successful description of fundamental physics. It is also the preeminent language for quantum many body systems \cite{as}. Thus, the techniques and insights developed to study renormalons in integrable relativistic field theories can be extended to the study of the ground state of integrable quantum gases. For gases of fermions with an attractive interaction, the vacuum is formed by bound states. This induces an energy gap, from breaking one of them, which is a non-perturbative scale of the theory. As we review in the third part of this thesis, this gap appears in perturbation theory in the form of a renormalon effect. For spin $1/2$ fermions, we identify it as a manifestation of superconductivity \cite{bcs}.

The work reviewed in this thesis is a case study for the physics of renormalons. From the particular models and observables, we found novel insights with general implications. However, among other limitations, the methods used are specific to integrable models and impossible to generalize. The microscopic theory of renormalons remains a distant goal. 
\section{Structure of the thesis}

This thesis is divided into three parts.
Part \ref{part-intro} serves as the extended introduction and reviews some important background. It is divided into three chapters:
\begin{itemize}
\item  Chapter \ref{cha_resurgence} is an introduction to resurgence. It is a brief explanation of Borel summation, trans-series and the framework of resurgence. It contains all the resurgence tools we use in this thesis.
\item Chapter \ref{cha-integrability} introduces integrability in general and the specific models studied in the thesis. We start with a pedagogical construction of the Bethe ansatz for the Lieb--Liniger model, followed by a review of the Gaudin--Yang model with arbitrary spin and the one dimensional Hubbard model. 
Then we review the application of the Bethe ansatz to relativistic field theories, where we rederive some reference equations. We present the relativistic models studied and finally we review the physics of the topological $\vartheta$ angle in integrable sigma models.
\item  Chapter \ref{sec_renormalons} introduces renormalons. It reviews some of the standard knowledge about this non-perturbative phenomenon. 
It ends with a schematic review of large $N$ renormalons as seen from ring diagrams in the $O(N)$ non-linear sigma model. This case serves as a paradigmatic example of renormalons, by showing how they connect to Feynman diagrams and discussing some relevant concepts. This last section is a summary of \cite{mmr} with some mention of \cite{mr-2dren}. 
\end{itemize}
The advanced reader can mostly skip or skim chapters \ref{cha_resurgence} and/or \ref{cha-integrability}, depending on their background, but we recommend reading section \ref{cha_largeN} for context on the problem of renormalons.

Part \ref{part-iqft} is dedicated to the study of renormalons and trans-series in integrable relativistic field theories. It is divided into two chapters,
\begin{itemize}
\item Chapter \ref{cha_antrans} is a comprehensive review of how to extract trans-series from the Bethe ansatz integral equations. We start with a detailed introduction to the case of Gross--Neveu followed by supplementary technical details. Then we extend the analysis to bosonic models, followed by a list of concrete results in specific models. We then compare the large $N$ limit of the results with previous literature. Finally, we study the exceptional case of the $O(3)$ non-linear sigma model, followed by models with a $\vartheta=\pi$ topological angle. This chapter reviews the analytic results of \cite{mmr-antrans, mmr-theta}.
\item Chapter \ref{cha_volin} includes both an in-depth review of Volin's method and a review of how to use exact series to test the trans-series results of chapter 5.
The first half of this chapter includes the details of the derivation of Volin's method and the particularities of applying it to relativistic field theories, expanding on the presentation of \cite{mr-ren,mr-long}.
The second half contains the details of numeric calculations that use the output of Volin's method to verify the analytic trans-series. This reviews the numeric analysis of \cite{mr-ren, mmr-antrans} with some additional details, updating the discussion of the former in light of the latter.
\end{itemize} 
For a reader focused on the overall picture and results, we recommend reading sections \ref{sec-summary}, \ref{sec-gn-1} and \ref{sec-bos}.

Part \ref{part-qmb} is dedicated to the study of renormalons in one dimensional quantum many body systems. 
This part remixes elements of \cite{mr-long, mr-ll, mr-hubbard, mr-three} into a single story. It is divided into two chapters,
\begin{itemize}
\item Chapter \ref{cha_GY} is a review of the relationship between superconductivity and resurgence. It focuses on the Gaudin--Yang model and the Hubbard model, in the case of spin $1/2$ fermions. We review some background on the BCS gap. Then we introduce a conjecture tying the gap to singularities in the Borel plane of the Borel transform of the ground state energy. We use Volin's method to test this conjecture in the Gaudin--Yang model. We verify the conjecture by also looking at the repulsive Gaudin--Yang model and at the Hubbard model in the small density limit.
\item Chapter \ref{cha_spin} focuses on the same models but with $\kappa$-components. It reviews the derivation of the energy gap in the Gaudin--Yang model both from the Bethe ansatz and from a relativistic effective theory. We then use ring diagrams in the large $\kappa$ limit to relate the conjecture of chapter 7 to renormalons, both in the Gaudin--Yang and Hubbard models.
\end{itemize} 
This part is conceptually independent from part \ref{part-iqft}, 
although some technical results are used.
Particularly, series obtained with Volin's method, described in chapter \ref{cha_volin}, are heavily discussed.

Lastly, part \ref{part-final} contains a brief closing discussion followed by two small appendices. Appendix \ref{app_volin} contains details on the implementation of Volin's method. Appendix \ref{Lieb--Liniger} describes the application of Volin's method to the Lieb--Liniger model and the disk capacitor problem, adding some details to the analysis presented in \cite{mr-ll}.

This thesis is based on the papers \cite{mr-ll,mr-long,mr-ren,mr-2dren,mr-hubbard,mr-three,mmr,mmr-antrans,mmr-theta}, of which I am a coauthor. In general, these results were the product of active collaboration between all the respective authors.
While the goal of the text is to review the results and methods of these papers, I tended to emphasize calculations to which I contributed in a substantial way.  
However, some crucial calculations were mostly the work of my collaborators with a less significant contribution of my own. 
In particular, the perturbative calculation for bosonic models in section \ref{sec-pert-bos} and the analytic trans-series for the Gaudin--Yang model reviewed in section \ref{sec-gy-antrans} were almost entirely done by my co-authors.

\chapter{Asymptotics and resurgence}
\label{cha_resurgence}
\epigraph{
\raggedleft
Why does it disturb us that the map be included in the map and the thousand and one nights in the book of the Thousand and One Nights? 
}
{Jorge Luis Borges, \\
Labyrinths: Selected Stories \& Other Writings (1962)}

In this first introductory chapter, we will motivate and introduce the framework of resurgence. The goal of this short review is that a reader with no previous exposure to resurgence can become familiar with it to the extent which is used in the main text. The presentation here generally follows \cite{mmlargen}, with some elements from \cite{dorigoni-resurgence}, both of which are good introductions to the topic.
For other in-depth reviews on resurgence and asymptotics, see also  \cite{abs,costinbook,sauzin}.

\section{Borel summation}

In physics, particularly in quantum physics, exact solutions are often out of reach. We then do perturbation theory, where we take an easy solution and ``perturb'' it with some small parameter $g$. The result of such process is a series which approximate the exact value of some observable $f(g)$,
\begin{equation}
\Big| f(g) - \sum_{n=0}^N a_n g^n\Big| \sim a_{N+1} g^{N+1}+\CO(g^{N+2}),\quad g\ll 1.
\label{asym-series}
\end{equation}
One could expect that the series is convergent and taking infinite terms would resolve $f(g)$. This is rarely the case. Rather, the series is asymptotic, which means the sum is divergent. It is still a good approximation, but, as we take more and more terms, the range of $g$ where \eqref{asym-series} holds becomes smaller and smaller. Conversely, if we hold $g$ fixed, the first few terms of the series improve the approximation of $f(g)$ but at some point the divergence starts to dominate and the approximation worsens.

This behavior often has a physical cause.
For example, in \cite{dyson-divergence-qed}, Dyson argued that this is to be expected in QED, and his argument can be projected into other problems. He posits that if the series in $e^2$ were to be convergent, then it would be convergent in a disk in $\mathbb{C}$, including negative $e^2$. But negative $e^2$ results in completely different physics, e.g. the vacuum would be unstable, and we should not expect the analytic continuation of QED to make sense . Thus, the series should not converge.

On a more mathematical ground, series in the quantum problems diverge because typically they are of the form
\begin{equation}
a_n\sim A^{-n} n!.
\end{equation}
At the end of this chapter, we will peek into why this is the case. The radius of convergence of such a series is $0$ but a perfectly finite function can be approximated in the sense of \eqref{asym-series}. At fixed $g$ one could find the minimal term $a_n g^n$, after which the divergence spoils the approximation,
\begin{equation}
\partial_n  |A^{-n} n! g^n| \big|_{n=N^*}=0\Rightarrow N^* = \Big|\frac{A}{g}\Big|, \quad |g|\ll 1
\end{equation}
where we use the Stirling approximation for the factorial $n!\sim \re^{n(\log n-1)}$. The truncation $N^*$ is the optimal truncation of the series and using \eqref{asym-series} we get
\begin{equation}
\Big| f(g) - \sum_{n=0}^{N^*} a_n g^n\Big| \sim \re^{-|A/g|},\quad g\ll 1.
\label{optimal-trunc}
\end{equation}
The error bound of \eqref{optimal-trunc} is concerning but enlightening. Concerning because it suggests that even if we have access to the infinite sequence of $a_n$ we can only approximate $f(g)$ up to some fixed error. Enlightening because it implies that limits of perturbation theory are of the same order as non-perturbative effects $\re^{-|A/g|}$. Furthermore, we see that the coefficient $A$ that we find in the large order behavior of the perturbative coefficients $a_n$ gives a clue about the form of the non-analytic terms that perturbation theory ignores. Luckily, the concerning aspect can be improved upon with more sophisticated methods, and this ``clue'' of non-perturbative physics can be made more rigorous.

Let us define a series as Gevrey-1 if two constants $A$ and $C$ exist such that
\begin{equation}
|a_n| \leq C A^n n!, \quad n\gg1.
\label{gevrey}
\end{equation}
From this point on, we assume that the power series we are concerned with are of this class.
What we need is a theory that relates asymptotic series, which are formal power series, to actual functions, objects that take values in $\IR$ or $\mathbb{C}$, and in the process incorporates non-perturbative contributions. This is the domain of Borel summation.

Let a formal power series be of the form
\begin{equation}
\varphi(g) = \sum_{n\geq 0} a_n g^n,
\label{typical-series}
\end{equation}
with $a_n$ a Gevrey-$1$ series, as defined in \eqref{gevrey}.
We then define the Borel transform as
\begin{equation}
\mathcal{B} \varphi(\zeta) = \hatvphi(\zeta) = \sum_{n\geq 0} \frac{a_n}{n!} \zeta^n.
\label{borel-transform}
\end{equation}
For a Gevrey-1 series, this series must be convergent in a small radius around the origin. Therefore, the Borel transform takes us from the strange netherworld of formal power series to the crystalline realm of analytic functions. If the radius of convergence of the series $\varphi$ was finite to begin with, then $\hatvphi$ would be entire. The question now is what does $\hatvphi$ tells us about $f$.

Through a modified Laplace transform, we can obtain a function of $g$ from $\widehat{\varphi}$, given by
\begin{equation}
s(\varphi)(g) = \CL \hatvphi(\zeta) = \int_0^\infty \re^{-\zeta} \hatvphi(g\zeta)\rd \zeta,
\label{borel-sum}
\end{equation}
this the Borel sum, or summation, of the power series $\varphi$. It is a simple exercise to check that
\begin{equation}
s(\varphi)(g) - \sum_{n= 0}^N a_n g^n = \CO(g^{N+1}),\quad g\ll 1.
\end{equation}
So $s(\varphi)(g)$ also has $\varphi$ as its asymptotic series. It is often convenient to rewrite \eqref{borel-sum} as
\begin{equation}
s(\varphi)(g) =  \frac{1}{g}\int_0^\infty \re^{-\zeta/g} \hatvphi(\zeta)\rd \zeta,
\label{borel-sum-alt}
\end{equation}
so that the dependency in $g$ lies outside of $\hatvphi$.

The Borel summation procedure takes us from an asymptotic series to an analytic function and then to a function of $g$ which is approximated by the  original asymptotic series. Schematically,
\begin{equation}
\begin{tikzpicture}[baseline=(current  bounding  box.center)]
\node (asym) {$\varphi(g) = \sum a_n g^n$};
\node (borel) [right=of asym] {$\hatvphi(\zeta)=\sum \frac{a_n}{n!} \zeta^n$};
\node (sum) [below=of borel] {$s(\varphi)(g) = \int_0^\infty \re^{-\zeta} \hatvphi(g\zeta)\rd \zeta$};
\node (function) [left=of asym] {$f(g)$};
\draw[->] (asym.east) -- node[above] {$\CB$} (borel.west);
\draw[->] (borel.south) -- node[right] {$\CL$} (sum.north);
\draw[->] (sum.west) -- node[above right] {$\approx$}(asym.south);
\draw[->] (function.east) -- node[above] {$\approx$}(asym.west);
\draw[->, dashed] (sum.west) to[out=160,in=-90] node[above] {?}(function.south);
\end{tikzpicture}.
\label{borel-sum-scheme}
\end{equation}
We have not answered whether this recovers the object we were trying to calculate. Two functions with the same same asymptotic approximations are either the same or differ by non-analytic contributions. For example, a term $\re^{-1/g}$ which is smaller than $g^N$ for any $N$. So the function $f(g)$ and $f(g)+\re^{-1/g}$ both have the same asymptotic series.

In some cases one can conclusively show that the Borel summation is the unique function approximated by the asymptotic series, as is the case of Phragmén--Lindel\"of theorem, by imposing additional constraints. Roughly, if the asymptotic series approximates a function $f(g)\approx\varphi(g)$ for $g$ spanning at least the sector $\arg g\in [-\pi/2,\pi/2]$, then the function is identical to the Borel summation, $f=s(\varphi)$, for $g>0$. The idea is that the function could differ from the resummation of the asymptotic series by exponentially suppressed factors $\sim\re^{-1/g}$. However, as $\arg g$ goes to $\pm \ri\pi/2$, these factors would become $|\re^{\pm\ri /|g|}|\sim \CO(1)$ which would obstruct the asymptotic approximation to any order $\CO(g^N)$. Then they cannot be there and the Borel summation must be exact. The Nevanlinna--Sokal\footnote{Nevanlinna proved this theorem in 1918 \cite{nevanlinna}, while Sokal brought to it to attention and rederived it in the 1980 \cite{sokal-nev}.} theorem further refines this statement.  For detailed presentations, see for example \cite{ramon-resurgence,stingl,costinbook}. However, in the problems we will be concerned with, this almost never happens.

\subsection{Large order behavior and Borel summation}

In order to better understand the relationship between Borel summation and large order behavior, let us start with a very elementary example, the exponential integral function. It is defined by
\begin{equation}
E_1(z) = \int_z^\infty\frac{\re^{-t}}{t}\rd t.
\end{equation}
Let $f_{-1}$ be
\begin{equation}
f_{-1}(z) = \re^{z} E_1(z),
\label{f-def}
\end{equation}
its asymptotic series when $z\gg 1$ is given by
\begin{equation}
f_{-1}(z) \approx \frac{1}{z}\sum_{n\geq 0} \frac{(-1)^n n!}{z^n} = \frac{1}{z}\phi_{-1}(1/z).
\end{equation}
The subscript $(-1)$ will become evident later.
  $1/z$ is our small parameter as in \eqref{typical-series}, and we can define the Borel transform \eqref{borel-transform} of $\phi(1/z)$
\begin{equation}
\widehat{ \phi}_{-1}(\zeta) = \sum_{n\geq 0} (-1)^n \zeta^n = \frac{1}{1+\zeta}.
\end{equation}
The Borel resummation \eqref{borel-sum} is easy to perform in this case,
\begin{equation}
s(\phi)(1/z) = \int_0^\infty \frac{\re^{-\zeta}}{1+\zeta/z}\rd \zeta = z \re^{z} \int_{z}^\infty \frac{\re^{-t}}{t}\rd t,
\label{phi-bs}
\end{equation}
which with the substitution $\zeta = t-z$ recovers the original definition
\begin{equation}
f_{-1}(\phi) =  \frac{1}{z} s(f)(z),
\end{equation}
as intended.

Let us now suppose that $z$ is negative, then there is a pole in the argument of
\eqref{phi-bs}. Consider instead the Borel sum performed at $\re^{\mp\ri 0}z$ where $\ri 0$ means a number with a small positive imaginary part. Then, we have
\begin{equation}
s(\phi)\left(\frac{\re^{\pm\ri 0}}{z}\right) = \int_0^{\re^{\pm\ri 0}\infty} \frac{\re^{-\zeta}}{1+\zeta/z}\rd \zeta.
\end{equation}
We can see that there is a branch cut because of the pole at $\zeta=-z$, which leads to a residue contribution
\begin{equation}
s(\phi)\left(\frac{\re^{+\ri 0}}{z}\right) - s(\phi)\left(\frac{\re^{-\ri 0}}{z}\right) = - 2\pi\ri\, z \re^{z}.
\label{dif-s-phi}
\end{equation}
However this is precisely as expected, since the exponential integral function has logarithmic branch cut from $-\infty$ to $0$,
\begin{equation}
\Im E_1(\re^{\mp \ri 0}z^{-1}) = \mp \ri \pi, \quad z<0.
\end{equation}

\begin{figure}
\begin{center}
\begin{tikzpicture}[scale=1.25]
\draw[->, thick] (-1,0)--(5,0) node[right]{$\textrm{Re}\,\zeta$};
\draw[->, thick] (0,-0.5)--(0,3) node[left]{$\textrm{Im}\,\zeta$};
\begin{scope}[rotate=30]
\draw[black, dashed] (0,0) -- (5,0);
\filldraw[cyan, fill=cyan] (2.5,0) circle (2pt) node[below,black]{$A$};
\begin{scope}[decoration={markings,
    mark=at position 0.75 with {\arrow{>}}}]
\path[draw,line width=1pt,color=cyan,postaction=decorate] 
    (0,0) -- (2,2*0.11043) arc (160+6.3401:20+6.3401:0.5) -- (4.5,0.5);
\path[draw,line width=1pt,color=cyan,postaction=decorate] 
    (0,0) -- (2,-2*0.11043) arc (-160-6.3401:-20-6.3401:0.5) -- (4.5,-0.5);
\end{scope}
\draw[] (4,0.5) node[above]{$s_{\theta+}$};
\draw[] (4,-0.5) node[below]{$s_{\theta-}$};
\end{scope}
\draw[black, thick] (1,0) arc (0:30:1);
\draw[] (1.2,.25) node[]{$\theta$};
\end{tikzpicture}
\end{center}
\caption{The countours $s_\pm$ and the pole of $\hatvphi$ in the Borel plane.}
\label{fig-Borel}
\end{figure}
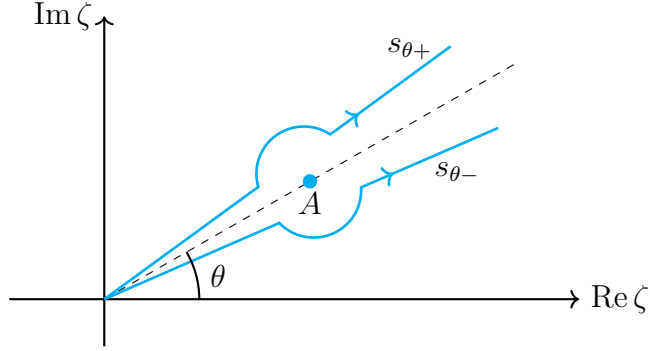

While this example is quite simple, the behavior seen here is quite paradigmatic. To make it more evident let
\begin{equation}
\phi_A(g) = \sum_{n\geq 0} A^{-n} n! g^n + 
\Big\{\begin{array}{c}
\text{convergent}\\
\text{series}
\end{array}\Big\}
 \rightarrow \widehat{\phi}_A(\zeta) = \frac{A}{A-\zeta} + \Big\{\begin{array}{c}
\text{entire}\\
\text{function}
\end{array}\Big\},
\label{phi-A}
\end{equation}
for some complex number $A$.
Let us also deform definition \eqref{borel-sum-alt} so that we integrate a ray in the complex plane with argument $\theta$,
\begin{equation}
s_\theta(\varphi)(g) = \frac{1}{g}\int_0^{\re^{\ri\theta}\infty}\re^{-\zeta/g} \hatvphi(\zeta)\rd\zeta.
\end{equation}
We also introduce the integrals slightly above and below this one, see figure \ref{fig-Borel},
\begin{equation}
s_{\theta\pm}(\varphi)(g) = \frac{1}{g}\int_0^{\re^{\ri\theta\pm\ri 0}\infty}\re^{-\zeta/g} \hatvphi(\zeta)\rd\zeta,
\end{equation}
and their discontinuity
\begin{equation}
\disc s_{\theta}(\varphi)(g) = s_{\theta+}(\varphi)(g) -  s_{\theta-}(\varphi)(g).
\end{equation}
This is a natural expression of the analytically continued Borel sum \eqref{borel-sum} for complex $g$. If $\arg g =\theta$, then
\begin{equation}
s(\phi)(g) = \int_0^\infty \re^{-\zeta}\hatvphi(g\zeta)\rd\zeta = \frac{1}{g}\int_0^{\re^{\ri\theta}\infty}\re^{-\zeta/g} \hatvphi(\zeta)\rd\zeta.
\end{equation}
Let us now choose $\theta=\arg A$. From \eqref{phi-A}, we see that this places the ray just so that it hits the singularity at $\zeta = A$, see figure \ref{fig-Borel}.
With these definitions, we can write \eqref{dif-s-phi}
\begin{equation}
s_{\theta+}(\phi_A)(g) -  s_{\theta-}(\phi_A)(g) = 2\pi\ri A g^{-1} \re^{-A/g}.
\label{phiA-disc}
\end{equation}
From this example, we see that from the large order behavior of the series we identify a singularity $\zeta=A$ and the Borel sum has a discontinuity along the ray where that singularity lies. Importantly, this discontinuity has a form reminiscent of a non-perturbative term.

Let us consider more general cases than $\phi_A$. First, the extension
\begin{equation}
\phi_{A,b} = \sum_{n\geq 0} A^{-n} \Gamma(n+b) g^n,
\end{equation}
for $b$ non-integer.
A simple calculation shows that 
\begin{equation}
\widehat{\phi}_{A,b} = \Gamma(b)\left(1-\frac{\zeta}{A}\right)^{-b}.
\end{equation}
The discontinuity of the Borel sum is no longer a simple residue but rather the integral of the discontinuity along the branch cut of rational power $-b$, which lies at $[A,\re^{\ri\theta}\infty)$, where $\theta =\arg A$. This is given by
\begin{equation}
\disc s_{\theta}(\phi_{A,b})(g) = \frac{\Gamma(b)}{g}\int_A^{\re^{\ri\theta}\infty}\re^{-\zeta/g}\left\{\left(1-\frac{\re^{+\ri 0}\zeta}{A}\right)^{-b}-\left(1-\frac{\re^{-\ri 0}\zeta}{A}\right)^{-b}\right\}\rd\zeta.
\end{equation}
Recalling that $(-x\re^{\pm\ri 0})^{-b} = \re^{\pm\ri\pi b} x^{-b}$ for $x>0$, the change of variables $\zeta = A+\re^{\ri\theta}x$ leads to
\begin{equation}
\ba
\disc s_{\theta}(\phi_{A,b})(g) 
&= \frac{\Gamma(b)|A|^{b}\re^{-A/g}}{|g|}\int_0^{\infty} \re^{-x/|g|} \left\{2\ri\sin(\pi b) |x|^{-b}\right\}\rd x ,
\\
&= 2\pi\ri |A|^{b} |g|^{-b} \re^{-A/g},
\ea
\label{proto-resurgence-relation}
\end{equation}
where we used the reflection formula $\Gamma(b)\Gamma(1-b)=\pi/\sin(\pi b)$.
If we consider a series 
\begin{equation}
\varphi \sim \sum_{n\geq 0} A^{-n} C_0 \Gamma(n+b) \left(1+\frac{\mu_1}{n+b-1}+\CO(n^{-2}) \right) g^n,
\label{prototype-corrections-0}
\end{equation}
then using \eqref{proto-resurgence-relation} term by term we would find
\begin{equation}
\disc s_{\theta}(\varphi)(g) \sim 2\pi\ri |A|^{b} C_0 |g|^{-b} \re^{-A/g}  \left(1+ \frac{\mu_1}{A} g + \CO(g^2)\right).
\end{equation}
We can now write a very useful relation, between the leading order behavior of a series and the discontinuity of its Borel transform
\begin{equation}
\ba
c_n &\sim A^{-n} C_0 \Gamma(n+b),\quad n\gg 1\\
\disc s_{\theta}(\varphi)(g) &\sim 2\pi\ri |A|^{b} C_0 |g|^{-b} \re^{-A/g}, \quad g\ll 1.
\ea
\label{resurgence-relation}
\end{equation}
Relation \eqref{resurgence-relation} also holds for integer $b$, although the proof follows from a logarithm branch cut instead of that of a rational function.

The attentive reader is now probably worried that we have added one more series, the series of correction to large order behavior or, equivalently, the series accompanying the discontinuity of the Borel transform. This series could itself be asymptotic. Fortunately, the framework includes the solution to our woes. Let us be a bit more careful and think of a generic Borel transform $\hatvphi$ which has a rational function branch cut starting at $A$, that is
\begin{equation}
\hatvphi(\zeta) = \text{holomorphic function} + \frac{\mathsf{S}_A}{2\sin(\pi b)}\left(A-\zeta\right)^{-b} \sum_{k\geq 0} c_k (\zeta-A)^k+\cdots.
\end{equation}
One can see that this is the natural consequence of asymptotic behavior of the type \eqref{prototype-corrections-0}.
Then the discontinuity will be given by
\begin{equation}
\disc s_\theta(\varphi)(g) = \frac{\ri \mathsf{S}_A \re^{-A/g}}{g} \int_0^{\re^{\ri\theta}\infty}\re^{-\xi/g} |\xi|^{-b}\left\{\sum_{k\geq 0} c_k(\re^{-\ri 0}\xi)^{k}\right\}\rd\xi.
\label{disc-of-s}
\end{equation}
One can eventually reorganize\footnote{To check that this is true asymptotically note that 
\be
\int_{\IR_+} \re^{-x/g} x^{-b+n}\rd x = \Gamma(n+1-b) g^{n+1-b}.
\ee
A more rigorous argument for getting $s(\varphi_1)$ is to introduce an equivalent, modified, Borel transform that removes the factors of $b$. See \cite{dorigoni-resurgence,abs,costinbook} for details.
} \eqref{disc-of-s} into 
\begin{equation}
\ba
\disc s_\theta(\varphi)(g) &= \ri \mathsf{S}_A |g|^{-b} \re^{-A/g}  \int_0^{\re^{\ri\theta}\infty}\re^{-\xi/g} \hatvphi_1(\re^{-\ri 0}\xi) \rd\xi \\
&= \ri \mathsf{S}_A |g|^{-b} \re^{-A/g} s_-(\varphi_1)(g),
\ea
\label{disc-to-stokes}
\end{equation}
where $\varphi_1$ is the formal power series
\begin{equation}
\varphi_1 = \sum_{n\geq 0} \Gamma(n+1-b) c_n g^n.
\end{equation}
Thus, in general, the discontinuity of the Borel sum of a power series is given by the Borel sum of a new power series. 
The overall constant $\mathsf{S}_A$ we call the Stokes constant associated with the singularity at $\zeta=A$.
To see how this relates to non-perturbative effects, we need to introduce trans-series.

\section{Trans-series}

Instead of introducing the theory of trans-series from the get-go, we will motivate it with three examples of increasing complexity, so that when we get to the general case the idea seems natural.

\subsection{A crescendo of trans-series examples}

We start with an\footnote{There are so many equations named after Euler, one wonders if the Axiom of Choice might be needed to name them.} Euler equation with for some number $A$,
\begin{equation}
f'(z) + A f(z) = \frac{A}{z}.
\end{equation}
If we plug an asymptotic ansatz of $f\sim \sum c_n/z^{n+1}$ for $z\gg 1$, we get a simple recursion relation for the $c_n$,
\begin{equation}
-n c_{n-1} + A c_n = A \delta_{n,0}.
\end{equation}
The solution of this recursion relation is given by \eqref{phi-A},
\begin{equation}
f_A \approx \frac{\phi_A(1/z)}{z}.
\end{equation}
For $A=-1$, the solution is the function $f_{-1}$ whose asymptotic behavior and ressummation we just discussed. However the solution to this equation is not unique. In general we can sum the homogeneous solution $\re^{-A/z}$. For $A<0$, the general solution is
\begin{equation}
f(z) = \frac{1}{z}s(\phi_A)(1/z) + C_0 \re^{-A z}.
\end{equation}
Thus, our perturbative solution does not fully capture the solution of the problem.
If we were interested in the solution when $z>0$, then it should be easy to see if the additional term is there, since it will be dominant. But what if we were interested in the case $z<0$ where $s(\phi_A)(1/z)$ is discontinuous and the added term is suppressed? Or rather, what if we were interested in the case where $A>0$?

For $A>0$, an explicit general solution can be written,
\begin{equation}
f(z) = A \re^{-A z}\int_{-\infty}^{Az} \frac{\re^t}{t}\rd t + C_0 \re^{-A z}.
\label{gen-A-eq}
\end{equation}
For $1/z\in \IR_+$, this is a well defined smooth function with one free parameter. Can we capture it from asymptotic analysis? We saw that the Borel sum is discontinuous for $z>0$ if $A>0$,
\begin{equation}
\frac{1}{z}\disc s(\phi_A)(1/z) = 2\pi\ri A \re^{- A z},
\end{equation}
but at the same time its real part actually gives the correct inhomogeneous solution
\begin{equation}
\frac{1}{z}\Re s_\pm(\phi_A)(1/z) =A \re^{-A z}\int_{-\infty}^{Az} \frac{\re^t}{t}\rd t,
\end{equation}
which can be shown from simple properties of the exponential integral. So we need something that cancels the imaginary ambiguity of $s_\pm$. Since we know that the general solution has exponentially suppressed terms, let us pick them differently in different parts of the complex plane as follows
\begin{equation}
\Phi^\pm(g) = \phi_A(g) + \big(C_0\mp \ri \pi A  \big) g^{-1} \re^{-A/g}.
\label{trans-series-euler}
\end{equation}
Then the general solution can be written by the unambiguous real function
\begin{equation}
f(z) = \frac{1}{z} s_\pm\left(\Phi^\pm\right)(1/z).
\label{gen-A-trans}
\end{equation}

What we wrote in \eqref{gen-A-eq} is the Borel summation of a trans-series, a series that includes regular power series terms and non-analytic terms like $\re^{-A/g}$, which we call trans-monomials. These terms come multiplied by trans-series parameters, which, as we just saw, might differ in different parts of the complex plane. In our example, we can label the trans-series parameter as
\begin{equation}
\CC_A^\pm = C_0\mp \ri \pi A.
\end{equation}
There is a deep relation between trans-series parameters and the Stokes constants of perturbation theory. In this simple example we saw that
\begin{equation}
\CC_A^+ = \CC_A^- - \ri (2\pi A).
\end{equation}
We can read from \eqref{phiA-disc},
\begin{equation}
\mathsf{S}_A = 2\pi A.
\end{equation}
We can thus start to squint at the pattern that the Stokes constans are connected to the jumps in the trans-series parameters,
\begin{equation}
\mathsf{S}_A = -\ri(\CC_A^--\CC_A^+),
\end{equation}
but the trans-series in this example was too simple. Let us consider a more interesting case.

The Airy equation is a second order linear ODE,
\begin{equation}
y''(x) = x y(x).
\label{airy-ode}
\end{equation}
Its solution is the linear combination of the two Airy functions
\begin{equation}
y(x) = \sigma_1 \Ai(x) +\sigma_2\Bi(x).
\label{Airy-gen-sol}
\end{equation}
For $x\gg 1$, both of these solutions admit asymptotic expansions
\begin{equation}
\Ai(x) \approx \frac{\re^{-\frac{2x^{3/2}}{3}}}{2x^{1/4}\sqrt{\pi}}  \varphi_\Ai\left(x^{-3/2}\right),\quad \Bi(x) \approx \frac{\re^{\frac{2x^{3/2}}{3}}}{2x^{1/4}\sqrt{\pi}}  \varphi_\Bi\left(x^{-3/2}\right),
\label{Ai-Bi-asym}
\end{equation}
given by the power series
\begin{equation}
\varphi_\Ai(z) = \frac{1}{2\pi}\sum_{n\geq 0} \left(-\frac{3}{4}\right)^n \frac{\Gamma\left(n+\frac{5}{6}\right)\Gamma\left(n+\frac{1}{6}\right)}{\Gamma\left(n+1\right)}z^n,\quad \varphi_\Bi(z) =\varphi_\Ai(-z).
\label{Ai-Bi-ps}
\end{equation}
Perhaps the most iconic property of the Airy function $\Ai(x)$ is that for large positive $x$ it is exponentially damped, while for large \textit{negative} $x$ it is oscillatory. The former is already visible in the asymptotic expansion \eqref{Ai-Bi-asym}, but can we also derive the latter from asymptotic series?

In order to extend $\Ai(x)$ for negative $x$, we should first resum it for positive $x$, that is positive $z$. The Borel transform of the power series in \eqref{Ai-Bi-ps} is
\begin{equation}
\hatvphi_\Ai(\zeta)=\, _2F_1\left(\frac{1}{6},\frac{5}{6};1;-\frac{3}{4} \zeta\right),\quad \hatvphi_\Bi(z) = \hatvphi_\Ai(-\zeta).
\label{phihat-Ai}
\end{equation}
Since the hypergeometric function is analytic over the negative reals, $\varphi_\Ai$ is Borel summable. We then have
\begin{equation}
\Ai(x) = \frac{\re^{-\frac{2x^{3/2}}{3}}}{2x^{1/4}\sqrt{\pi}}  s(\varphi_\Ai)\left(z\right), \quad x>0, z= x^{-3/2}.
\end{equation}

To find $\Ai$ with a negative argument, let's take $x$ from the positive real axis to the negative real axis by continuing through the lower half plane. This implies the argument of $z=x^{-3/2}$ will follow in the upper half plane from $0$ to $3\pi/2$, which we can visualize as rotating the ray of Borel summation anti-clockwise. During this continuation, this ray will cross the negative real axis. We then hit a difficulty because \eqref{phihat-Ai} is discontinuous for $z<0$. In particular, using the formula for the discontinuity of the hypergeometric function,
\begin{equation}
\hatvphi_\Ai(\re^{+\ri 0}\zeta)-\hatvphi_\Ai(\re^{-\ri 0}\zeta) = \ri \left\{\, _2F_1\left(\frac{1}{6},\frac{5}{6};1;\frac{3 \zeta}{4}+1\right)\right\}, \quad \zeta\in (-\infty,-4/3].
\end{equation}
With the change of variable $\zeta=-4/3+\xi$ and the formulae in \eqref{phihat-Ai}, we see that this discontinuity is precisely the Borel transform of $\hatvphi_\Bi$. We then find
\begin{equation}
\disc s_{\pi}(\varphi_\Ai)(z) =  \ri \re^{\frac{4}{3z}} s_{\pi}(\varphi_\Bi)(z),
\label{disc-Ai}
\end{equation}
where $\varphi_\Bi$ is Borel summable along the negative real axis. 
This is a relation of the form \eqref{disc-to-stokes} with
\begin{equation}
A = -\frac{4}{3},\quad b=0,\quad \mathsf{S}_{-4/3} = 1.
\end{equation}
But the Airy function does not have a branch cut, in fact it is entire. To reconstruct this from it asymptotics we then need need to write it as a trans-series
\begin{equation}
\Phi_\Ai^\pm(z) = \varphi_\Ai(z) + \CC^\pm_{-4/3} \re^{\frac{4}{3z}} \varphi_\Bi(z),
\end{equation}
with a trans-series parameter
\begin{equation}
\CC^-_{-4/3} =0,\quad \CC^+_{-4/3} = \CC^-_{-4/3}-\ri\mathsf{S}_{-4/3},
\end{equation}
such that
\begin{equation}
\ba
\Ai(x) &= \frac{\re^{-\frac{2x^{3/2}}{3}}}{2x^{1/4}\sqrt{\pi}}  s_{\pi\pm}\left(\Phi_\Ai^\pm\right)(z),\qquad z={x^{-3/2}},\\
&= \frac{\re^{-\frac{2 x^{3/2}}{3}}}{2x^{1/4}\sqrt{\pi}}  \Big( s_{\pi-} (\varphi_\Ai)(z)\Big),\quad  \arg z \leq \pi\\
&= \frac{\re^{-\frac{2 x^{3/2}}{3}}}{2x^{1/4}\sqrt{\pi}}  \left( s_{\pi+} (\varphi_\Ai)(z) -\ri \re^{\frac{4 x^{3/2}}{3}} s_{\pi+} (\varphi_\Bi)(z)\right), \quad  \arg z \geq \pi.
\ea
\label{Ai-as-trans-series-neg-z}
\end{equation}
With \eqref{disc-Ai} one can see that both $\pm$ cases give the same unambiguous answer when $\arg z =\pi$.

From $\arg x= - 2\pi/3$ to $\arg x = -\pi$, that is from $\arg z=\pi$ to $\arg z = 3\pi/2$, we don't cross any further discontinuity of $\hatvphi_{\Ai,\Bi}$. We can then continue \eqref{Ai-as-trans-series-neg-z} after the jump. Accounting for the branch of square root by being judicious with the phases, for $\arg z>\pi$ we have
\begin{equation}
\Ai(-x) = \frac{\re^{\frac{\ri\pi}{4}-\frac{2\ri x^{3/2}}{3}}}{2x^{1/4}\sqrt{\pi}}  \left( s_{3\pi/2} (\varphi_\Ai)(-\ri x^{-3/2}) + \re^{-\frac{\ri\pi}{2}+\frac{4\ri x^{3/2}}{3}} s_{3\pi/2} (\varphi_\Bi)(-\ri x^{-3/2})\right),
\label{Ai-neg-x}
\end{equation}
where $x>0$. From the asymptotic series \eqref{Ai-Bi-ps}, we have that $\varphi_\Ai(z)\sim 1+\CO(z)$ and similarly for $\Bi$. Then the leading behavior of \eqref{Ai-neg-x} is 
\begin{equation}
\Ai(-x) =  \frac{\cos\left(\frac{2 x^{3/2}}{3}-\frac{\pi}{4}\right)}{x^{1/4}\sqrt{\pi}},\quad x\gg 1,
\end{equation}
which is the well known oscillatory behavior of the Airy function.

There are several lessons to extract from this example. First and foremost, we see that very natural functions are expressed as the Borel summation of a trans-series, when we study them using asymptotic series. Furthermore, trans-series can have asympotic series multiplied by the exponentially suppressed trans-monomials. Another, deeper, insight is that when we built this trans-series we started with a power series $\varphi_\Ai$ but found a second power series $\varphi_\Bi$ in its discontinuity. From the cancellation of ambiguities in Borel summation, we see that we must include this power series $\varphi_\Bi$ in the trans-series. Or rather, by inspecting the discontinuity of the Borel transform, which is to say the large order behavior of the series, we discovered that the trans-series had a higher sector we had not been privy to. The series $\varphi_\Bi$ ``resurges'' in the large order behavior of $\varphi_\Ai$.

That the solution of the Airy equation is a trans-series with only two formal power series is a particularity of the fact that the equation \eqref{airy-ode} is linear. If we take a non-linear equation, like a Riccati equation,
\begin{equation}
f'(z) + A f(z) + \frac{f(z)^2}{z^2} = - \frac{1}{z},
\end{equation}
then the solution is given by a trans-series with infinite sectors,
\begin{equation}
\Phi(z,\sigma) = \phi_0(z) + \sum_{n\geq 1} \sigma^n \re^{- n A z} \phi_n(z).
\label{Ric-trans-series}
\end{equation}
There is still a single trans-series parameter which, for $A>0$, jumps across the positive real axis as
\begin{equation}
\Phi^+(z,\sigma) = \Phi^-(z,\sigma+\ri\mathsf{S}_A).
\end{equation}
This happens because the Borel transform of $\phi_0$ has singularities at all integer multiples of $A$, $nA$ as suggested in \eqref{Ric-trans-series}, and the equation limits the trans-series to its specific form. In principle, one could calculate $\mathsf{S}_A$ from the large order behavior of $\phi_0$ as we discussed in \eqref{resurgence-relation}. Furthermore, this means all the $\phi_n$ themselves are encoded in the large order behavior of $\phi_0$. We see the power of the resurgence framework: an infinite trans-series, which includes non-perturbative effects, can then be found by inspecting the asymptotic analysis of the asymptotic series. For an extensive analysis of this equation, see \cite{dorigoni-resurgence}.

\subsection{General theory}

Let us summarize the framework discussed so far. First, we introduced trans-monomials, which are the non-analytic terms that can figure in trans-series, e.g. 
\begin{equation}
z \re^{z}, g^b \re^{-A/g}, \re^{-\re^{1/g}}, \log \log z, \cdots.
\end{equation}
One can keep composing exponentials or logarithms, but fortunately such terms do not usually appear in QFT. We will only care about terms of the form
$g^b \re^{-A/g}$
for the moment, where we assume $|g|\ll 1$. Trans-monomials have an order relationship
\begin{equation}
g^{b_1} \re^{-A_1/g} > g^{b_2} \re^{-A_2/g} \quad\text{if}\quad 
\begin{cases} \Re \frac{A_1}{g} < \Re \frac{A_2}{g}&\\
 \Re \frac{A_1}{g} = \Re \frac{A_2}{g}, & b_1<b_2
\end{cases}
\end{equation}
A ``good'' trans-series is then a partially ordered infinite sum, i.e. the next term is never larger than the preceding one, of the form
\begin{equation}
T(g) = \sum_{A_i,b_j} c_{A_i,b_j} \re^{-A_i/g} g^{b_j} .
\end{equation}
These are effectively series in two small parameters, $g$ and $\re^{-1/g}$, which can be treated as independent.
For more details on the mathematical construction of generic trans-series, see \cite{trans-series-beginners}.

The trans-series that we will need are of the form
\begin{equation}
\Phi^\pm(g) = \varphi_0(g) + \sum_{i\geq 1} \CC_i^\pm \re^{-\frac{A_i}{g}} g^{-b_i} \varphi_i(g),
\label{general-trans-series}
\end{equation}
where the $\varphi_i$ are asymptotic series
\begin{equation}
\varphi_i(g) = \sum_{k\geq 0}\varphi_{i,k} g^k,
\end{equation}
and the sequences $A_i$ and $b_i$ are arbitrary sequences of numbers. The $A_i$ obey the partial ordering $\Re (A_i/g) \leq \Re (A_{i+1}/g)$. We will be particularly interested in the case of $g>0$ and real $A_i$ such that $0 < A_1 \leq A_2 \leq A_3 \leq \cdots$. One can reorganize the trans-series such that the $A_i$ are distinct, but then one might need different $\varphi^\pm_k$. 

The trans-series \eqref{general-trans-series} is the object we naturally obtain if the Borel transform of $\varphi_0$ has singularities along the positive real axis. The $\CC_i^\pm$ are the trans-series parameters. As we saw in our examples, they play the important role of ensuring a well defined resummation,
\begin{equation}
f(g)=s_\pm\big(\Phi^\pm)(g),
\label{sum-trans-series}
\end{equation}
but at the same time encode information not a priori specified by the asymptotic expansion.
Terms in the trans-series can be formally imaginary and/or ``ambiguous'', i.e. their values depend on the location of the complex plane. This does not mean that there must be imaginary, much less ambiguous, contributions to $f(g)$. The trans-series is a formal object which becomes a function only after resummation. The resummed series \eqref{sum-trans-series} is a real unambiguous finite number. When we refer to imaginary and real terms in the trans-series, we mean in this formal sense.

The cornerstone of resurgence is that the $\varphi_i$ are connected. As we discussed, the discontinuity of the Borel sum of $\varphi_0$ will probe the higher $\varphi_0$. To leading order we must have,
\begin{equation}
\disc s(\varphi_0)(g) = -\left(\CC^+_1-\CC^-_1\right)  \re^{-\frac{A_1}{g}} g^{-b_1} s(\varphi_1)(g)+\cdots.
\end{equation}
Which implies that
\begin{equation}
\mathsf{S}_1 = -\ri (\CC^-_1-\CC^+_1),
\end{equation}
where we use the definition of Stokes constants in \eqref{disc-to-stokes}. Furthermore, using \eqref{resurgence-relation}, we can see this directly in the large order behavior of the series $\varphi_0$,
\begin{equation}
\varphi_{0,n} \approx \frac{\mathsf{S}_1}{2\pi}\sum_{k\geq 0} A_1^{k-b_1-n} \Gamma(n+b_1-k) \varphi_{1,k}
\label{res-relation-varphi}
\end{equation}
Of course, $\varphi_1$ itself could be non Borel summable along the positive real axis. Its discontinuity would then need to be fixed by $\varphi_2$ or higher $\varphi_k$, which also has to counter-act the singularity of $\hatvphi_0$ at $\zeta=A_2$. The emerging relations between the $\varphi_k$ and their Borel singularities become quite intricate and can unveil beautiful mathematical patterns. It is often at this point that one would start using terms like ``Alien derivative'' or ``Stokes automorphism''. Since in this thesis we will only study the leading sector, we will not introduce higher arcana and we refer to \cite{abs, dorigoni-resurgence} for introductions to the proper formalism.

Nevertheless, let us metion the concept introduced in \cite{mg-peacock} of \textit{minimal resurgent structure} of $\varphi_0$. This is a set of formal power series which includes all series necessary to characterize the discontinuities of the Borel sum of $\varphi_0$. This would include $\varphi_1$, and, if $\varphi_1$ is discontinuous along the positive real axis, also the series that shows up in its discontinuity and so forth. In a sense, these are the series which are ``discoverable'' by $\varphi_0$ by looking at the discontinuities of Borel summation. Often, this is simply the set $\varphi_k$.

Our previous examples were very particular cases of \eqref{general-trans-series}. In general, the trans-series for an $n$-th degree ODE can be characterized by $n$ trans-series parameters. For linear ODEs, the trans-series is a finite sum of asymptotic series with their respective trans-monomials, like in the Airy function, but in general they are infinite, like in the, non-linear, Riccati equation. 

There are two aspects to the parameters of such trans-series. On one hand, they ``jump'' according to the Stokes constants. These are invariants of the problem and can be inspected from the large order behavior of the multiple series in the trans-series. 
On the other hand, there is a real ``invisible'' part. In our examples, we had $C_0$ in \eqref{trans-series-euler} or the real initial values of $\sigma_{1,2}$ in \eqref{Airy-gen-sol}. Roughly, in the theory of ODEs these values represent the initial conditions to which the asymptotic behavior $z\rightarrow \infty$ is blind. They are genuinely unfixed in the general solution. If a strange deity gave us a specific solution of the ODE, and we had access to the trans-series solution, we could in principle calculate the initial conditions by estimating the non-perturbative effects numerically. This is akin to what happens in QFT, where all the trans-series parameters are a priori fixed by the path integral, but we might only be able to reconstruct the trans-series.

But QFT is not an ODE, much less a linear ODE. In principle, we have to assume there are infinite series and infinite trans-series parameters. This is similar to what happens in finite difference equations. Intuitively one can think from the point of view of Schwinger-Dyson equations, a QFT would be a sequence of infinite equations. One could even speculate that the infinite sequence of boundary conditions of the path integral must specify infinite trans-series parameters. Truth is, very little is known about trans-series in a general QFT so we must be epistemologically humble. For our case, we think \eqref{general-trans-series} is a sufficiently general structure. To better understand why, let us recall why trans-series are natural in QFT and thus resurgence such an interesting tool.

\section{Resurgence in QFT}
\label{sec-res-qft}

One of the reasons one might think about the Borel summation of trans-series as an important object for QFT is that they naturally appear in many ``path integral-like'' problems. For example, the Airy function $\Ai$ can be written as a ``0d path integral'' in the form
\begin{equation}
\Ai(x) = \frac{1}{2\pi\ri}\int_\gamma \re^{x z - \frac{z^3}{3}}\rd z,
\end{equation}
where $\gamma$ is path which starts from infinity above the ray $\re^{-2\pi\ri/3}\IR_+$ , approaches the origin and deflects to infinity asymptotic to the ray $\re^{2\pi\ri/3}\IR_+$. This integral has two saddle points, but for $x>0$ only one of them contributes. As we continue $x$ analytically, the other saddle point starts contributing to the steepest descent deformation precisely when the Stokes jump happens, see \cite{mmlargen}. We then see $\varphi_{\Ai,\Bi}$ as perturbations around the different saddles. We can think of $\varphi_{\Ai}$ as regular perturbation theory and $\varphi_\Bi$ as non-perturbative physics which ``resurges''.

Schematically, in QFT we have Euclidean path integrals of the form
\begin{equation}
\mathcal{Z} = \int \mathcal{D} \phi \, \re^{- S(\phi)/g}.
\end{equation}
We can expand around the saddle points which solve the equations of motion,
\begin{equation}
\frac{\delta S}{\delta \phi}\Big|_{\phi_0}=0,\quad S(\phi_0) = S_0,
\end{equation}
to find an expansion of the form
\begin{equation}
\mathcal{Z} \approx \big(\text{perturbation theory}\big) + \re^{-\frac{S_0}{g}}\big(\text{perturbation around }\phi_0\big)+\cdots.
\end{equation}
This is, briefly, an instanton effect. We can then identify a emergent trans-series of the form \eqref{general-trans-series} with $\varphi_0$ playing the role of perturbation theory, the $A_i$ are the instanton actions and $\varphi_{k\geq 1}$ are perturbative correction around instanton and multi-instanton solutions. For unstable instantons, there are modes with negative mass. Since the one-loop action will lead to an overall square root of the determinant of the gaussian term in the action, this would lead to ambiguous imaginary trans-series parameters, as expected. 

The strategy of pertubartion around saddles has had remarkable success in $0+1$ dimensional QFT, i.e. quantum mechanics \cite{bw1,bw2,bender-wkb,bender-resonance,voros,voros-quartic,mmbook}; as well as in Chern Simons theories \cite{witten-acs,peacock-ggm,mg-peacock}, which are $2+1$ TQFT; among other indirect applications such as supersymmetric gauge theories, see e.g. \cite{ggm,mm-is}. However these are very special limits of QFT. In practice, for renormalizable QFT, instanton calculations are plagued with technical obstructions, as pointed out back in \cite{coleman-as}.

As we discuss in chapter \ref{sec_renormalons}, due to the existence of renormalons, saddle points of the actions do not even capture all non-perturbative corrections in a QFT. Nevertheless, resurgence remains a useful if not key tool in QFT. On one hand, it provides us with a precise mathematical formalism to make sense of the perturbative series and incorporate non-perturbative corrections. The Borel summation of trans-series \eqref{sum-trans-series} is a clear map from divergent asymptotic series to finite testable numbers. On the other hand, it allows us to explore non-perturbative physics from the large order behavior of perturbation theory, using relations such as \eqref{res-relation-varphi}.

It is useful to distinguish two levels of application of the resurgence framework to QFT, as outlined in \cite{dpmss}. The first one is \textit{weak resurgence}, which is the statement that observables in QFT can be written as unambiguous Borel summed trans-series, of the form \eqref{sum-trans-series}. This is not a statement about how accessible this trans-series is, simply that there is such a trans-series. 

The statement of \textit{strong resurgence} is that perturbation theory fixes the structure of the trans-series by requiring the cancellation of ambiguities in Borel summation. The overall constants are not fixed from perturbation theory alone: we know the Stokes jumps but not the unfixed part of the trans-series coefficients. More technically, the trans-series is given by the infinite linear combination of the minimal resurgent structure of perturbation theory, with the linear coefficients being the trans-series parameters and the appropriate trans-monomials. We are left with the additional exercise of fixing the ``free'' or invisible part of the trans-series parameters, for example with other limits, lattice calculations, or even experiments. This is similar to the exercise in ODEs of knowing the asymptotic expansion but not knowing the initial conditions, which we previously discussed. 

An example of the breakdown of weak resurgence is a divergent sum over the Borel summation of all $\varphi_k$. To our knowledge, there is yet no such example in QFT. An example of the breakdown of strong resurgence is a theory whose perturbation is Borel summable, thus suggesting that we do not need a trans-series, but where there are in fact non-perturbative effects. This happens, for example, in the ``Chesire cat resurgence'' of some supersymmetric theories \cite{chesire-unsal} and in many large $N$ calculations \cite{dpmss,beneke-braun}. One could even have a series that is non-Borel summable but whose ambiguity does not capture all sectors of the trans-series; we will see such an example with the $O(3)$ non-linear sigma model.

One could add an additional level, a ``strongest resurgence'', which would be the applicability of median ressumation. Median ressumation is a precise sumation method, see section 7 of \cite{dorigoni-resurgence}, but  it can be briefly summarized as assuming that any term which is not fixed by the cancellation of ambiguities of the perturbative series is zero. Perturbation theory, once promoted to the minimal trans-series which can be unambiguously Borel summed, provides the exact solution. This can happen in very non-trivial examples, see for example \cite{abbh2}, and it might be applicable in some examples we will study. However it is certainly not generic, as we will amply show.

We can summarize the different version of the resurgence program by upgrading our previous scheme \eqref{borel-sum-scheme},
%
%
\begin{equation}
\begin{tikzpicture}[baseline=(current  bounding  box.center)]
\node (function)  {$f(g)$};
\node (asym) [right=of function] {$\varphi(g) = \sum a_n g^n$};
\node (borel) [right=of asym] {$\hatvphi(\zeta)=\sum \frac{a_n}{n!} \zeta^n$};
\node (sum) [below=of borel] {$s(\varphi)(g) = \int_0^\infty \re^{-\zeta} \hatvphi(g\zeta)\rd \zeta$};
\node (stokes) [below=of sum] {$\{A_k, \mathsf{S}_k,\varphi_k\}$};
\node (trans) [below=of function] {$\Phi^\pm(g) = \varphi(g) + \sum \CC_i^\pm \re^{-\frac{A_i}{g}} \varphi_i(g),$};
\node (other) [below =of trans] {$
\begin{cases} 
\text{if weak resurgence} &  \text{other}\, A_i,\, \varphi_i,
\\
\text{if strong resurgence} & \CC_i,
\\
\text{if median resum.} & 0,
\end{cases}
$};
\draw[->] (asym.east) -- node[above] {$\CB$} (borel.west);
\draw[->] (borel.south) -- node[right] {$\CL$} (sum.north);
\draw[->] (function.east) -- node[above] {$\approx$}(asym.west);
\draw[->] (sum.south) -- node[right] {$\disc$}(stokes.north);
\draw[->] (stokes.west)  
to[out=180,in=-0] node[below left = -0.15cm] {$\disc s(\Phi) = 0$}(trans.east);
\draw[->] (other.north) --node[above right] {}(trans.south);
\draw[->] (trans.north) -- node[left] {$s_\pm(\Phi^\pm)$}(function.south);
\draw[->,dashed] (stokes.south) to[out=-90,in=0] node[below] {what else?}(other.east);
\end{tikzpicture}.
\label{resurgence-scheme}
\end{equation}

The application of resurgence to renormalizable QFT has now developed through many avenues. Some recent examples include: higher loop calculations of perturbation theory \cite{dunne-phi4}, large $N$ calculations \cite{dpmss}, using the Schwinger-Dyson equations describing a specific subset of diagrams \cite{dyson-dunne,broadhurst-diagrams} or even twisted compactification combined with an effective QM model which approximates some aspects of $1+1$ models \cite{cherman-dorigoni-dunne-unsal,dunne-unsal}. An advantage of the methods we will review in this thesis is that we obtain results for finite $N$ and infinite volume, without any auxiliary approximation or truncation. Furthermore, we will be able to specify the full trans-series parameters, both the part specified by perturbation theory and the unambiguous part.

\chapter{Integrable models}
\label{cha-integrability}
\epigraph{
\raggedleft
Se eu não morresse, nunca! E eternamente\\
Buscasse e conseguisse a perfeição das cousas!\footnotemark}
{Cesário Verde,\\ O Sentimento dum Ocidental -- ``Horas Mortas'' (1887)}
\footnotetext{``If I were to never die! And forever// Seek and find perfection in all things!''}

In this chapter, we review some background on integrable systems and lay out important information about the models studied in the thesis. Section \ref{sec_iqmb} contains a pedagogical review of the ground state of the Lieb--Liniger model and a presentation of the Gaudin--Yang and Hubbard models. Section \ref{sec_iqft} is an overview of the application of integrability to QFT which includes some important general definitions and a review of the derivation of the Bethe ansatz integral equations for the ground state with a chemical potential. Section \ref{sec-models} includes the important data of the relativistic models studied in this thesis. Lastly, section \ref{sec-theta-intro} contains a review of the $\vartheta$ topological angle in sigma models and a quick presentation of the Bethe ansatz in gapless theories.

\section{Integrable quantum many-body systems}
\label{sec_iqmb}

Quantum integrable systems are a class of systems which can be solved exactly. They include many non-relativistic systems of many particles, ranging from one dimensional spin chains and quantum gases to two dimensional lattices at thermal equilibrium. They also include relativistic field theories in two dimensional space-times. They are fully interacting theories and have complex bound states in their spectra. But particles always scatter in two to two collisions, which resembles an ``almost free theory''. They are thus ideal toy models, both rich and tractable. 

In this section, we start with non-relativistic one dimensional models. We do an in-depth review of the Bethe ansatz solution of the ground state of the Lieb--Liniger model, followed by a quick review of the ground state of the Gaudin--Yang model with spin $1/2$ and arbitrary spin. Lastly we quickly lay out the case of the Hubbard model.

\subsection{Bethe ansatz for the Lieb--Liniger model}

The simplest integrable system is a gas of spinless bosons with a $\delta$-function interaction potential living on a circle of length $L$. The Bethe ansatz solution of this system was originally found by Lieb and Liniger in \cite{ll}. A detailed treatment of this system, as well as a comprehensive introduction to integrability, can be found in \cite{sb-book}. For a review on the physics of one dimensional bosons, see \cite{bthierry}, or \cite{giamarchi} for a reference on one dimensional models in general. Here we revise how to obtain the integral equation which characterizes the ground state, starting from the Bethe ansatz.

The Hamiltonian for $N$ bosons is given by
\be
H=-\sum_{i=1}^N  \frac{\partial^2}{\partial x_i^2}+ 2c \sum_{1\le i<j \le N} \delta(x_i-x_j).
\label{delta-LL}
\ee
In this convention, $c>0$ is a repulsive interaction. We also choose the units
\begin{equation}
\hbar = 2m=1,
\label{qmb-units}
\end{equation}
which we will conserve in all discussions of non-relativistic systems. 
We want first to find the eigenfunctions of the Hamiltonian \eqref{delta-LL}. When no particle overlaps, the Hamiltonian is simply the free Hamiltonian, so we must have a sum of plane waves. However, this is still an interacting theory, so there is more to it. In order to find the ground state, we first construct the entire spectrum.

The key idea of the Bethe ansatz \cite{bethe} is to propose that there is a single set of momenta $k_j$ and we need only sum over permutations,
\begin{equation}
\psi(x_1,\cdots, x_N) = \sum_{\sigma\in S_N} 
A(\sigma)  \re^{\sum_{i=1}^N \ri k_{\sigma_i} x_i},\quad x_j < x_{j+1},
\label{bethe-ansatz-wf}
\end{equation}
where $A(\sigma)$ are constants\footnote{It is conventional to factor the sign of the permutation $(-1)^\sigma$ out of $A(\sigma)$, as done in \cite{sb-book}. To connect with the discussion in the relativistic case, we chose not to do so.} and $S_N$ is the set of permutations of $N$ elements. 
 Because identical bosons commute, if we need $\psi$ with unordered arguments we can simply exchange the arguments until $0\leq x_1< \cdots < x_N \leq L$, and then use \eqref{bethe-ansatz-wf}.

In order to find the spectrum of the problem, we need the $k_j$ to solve $H\psi = E\psi$.  By evaluating it when all coordinates $x_j$ are different, we obtain the usual
\begin{equation}
E = \sum_{i=1}^N k_i^2.
\label{E-discrete}
\end{equation}
 However, at the contact points,
the $\delta$-function interaction imposes a familiar condition on $\psi$,
\begin{equation}
\left(\frac{\partial}{\partial x_{j+1}}-\frac{\partial}{\partial x_j}\right)\psi\Big|_{x_{j+1}-x_j\rightarrow 0^+} = c \psi\Big|_{x_j=x_{j+1}}.
\label{delta-constraint}
\end{equation}
We can use the constraint \eqref{delta-constraint} to compare the function when $x_{j+1} - x_{j}\rightarrow 0^+$, where we use \eqref{bethe-ansatz-wf}, with the function when $x_{j+1} - x_{j}\rightarrow 0^-$, where we must swap the two coordinates and then plug them into  \eqref{bethe-ansatz-wf}. 
Let $\sigma_{i\leftrightarrow j}$ be the permutation $\sigma$ with the entries $i$ and $j$ exchanged. We can compensate the swapping of the $x_j$ and $x_{j+1}$ when $x_{j+1} - x_{j}\rightarrow 0^-$, by swapping the respective entries in all permutations $\sigma$. Then we have
\be
\sum_{\sigma\in S_N} \left(\ri k_{\sigma_{j+1}} - \ri k_{\sigma_{j}} \right) 
 A(\sigma)
  \re^{\sum_{i=1}^N \ri k_{\sigma_i} x_i} 
    =
c \sum_{\sigma\in S_N}
A(\sigma_{j\leftrightarrow j+1})
  \re^{\sum_{i=1}^N \ri k_{\sigma_i} x_i}
\ee
Because only $x_j = x_{j+1}$ are identified, for each permutation $\sigma$ only $\sigma_{j\leftrightarrow j+1}$ has the same exponential factor in each of the sums. For the equation to hold for all values of $x_1,\cdots, x_N$, it must hold separately for each such subset. We isolate those terms on each side and
solve
\begin{equation}
\frac{A(\sigma_{j\leftrightarrow j+1}) }{A(\sigma)} = \frac{k_{\sigma_{j}} - k_{\sigma_{j+1}}-\ri c}{k_{\sigma_{j}}-k_{\sigma_{j+1}}+\ri c}.
\label{A-sigma}
\end{equation}

In light of \eqref{A-sigma}, we can read $A(\sigma_{j\leftrightarrow j+1})$ as $A(\sigma)$ after the elastic scattering of particles $j$ and $j+1$, 
\begin{equation}
A(\sigma_{j\leftrightarrow j+1}) = S(k_{\sigma_{j}}-k_{\sigma_{j+1}})A(\sigma), 
\label{A-SA}
\end{equation}
with the S-matrix
\begin{equation}
S(k) = \frac{k-\ri c}{k+\ri c} = - \re^{2\ri \tan^{-1}(k/c)}.
\label{ll-S}
\end{equation}
Then we can write any $A(\sigma)$ by starting from a reference $A(1\cdots N)$ and scattering particles two by two, picking up factors of $S$, until the particles are ordered as $\sigma$. Schematically each $A(\sigma)$ is
\begin{equation}
A(\sigma) = \prod S(k_{\sigma_{i}}-k_{\sigma_{j}}) A(1\cdots N).
\label{A-prodS}
\end{equation}
The overall constant $A(1\cdots N)$ is left unfixed.

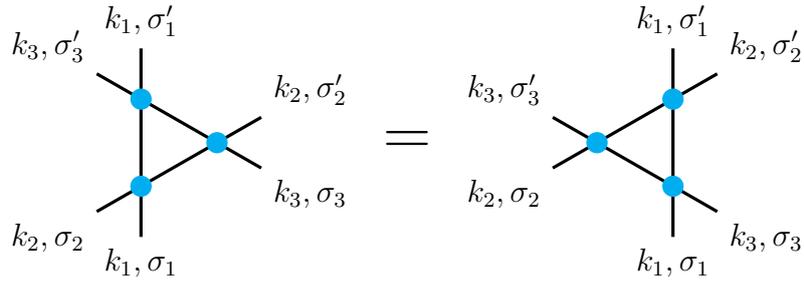
\begin{figure}
\centering
\begin{tikzpicture}[scale=1]
\begin{scope}[shift={(-1.5,0)}]
\draw[very thick] (-2,1.25) node[above]{$k_1,\sigma'_1$} -- (-2,-1.25) node[below]{$k_1,\sigma_1$};
\draw[very thick] (-0.41746, 0.33632) node[above right]{$k_2,\sigma'_2$} -- (-2.58253, -0.91367) node[below left]{$k_2,\sigma_2$};
\draw[very thick] (-0.41746, -0.33632) node[below right]{$k_3,\sigma_3$} -- (-2.58253, 0.91367) node[above left]{$k_3,\sigma'_3$};
\fill[fill=cyan] (-1,0) circle (4pt);
\fill[fill=cyan] (-2,0.57735) circle (4pt);
\fill[fill=cyan] (-2,-0.57735) circle (4pt);
\end{scope}
\begin{scope}[shift={(+1.5,0)}]
\draw[very thick] (2,1.25) node[above]{$k_1,\sigma'_1$} -- (2,-1.25) node[below]{$k_1,\sigma_1$};
\draw[very thick] (0.41746, -0.33632) node[below left]{$k_2,\sigma_2$} -- (2.58253, 0.91367) node[above right]{$k_2,\sigma'_2$};
\draw[very thick] (0.41746, 0.33632) node[above left]{$k_3,\sigma'_3$} -- (2.58253, -0.91367) node[below right]{$k_3,\sigma_3$};
\fill[fill=cyan] (1,0) circle (4pt);
\fill[fill=cyan] (2,0.57735) circle (4pt);
\fill[fill=cyan] (2,-0.57735) circle (4pt);
\end{scope}
\draw (0,0) node{\huge $=$};
\end{tikzpicture}
\caption{Visual representation of the Yang-Baxter equation. The factorization of the S-matrix requires that the two sequences of 2-to-2 collisions be the same. The $k_i$ are the momenta and the $\sigma_i$ represent internal degrees of freedom, which are absent in the Lieb--Liniger model.}
\label{fig-ybe}
\end{figure}

Equation \eqref{A-prodS} imposes strong consistency conditions on the S-matrix. If $N=3$, the factor for the permutation $A(312)$ can be obtained by permuting either $(1,2)\rightarrow (1,3) \rightarrow (2,3)$ or $(2,3)\rightarrow (1,3) \rightarrow (1,2)$. In the Lieb--Liniger case, this is trivially satisfied, but if our particles had internal degrees of freedom, such as spin or colour, this would be a non-trivial matricial constraint called the Yang Baxter Equation, see figure \ref{fig-ybe}.  Notice as well that $S(0)=-1$. Even though the elemental particles are physically bosons, the $k_i$ carrying excitations obey Fermi statistics. This ensures that there can be no two identical $k_i$, otherwise the wavefunction would vanish.

Back to the wave function, it must also obey the boundary condition
\begin{equation}
\psi(x_1=0,x_2,\cdots,x_N) = \psi(x_2,\cdots,x_N,x_1=L),
\label{periodic-bc}
\end{equation}
where we shift the position of $x_1$ to keep the arguments ordered. To compare both sides of the equation requires doing a cyclic permutation of $\sigma$ which, using \eqref{A-SA}, can be written simply as
\begin{equation}
A(\sigma_2\cdots \sigma_N \sigma_1) = \prod_{j=2}^N S(k_{\sigma_{1}}-k_{\sigma_{j}}) A(\sigma).
\end{equation}
Plugging this into \eqref{periodic-bc} we retrieve the following set of equations:
\begin{equation}
\re^{\ri k_i L} = \prod_{j\neq i} S(k_j-k_i),\quad k_i =1,\cdots N,
\label{gen-ba}
\end{equation}
which are the Bethe ansatz equations.
Physically these equations tell us that if we take an excitation, scatter it with all the others and go around the circle, we return to the initial position. Thus, the phase from the circle must cancel the phase shifts from scattering. 

For Lieb--Liniger, we restore the S-matrix in \eqref{ll-S},
\begin{equation}
\re^{\ri k_i L} = \prod_{j\neq i} \frac{k_i-k_j+\ri c}{k_i-k_j-\ri c},\quad k_i =1,\cdots N.
\label{BA-ll}
\end{equation}
Each set of solutions, or ``Bethe roots'', $\{k_i\}$  that solves these equations provides an eigenfunction of $H$ with eigenvalue \eqref{E-discrete}.
If we take the logarithm of \eqref{BA-ll}, using \eqref{ll-S}, we find
\begin{equation}
\ri k_i L + \sum_{i\neq j} 2\ri \tan^{-1}\left(\frac{k_i-k_j}{c}\right)=2\pi\ri\, n_i, \quad i=1,\cdots N.
\label{log-BA}
\end{equation}
where the $n_i$ are integers or half integers depending on whether N is odd or even, respectively. Now each set of $n_i$ specifies a set of $k_i$, and due to Fermi statistics no two $n_i$ are the same. In order to find the ground state, we need to find the set of $n_i$ that minimizes \eqref{E-discrete} after solving \eqref{log-BA}. We can think of this as minimizing $E_i$ with respect to $k_i$ and $n_i$ with constraints, but equations \eqref{log-BA} are hard to solve in the discrete case. 

We take the ``thermodynamic'' limit
\be
\label{thl}
L\rightarrow \infty, \qquad N\rightarrow\infty, \qquad \frac{N}{ L}=n\quad \text{finite}.
\ee
The Bethe roots becomes a continuous variable $k_j\rightarrow k$ and the state numbers a function $n_j \rightarrow n(k)$.
 We introduce $\rho(k)$, the density of roots, such that $\sum(\cdots)\rightarrow L\int (\cdots) \rho(k)\rd k$, and write
\begin{equation}
k + \int_\IR 2 \tan^{-1}\left(\frac{k-k'}{c}\right)\rho(k')\rd k' = 2\pi \frac{n(k)}{L}.
\label{pre-pre-ll}
\end{equation}
It is useful to introduce the density of available states,
\begin{equation}
r(k) = \frac{1}{L} \frac{\rd n}{\rd k}.
\end{equation}
Since the Pauli exclusion principle imposes $n_{i+1}-n_i\geq 1$, we must have, in the thermodynamic limit,
\begin{equation}
r(k) \geq \rho(k).
\label{r-ineq}
\end{equation}
We can assume that both $\rho(k)$ and $r(k)$ are even functions, which is consistent with \eqref{pre-pre-ll}. Without loss of generality we assume that $\rho(k)$ is compactly supported on $[-Q,Q]$, with $Q\in[0,\infty]$. Then taking a derivative of \eqref{pre-pre-ll}
we find
\begin{equation}
1 + \int_{-Q}^Q \frac{2c \rho(k')}{c^2+(k'-k)^2}\rd k' = 2\pi r(k).
\label{pre-ll}
\end{equation}
The density, which is a parameter of the problem, is given by
\begin{equation}
n = \int_{-Q}^Q \rho(k')\rd k',
\label{n-def}
\end{equation}
and the energy is 
\begin{equation}
\frac{E}{L} = \int_{-Q}^Q k^2 \rho(k')\rd k'.
\end{equation}

To find the ground state, we must minimise $E$ with respect to our unknown functions $\rho(k)$, $r(k)$ and parameter $Q$ while obeying the constraints \eqref{r-ineq}, \eqref{pre-ll} and \eqref{n-def}. This might seem complicated but the optimal solution is simply to occupy all the available states up to some momentum $Q$, i.e. to saturate \eqref{r-ineq}. To see that, we can vary $\rho = r - \eta$ and $Q = Q + q$, with $r$ fixed by the constraint \eqref{r-ineq}. Because $\rho(k) \leq r(k)$, we must have $1\gg \eta(k) \geq 0$ and $1\gg |q|>0$. The integral constraint \eqref{n-def} relates the two variations
\begin{equation}
\delta \int_{-Q}^Q \rho(k)\rd k = 0 \Rightarrow 2q\, r(Q)  = \int_{-Q}^Q \eta(k)\rd k.
\label{eta-q}
\end{equation}
Varying the energy and plugging in \eqref{eta-q} we have
\begin{equation}
\delta E \propto  2  q  Q^2 r(Q)
 -   \int^Q_{-Q} k^2 \eta(k)\rd k =  \int_{-Q}^Q \big(Q^2-k^2\big) \eta(k) \rd k \geq 0.
\end{equation}
Therefore, the optimal solution is $\rho=r$. 

We could have guessed that the ground state energy is found by filling the lowest available states by analogy with the free Fermi gas. In this case, which is found when $c\rightarrow \infty$, the optimal solution is simply to take the $n_i$ sequentially, without any gaps. The variational problem confirms that the ground state at finite $c$ is also found by taking the set of smallest possible $n_i$,
\begin{equation}
n_i = -\frac{N-1}{2} + i - 1,\quad i = 1,\cdots, N,
\label{gs-n}
\end{equation}
which are integers for odd $N$ and half integers for even $N$ as expected.

Returning to the thermodynamic limit, let us define 
\be 
x = \frac{k}{c},\quad B = \frac{Q}{c}, \quad f(x) = 4\pi \rho(c x).
\ee
And let us introduce the dimensionless coupling $\gamma=c/n$. Then, in terms of the new variables, we have that $f$ must satisfy
\begin{equation}
\frac{f(x)}{2}-\frac{1}{2\pi}\int_{-B}^B \frac{f(y)}{1+(x-y)^2}\rd y =1,
\label{ll-inteq}
\end{equation}
and
\begin{equation}
\frac{1}{\gamma} = \frac{1}{4\pi}\int_{-B}^B f(x)\rd x,\quad e= \frac{E}{n^3L} = \frac{\gamma^3}{4\pi}\int_{-B}^B x^2f(x)\rd x.
\label{ll-rho-e}
\end{equation}

These equations provide an exact description of the ground state. However, equation \eqref{ll-inteq} is deceptively simple. In the weak coupling limit $c\ll 1$, where $B\gg 1$, the kernel is singular. The first few orders of this expansion were found in \cite{popov, tw-ll} with significant effort. In chapters \ref{cha_volin} and \ref{cha_GY}, we will see how to attack this expansion systematically.
A curious spin-off of such calculation is that the integral equation \eqref{ll-inteq} is Love's equation, which originally appeared in study of the circular plate capacitor \cite{loveeq}. The problem of calculating the capacitance of the circular plate capacitor at small plate separation dates even further back to work of Kirchhoff in 1877 \cite{kirchhoff}. For more than a century, only the first terms of the asymptotic expansion for small plate separation were known \cite{hutson}. Then the perturbative solution of the Lieb--Liniger model also solves this long standing problem for free. We analyze the integral equation \eqref{ll-inteq} and the disk capacitor problem in appendix \ref{Lieb--Liniger}.

While the Lieb--Liniger model will not be the main focus of our study, it serves as an illustrative example of integrable systems. All the following systems will be more complicated, but conceptually similar.

\subsection{Gaudin--Yang model}

The Gaudin--Yang \cite{gaudin,yang} model is another foundational model in the study of integrable systems. This is a model of spin $1/2$ fermions with a $\delta$-potential interaction on a circle of length $L$. The Hamiltonian for this model is 
\be
H=-\sum_{i=1}^N  \frac{\partial^2}{\partial x_i^2}- 2c \sum_{1\le i<j \le N} \delta(x_i-x_j),
\label{delta-GY}
\ee
so that when $c>0$ the interaction is attractive.
The $\delta$-potential maintains our plane wave intuition for the wave function, but now the spin adds an internal degree of freedom of dimension $2$. The dimensionless parameter for this model is the coupling
\begin{equation}
\gamma=\frac{c}{n},
\label{gammadef}
\end{equation}
and we are interested in the weak coupling limit $\gamma\ll 1$.
The dilute limit $n\ll 1$ is strongly coupled, which is a particularity of one dimension.

The construction of the Bethe ansatz we reviewed for the Lieb--Liniger model, sometimes called the ``coordinate Bethe ansatz'', is lackluster for models with internal degrees of freedom. Instead one should use the ``algebraic Bethe ansatz'', which is found with the powerful method of ``quantum inverse scattering''. We will not review this approach and we refer to \cite{sb-book} for a detailed introduction or to section 2 of \cite{egsv} for a short but pedagogical introduction. Schematically, in the algebraic Bethe ansatz, for each additional internal degree of freedom one needs to solve for an additional set auxiliary Bethe roots. This leads to the ``nested bethe ansatz'' equations, the Gaudin--Yang model being the inaugural example. For $M$ total particles of which $N$ are spin down, these equations are
\begin{equation}
\ba
\re^{\ri k_j L} &= \prod_{\alpha=1}^{M_1} \frac{k_j-\lambda_\alpha+\ri c/2}{k_j-\lambda_\alpha-\ri c/2},\quad j=1,\cdots,M,\\
\prod_{\beta=1}^{N} \frac{\lambda_\alpha-\lambda_\beta+ \ri c}{\lambda_\alpha-\lambda_\beta- \ri c} &=-\prod_{j=1}^{M}\frac{\lambda_\alpha-k_j+\ri c/2}{\lambda_\alpha-k_j-\ri c/2},\quad \alpha=1,\cdots, N,
\ea
\label{NBA-GY}
\end{equation}
where now we must solve for the $k_i$ and the auxiliary $\lambda_\alpha$. Like in Lieb--Liniger, these equations follow from the periodic boundary conditions, given the S-matrix originating in the $\delta$-potential.
The energy is still given solely through the momenta
\begin{equation}
E= \sum_{j=0}^M k_j^2.
\end{equation}
This problem was generalized to fermions with $\kappa$ identical components, i.e. arbitrary spin or ``flavour'', in \cite{sutherland,taka}. For the rest of this section, we follow the general case. For a review of the physics of one-dimensional fermions, including the $\kappa$-component Gaudin--Yang model, see for example \cite{guan-review}.

To write the Bethe ansatz equation in the multi-component case,
we label the number of fermions in the state $|i \rangle$ by $\mathsf{N}^i$. We choose the label of the states, $i=1,\cdots,\kappa$,  such that the populations are ordered as $\mathsf{N}^1\geq \mathsf{N}^2 \geq \cdots \geq \mathsf{N}^\kappa$. The we introduce the sums,
\be
M_i= \sum_{j=i}^{\kappa-1} \mathsf{N}^{j+1},\quad i =0,\cdots, \kappa-1
\ee
where $M_0$ is the particle total, $M_1$ is the total of particles in a state with label higher than $1$ and so forth.
The solution can be characterized by a system of $\kappa$ nested Bethe ansatz equations,
\begin{equation}
\ba
\re^{\ri k_j L} &= \prod_{\alpha=1}^{M_1} \frac{k_j-\lambda^{(1)}_\alpha+\ri c'}{k_j-\lambda^{(1)}_\alpha-\ri c'},\quad j=1,\cdots,M_0,\\
\prod_{\eta=1}^{M_l} \frac{\lambda^{(l)}_\alpha-\lambda^{(l)}_\eta+2 \ri c'}{\lambda^{(l)}_\alpha-\lambda^{(l)}_\eta-2 \ri c'} &=-\prod_{\beta=1}^{M_{l-1}}\frac{\lambda_\alpha^{(l)}-\lambda_\beta^{(l-1)}+\ri c'}{\lambda_\alpha^{(l)}-\lambda_\beta^{(l-1)}-\ri c'}\prod_{\delta=1}^{M_{l+1}}\frac{\lambda_\alpha^{(l)}-\lambda_\delta^{(l+1)}+\ri c'}{\lambda_\alpha^{(l)}-\lambda_\delta^{(l+1)}-\ri c'},\\
&\quad \alpha=1,\cdots, M_l, \quad l= 1,\cdots, \kappa-1,
\ea
\label{NBA}
\end{equation}
where $c'=c/2$. 
We now have the quasi-momenta $k_j$ and $\kappa-1$ ``levels'' of auxiliary Bethe roots $\lambda_\alpha^{(l)}$. In \eqref{NBA}, we consider $\lambda_\alpha^{(0)}=k_\alpha$.
Each set of $\{k_i,\lambda_\alpha^{(l)}\}$ determines a wavefunction with energy eigenvalue 
\be
E= \sum_{j=1}^{M_0} k_j^2.
\label{E-M0}
\ee
We are interested in obtaining the ground-state among these states.

The physics of this problem with an attractive interaction, $c>0$, is quite rich. The fermions can form bound states of $m$ particles, with $2\leq m\leq \kappa$, corresponding to irreducible representations of the $\mathfrak{su}(\kappa)$ algebra. At the level of solutions to the Bethe ansatz, these are seen as solutions with complex quasi-momenta $k_i$. At large $L$, the quasi-momenta of each particle $q$ in the bound state $j$ of size $m$ lies in a ``string" in the complex plane
\be
k_j^{m,q}=\lambda_j^{m}+\ri(m+1-2q)c'+\CO(\re^{-L}),\quad q=1,\cdots,m.   
\label{kstring}
\ee
At each of the lower levels up to $l=m-1$, there are $m-l$ auxiliary roots associated with each string $j$, $\lambda^{(l)m,q}_j$. They obey
\be
\lambda^{(l)m,q}_j = \lambda_j^{m}+\ri (m-l+1-2q)c'+\CO(\re^{-L}),\quad q = 1,\cdots m-l, \quad l = 1,\cdots, m-1.
\ee
Ultimately, all the roots associated with the bound state $j$ of size $m$ are encoded in the single real variable $\lambda_j^{m}$, which is identical to the single root at the lowest level $m-1$, $\lambda^{(m-1)m,1}_j$.

We can ``reparameterize'' the Bethe ansatz equation in terms of the bound state roots $\lambda_j^{m}$ and the number of bound states of size $m$. Because the $\mathsf{N}^i$ are ordered, a $m$ size bound state must have one particle with label $|m \rangle$ and an $m-1$ bound state must have none. Then
\be
N_m= \mathsf{N}^{m}-\mathsf{N}^{m+1}, \quad m=1, \cdots, \kappa-1.
\ee
We include $N_1$, the number of ``$1$ particle bound states'' or unbound fermions, and $N_\kappa=\mathsf{N}^{\kappa}$.
Plugging these solutions into \eqref{NBA} and taking the logarithm, see e.g. \cite{Lee_2011}, we find
\begin{multline}
 m \lambda_j^m L = 2\pi K_j^m + \sum_{p=1}^{m-1}\sum_{q=p}^\kappa\sum_{l=1}^{N_q} 2\tan^{-1}\left(\frac{\lambda_j^m-\lambda^q_l}{(q+m-2p)c'}\right)
 \\
 +\sum^\kappa_{q=m+1}\sum_{l=1}^{N_q} 2\tan^{-1}\left(\frac{\lambda_j^m-\lambda^q_l}{(q-m)c'}\right),
 \label{leeNBA}
\end{multline}
where
\be
 m = 1,\cdots,\kappa\quad j = 1,\cdots,N_m.
\ee
Where the $K_j^{m}$ are integers or half integers, depending on $N_m$, and are the analogue of the $n_i$ in the Lieb--Liniger equations.

For a fixed number of bound state of each size $\{N_1,\cdots,N_\kappa\}$, the minimum energy state chooses the lowest available $K_j^m$ in each set of bound states,
\be
K_j^m = - \frac{N_m-1}{2}+j-1,
\label{Kjm}
\ee
which follows from identical arguments to the Lieb--Liniger case.
Inserting \eqref{kstring} into \eqref{E-M0}, the energy of the resulting wavefunction is
\begin{equation}
E(N_1,\cdots,N_\kappa)  =\sum_{m=1}^\kappa \sum_{j=1}^{N_m}m\left( (\lambda_j^m)^2-\frac{\left(m^2-1\right) c ^2}{12}\right).
\label{energynba}
\end{equation}
Where we can think of $\lambda_j^m$ as the kinetic energy and the constant term as the potential energy of the bound state of size $m$.
The ground state of the system is the configuration of bound states with minimal energy. Intuitively, we can see in \eqref{energynba} that the binding energy is highest for the $\kappa$ particles bound state, so it is not surprising that the energy is minimized when all fermions are in such bound states (in the $\kappa=2$ case, these are Cooper pairs). From the point of view of the algebra $\mathfrak{su}(\kappa)$, this state is in the fully anti-symmetric singlet representation.  Then \eqref{leeNBA} reduces to
\begin{equation}
\kappa \lambda^\kappa_j L = 2\pi K_j^\kappa + \sum_{l=1}^{N_\kappa} \sum_{p=1}^{m-1}2\tan^{-1}\left(\frac{\lambda^\kappa_j-\lambda^\kappa_l}{(2\kappa-2p)c'}\right),\quad j=1,\cdots, N_\kappa,
\label{NBAgs}
\end{equation}
where we are left with solving for the $\lambda^\kappa_j$.

Let $N$ be the total number of particles, so that $N=\kappa N_\kappa$ and take the thermodynamic limit \eqref{thl}. Now the continuous variable is $\lambda$, the limit of 
$\lambda^\kappa_j$.
Since the $K_j^\kappa$ are fully filled according to \eqref{Kjm}, the density of roots and the density of states are the same, which we define as
$f(\lambda) = L^{-1}\rd K/\rd\lambda$.  After a derivative, \eqref{NBAgs} becomes
\begin{equation}
\frac{\kappa}{2\pi}= f(\lambda)+ \frac{1}{2\pi}\int_{-Q}^Q \rd \lambda' f(\lambda') \sum_{p=1}^{\kappa-1} \frac{2 p c}{(pc)^2 + (\lambda-\lambda')^2}.
\label{pregsc}
\end{equation}
Like in the Lieb--Lininger model, $Q$ is fixed by the density
\begin{equation}
\int_{-Q}^Q f(\lambda) \, \rd \lambda  = {n \over \kappa}.
\end{equation}
The energy density per unit length \eqref{energynba} is a simple integral,
\be
\label{e-ground}
E = \kappa \int_{-Q}^Q \left( \lambda^2 - {\kappa^2-1 \over 12} c^2 \right) f(\lambda) \rd \lambda. 
\ee
We now have a complete description of the ground state of the multi-component Gaudin--Yang model in thermodynamic limit, as originally found in \cite{taka}.

It is useful to change variables as
\be
\theta={ \lambda\over c}, \qquad B={Q \over c}, \qquad \chi(\theta)= 2\pi f(\lambda). 
\label{multi-gy-convention}
\ee
The integral equation (\ref{pregsc}) then reads
\be
\chi(\theta)-\int_{-B}^B  K(\theta-\theta')\chi(\theta') \rd \theta'= \kappa,
\label{gscont-mgy}
\ee
with the kernel
\be
K(\theta)=-\frac{1}{2\pi} 
\left( \psi^{(0)}(\kappa +\ri\theta)+\psi^{(0)} (\kappa -\ri\theta)-\psi^{(0)} (1-\ri \theta )-\psi^{(0)} (1+\ri \theta ) \right).
\label{conteq}
\ee
With this normalization we have that
\begin{equation}
\frac{1}{\gamma} = \frac{n}{c} = \frac{\kappa}{2\pi}\int_{-B}^B \chi(\theta)\rd\theta,\quad \frac{E}{n^3} = \frac{\kappa\gamma^3}{2\pi} \int_{-B}^B \theta^2\chi(\theta)\rd\theta-\frac{\kappa^2-1}{12}\gamma^2.
\end{equation}
This is the convention used in \cite{mr-three}.
For completeness, we define another useful physical quantity, the Fermi momentum
\begin{equation}
k_F = \frac{\pi n}{\kappa},
\label{kfermi}
\end{equation}
which is related to the Fermi velocity by $k_F = \frac{1}{2} v_F$, see \eqref{qmb-units}.

When $\kappa=2$ we recover the standard Gaudin--Yang model. It is more conventional in this case to write $x$ instead of $\theta$ and $f(x)$ instead of $\chi(\theta)$ (do not confuse with $f(\lambda)$). Then the integral equation is
\begin{equation}
\frac{f(x)}{2}+\frac{1}{2 \pi}\int_B^B\frac{f(y)}{1+(x-y)^2}\rd y = 1\,.
\label{volin_eq_TBA_GY}
\end{equation}
While \eqref{volin_eq_TBA_GY} looks very similar to the Lieb--Liniger equation \eqref{ll-inteq}, they have remarkably different solutions, as we shall see in chapter \ref{cha_GY}.

It is also sometimes useful to use a different change of variables,
\begin{equation}
x = \frac{2\lambda}{\kappa c},\quad \mathsf{B}= \frac{2Q}{\kappa c},\quad \mathsf{f}(x) = 2\pi f(\lambda).
\label{mr-long-convention}
\end{equation}
With this convention, the integral equation is
\be
\mathsf{f}(x)-\int_{-\mathsf{B}}^\mathsf{B}  K(x-x')\mathsf{f}(x')\rd x' = \kappa,
\label{gscont}
\ee
where the kernel $K(x)$ is such that its Fourier transform obeys
\begin{equation}
1-\tilde K(\omega) = \frac{1-\re^{-2|\omega|}}{1-\re^{-\frac{2}{\kappa}|\omega|}}.
\label{K-multigy}
\end{equation}
The observables are given by
\begin{equation}
\frac{1}{\gamma}=\frac{\kappa^2}{4\pi}\int_{-\mathsf{B}}^\mathsf{B}\mathsf{f}(x)\rd x, \quad \frac{E}{n^3} = \ \frac{\gamma^3\kappa^4}{16\pi} \int_{-\mathsf{B}}^\mathsf{B} x^2\mathsf{f}(x)\rd x-\frac{\kappa^2-1}{12}\gamma^2.
\label{observables-notebook}
\end{equation}
This is the convention used when treating the multi-component model in \cite{mr-long}. 
When $\kappa=2$, \eqref{multi-gy-convention} and \eqref{mr-long-convention} are identical and both \eqref{gscont-mgy} and \eqref{gscont} reduce to \eqref{volin_eq_TBA_GY}.

\subsection{Hubbard model}

As the final non-relativistic model, we present the one dimensional Hubbard model. The solid state system \textit{par excellence}, it is a model of $N$ fermions on a lattice of $L$ sites, a periodic chain. Like in the Gaudin--Yang model we let the fermion have $\kappa$-components, or flavours. 

The Hamiltonian can be separated into two parts,
\be
\mH=\mH_0+\mH_I.
\ee
The first is the kinetic term, given by a next-neighbor hopping interaction
\be
\mH_0=\sum_{i,j, \sigma} t_{ij} c^\dagger_{i \sigma} c_{j \sigma}= \sum_{k, \sigma} \epsilon_k a_{k \sigma}^\dagger a_{k \sigma}, 
\ee
where $c_{i\sigma}^\dagger$ is the creation operator for a fermion with flavour $\sigma$ at site $i$, $a_{k \sigma}^\dagger$ its Fourier transform and $\epsilon_k$ is given by
\be
\label{eps-k}
\epsilon_k=- 2\cos(k).
\ee
The interacting term, $\mH_I$, is a four-site interaction
\be
\label{sig-int}
\mH_I= -u \sum_j \sum_{\sigma,\tau} c^\dagger_{j \sigma} c_{j \sigma} c^\dagger_{j \tau} c_{j \tau}=- {u\over L} \sum_{k, k', q} \sum_{\sigma, \tau} a^\dagger_{k +q, \sigma} a_{k, \sigma} a^\dagger_{k' -q, \tau} a_{k', \tau}. 
\ee
with coupling constant $u$. We assume that the interaction is attractive, which means $u>0$ in the present convention.

In the thermodynamic limit, it is preferable to organize $N$ and $L$ into the density
\be
n={N \over L}
\ee
which must be $n \in (0,1]$. This is the parameter is also called the ``filling'', and for $\kappa=2$ the value $n=1$ is called ``half-filling''.

The $\kappa=2$ case has the particularity of being integrable, as found by \cite{lw}.
In the limit \eqref{thl}, we find that the ground state is given by an integral equational analogous to \eqref{volin_eq_TBA_GY},
\be
\label{inteq-hub}
{f(x) \over 2}+{1\over 2\pi} \int_{-B}^{B} {f(x') \over 1+ (x-x')^2}= {\rm Re} {1\over  {\sqrt{1-(x- \ri/2)^2 u^2}}}. 
\ee
The density $f(x)$ and the parameter $B$ are related to the density through
\be
 {n \over u}={1\over \pi} \int_{-B}^{B} f(x) \rd x, 
 \label{hub-dense}
\ee
and to the ground state energy through
\be
\label{e-hub}
E(u,n)= -{2 u \over \pi}  \int_{- B}^{ B} {\rm Re}\,   {\sqrt{1-(x- \ri/2)^2 u^2}} \, f(x) \rd x. 
\ee
We can see that, naively, when $u\rightarrow 0$ equation \eqref{inteq-hub} becomes \eqref{volin_eq_TBA_GY}. One has to be careful and scale $n$ in \eqref{hub-dense} as well. We will discuss the double limits of the Hubbard model in chapters \ref{cha_GY} and \ref{cha_spin} in more detail.

\section{Integrable asymptotically free QFT}
\label{sec_iqft}

Asymptotically free quantum field theories are an important family of QFT. Gauge theories in $3+1$ dimensions are asymptotically free \cite{asym-freedom,asym-freedom2}, which lies at the base of many of their key features. While asymptotically free theories in $1+1$ dimensions do not describe the fundamental forces of nature they are useful toy models. Particularly since the study of large order behavior of perturbation theory, and non-perturbative physics through resurgence, in realistic $3+1$d gauge theories is an herculean endeavor. Low dimension toy models provide an ideal laboratory: they are sufficiently intricate to capture subtle aspects of QFT while being sufficiently simple to be solvable. Particularly if they are integrable.

Integrabilitiy can be found in the context of relativistic field theories, including asymptotically free theories. This is a well established field with very rich literature. They can be seen as the relativistic generalization of the quantum gases and spin chains we just discussed, to the point that the non-relativistic models can be obtained as the limit of the relativistic theories  (and vice-versa, as we shall discuss in chapter \ref{cha_spin}). Like the non-relativistic models, they can only exist in $1+1$ space-time dimensions\footnote{$\CN=4$ SYM, which is a $3+1$ dimensional theory, is often described as ``integrable'' but it is not an integrable field theory per se, rather one can use integrability to calculate correlation functions, see \cite{integrability-adscft}.}.

The conceptual application of integrability is similar to the non-relativistic systems.
When we reviewed the Lieb--Liniger model, we saw that the application of the Bethe ansatz was contingent on the scattering process factorizing into a sequence of 2-to-2 collisions. 
This is the defining property of integrable models.
In terms of symmetries, integrable models have an infinite tower of conserved charges which make the theory ``almost free''. One can think of these charges as conserving not just momentum and energy in a collision, $p$ and $p^2$, but all integer powers $p^k$. These charges highly restrict scattering: there can be no particle production and the set of momenta cannot be altered and, ultimately, we recover that an $n$-particle collision must factor into 2-to-2 collisions. 

One of the important features of integrable field theories is that they are so constrained that it is often possible to obtain the S-matrix.
A first simplification comes simply from the kinematics of 
 $1+1$ dimensional space-time. An on-shell particle of mass $m$ can have its single momentum component $p$ re-expressed with the rapidity variable $\theta$ such that
\begin{equation}
E = m \cosh\theta, \quad p = m \sinh \theta.
\end{equation}
In a 2-to-2 scattering process in $1+1$d, 
there is a single Mandelstam variable:  $s=(p_1+p_2)^2$. In terms of $\theta$, $s$ depends only on the difference of rapidities $\theta_1-\theta_2$. Thus, we can write S-matrix simply as a function of the difference of rapidities, $\mathcal{S}(\theta_1-\theta_2)$. To bootstrap the full S-matrix one then needs to use:
\begin{itemize}
\item unitarity, which implies
\begin{equation}
\mathcal{S}(\theta)\mathcal{S}(-\theta)=1,
\label{unitarity}
\end{equation}
\item crossing symmetry, 
\begin{equation}
S_{ij}^{kl}(\theta)=S_{il}^{kj}(\ri\pi-\theta),
\end{equation}
\item the symmetries of theory, e.g. in the $O(N)$ non-linear sigma model $O(N)$ symmetry fixes the index structure
\begin{equation}
S_{ij}^{kl}(\theta)\propto\left(\delta_i^k \delta_j^l - P(\theta) \delta_i^k \delta_j^l + K(\theta) \delta_{ij} \delta^{kl}\right),
\end{equation}
\item the integrability prohibition on particle production,
\item the spectrum of the theory, which restricts the poles of S-matrix,
\item factorization, encapsulated in the Yang-Baxter equation which we represent in figure \ref{fig-ybe},
\begin{equation}
\mathcal{S}_{12}(\theta_1-\theta_2)\mathcal{S}_{13}(\theta_1-\theta_3)\mathcal{S}_{23}(\theta_2-\theta_3)=\mathcal{S}_{23}(\theta_2-\theta_3)\mathcal{S}_{13}(\theta_1-\theta_3)\mathcal{S}_{12}(\theta_1-\theta_2).
\end{equation}
\end{itemize}
Sometimes it is also useful to compare with perturbation theory to check guesses that might be made in the bootstrap. 

In the $O(N)$ non-linear sigma model, for example, the exact S-matrix is given by
\begin{equation}
S^{kl}_{ij}(\theta) = s\left(\frac{\theta}{2\pi\ri}\right)\left(\delta_i^k \delta_j^l -\frac{2\pi\ri\Delta}{\theta} \delta_i^k \delta_j^l + \frac{2\pi\ri\Delta}{\theta-\pi\ri} \delta_{ij} \delta^{kl}\right),
\label{full-S}
\end{equation}
with 
\begin{equation}
s(x)= \frac{x}{\Delta -x}\frac{\Gamma (1-x) \Gamma \left(\frac{1}{2}+x\right)   \Gamma (\Delta +x)\Gamma\left(\frac{1}{2}+\Delta-x\right)}{\Gamma (1+x) \Gamma \left(\frac{1}{2}-x\right) \Gamma (\Delta -x) \Gamma \left(\frac{1}{2}+\Delta +x\right)},
\label{s-on}
\end{equation}
where
\begin{equation}
\Delta = \frac{1}{N-2}.
\end{equation}
This was obtained in the foundational work of \cite{zamo-zamo}.
For a pedagogical presentation of the bootstrap of this S-matrix, we refer to section 7 of \cite{zarembo-lectures}, from which \eqref{full-S} is read.

Importantly, thanks to factorization, the exact wavefunctions of multi-particle states obey the Bethe ansatz equations.
With the S-matrix at hand,  we can write such equations explicitly.
Of course, one should also find a way of extracting interesting observables from these wavefunctions.
A particularly prosperous approach, initiated in the pioneering work of \cite{pw,wiegmann2}, is to introduce a chemical potential $h$ which couples a conserved charge $\mathsf{Q}$,
\begin{equation}
\mathsf{H}-h\mathsf{Q}.
\label{H-hQ}
\end{equation}
Motivated by condensed matter applications, one can also think of this term as applying an external magnetic field.  The potential $h$ also plays the important rôle of providing an external tuneable parameter, which would otherwise be nonexistent in a renormalizable theory. 

The impact of \eqref{H-hQ} is that the creation of particles charged under $\mathsf{Q}$ is energetically favourable, and thus the ground state of the modified Hamiltonian is populated by a finite number of particles. Then, we can treat the resulting state using the Bethe ansatz.
Furthermore, if $h$ is much larger than the mass scale of the theory, $h\gg m$, only particles that optimize the trade-off of charge, $-hq$, to energy, $+m$, are created. These would be the particles whose charge is the largest eigenvalue of $\mathsf{Q}$. Bound states of such particles would have at most the sum of charges but at least the more than the sum of the masses, being disfavored. With a clever choice of $\mathsf{Q}$, one can engineer a ground state populated by a single particle species\footnote{For further discussion on the particle content in the limit $h\gg m$, see the detailed description of full finite temperature TBA in \cite{fendley2} from which one can obtain the limit $T\rightarrow 0$.}, or a tractable number of them. 

Our observable of interest will be the free energy. In the Euclidean theory on a compact space, with $L$ the size of the spatial dimension and $\beta$ for the Wick-rotated time dimension, it can be defined as
\begin{equation}
F(h) = - \lim_{L,\beta\rightarrow \infty} \frac{1}{\beta L}\log\Tr\left[\re^{-\beta (\mathsf{H}-h\mathsf{Q})}\right].
\end{equation}
As we will show in the next section, and originally found in \cite{pw}, we can use the Bethe ansatz solution of the ground state to calculate
\begin{equation}
\CF(h)=F(h)-F(0).
\label{CF-def}
\end{equation}
An important subtlety in this quantity is that $F(0)$ is the energy of the original ground state of the theory, the vacuum per se, while $F(h)$ is the energy, including the chemical potential contribution, of the excited state populated with particles, which becomes the ground state under the modified Hamiltonian. Because these are distinct physical quantities, it is not a priori true that $\CF(0)$ should be $0$. In fact, it is when $h\rightarrow m$ that the chemical potential becomes insufficient to create any particles and $\CF(h)\rightarrow 0$.

The free energy $\CF(h)$ can also be calculated from perturbation theory, as a power series in the coupling $g$ in the weak coupling limit. Since these are asymptotically free theories, weak coupling requires high energy, thus $h$ very large, which matches the regime of validity of the Bethe ansatz. Note that the coupling $g$ must be renormalised.
In an asymptotically free theory, the $\beta$-function can be written as
\begin{equation}
\beta(g) = \mu \frac{\rd g(\mu)}{\rd\mu} = - \beta_0 g^3 - \beta_1 g^5 +\cdots,
\label{beta-def}
\end{equation}
where $\beta_0$ and  $\beta_1$ are scheme independent, with $\beta_0>0$. It is also useful to define
\begin{equation}
\xi = \frac{\beta_1}{2\beta_0^2}.
\label{xi-def}
\end{equation}
In these theories, there is a dynamically generated scale given by
\begin{equation}
\ba
\Lambda &= \mu  \re^{-\frac{1}{2\beta_0 g(\mu)^2}}\left(2\beta_0 g(\mu)^2\right)^{-\xi}\exp\left\{-\int_0^{g(\mu)} \left(\frac{1}{\beta(g')}+\frac{1}{\beta_0 g'^3}-\frac{\beta_1}{\beta_0^2g'}\right)\rd g'\right\}\\
&\sim \mu \re^{-\frac{1}{2\beta_0 g^2}}\left(2\beta_0 g^2\right)^{-\xi}\left(1+\CO(g)\right),
\ea
\label{Lambda-def}
\end{equation}
where $g(\mu)$ is the running coupling at scale $\mu$ and $\Lambda$ is independent of $\mu$. One can also define the RG-invariant coupling $\bar g (h)$,
\begin{equation}
\log\left(\frac{h}{\mu}\right) = \int_{g(\mu)}^{\bar g (h)} \frac{\rd g'}{\beta(g')},
\label{gbar-def}
\end{equation}
which can be written as a function of $h$ and $\Lambda$,
\begin{equation}
\frac{1}{2\beta_0\bar g^2(h)}= \log\left(\frac{h}{\Lambda}\right) + \xi \log\log\left(\frac{h}{\Lambda}\right)+\cdots.
\label{g-to-L}
\end{equation}
One can then apply the usual techniques of renormalizable QFT to integrable theories, ignoring their more rigid structure.

In perturbative calculations, the result is naturally expressed as a function of $h/\Lambda$, due to \eqref{g-to-L}, while in the Bethe ansatz it is naturally expressed as a function of $h/m$. Comparing both is enough to obtain the exact mass gap $m/\Lambda$, without more than a two loop calculation. Physically, it connects the IR information which is incorporated in the Bethe ansatz, including the spectrum and exact matrix, with the purely an UV description of perturbation theory. This strategy was started by \cite{hn,hmn} and applied in many different models, see e.g.  \cite{fnw1,fnw2,pcf,eh-ssm} and \cite{eh-review}
for a review. Our objective in chapter \ref{cha_antrans} will be to derive a trans-series description of this observable by building on their techniques.

\subsection{Bethe ansatz integral equations in QFT}

Let us now construct the appropriate Bethe ansatz to describe the ground state of the modified Hamiltionian \eqref{H-hQ} in the infinite volume limit.
We start by considering the spatial dimension to be a finite circle of length $L$. We can then write a Bethe ansatz of the form \eqref{bethe-ansatz-wf} for the ground state wave-function with $N$ particles created by the $h$ field.
Like in Lieb--Lininger, the periodic boundary conditions require
\begin{equation}
\re^{-\ri m L \sinh(\theta_i)} = \prod_{i\neq j} S(\theta_i-\theta_j), \quad i=1,\cdots,N.
\label{bethe-rapidity}
\end{equation}
Here we assume that the particles ``behave as fermions'' in rapidity space. That is, either they are genuine fermionic fields or their S-matrix is $-1$ when $\theta_i-\theta_j=0$.

Taking the logarithm of \eqref{bethe-rapidity}, we find
\begin{equation}
\ri m L \sinh\theta_i + \sum_{i\neq j} \log S(\theta_i-\theta_j) = 2\pi\ri n_j
\label{log-of-ba}
\end{equation}
where the $n_j$ are integers. From the equation \eqref{log-of-ba}, one can see that the choice of the $n_j$ will specify a set of $\theta_i$ in a complicated way. Because the particles are anti-commuting in rapidity space, they obey the Pauli exclusion principle with respect to $n_j$. So a choice of $N$ different $n_j$ specifies the exact wave function of \textit{a} state of the theory with $N$ particles, and vice-versa, but to identify \textit{the} ground-state we would need to minimize the energy 
with respect to the $n_j$. 
The energy is calculated with the modified Hamiltonian \eqref{H-hQ},
\begin{equation}
E = \sum_{i} \big(m\cosh(\theta_i) - h),
\label{E-def}
\end{equation}
where we assume $\mathsf{Q}$ is normalized such that its largest eigenvalue is $+1$. 

We now take the thermodynamic\footnote{In some sources, the resulting Bethe ansatz is called the ``TBA'', and, indeed, it is the thermodynamic Bethe ansatz at $T=0$. However, we avoid this term to not confuse it with the finite temperature Bethe ansatz.} limit $L\rightarrow\infty$, with $N/L=\rho$ finite.  Equation \eqref{log-of-ba} becomes
\be
\ri m \sinh(\theta) + \int_\IR \log S(\theta-\theta') \lambda(\theta)\rd\theta = \frac{2\pi\ri}{L} n(\theta),
\label{cont-limit}
\ee
where $\lambda(\theta)$ is the density of roots, or the density of occupied states. The 
total particle density imposes
\begin{equation}
\int_{-B}^B \lambda(\theta)\rd\theta= \rho,
\label{lambda-int}
\end{equation}
with some $B\in [0,\infty]$. The energy is obtained from
\begin{equation}
\frac{E}{L} = \int_{-B}^B \big(m \cosh \theta - h \big)\lambda(\theta)\rd\theta.
\label{E-cont}
\end{equation}

Let
$
L^{-1}\rd n/\rd \theta$ 
be the density of available states in rapidity space, which
must be greater or equal than $\lambda$ due to the Pauli exclusion principle in rapidity space.
As we discussed in the case of the Lieb--Liniger model, if we minimize the energy \eqref{E-cont} with fixed density $\rho$, the optimal solution is to fill all the available states, i.e. $\lambda(\theta)= L^{-1}\rd n/\rd \theta$. 
We then define the kernel $K$
\begin{equation}
K(\theta) = \frac{1}{2\pi\ri}\frac{\rd}{\rd\theta}\log S(\theta).
\label{K-def}
\end{equation}
and take the derivative of \eqref{cont-limit},  finding the integral equation
\begin{equation}
\lambda(\theta) - \int_{-B}^B K(\theta-\theta') \lambda(\theta')\rd\theta' = \frac{m}{2\pi}\cosh(\theta).
\label{chi-to-lambda}
\end{equation}
Because of the unitarity requirement \eqref{unitarity}, $K(\theta)$ must be an even function of $\theta$ and, consequently, so must $\lambda(\theta)$.
Equations \eqref{chi-to-lambda} and \eqref{lambda-int} specify $\lambda(\theta)$ and $B$ given $\rho$.

Recall the physics of the problem: the only external parameter is $h$, the chemical potential, that induces the appearance of $L\rho$ many particle such that they minimise the energy \eqref{E-def}, or in the continuum limit, \eqref{E-cont}.
We have already found $\lambda$ and $B$ such that they minimise the energy at fixed $\rho$, so we are left with finding the optimal $\rho$.
It is useful to introduce the energy function
\begin{equation}
e(\rho) = \int_{-B}^B m \cosh(\theta)\chi(\theta)\rd\theta,
\end{equation}
such that
\begin{equation}
\frac{1}{L}\frac{\delta E}{\delta \rho} = e'(\rho) - h = 0.
\label{energy-min}
\end{equation}
While this does not seem very practical, we have in essence solved the problem. Equation \eqref{energy-min} tells us that we can think of the energy of ground state 
 as the Legendre transform of $e(\rho)$, with $\rho$ and $h$ as the Legendre conjugate variables. $e(\rho)$ is specified by solving the constraints for $\lambda(\theta)$ and $B$ with a fixed $\rho$.

The energy density \eqref{E-cont} should be identified with $\CF(h)$, which we discussed in field theoretical terms. Both come from the expectation value of $\mathsf{H}-h\mathsf{Q}$ and are averaged over volume. Furthermore, \eqref{E-def} only counts the energy of the particles \textit{on top of} the vacuum, it is zero in the absence of particles, and thus implicitly subtracts the contribution of the vacuum itself, as does the definition of $\CF(h)$ in \eqref{CF-def}. 

The Legendre transform of the free energy, $e(\rho)$, can be interpreted as an ``internal energy'', since it measures only the energy of the particles under $\mathsf{H}$ rather than the full modified Hamiltonian $\mathsf{H}-h\mathsf{Q}$. In physics, one often thinks of the chemical potential as ``selecting'' an excited state. In this mindset, $e(\rho)$ would be the energy of the excited state with $L\rho$ particles selected by the corresponding $h$. Since it is the Legendre transform of the free energy with respect to the chemical potential, one can also identify it with the grand potential, or a potential density. In this thesis\footnote{It is not uncommon in the literature, including the works this thesis is based on, to label this energy also as the ``ground state energy''. While $e$ is ultimately only a different way of encoding the energy of the ground state, we will avoid this terminology in this thesis. Confusingly, in non-relativistic there is only one notion of energy, the ground state energy, which we denote with the symbol $e$. }, we will prefer ``internal energy'' and also use ``normalised energy density'', or simply ``energy density'', when referring to the ratio $e/\rho^2$. 

\subsection{The integral equations}

We can summarize the solution of the variational problem in what we call the ``canonical problem''. Let $\chi(\theta) = 2\pi \lambda(\theta)$. Given $K(\theta)$ defined from the appropriate S-matrix in \eqref{K-def}, we need to solve the integral equation
\begin{equation}
\chi(\theta) - \int_{-B}^B K(\theta-\theta')\chi(\theta')\rd\theta' = m \cosh(\theta),
\label{iqft_geneq}
\end{equation}
where $B$ is indirectly fixed by $\rho$ through
\begin{equation}
\frac{1}{2\pi}\int_{-B}^B \chi(\theta)\rd\theta = \rho,
\end{equation}
to calculate $e(\rho)$ with
\begin{equation}
e(\rho) = \frac{m}{2\pi}\int_{-B}^B \cosh(\theta)\chi(\theta)\rd\theta.
\end{equation}
The free energy can then be obtained through the Legendre transform
\begin{equation}
\CF(h) = \min_\rho[e(\rho) - h\rho],
\end{equation}
or, more conveniently,
\begin{equation}
\ba
\CF(h) &= e(\rho) - h\rho,\\
h &= e'(\rho).
\ea
\end{equation}
This wraps up a well defined procedure to obtain the free energy. But one can also avoid the variable $\rho$ entirely.

Define $\epsilon(\theta)$ such that it satisfies, by fiat,
\begin{equation}
\epsilon(\theta) - \int_{-B}^B K(\theta-\theta')\epsilon(\theta')\rd\theta' = h-m \cosh(\theta).
\label{iqft_geneq_hm}
\end{equation}
From this equation, $\epsilon(\theta)$ must be an even function.
Then we can plug it into \eqref{E-cont} and use equation \eqref{iqft_geneq} to remove $\chi(\theta)$,
\begin{equation}
\ba
\frac{E}{L} &= \frac{1}{2\pi}\int_{-B}^B (m\cosh\theta-h)\chi(\theta)\rd\theta,\\
&= - \frac{1}{2\pi}\int_{-B}^B \left(\epsilon(\theta) - \int_{-B}^B K(\theta-\theta')\epsilon(\theta')\rd\theta'\right)\chi(\theta)\rd\theta,\\
&= - \frac{m}{2\pi}\int_{-B}^B \epsilon(\theta)\cosh(\theta)\rd\theta.
\ea
\label{E-func-epsilon}
\end{equation}
The energy must still be minimal with respect to $\rho$, but in the final line of \eqref{E-func-epsilon} it depends on $\rho$ only indirectly through $B$. So we can instead minimise with respect to $B$. As for the dependence on $B$, $E$ depends on $B$ through the limits of the integral and indirectly through $\epsilon(\theta)$ itself, due to \eqref{iqft_geneq_hm}. It is a simple exercise to show that
\begin{equation}
\frac{\delta E}{\delta B} \propto \epsilon( B).
\end{equation}
Then the extremisation of the ground state energy requires
\begin{equation}
\epsilon(\pm B) = 0.
\label{iqft-epsilon-bc}
\end{equation}

Equations \eqref{iqft_geneq_hm} and \eqref{iqft-epsilon-bc} are enough to determine $\epsilon(\theta)$. We then obtain the free energy directly from
\begin{equation}
\CF(h) = - \frac{m}{2\pi}\int_{-B}^B \epsilon(\theta)\cosh(\theta)\rd\theta.
\label{fh-cosh}
\end{equation}
Should we be interested we can recover the variable $\rho$ and $e$ by the running the Legendre transform in reverse,
\begin{equation}
\ba
\rho &= - \CF'(h),\\
e(\rho)  &= \CF(h)  + h\rho.
\ea
\end{equation}

\section{Notable integrable models}
\label{sec-models}
In this section, we present the concrete asymptotically free integrable QFT studied in this thesis. These are mostly the models for which the Bethe ansatz integral equations had been studied to find the exact mass gap.

\subsection{\texorpdfstring{$O(N)$}{O(N)} non-linear sigma model}

The $O(N)$ non-linear sigma model is a model with a $N\geq 3$ boson fields $\sigma_a$ satisfying 
\be 
\boldsymbol{\sigma}^2=1
\label{on-one}
\ee
 It is thus a sigma model with the $N$-sphere, $\mathbb{S}^N$, as its target space. Its Lagrangian is
\begin{equation}
\CL = \frac{1}{2g_0^2} (\partial_\mu \boldsymbol{\sigma})\cdot(\partial_\mu \boldsymbol{\sigma}),
\label{ON-lag}
\end{equation} 
whose interactions are very non-linear due to the constraint \eqref{on-one}. It is useful to characterize this model with the parameter,
\begin{equation}
\Delta = \frac{1}{N-2}.
\label{Delta2}
\end{equation}
The $\beta$-function coefficients \cite{bzj} are given by
\begin{equation}
\beta_0 = \frac{1}{4\pi\Delta},\quad \beta_1 = \frac{1}{8\pi\Delta}.
\end{equation}
This theory is gapped, and its mass gap was obtain with integrability techniques in \cite{hn,hmn},
\begin{equation}
\frac{m}{\Lambda}=\left(\frac{8}{\re}\right)^{\Delta }\frac{1}{\Gamma (\Delta +1)}.
\end{equation}

For the coupling with the $h$ chemical potential, we follow \cite{hn,hmn} and couple the model to the $O(N)$ charge $Q_{12}$. The $Q_{ab}$ charges are the generator of rotations in field space, $\mathbb{S}^N$. This charge acts on the first two components of the field $\sigma_1,\sigma_2$ as the matrix
\begin{equation}
Q_{12}= \left[
\begin{array}{cc}
 0 & -\ri\\
 \ri & 0 \\
\end{array}
\right].
\end{equation}
There are then two charged states with charge $\pm 1$. The $h$ fields triggers the presence of the $+1$ charge state,
\begin{equation}
\phi = \frac{1}{\sqrt{2}} \left(\sigma_1+\ri\sigma_2\right).
\end{equation}
In order to construct the Bethe ansatz \eqref{bethe-rapidity} for the gas of the $\phi$ particles, we extract the relevant S-matrix from \eqref{full-S},
\begin{equation}
\mathcal{S}_{\phi\phi\rightarrow\phi\phi}(\theta)= -
\frac{
\Gamma \left(\frac{1}{2}-\frac{\ri \theta }{2 \pi }\right) 
\Gamma \left(1+\frac{\ri \theta }{2 \pi }\right) 
\Gamma \left(\Delta -\frac{\ri \theta }{2 \pi }\right) 
\Gamma \left(\frac{1}{2}+\Delta +\frac{\ri \theta }{2 \pi }\right)}
{ 
\Gamma \left(\frac{1}{2}+\frac{\ri \theta }{2 \pi }\right) 
\Gamma \left(1-\frac{\ri \theta }{2 \pi }\right)
\Gamma \left(\Delta +\frac{\ri \theta }{2 \pi }\right)
\Gamma \left(\frac{1}{2}+\Delta-\frac{\ri \theta }{2 \pi }\right) }
\label{S-ON}
\end{equation}
As pointed out in \cite{hn}, $S(0) = -1$ which means that the Bethe ansatz solution is analogous to a gas of fermions.
With \eqref{K-def} we obtain the kernel, whose Fourier transform is given by
\be
1-\tilde K(\omega)=\frac{1-\re^{-2\pi\Delta|\omega|}}{1+\re^{-\pi|\omega|}}.
\ee

For perturbative calculations, it is necessary to write the Euclidean Lagrangian coupled to $h$. The currents associated with the charges $Q^{ab}$ are
\begin{equation}
Q_{ab}=\int J_0^{ab}\rd^2x,\quad J_\mu^{ab}=\frac{1}{g_0^2}\left(\sigma_a\partial_\mu \sigma_b-\sigma_b\partial_\mu \sigma_a\right),
\end{equation}
which leads to a Hamiltonian density
\begin{equation}
\mathcal{H}-hJ^{12}_0= \frac{g_0^2}{2} \Pi_a \Pi_a+\frac{1}{2 g_0^2}\left(\partial_1 \sigma_a\partial_1 \sigma_a\right) - h \big(\sigma_1\Pi_2 -\sigma_2\Pi_1\big).
\end{equation}
The Legendre transform into the Lagrangian produces an additional quadratic term. After a Wick rotation, one finds the Euclidean action,
\begin{multline}
\CL_h =
\frac{1}{2g_0^2}\bigg\{ \partial_\mu \boldsymbol{\sigma}\cdot \partial^\mu\boldsymbol{\sigma}
+2\ri h \left(\sigma_1\partial_0 \sigma_2-\sigma_2 \partial_0 \sigma_1\right)+h^2(\sigma_3^2 + \cdots + \sigma_N^2 - 1) \bigg\}.
\label{ON-euc}
\end{multline}

Everything described so far applies to $N=3$ but this turns out to be a very special case. In most of this thesis, we distinguish between the $N\geq 4$ case and the $N=3$ case, often referred to simply as the $O(3)$ model. When we refer to the fields of the $O(3)$ model specifically, we use $\mathbf{n}$ instead of $\boldsymbol{\sigma}$.

\subsection{Gross--Neveu model}

The Gross--Neveu model is a model of $N>4$ Majorana fermions with Lagrangian
\begin{equation}
\CL = \frac{\ri}{2}\bar{\boldsymbol{\chi}}\cdot\slashed{\partial}\boldsymbol{\chi}+\frac{g^2}{8}\left(\bar{\boldsymbol{\chi}}\cdot\boldsymbol{\chi}\right)^2.
\end{equation}
It is an asymptotically free theory. In fact, it is the first theory where a renormalon effect was identified, right in the same paper where the model was introduced \cite{gross-neveu}.
Its $\beta$-function coefficients are
\be
\beta_0= {1 \over 4 \pi \Delta}, \qquad \beta_1=- {1\over 8 \pi^2 \Delta}, 
\ee
with $\Delta$ also given by (\ref{Delta2}). 
Much like the $O(N)$ NLSM, this theory is gapped, with
\begin{equation}
\frac{m}{\Lambda}=\frac{(2 \re)^\Delta}{\Gamma(1-\Delta)}\,,
\end{equation}
calculated in \cite{fnw1,fnw2}.

The exact matrix of this theory was also found in \cite{zamo-zamo}. Like in the $O(N)$, the ideal charge is the $O(N)$ charge $Q_{12}$ as done in \cite{fnw1,fnw2}. For the $+1$ charged state under this $Q$, we have the S-matrix
 \be
 S(\theta)= {\Gamma\left(1+ \frac{\ri\theta}{2\pi}\right) \Gamma\left({1\over2}-\frac{\ri\theta}{2\pi}\right) \Gamma\left(1-\Delta -\frac{\ri\theta}{2\pi}\right) \Gamma\left({1\over 2}-\Delta + \frac{\ri\theta}{2\pi} \right)
 \over 
 \Gamma\left(1-\frac{\ri\theta}{2\pi}\right) \Gamma\left({1\over2}+\frac{\ri\theta}{2\pi}\right) \Gamma\left(1-\Delta +\frac{\ri\theta}{2\pi}\right) \Gamma\left({1\over 2}-\Delta - \frac{\ri\theta}{2\pi}\right)}.
 \ee
Notice that $S(0)=1$: because the massive particles are fermions themselves, the Bethe ansatz solution describes a Fermi gas without requiring the S-matrix to be ``anti-commuting''.
The Fourier transform of \eqref{K-def}
leads to
\begin{equation}
1-\tilde K(\omega)=\frac{1+\re^{-2\pi  \left(\frac{1}{2}-\Delta\right)|\omega|}}{1+\re^{-\pi  |\omega|}}. 
\end{equation}

The Gross--Neveu model with $N=2,3,4$ has completely different  physics and is outside of the scope of this thesis. The $N=4$ is an interesting limiting case, but kink solutions become the stable particle instead of the fundamental fermions. See \cite{fnw1,fnw2,dpmss} for discussions.

\subsection{\texorpdfstring{$\CN=1$}{N=1} supersymmetric \texorpdfstring{$O(N)$}{O(N)} non-linear sigma model}

The last of our $O(N)$ symmetric models, the $\CN=1$ supersymmetric $O(N)$ non-linear sigma model contains both $N$ bosons $\sigma_a$ and $N$ Majorana fermions $\chi_a$. It can be seen as the coupling of an $O(N)$ non-linear sigma model with a $O(N)$ Gross--Neveu model. Originally presented in \cite{witten}, it has the Lagrangian,
\be
\label{susyL}
\CL={1\over 2 g_0^2} \left\{  \partial_\mu {\boldsymbol{\sigma}} \cdot  \partial^\mu {\boldsymbol{\sigma}}+ \ri \,  \overline {\boldsymbol{\chi}} \cdot \slashed{\partial}\boldsymbol{\chi}+ {1\over 4} \left(\overline{\boldsymbol{\chi}} \cdot \boldsymbol{\chi}  \right)^2 \right\}, 
\ee
where the fields obey the constraints
\be
\boldsymbol{\sigma}^2=1,\quad \boldsymbol{\sigma} \cdot \boldsymbol{\chi}=0. 
\ee
Its $\beta$-function coefficients are
\be
\beta_0={1\over 4 \pi \Delta}, \qquad \beta_1=0, 
\ee
and again it is a gapped theory,
\begin{equation}
\frac{m}{\Lambda}=2^{2\Delta}\frac{\sin(\pi\Delta)}{\pi\Delta}\,,
\end{equation}
as found in \cite{eh-ssm}.

The exact S-matrix for this model was found in \cite{Shankar1978}. Its Bethe ansatz solution when coupled to $h$ was developed in \cite{eh-ssm} to find the mass gap. It is a subtler case than the ones we have so far described. Because of supersymmetry, any conserved charge will couple to a supermultiplet, and thus the ground state will always be populated by the boson and fermion superpartners. This leads not to a single Bethe ansatz integral equation but two coupled equations. As explained in \cite{eh-ssm}, one can eventually reduce the two integral equations to a single one of the form \eqref{iqft_geneq} with the kernel specified by
\begin{equation}
\label{rthetasusy}
1-\tilde K(\omega)= 
\re^{\pi|\omega|/2}{\cosh\left( (1-2\Delta) \frac{\pi \omega}{2} \right) \sinh(\pi \Delta |\omega|) \over  \cosh^2\left(\frac{\pi \omega}{2} \right)}.
\end{equation}

\subsection{\texorpdfstring{$SU(N)$}{SU(N)}  principal chiral field}

The $SU(N)$ principal chiral field is a sigma model where the fields take values in the group manifold $SU(N)$. It is thus a ``matrix-valued'' QFT.
The Lagrangian is
\be
\CL= {1\over  g_0^2} \tr \left(\partial_\mu U^\dagger \,  \partial^\mu U\right).
\label{lag-pcf}
\ee
with $U$ a determinant $1$ unitary $N\times N$ matrix.
For this model, it is more convenient to define
\be
\label{oDelta}
 \Delta ={1\over N}. 
\ee
In terms of this parameter, the $\beta$-function coefficients, see \cite{McKane1980}, are
\be
\beta_0={1\over 16  \pi  \Delta}, \qquad \beta_1={1\over 256 \pi^2  \Delta^2}.
\label{pcf-beta}
\ee
As shown in \cite{pcf}, this model is gapped with mass gap
\begin{equation}
\frac{m}{\Lambda}=\sqrt{\frac{8\pi}{\re}}\frac{\sin(\pi \Delta)}{\pi \Delta}\,.
\end{equation}
The full spectrum is composed of $N-1$ particle species, each transforming under $SU(N)$ as a fully anti-symmetric tensor of rank $r$, i.e. a fundamental representation of $SU(N)$. Their masses are given by
\begin{equation}
m_r = m \frac{\sin(\pi r\Delta)}{\sin(\pi\Delta)},
\end{equation}
where $m_1=m$ is the mass of the lowest particle, corresponding to the vector representation. One can think of the higher mass particles as bound states of the vector particle. The exact S-matrix is also known for $SU(N)$, and other Lie groups, as target space \cite{Wiegmann1984,wiegmann-pcf,Abdalla1984,pcf-all-lie}.
For different Lie groups, see \cite{h-pcf} for a mass-gap calculation. 

The conserved charge $Q$ used as a magnetic field is now an element of the Lie algebra $\mathfrak{su}(N)$.  The Euclidean action can be written with a ``covariant derivative''
\be
S={1\over g_0^2} \int \rd^2 x \Tr \left(\overline D_\mu U^\dagger (x) D_\mu U(x) \right), 
\label{pcf-euc}
\ee
where 
\be
D_\mu U = \partial_\mu U - h \delta_{\mu 0} (Q U + U Q),
\quad
\overline D_\mu U^\dagger =  \partial_\mu U^{\dagger} + h \delta_{\mu 0} (U^{\dagger} Q + Q U^{\dagger}).
\ee
One can assume $Q$ is part of the Cartan subalgebra. It can then be characterized by a normalized $N$-component vector of abelian charges $q_i$ such that $\sum_i q_i=0$. With the choice of \cite{pcf},
\begin{equation}
q_1 = \frac{1}{2},\quad q_{i\geq 2} = - \frac{1}{2(N-1)},
\label{q-pcf}
\end{equation}
there is a single particle species, the least massive particle which transforms as a vector under $SU(N)$. The S-matrix for such particles is
\be
\label{spcf}
S(\theta)=-\frac{\Gamma^2\left(1+ \frac{\ri\theta}{2\pi}\right) \Gamma\left( \Delta- \frac{\ri\theta}{2\pi}\right) \Gamma\left(1-  \Delta -\frac{\ri\theta}{2\pi}\right)  }{ 
 \Gamma^2\left(1- \frac{\ri\theta}{2\pi}\right) \Gamma\left( \Delta + \frac{\ri\theta}{2\pi}\right) \Gamma\left(1-  \Delta +\frac{\ri\theta}{2\pi}\right)  }, 
 \ee
with which we can write the habitual Bethe ansatz integral equations. We will call this the ``standard charge choice''.

In \cite{fkw1,fkw2}, an alternative choice of charges was introduced which we refer to as \textbf{FKW charges}. There, the charge vector is given by
\begin{equation}
q_k = \frac{\cos\left(\pi\Delta k -\frac{\pi\Delta}{2}\right)}{\cos\left(\frac{\pi\Delta}{2}\right)}.
\end{equation}
This choice creates all species of particle, with the specific particle densities guaranteeing that the Fermi rapidities are all identical to $B$. The resulting system of $N-1$ coupled integral equations can then be reduced to a single one of the form \eqref{iqft_geneq}, with
\begin{equation}
1- \tilde K(\omega) =  \frac{\pi \Delta}{2}\frac{\sinh(\pi\Delta|\omega|)}{\cosh(\pi\Delta\omega)-\cos(\pi\Delta)}.
\label{kernelfkw}
\end{equation}
However, the presence of multiple particles requires an adaptation of the observables. First we use 
\be
\ba
\rho=\frac{1}{4}\int_{-B}^B \chi(\theta)\rd\theta,\quad 
e=\frac{m}{8\sin^2(\pi\Delta)}\int_{-B}^B \chi(\theta)\cosh\theta\rd\theta ,
\label{rhoe-fkw}
\ea
\ee
instead. For the free energy, we must change the Legendre transform as follows
\be
h = 2\sin^2(\pi {\Delta})\, e'(\rho),\qquad \mathcal{F}(h)= e-\frac{h\rho}{ 2\sin^2(\pi {\Delta})}. 
\ee
This choice was introduced in \cite{fkw1,fkw2} because it permits an analytic solution in the large $N$ limit. For our purposes, it serves, on one hand, as an additional test of our results and, on the other, as a case where a different choice of $Q$ is readily available.

\subsection{Coset sigma models}

Coset sigma models are sigma models whose target space is $G/H$, a quotient of two groups\footnote{To form the quotient of two groups we identify elements $g$ of $G$ as equal $g=g'$ if $g'=gh$ for some element $h$ of $H$.}. Usually $G$ is a Lie group and $H$ is maximal subgroup of $G$ so that the space is symmetric. The $O(N)$ sigma model is a symmetric coset sigma model, since $\mathbb{S}^N=SO(N+1)/SO(N)$. The principal chiral field can also be thought of the degenerate case where $H$ is trivial. We will be studying two families of symmetric coset spaces, $SU(N)/SO(N)$ and $O(2P)/O(P)\times O(P)$, which were introduced by Fendley in \cite{fendley}. We will refer to them as ``Fendley's coset models''.

There are multiple ways of constructing coset sigma models. One way is to take the principal chiral model for $G$ and gauge the action of $H$. An introductory presentation to this approach is done in \cite{zarembo-lectures}. Another way is to impose further conditions on the matrices $U$ of \eqref{lag-pcf}, as done in \cite{fendley}. The $SU(N)/SO(N)$ model is obtained by imposing that $U$ be a symmetric unitary matrix with determinant $1$. Meanwhile for the $O(2P)/O(P)\times O(P)$ one starts with an orthogonal $2P\times 2P$ matrix and further imposes that it is symmetric and traceless.

Like the regular principal chiral field, these models are asymptotically free. Here the requirement that the coset is a symmetric space is crucial, because it ensures that there is only coupling constant. This is often phrased as saying that the target space preserves its shape, so that the single coupling $g_0^2$ corresponds to the size of target space. This interpretation is also valid in $O(N)$ NLSM and PCF model\footnote{If one thinks of a sigma model as a string living on target space, we could instead call this the inverse size, or tension, of the string.}. See \cite{zarembo-lectures, fendley, foz} for further discussions of the renormalization of these models. For both of Fendley's models, the $\beta$-function is given by
\begin{equation}
\beta_0 = \frac{1}{16\pi\Delta},\quad \beta_1 = \frac{1}{256\pi^2\Delta^2}\left(1+2\Delta\right),
\label{beta-coset}
\end{equation}
where
\begin{equation}
\Delta = \begin{cases}
\frac{1}{N}, & SU(N)/SO(N),\\
\frac{1}{2P-2}, & O(2P)/O(P)\times O(P),
\end{cases}.
\label{delta-coset}
\end{equation}
Note that \eqref{beta-coset} become the coefficients of the PCF model \eqref{pcf-beta} in the small $\Delta$ limit.

When coupled to a charge $Q$, which must be an element of the Cartan subalgebra of $G$, the Euclidean action is again \eqref{pcf-euc}, but with appropriate restrictions on $U$. For the $O(2P)/O(P)\times O(P)$ model, the choice of charge proposed in \cite{fendley} is proportional to the matrix
\begin{equation}
Q\propto (\delta_{i,1}\delta_{j,2}-\delta_{i,2}\delta_{j,1}).
\end{equation}
This charge creates only one species of particle, charged with its largest eigenvalue and, from the exact S-matrix calculated in the same reference, the usual Bethe ansatz integral equation is derived. It has the kernel specified by
\begin{equation}
1-\widetilde K(\omega) = \re^{-\pi\Delta|\omega|} \frac{2\cosh\big( \tfrac{1}{2}(1-2\Delta)\pi\omega \big)
\sinh(2\pi\Delta|\omega|)}{\cosh\left(\frac{\pi\omega}{2}\right)}. 
\label{eq_kernel_fend}
\end{equation}
This can also be used to calculate its mass gap, which is
\be
{m \over\Lambda}= {\sqrt { \pi}}  {2^{5 \Delta+2 } \re^{-{1\over 2}-\Delta } \over \Gamma(1- \Delta) \Gamma(1+ 2 \Delta)}. 
\ee

For the $SU(N)/SO(N)$, one could choose \eqref{q-pcf} which again only creates one species of particle. However, it is more useful to pick
\begin{equation}
 q\propto (1,-1,0,\cdots,0),
\end{equation}
which creates two types of particles described by identical Bethe ansatz integral equations of the form \eqref{iqft_geneq} with the same kernel \eqref{eq_kernel_fend} as the other coset model.
The mass gap for this model is
\be
{m \over\Lambda}= {\sqrt { \pi}}  {2^{3\Delta+2 } \re^{-{1\over 2}-\Delta } \over \Gamma(1- \Delta) \Gamma(1+ 2 \Delta)}. 
\ee
Because they share the same integral equation, we will in general treat Fendley's coset models together, with the only distinction coming from \eqref{delta-coset}.

\subsection{Bosonic models and fermionic models}

In this thesis, we will often distinguish between ``bosonic'' models and ``fermionic'' models. 
In the context of mass gap calculations, this distinction originally referred simply to the field content, ``models with bosons'' and ``models with fermions'', see for example in \cite{pcf,eh-ssm,eh-review}. 
It had been observed that the Bethe ansatz integral equation has different properties depending on the statistics. However, with the introduction of models with both bosons and fermions, like supersymmetric models, it is useful to categorize the type of Bethe ansatz integral equation by whether it is of the type that originated in models with bosons, ``bosonic'', or in models with fermions, ``fermionic''. Then the supersymmetric $O(N)$ NLSM, while having both bosons and fermions in the ground state in equal footing, is called a ``bosonic'' model because its Bethe ansatz integral equation is more similar to that of the $O(N)$ NLSM than to that of the GN model.

Precisely, we define the dichotomy as
\begin{description}
\item[Bosonic models] satisfy
\begin{equation}
1-\int_\IR K(\theta) \rd\theta  =0.
\label{K-bos-cond}
\end{equation}
They include the $O(N)$ NLSM, its $\CN=1$ supersymmetric extension, the PCF model and Fendley's coset models. Their non-relativistic analogue is the Lieb--Liniger model.
\item[Fermionic models] satisfy
\begin{equation}
1-\int_\IR K(\theta) \rd\theta \neq 0.
\label{K-ferm}
\end{equation}
The only relativistic fermionic model that we study is the $O(N)$ Gross--Neveu model, for which $\int K(\theta) =0$. Other models in this category not studied in this thesis include the chiral GN model and Thirring models, see \cite{eh-review}. Its non-relativistic analogues are the Gauding-Yang model and its variants.
\end{description}

We can check that this definition reduces to the statistics of the field content when there is a single particle species in the ground state. In such cases, we have that the kernel is given by  \eqref{K-def}. Furthermore, since asymptotically free relativistic theories are non-interacting in the UV limit, we have that $S(\pm\infty)=1$.
Using this, we can integrate \eqref{K-bos-cond} to become
\begin{equation}
S(0) = -1.
\label{S-minus-1}
\end{equation}
Recall that for the Bethe ansatz to hold we need the particles to be anti-commuting with respect to their rapidities. This means that either the fields are themselves fermions, or their S-matrix is ``fermion-like'' with respect to rapidities. \eqref{S-minus-1} implies the latter, so the fields must be bosons. In contrast, for theories like the supersymmetric $O(N)$ NLSM \eqref{K-def} does not hold, because the final integral equation comes from reducing two coupled integral equations, one for the fermions and other for the bosons.  Meanwhile for a theory with only fermions one has  $S(0) = 1$ and thus $\int_\IR K(\theta)\rd\theta=0$ which automatically satisfies \eqref{K-ferm}.

Nonetheless, there is a physical interpretation which matches the definitions \eqref{K-bos-cond} and \eqref{K-ferm} even in theories with both bosons and fermions, as remarked in \cite{eh-review}.  In ``bosonic'' models, there is a classical contribution to the free energy, which is usually the case of theories that include bosons. Meanwhile, the ``fermionic'' solution describes the models for which there is no classic contribution to the free energy, which is the case of models that contain only fermions. This is a direct consequence of calculating the free energy with the Bethe ansatz but can also be found from perturbation theory.

\section{The topological \texorpdfstring{$\vartheta$}{theta} angle in integrable models}
\label{sec-theta-intro}

A final tool in the study of non-perturbative physics in integrable models is topological $\vartheta$ angle. If we think of our models as toy models for QCD, then this will be toy model for the $\vartheta$ angle in QCD, see \cite{mmbook} for a nice introduction. However this type of term is also interesting from a condensed matter perspective: in the $O(3)$ sigma model it was originally introduced by Haldane \cite{haldane} to set up relativistic realizations of the Heisenberg spin chain. These are also interesting as integrable models in themselves. In this final part of the section, we will review the construction of the $\vartheta$ angle in the $O(3)$ model and Fendley's coset models.

\subsection{The \texorpdfstring{$\vartheta$}{theta} angle in sigma models}

As the name suggests, the distinct feature of the $\vartheta$ angle is that it is topological. It is topological first in the sense that it does not depend on the metric of the base space, in our case the $1+1$ space-time. Furthermore, $\vartheta$ terms are a particular class of topological terms that sense the topology of the \textit{field configuration}. In this section, we will mostly refer to this second stronger notion of topological invariance. 
Physically, a $\vartheta$ angle gives a complex weight to field states depending on their topological charge. Mathematically, they characterize the topology of maps (fields) from the base space (space-time) to the target space (field values). For a pedagogical introduction to topology and $\vartheta$ terms in the context of field theory in condensed matter, we refer to \cite{abanov-topology}.

Let us look into the $O(3)$ sigma model, with the Lagrangian in \eqref{ON-lag}. An $O(3)$ vector is simply a point on the sphere $\mathbb{S}^2$. On the other hand, finiteness of the action requires that fields go to a constant at $\infty$ in $\IR^2$. Therefore, by identifying that constant as the value at $\infty$ we can think of these fields as living on the sphere as well (for example by identifying the North pole with the origin, and infinity as the South pole). A field configuration is then a map $\mathbb{S}^2\rightarrow\mathbb{S}^2$. 

There is a natural topological invariant of such a map which is its degree. One can think of the degree as a generalization for spheres of the winding number in maps from the circle to the circle: it captures how many times the map from the base sphere wraps the target sphere. Like the winding number, there is a notion of addition: if image of the northen hemisphere wraps the target $n$ times, the image of the southern hemispheres to wrap the target $m$ times, and the image at the equator is constant, then the degree of the map will be $n+m$. 

The degree is topologically invariant because if a map is homotopically equivalent to another, that is if one can continuously deform one into the other, then they must share their degree. Thus, this topological invariant encodes the group of homotopy classes of maps from the (base) sphere to the (target) sphere. The technical name for this object is the second homotopy group of the (target) sphere, or $\pi_2(\mathbb{S}^2)$ for friends and mathematicians. More generally, $\pi_n(\mathcal{M})$ is the group formed by the homotopy classes of maps from $\mathbb{S}^n$ to $\mathcal{M}$. For the sphere, $\pi_2(\mathbb{S}^2)$ is isomorphic to $\mathbb{Z}$, which we can think of as the ``group of wrapping numbers''.


Let then $\mathbf{n}$ be a specific configuration of the $O(3)$ sigma model. We can define the action
\begin{equation}
S_\vartheta[\mathbf{n}] = \ri \vartheta n,
\label{action-O3}
\end{equation}
where $n$ is the degree of $\mathbf{n}$. Since it is an integer, $\vartheta$ is periodic and we restrict it to take values in $[0,2\pi]$. This action can be implemented with a concrete local Lagrangian. Define the topological charge density to be
\begin{equation}
q[\mathbf{n}] = \frac{1}{8\pi} \epsilon^{abc}\epsilon_{\mu\nu} n^a \partial_\mu  n^b\partial_\nu n^c,
\label{topo-density}
\end{equation}
where we work directly in Euclidean signature. 
 The topological charge $\mathcal{Q}$ is given by
\begin{equation}
\mathcal{Q}[\mathbf{n}] = \int q[\mathbf{n}](x) \rd^2 x,
\label{topo-charge}
\end{equation}
and the action is defined as
\begin{equation}
S_\vartheta[\mathbf{n}] = \ri\vartheta\mathcal{Q}[\mathbf{n}].
\end{equation}

To see that this Lagrangian results in a \textit{topological} charge, consider a field configuration $\mathbf{n}_0$ and smoothly deform it into $\mathbf{n}_1$. To prove this rigorously, we would have to show that it holds for any continuous deformation, i.e. homotopy, and could not assume smoothness, but this example suffices for a ``small variation''. We write the deformation as $\mathbf{N}(t)$ such that $\mathbf{N}(0) = \mathbf{n}_0$ and $\mathbf{N}(1)=\mathbf{n}_1$. We can think of $\mathbf{N}(x,y,t)$ as a map from $\mathbb{S}^2\times[0,1]$ to the sphere. Using Stokes theorem on this ``cylinder'', we have that
\begin{equation}
\ba
\mathcal{Q}[\mathbf{n}_1]-\mathcal{Q}[\mathbf{n}_0] &= \int_{\mathbb{S}^2\times [0,1]} \partial_t q[\mathbf{N}](x,y,t)\,\rd t\rd^2 x  \\
&= \frac{1}{4\pi}\int_{\mathbb{S}^2\times [0,1]} \epsilon^{abc} \partial_t N^a \partial_x N^b \partial_y N^c \,\rd t \rd^2 x = 0.
\ea
\label{is-topo}
\end{equation}
We first used the anti-symmetric structure of the topological density \eqref{topo-density} to simplify the derivative of $q$. The last integrand is the volume element spanned by the three dimensional vectors $\partial_i \mathbf{N}(x,y,t)$. However, since $\mathbf{N}(x,y,t)^2 = 1$,  each of these vectors satisfy $\mathbf{N}\cdot \partial_i \mathbf{N} =0$ and thus the three vectors lie in the same 2 dimensional plane orthogonal to $\mathbf{N}(x,y,t)$, and the volume element vanishes.

This shows that the charge \eqref{topo-charge} is topological, but does it capture the degree? Indeed, we can write the charge density \eqref{topo-density} as a differential form
\begin{equation}
q[\mathbf{n}] \rd^2 x = \frac{1}{4\pi} \epsilon^{abc} n^a \rd n^b\wedge \rd n^c = \frac{1}{4\pi}i_\mathbf{n}(\rd \mathbf{n}\wedge \rd \mathbf{n}\wedge\rd \mathbf{n}),
\label{form-volume}
\end{equation} 
which makes it evident that it is also topological in the sense of not depending on the base space metric.
Notice that $\mathbf{n}$ is a point in $\IR^3$ constrained to the surface of the sphere, and it is also the normal vector to the surface of the sphere at that same point. Thus, the last equality in \eqref{form-volume} is the three dimensional volume form at point $\mathbf{n}$ contracted with the normal vector to the surface: the surface element of the target sphere. Hence \eqref{topo-charge} is the volume of the target space spanned by the map divided by the volume of the sphere, or ``how many times the field wraps the sphere''.

The reason we are interested in a topological angle is because of its interplay with non-perturbative physics. Let us see how the usual action is related to the $\vartheta$ term by comparing their Lagrangians. Observing that
\begin{equation}
n^a n^a =1\Rightarrow n^a \partial_\mu n^a =0\Rightarrow (\epsilon^{abc} n^b \epsilon_{\mu \nu}\partial_\nu n^c)^2 = (\partial_\mu n^a)^2,
\end{equation}
we can relate the usual Lagrangian to the topological charge density as
\begin{equation}
\ba 
\partial_\mu n^a \partial_\mu n^a -  \epsilon^{abc}\epsilon_{\mu\nu} n^a \partial_\mu  n^b\partial_\nu n^c &= \frac{1}{2}(\partial_\mu n^a)^2 + \frac{1}{2}(\epsilon^{abc}\epsilon_{\mu \nu} n^b \partial_\nu n^c)^2
\\
&\qquad-  \epsilon^{abc}\epsilon_{\mu\nu} n^a \partial_\mu  n^b\partial_\nu n^c,\\
&= \frac{1}{2}\left(\partial_\mu n^a-\epsilon^{abc}\epsilon_{\mu\nu} n^b \partial_\nu n^c\right)^2\geq 0.
\ea
\end{equation}
We can write a similar inequality with the opposite sign in the topological charge.
This inequality implies
\begin{equation}
g^2 S[\mathbf{n}]\geq 4\pi |\mathcal{Q}|.
\end{equation}
There are solutions that saturate this bound, and are thus local minima of the action.  They correspond to combinations of instantons and anti-instantons, and can even be explicitly written. A single instanton solution, for example, is
\begin{equation}
\mathbf{n}(x,y) = \left\{\frac{2 x}{1+x^2+y^2},\frac{-2 y}{1+x^2+y^2},\frac{-1+x^2+y^2}{1+x^2+y^2}\right\},
\end{equation}
which has charge $1$. In fact, this is simply the identity map of $\mathbb{S}^2$ written in stereographic coordinates. One can find other equivalent solutions by translating, stretching and rotating this one. In the presence of the topological $\vartheta$ term, the action of a general multi-instanton solution  is then given by
\begin{equation}
S =\frac{ 4\pi |\mathcal{Q}|}{\bar{g}^2(\mu)}+\ri\vartheta\mathcal{Q}.
\label{instanton-action-O3}
\end{equation}
Thus, the $\vartheta$ term is supposed to interact in a very clean way with non-perturbative effects coming from instantons. Which is not to say this is the only avenue through which the $\vartheta$ term acts. In Yang--Mills theory, for example, it is known from lattice \cite{giusti} and large $N$  calculation \cite{witten-largeN} that there can be consequences of the $\vartheta$ term beyond its effect through instantons.

One can wonder what other sigma models admit a non-trivial topological $\vartheta$ angle. This is obtained by inspecting the aforementioned second homotopy group. The target space of the $O(N\geq 4)$ sigma model is a higher dimensional sphere, but $\pi_2(\mathbb{S}^n)$ is the trivial group for $n>2$. As for the principal chiral field, the target space is a Lie group $SU(N)$, but $\pi_2$ of any Lie group must be trivial (for $SU(N)$ one can obtain this from the fact that it is simply connected).

As for the topology of coset spaces, it depends crucially on $H$. If we have a coset space $G/H$ where $G$ is simply connected, like $SU(N)$, then $\pi_2(G/H)=\pi_1(H)$. $\pi_1$ is the first homotopy group being the homotopy group of maps from the circle to $H$ or the ``group of winding numbers of a circle on $H$''. In general, $\pi_2(G/H)\subset\pi(H)$. The fundamental group $\pi_1$ is also trivial for most simple groups, with the exception of $U(1)$ (identically, $SO(2)\,$), where it is $\mathbb{Z}$, and $SO(N\geq 3)$, where it is $\mathbb{Z}_2$. So if $H=SO(2)$, $\pi_2(G/H)$ might be $\mathbb{Z}$, as it happens with the $O(3)$ model which can bee seen as the coset $SO(3)/SO(2)$. If $H=SO(N\geq 3)$, $\pi_2(G/H)$ can be $\mathbb{Z}_2$. However, if $G$ is not $SU(N)$ it might be trivial. E.g., the $O(N\geq 4)$ NLSM, which can be as seen as the coset $SO(N+1)/SO(N)$, has trivial $\pi_2$. 

Fendley's coset models were in part engineered in \cite{fendley} to admit a topological angle. As we just discussed, because $\pi_2(SU(N)/SO(N))=\mathbb{Z}_2$, in $SU(N)/SO(N)$ coset models there is a topological invariant of a field configuration, but it can only take the values $0$ and $1$. For $O(2P)/O(P)\times O(P)$, it is found that the $\pi_2$ is also $\mathbb{Z}_2$, then there also is a topological charge with values $0,1$.
We can  then write an action in the style of \eqref{action-O3} where $n$ is the $\mathbb{Z}_2$ invariant, $0$ or $1$. It turns out that the only consistent angles for $\vartheta$ would be then $0$ or $\pi$.
In other words, for Fendley's coset model the only non-trivial topological angle possible is $\vartheta=\pi$. 

It is hard to find a local topological charge, like in the $O(3)$ model, whose integral is by itself $\mathbb{Z}_2$ valued. The problem being that if we take a field configuration with charge $1$ we can deform it to be supported only in some subregion of the base space, like a disk around the North pole. Then we can put a copy in some other subregion, like for example a disk around the South pole. The integral of a local Lagrangian would output twice the integral in one of the subregions, and thus $2$. Considering that a trivial field has charge $0$, we have three values, not $\mathbb{Z}_2$. However, the action \eqref{action-O3} defined with $\vartheta=\pi$ naively obeys this composition, even if it is not written with a local Lagrangian. This is not unexpected, the connection between having only two quantum mechanically consistent values of $\vartheta$ and the non-locality of topological charges in the action was studied in $2+1$ dimensions with a $\mathbb{S}^2$ target space in \cite{fks-theta}, see also \cite{aw-theta}. While it might still be possible to write a Lagrangian whose integral is always an integer $n$, such that $n$ is not topologically invariant but $\ri \pi n$ is, 
this is not strictly necessary to study the physics of these models.

\subsection{The gapless Bethe ansatz at \texorpdfstring{$\vartheta=\pi$}{theta=pi}}

In general, adding a $\vartheta$ angle will break integrability. However, as has been shown by \cite{zz-theta,foz}, if $\vartheta=\pi$ in the $O(3)$ sigma model, integrability is preserved. Building on this result, Fendley showed in \cite{fendley} showed that the coset models too are integrable with the $\vartheta=\pi$ angle. 
Nevertheless, it radically alters the physics of the system. As had been conjectured by \cite{haldane} and shown in \cite{zz-theta, foz}, the $O(3)$ sigma model with $\vartheta=\pi$ is a theory with massless particles. In \cite{fendley}, the same was shown for the coset models at $\vartheta=\pi$. To finish this chapter, we present the Bethe ansatz integral equations in a gapless integrable theory. We will discuss the implications of their specific solutions at the end of chapter \ref{cha_antrans}.  For a review on massless integrable theories, see \cite{fs-rev}.

With massless particles, the kinematics are different. In particular, the relationship between the energy and momentum of the particles and their rapidity are given by
\begin{equation}
e= \pm p = \frac{M}{2} \re^{\pm\theta},
\end{equation}
where we must now distinguish between right-moving particles ($e=p$) and left-moving particles ($e=-p$). 
The mass parameter $M$ is an arbitrary mass scale which normalizes the definition of the rapidity. In \cite{zz-theta, foz}, it is recommended that one chooses $M$ such that ``interesting physics'' happens around $\theta \sim 0$. 
For example, in \cite{foz} it is chosen such that it corresponds to the point where the asymptotically free UV regime changes into the non-trivial IR CFT. For our purposes, it is simpler to choose $M=m$ where $m$ is the mass of the $\vartheta=0$ theory. 

The ground state, once the magnetic field $h_{\vartheta=\pi}=H$ is turned on, will be populated by a finite density of particles charged under $\mathsf{Q}$ with energies up to some bound. 
We now have two distinct scattering processes: a left-left scattering, which is identical to right-right, and a left-right scattering. Both can be obtained from the S-matrices derived in \cite{zz-theta,foz,fendley}. In order to connect with the $\vartheta=0$ case, we write the magnetic field $H$ in terms of $h$ as
\begin{equation}
H =  t h_{\vartheta=0}.
\end{equation}
Given our choice of $M$, this constant $t$ is fixed by requiring the $\vartheta=0,\pi$ cases to have the same perturbative parameter.

In the ground state, particles still completely fill the lower energy states up to some state of rapidity $B$, which is fixed by constraints. However,
because the dependency of the energy on the rapidity is exponential, a bounded interval of energies means a semi-infinite interval in rapidities, $(-\infty,B]$ for right moving particles and $[B,\infty)$ for left moving particles.
The Bethe ansatz solution then leads to
\begin{equation}
\begin{aligned}
\epsilon_1(\theta) - \int_{-\infty}^B \varphi_1(\theta-\theta')\epsilon_1(\theta')\rd \theta' - \int^{\infty}_{-B} \varphi_2(\theta-\theta')\epsilon_2(\theta')\rd \theta' &= th - \frac{m \re^\theta}{2},\\
\epsilon_2(\theta)  - \int_{-\infty}^B \varphi_2(\theta-\theta')\epsilon_1(\theta')\rd \theta' - \int^{\infty}_{-B} \varphi_1(\theta-\theta')\epsilon_2(\theta')\rd \theta' &= th - \frac{m \re^{-\theta}}{2} .
\end{aligned}
\label{orig_elr_inteq}
\end{equation}
where $\epsilon_{1,2}$ are the rapidity distributions for right/left moving particles,
and
\begin{equation}
\varphi_1(\theta) = \frac{1}{2\pi\ri}\frac{\partial}{\partial\theta} \log S_{LL}(\theta),\quad \varphi_2(\theta) = \frac{1}{2\pi\ri}\frac{\partial}{\partial\theta} \log S_{LR}(\theta),
\end{equation}
 with $S_{LL}(\theta)$ the left-left (or right-right) S-matrix, and $S_{LR}(\theta)$ the left-right (or right-left) S-matrix. 
The upper equation in \eqref{orig_elr_inteq} only holds for $\theta<B$ while the lower one only holds for $\theta>-B$.
Much like in the massive case, the boundary condition is
\begin{equation}
\epsilon_1(B) = \epsilon_2(-B) = 0,
\end{equation}
and the symmetry of the system imposes
\begin{equation}
\epsilon_1(\theta) = \epsilon_2(-\theta).
\end{equation}
The free energy is given by
\begin{equation}
\ba
\CF(h,\pi)& = -\frac{1}{2\pi}\left(\int_{-\infty}^B \frac{M\re^\theta}{2}\epsilon_1(\theta)\rd\theta + \int^{\infty}_{-B} \frac{M\re^{-\theta}}{2}\epsilon_2(\theta)\rd\theta\right)
\\
& = -\frac{M}{2\pi}\int_{-\infty}^B \re^\theta\epsilon_1(\theta)\rd\theta.
\ea
\label{free-en-massless}
\end{equation}

In the massless case, one can still work directly in the canonical formalism, where we parameterize the solution with the density $\rho$ rather than $h$. We introduce the normalized densities of states $\chi_{1,2}$ for right (left) moving particles, which satisfy
\begin{equation}
\begin{aligned}
\chi_1(\theta) - \int_{-\infty}^B \varphi_1(\theta-\theta')\chi_1(\theta')\rd \theta' - \int^{\infty}_{-B} \varphi_2(\theta-\theta')\chi_2(\theta')\rd \theta' &= \frac{m \re^\theta}{2}, \quad &\theta&<B,\\
\chi_2(\theta)  - \int_{-\infty}^B \varphi_2(\theta-\theta')\chi_1(\theta')\rd \theta' - \int^{\infty}_{-B} \varphi_1(\theta-\theta')\chi_2(\theta')\rd \theta' &=\frac{m \re^{-\theta}}{2}, \quad &\theta&> -B .
\end{aligned}
\label{orig_chi_inteq}
\end{equation}
Using the symmetry $\chi_1(\theta)=\chi_2(-\theta)$ and the parity of $\varphi_{1,2}$, one can characterize the ground state with a single integral equation,
\begin{equation}
\chi_1(\theta) - \int_{-\infty}^B \big( \varphi_1(\theta-\theta') + \varphi_2(\theta + \theta') \big) \chi_1(\theta') \rd \theta' =  m\re^\theta,
\label{merTBA2}
\end{equation}
from which follows
\be
e={ M \over 4 \pi} \int_{-\infty}^B \re^\theta \chi_1 (\theta) \rd \theta, \qquad \rho={ t \over 2 \pi } \int_{-\infty}^B \chi_1 (\theta) \rd \theta.
\label{erho_pi}
\ee
This representation is more convenient for numeric manipulation, for example.

\chapter{Renormalons}
\label{sec_renormalons}
\setlength{\epigraphwidth}{.65\textwidth}
\epigraph{
\raggedleft
``It's very hard to talk quantum using a language originally designed to tell other monkeys where the ripe fruit is.''}{Terry Pratchett,\\ Night Watch (2002)}

In this chapter, we start with a summary of the main ideas in the literature about renormalons, corresponding to section \ref{sec-ren-summary}. The presentation is neither very precise nor very sharp, but this is a topic riddled with conjectures and lacking an overall framework. 
In section \ref{cha_largeN}, we analyze the specific case of ring diagrams in the large $N$ limit of the $O(N)$ non-linear sigma model. This is a concrete realization of renormalons in integrable models, which serves as a template for how renormalons manifest at the level of Feynman diagrams. 

\section{Renormalons in asymptotically free theories}
\label{sec-ren-summary}

In section \ref{sec-res-qft}, we sketched the idea that non-perturbative sectors of the trans-series can come from saddle points of the action, i.e. instanton effects. In the early 1970's, the hope was that this would be the source of all non-perturbative sectors, the program of instanton dominance. However, this hope was dissolved by the observation of renormalon effects \cite{gross-neveu,thooft}.

\begin{figure}
\centering
\includegraphics[height=0.25\textwidth]{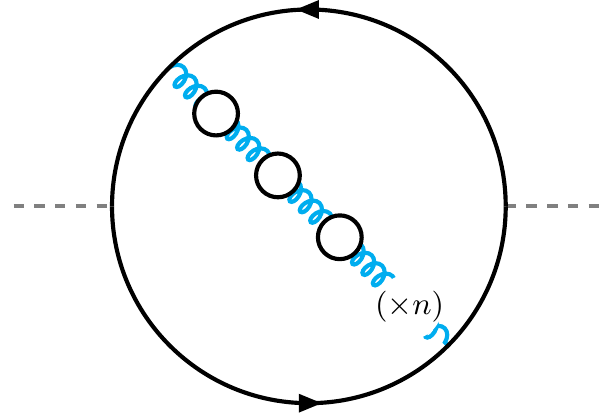}
\caption{An illustration of the typical renormalon diagram in particle physics. These diagrams, after renormalization, would contribute as $\propto n!$.}
\label{ren-diag}
\end{figure}

At the level of Feynman diagrams, the distinction lies in how the factorial behavior \eqref{gevrey} comes about. For instanton effects, they manifest as a combinatorial explosion of diagrams \cite{Bender1976,Lam1968}. At each order $g^n$ there are $\sim n!$ many diagrams, each valued $\CO(1)$ once appropriately renormalised. As for renormalons, they manifest as a family of diagrams, such that each order $g^n$ there is a single diagram, or a constant few, but which is valued $n!$ once renormalised. Typically, this is due to the integrand behaving as $(\log k)^n$ close to the UV or to the IR. This happens in diagrams with chains of ``bubbles'', like that of figure \ref{ren-diag}. It is expected that asymptotically free theories in general have IR renormalons which obstruct Borel summation. This includes the most important asymptotically free theory of all, QCD, which has been checked with lattice calculations \cite{pineda,pineda2}.

Let us focus for now on renormalizable asymptotically free theories.
In order to see what non-perturbative scale this effect would correspond to, it  is useful to consider the renormalization of the non-perturbative part, as argued in \cite{beneke,bly} for example. Let's say we are studying an observable with perturbative and non-perturbative contributions $F(g) = F_{\text{p}}(g) +  F_{\text{np}}(g)$, which we want to evaluate at some renormalised coupling $\bar{g}$ set by an external scale $h$, $\bar{g}(\mu=h)$. 
By construction, the perturbative part is RG-invariant.
Then, the non-perturbative contribution to an observable must satisfy an RG equation of the form
\begin{equation}
\left(\mu \frac{\partial}{\partial\mu}+\beta(g)\frac{\partial}{\partial g} + \gamma(g) \right) F_{\text{np}}\left(h/\mu,g(\mu)\right)=0,
\label{RG-F}
\end{equation}
where $d$ and $\gamma(g) \approx \gamma_1 g^2+\cdots$ are, respectively, the classical and anomalous dimensions of the observable. \eqref{RG-F} implies
\begin{equation}
F_{\text{np}}(1,\overline{g}) = C(h) \re^{-\frac{d}{2\beta_0 \overline{g}^2}}\left(\overline{g}^2\right)^{-d\xi + \frac{\gamma_1}{2\beta_0}}\left(1+\CO(\overline g^2)\right).
\label{RG-F-exp}
\end{equation}
Less cryptically, what RG-invariance requires is that any non-perturbative effect be proportional to some power $\ell$ of the dimensionless ratio between the dynamically generated scale \eqref{Lambda-def} and the external scale parameter,
\begin{equation}
\left(\frac{\Lambda}{h}\right)^\ell\left(1+\CO(\bar g^2)\right).
\label{power-Lh-p}
\end{equation}
However this is not specific to renormalons. Instantons too should obey \eqref{power-Lh-p}, see \cite{mmbook} for discussions.

In \cite{thooft,parisi1,parisi2,parisi-rg}, Parisi and 't Hooft argued that renormalons correspond to effects of the form \eqref{power-Lh-p} with $\ell$ integer. In terms of the corresponding singularities in the Borel plane, this implies singularities at
\begin{equation}
\zeta= \frac{\ell}{2\beta_0},\quad \ell \in\mathbb{N},
\label{IR-ren}
\end{equation}
for the Borel space $\zeta$ dual to $\bar g^2$. We call these conjectured singularities the conventional IR renormalons, and they would obstruct Borel summation. 't Hooft also argues for the existence of UV renormalons, 
\begin{equation}
\zeta= -\frac{\ell}{2\beta_0},\quad \ell \in\mathbb{N}.
\label{UV-ren}
\end{equation}
These do not obstruct Borel summation but do control the large order behavior of perturbation theory, as we explored in chapter \ref{cha_resurgence}. The UV and IR labels imply that the \eqref{IR-ren} come from $\log k$ behavior in the IR, and respectively \eqref{UV-ren} come from divergences in the UV. Physically, this fits our intuition of asymptotically free theories, the UV is under control but in the IR lies the breakdown of perturbation theory. In contrast, for QED the reverse is observed: UV effects manifest in the positive real axis, obstructing Borel summation.

When an observable is amenable to an OPE expansion, these integers have a canonical interpretation. Let us imagine we are inspecting an observable with an external momentum scale $P$, like a two point function, then
\begin{equation}
F_\mu(P,g) = F_{\mu,\text{p}}(P,g) + \sum_{\CO_i} P^{-d_i}\langle\CO_i\rangle_\mu C_i(P/\mu, g)
\end{equation}
where $\langle\CO\rangle_\mu$ are the non-perturbative vevs of the operators $\CO_i$ with classical dimension $d_i$. The RG argument from \eqref{RG-F-exp} can be applied to each of these contributions individually and we find
\begin{equation}
F_\mu(P,g) = F_{\mu,\text{p}}(P,g) + \sum_{\CO_i} C_i(P) \re^{-\frac{d_i}{2\beta_0 \overline{g}^2}}\left(\overline{g}^2\right)^{-d_i\xi + \frac{\gamma_{1,i}}{2\beta_0}}\left(1+\cdots\right),
\label{OPE-expansion}
\end{equation}
where $\gamma_{1,i}$ are the anomalous dimensions of the respective operators $\CO_i$. Then we would identify the integer classical dimensions $d_i$ with the integers $\ell$ in \eqref{IR-ren}. Because one might be able to restrict which operators can contribute, for example on symmetry grounds, the allowed $d_i$, in particular the leading one, can be specified. 

The study of renormalons in the two point function and their relation to the OPE has a long literature 
\cite{David1981,david1,david2,david3,itep,itep2d,beneke-lo,Beneke1997,bly,Shifman2015,shifman}. In most of these cases, the analysis is done in the large $N$ limit and always validates the conventional positions \eqref{IR-ren}. In some 2d models with twisted compactifications, the conventional IR renormalons at large $N$ are also matched by semi-classical effects in a quantum mechanical effective model \cite{dunne-unsal,cherman-dorigoni-dunne-unsal}, but it is unclear what can be extrapolated from such approximation. For canonical reviews on renormalons, see \cite{Shifman2015a,beneke}.

However, the conventional view turns out to be limited. In particular, not all observables in QFT can be expanded in an OPE. One could hope that \eqref{OPE-expansion} would still hold, but it we found in \cite{mmr-antrans} that it does not for the free energy, since there are renormalon effects which do not correspond to \eqref{IR-ren}. This leaves us with very little to specify what a renormalon is. The only general criterion is that a renormalon is a non-perturbative effect which is manifested at the level of perturbation theory as a series of factorially divergent diagrams. This is a somewhat unsatisfying definition, since it perturbatively defines a non-perturbative effect, but it is, at the moment, the best available description.

There is, nonetheless, one further feature of renormalons which helps us disentangle them from instantons. Renormalons, unlike instantons, should survive in the large $N$ limit. This is built in for the conventional renormalons, \eqref{IR-ren} and \eqref{UV-ren}. More generally, it follows from their diagrammatic description: at large $N$ most diagrams as sub-dominant and the combinatorical factors required to add up to $n!$ are broken up into lower orders in $1/N$. Meanwhile, a diagram with many bubbles can be dominant in the 't Hooft limit. In fact, they tend to be, since bubbles of field with $N$ components will contribute with factors of $N$. However, even this test only ``suggests'' an effect is an instanton or a renormalon and should not be taken as conclusive.

In the end, renormalons are still deeply ill-understood. A simplistic description would be that renormalons are the price paid by the perturbative series for being ignorant of non-perturbative effects, like $\Lambda_{\text{QCD}}$ in QCD and the Landau pole in QED. However, as we show in chapter \ref{cha_antrans}, renormalon effects do not merely cancel ambiguities of the Borel summation, they include real unambiguous contributions to observables. 
Furthermore, renormalons are not exclusive to renormalizable QFT. They can be found in super-renormalizable theories, like in \cite{mr-2dren}; in non-relativistic theories which do not require renormalization, which we inspect in part \ref{part-qmb}; and even in quantum mechanics, as recently found by \cite{Pazarbasi2019}.
Since the OPE seems to prima facie fail as a general theory of renormalons, the question of what form of ``renormalon calculus'' can reproduce these contributions from first principles in a general QFT remains completely unanswered.

To close off this deambulation through renormalon ``lore'', let us look at a concrete example of a renormalon effect in an integrable model.

\section{Large \titleN renormalons from ring diagrams}
\label{cha_largeN}

To finish this chapter and transition into the main text, we will look at renormalons in the $O(N)$ non-linear sigma model at large $N$. They turn out to be related to ring diagrams, a particular family of diagrams that constitute a divergent series. This discussion will schematically review some of the main results of \cite{mmr}, while avoiding technical details. 
The goal is to show why we find renormalons from a perturbative point of view. We will also contextualise the results of \cite{mmr} in light of posterior work in the literature, particularly the large $N$ analysis of \cite{dpmss} and sketch the overall conceptual picture.   

\subsection{\texorpdfstring{$O(N)$}{O(N)} non-linear sigma model}

We introduced the $O(N)$ non-linear sigma model in \ref{sec-models}. However, the Lagrangian \eqref{ON-euc} is not convenient for perturbation theory, due to its non-linear constraint. We start by adding an auxiliary field $X$ which implements $\mathbf{\sigma}^2=1$. The Lagrangian, coupled to $h$, becomes akin to a linear sigma model,
%
\begin{multline}
\CL_h =
\frac{1}{2g_0^2}\bigg\{ \partial_\mu \boldsymbol{\sigma}\cdot \partial^\mu\boldsymbol{\sigma}+ X(\boldsymbol\sigma^2 - 1) \\
+2\ri h \left(\sigma_1\partial_0 \sigma_2-\sigma_2 \partial_0 \sigma_1\right)+h^2(\sigma_3^2 + \cdots + \sigma_N^2 - 1) \bigg\}.
\end{multline}

Furthermore, we want to study this model in the large $N$ limit \cite{thooft-largen}. Some further modifications are then required.
We first introduce the 't Hooft coupling $\pi\lambda_0=(N-2) g^2_0$, which is preferable for the large $N$ analysis.
Then we parameterize $\boldsymbol\sigma$ and $X$ as fluctuations around a constant configuration $\sigma$, in the $\sigma_1$ direction, and $\chi$,
\begin{equation}
\ba
\boldsymbol\sigma(x) &= \Big(\sigma,0,\dots,0 \Big) + \sqrt{2\pi\Delta\lambda_0}\Big(\tilde\sigma_1(x),\tilde\sigma_2(x),\eta_1(x),\dots,\eta_{N-2}(x) \Big),\\
X(x) &= \chi +  \sqrt{2 \pi \Delta \lambda_0} \tilde{\chi}(x).
\ea
\end{equation}
Finally, we organize the $N-2$ $\eta$-fields into a vector $\boldsymbol\eta = (\eta_1,\dots,\eta_{N-2})$.
The resulting Lagrangian can be expanded into three pieces: the tree level which specifies the vevs, a quadratic kinetic term, and the interaction vertices for the dynamical fields. 
They are given by
\begin{equation}
\ba
\CL_\text{tree} &= \frac{\chi}{2}(\sigma^2-1) - \frac{h^2}{2},\\
\CL_\text{kinetic} &= \frac{1}{2}\boldsymbol\eta \cdot (-\partial^2 + \chi + h^2) \boldsymbol\eta + \frac{1}{2}
\left[
\begin{matrix}
\tilde\sigma_1\\
\tilde\sigma_2\\
\tilde\chi
\end{matrix}
\right]^T
\left(
\begin{matrix}
-\partial^2 + \chi & 2ih\partial_0 & \sigma\\
-2ih\partial_0 & -\partial^2 + \chi & 0\\
\sigma & 0 & 0
\end{matrix}
\right)
\left[
\begin{matrix}
\tilde\sigma_1\\
\tilde\sigma_2\\
\tilde\chi
\end{matrix}
\right],\\
\CL_\text{int} &= \frac{1}{2}\tilde\chi \boldsymbol\eta \cdot \boldsymbol\eta + \frac{1}{2}\tilde\chi \Big( \tilde\sigma_1^2 + \tilde\sigma_2^2 \Big).
\ea
\end{equation}
And the total Lagrangian is simply
\begin{equation}
\label{lag-nlsmh}
\CL_h = \frac{1}{2\pi\Delta\lambda_0}\CL_\text{tree} + \CL_\text{kinetic} + \sqrt{2\pi\Delta\lambda_0}\CL_\text{int}.
\end{equation}
The propagators of the dynamical fields in this theory are
\begin{equation}
\ba
D_{\tilde{\sigma}_1\tilde{\sigma}_1} &=  D_{\tilde{\sigma}_1\tilde{\sigma}_2} = D_{\tilde{\sigma}_2\tilde{\sigma}_1}= 0, &D_{\tilde{\sigma}_2\tilde{\sigma}_2} &= \frac{1}{k^2+\chi},\\
D_{\tilde{\sigma}_1\tilde{\chi}} &= D_{\tilde{\chi}\tilde{\sigma}_1} =\frac{1}{\sigma}, &D_{\tilde{\sigma}_2\tilde{\chi}} & = - D_{\tilde{\chi}\tilde{\sigma}_2} = \frac{2 h k_0}{\sigma(k^2+\chi)}, \\
D_{\tilde{\chi}\tilde{\chi}}&= -\frac{1}{\sigma^2}\left[k^2+\chi+ \frac{4 h^2 k_0^2}{k^2+\chi}\right], &
D_{\eta_i \eta_j} &= \delta_{ij} \frac{1}{k^2+h^2+\chi}, \quad i,j=1, \dots, N-2. 
\ea
\label{nlsmLkin}
\end{equation}

In the large $N$ limit, we can write the effective potential as a power series in $\Delta$, 
\be
\label{v-delta}
V(\sigma, \chi; h) = \sum_{\ell \ge 0} V_{(\ell)}(\sigma, \chi; h) \Delta^{\ell-1}. 
\ee
The vevs themselves, which minimize this potential, can be expanded order by order in $N$,
\be
\sigma= \sigma_{(0)}+ \CO\left(\Delta\right), \qquad \chi= \chi_{(0)}+ \CO\left(\Delta\right).
\ee
The minimisation condition at leading order is then
\be
{\partial V_{(0)}\over \partial \sigma_{(0)}}= {\partial V_{(0)}\over \partial \chi_{(0)}}=0.
\label{v-minima}
\ee
Due to \eqref{v-minima}, there is no direct contribution from $\sigma_{(1)},\xi_{(1)}$ to the free energy, which is given simply by the effective potential evaluated at the vev,
\begin{equation}
\CF(h) = V(\sigma,\xi;h)\big|_{\delta V=0}.
\end{equation}
Thus, the leading order vevs turn out to be enough to calculate up to NLO, 
\be
\begin{aligned}
\CF(h)= \frac{1}{\Delta} V_{(0)}(\sigma_{(0)}, \chi_{(0)};h)+ V_{(1)}(\sigma_{(0)}, \chi_{(0)};h)+ \CO\left(\Delta\right).
\end{aligned}
\ee
Of course, we still need to compute $V_{(0)}$ and $V_{(1)}$.

Let us start, naturally, with $V_{(0)}(\sigma, \chi;h)$. There is a tree level contribution from $\CL_\text{tree}$ and a one-loop contribtuion from the $\eta$ fields which gets multiplied by $1/\Delta = N-2$. In total we have,
\begin{equation}
V_{(0)} (\sigma,\chi;h) =  \frac{1}{4 \pi  \lambda_0 } \left(\chi(\sigma^2-1)-h^2 \right)+\frac{1}{2} \int \frac{\rd^d k}{(2\pi)^d}\log(k^2+h^2+\chi).
\end{equation}
This integral can be evaluated using dimensional regularization,
\be
V_{(0)}(\sigma, \chi;h)= {1 \over 4 \pi \lambda_0} \left( \chi (\sigma^2-1)- h^2 \right)+ {(h^2+ \chi)^{d/2} \over  (4 \pi)^{d/2}} {1 \over d} \Gamma\left({\epsilon \over 2} \right).
\label{v-leading-order}
\ee
which leads to $\sigma_{(0)}$ and $\chi_{(0)}$.

In the minima of \eqref{v-leading-order}, there is a symmetry broken classical vacuum with $\sigma=\sigma_1=1$. If we had $h=0$, then $\chi=0$ and $\tilde \sigma_2$, $\boldsymbol{\eta}$ would play the role of Goldstone bosons. However, in $1+1$ dimensions this is not quantum mechanically allowed, as formulated by Coleman-Mermin-Wagner theorem. Schematically, the $2$-point function of massless fields in $1+1$ dimensions is IR divergent. Nonetheless, in the presence of the $h$ coupling these field will acquire a mass and the phase becomes sensible. It should be noted that for $h=0$ the usual analysis is done in a different, gapped vacuum, where $\chi$ picks up a vev and $\sigma=0$. We can see that there is a similar minimum of \eqref{v-leading-order} with $h\neq 0$. However, this solution does not connect naturally to finite $N$ perturbation theory. We will discuss this ``phase'' again at the end of the chapter.

%
Choosing the minimum with vanishing $\chi_{(0)}$ we have
\be
\label{sigma0}
{\sigma^2_{(0)} \over \lambda_0} = {1\over \lambda_0}- \frac{1}{2} \left(\frac{h^2}{4\pi}\right)^{-\epsilon} \Gamma\left({\epsilon \over 2} \right).
\ee
The divergent terms on the r.h.s. of \eqref{sigma0} remind us that this theory needs to be renormalised. After all, it is asymptotically free.  Since it is not the focus of this chapter, we will skip over such complications and refer the interested reader to the detailed treatment in \cite{mmr}. Then, with $\lambda$, the renormalised coupling at scale $\mu=h$, we write
\be
{\sigma_{(0)}^2 \over \lambda_0 }= {1\over \lambda} + \CO\left(\Delta, \epsilon\right),
\label{sigma^2/lambda_limit}
\ee
which leads to the leading order, finite, free energy
\be
\CF_{(0)}(h)=-{h^2 \over 4 \pi} \left( {1 \over \lambda} - \frac{1}{2}  \right). 
\ee

\subsection{Ring diagrams}

\begin{figure}
\centering
\includegraphics[width=0.9\textwidth]{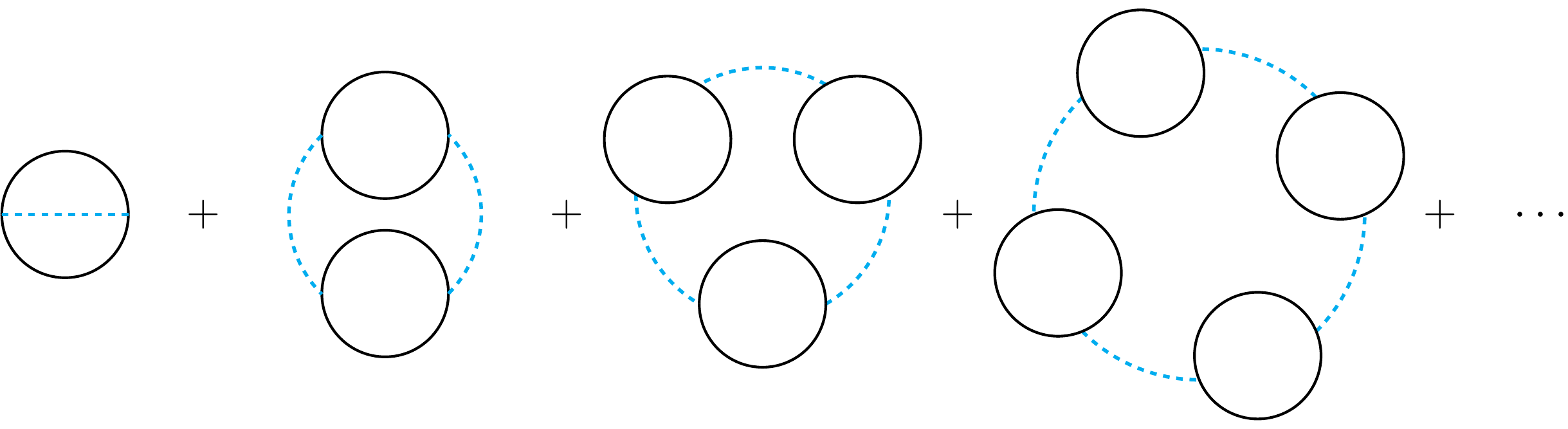}
\caption{Ring diagrams for the $O(N)$ non-linear sigma model. The black line is the propagator of the $\boldsymbol{\eta}$ field and the dashed line is the propagator of the $\tilde\chi$ field.}
\label{fig-ring-diag}
\end{figure}

The calculation of $V_{(1)}$ is substantially more ellaborate. As presented in \cite{root},  see also \cite{Jackiw1974,mr-2dren}, this term is given by the sum over ring diagrams. A ring diagram is composed of $m$ loops strung together by a chain of auxiliary fields, in this case $\tilde\chi$, see figure \ref{fig-ring-diag}. These diagrams contribute at order $\Delta^0$ because each bubble is connected to two vertices, picking up $(\sqrt\Delta)^2$, but then sums over $N-2= \Delta^{-1}$ identical fields, in this case the $\eta$ fields. So for any number $m$ of bubbles, the diagram is of order $\Delta^0$. At the same time, each vertex contributes with $\sqrt{\lambda_0}$ so each diagram is proportional to its respective power of $\lambda^m$. 

The sum over ring diagrams is given by,
\begin{equation}
-\sum_{m\geq 1} \frac{1}{2 m}  \int\frac{\rd^d k}{(2\pi)^d} \Big(2 \pi \lambda_0 D_{\tilde{\chi}\tilde{\chi}}(k) \Pi(k^2,h^2+\chi)\Big)^m.
\label{ring}
\end{equation}
The factor of $2m$ can be seen from the symmetries of the diagrams, $m$ shifts and one reflection but can also be derived straightforwardly from combinatorics.
The function $\Pi$ encodes the loop integral over the $\eta$ fields in a single bubbble and is given by
\begin{equation}
\Pi(k^2,M^2) ={1\over 2} \int {\rd^d q \over (2\pi)^d} {1\over (q^2 + M^2) \left( (k+q)^2 + M^2\right)}.
\end{equation}
This is often christened the ``polarization loop''. 

We are worried about the large order behavior of the series \eqref{ring}. Schematically, we can write a ring diagram in the form
\begin{equation}
 - \frac{\lambda^m}{2m}\int\frac{\rd^d k}{(2\pi)^d} \boldsymbol{\Pi}(k)^n\,,
\label{ren_prototype}
\end{equation}
where $\boldsymbol{\Pi}(k)\sim 2 \pi D_{\tilde{\chi}\tilde{\chi}}(k) \Pi(k^2,h^2+\chi)$ is the contribution of a single bubble and a single $\chi$ propagator. Let us assume the series of diagrams is asymptotic and define a formal power series,
\begin{equation}
\varepsilon (\lambda) = - \sum_{m\geq 1} \frac{\lambda^m}{2m}\int\frac{\rd^d k}{(2\pi)^d} \boldsymbol{\Pi}(k)^n.
\label{ren_series}
\end{equation}
To inspect its large order behavior, we define the Borel transform as 
\begin{equation}
\hat\varepsilon (\zeta) = - \sum_{m\geq 1} \frac{\zeta^m}{2m!m}\int\frac{\rd^d k}{(2\pi)^d} \boldsymbol{\Pi}(k)^n.
\end{equation}
Since the Borel transform is a convergent sum for sufficiently small $\zeta$, we are allowed to exchange the integral and the sum, 
\begin{equation}
\hat\varepsilon (\zeta) =  - \frac{1}{2} \int\frac{\rd^d k}{(2\pi)^d}\left\{-E_1\big(- \zeta  \boldsymbol{\Pi}(k)\big) - \log \left(-\zeta   \boldsymbol{\Pi}(k) \right)- \gamma_E \right\},
\end{equation}
where $E_1$ is the exponential integral function. The integrand is an entire function of $\zeta \boldsymbol{\Pi}(k)$ because the branch cuts of the $\log$ and $E_1$ cancel. 

One can then compute the Borel ressumation by exchanging the order of integration, 
\begin{equation}
s_\pm(\varepsilon)(\lambda) = \frac{1}{\lambda}\int_0^{\re^{\pm\ri 0}\infty}  \re^{-\zeta/\lambda} \hat\varepsilon(\zeta) \rd \zeta = \frac{1}{2}\int\frac{\rd^d k}{(2\pi)^d} \log\big(1-\re^{\mp\ri 0}\lambda \boldsymbol{\Pi}(k) \big)\,.
\label{prototype_resum}
\end{equation}
Thanks to the branch cut of the logarithm we see that the Borel ressumation might have a discontinuity for some values of $\lambda$, as we expected from the divergent asymptotics. We can also calculate the discontinuity as a function from the integral representation in \eqref{prototype_resum}, since 
\begin{equation}
\disc \varepsilon_{(0)}(\lambda) =  \frac{1}{2} \int _{M^d(\lambda)}  (2\pi\ri)\rd^d p
\end{equation}
where $M^d(\lambda)$ is the subregion of the positive half line where
$\boldsymbol{\Pi}(k) > \frac{1}{\lambda}$. The integral in \eqref{prototype_resum} is actually what one obtains by swapping the sum and the integral in \eqref{ren_series}, which in practice is a more simple, if heuristic, technique. No matter how we get it, we can relate the discontinuity of the Borel sum with the large order behavior of perturbation theory, as we saw in \eqref{disc-to-stokes}.

The recipe outlined here is but a schematic picture of what one needs to actually extract an asymptotic series and its summation from \eqref{ring}. In practice, there are many obstacles which are thoroughly presented in \cite{mmr}. First and foremost, the momentum integrals are in fact divergent and the coupling $\lambda_0$ must be renormalised accordingly. This is an elaborate computation that, using the all loops $\beta$-function at large $N$ calculated in \cite{bh}, retrieves a finite result at each order. These finite coefficients of the series in the renormalised coupling $\lambda$ can be resummed using the above heuristic of swapping the sum and the integral. 

In the end, is is found in \cite{mmr} that the resummed ring diagrams can be expressed as four functions such that
\begin{equation}
\CF_{(1)}(h) \approx -\frac{h^2}{2\pi}\big[W(\lambda) + X(\lambda) + Y(\lambda) + Z(\lambda) \big].
\end{equation}
The first function is a convergent function determined by $\beta$-function  read from \cite{bh},
\begin{equation}
\ba
W(\lambda)& = -\frac{1}{2}\int_0^\lambda\left(\frac{1}{u}+\frac{\beta_{(1)}(u)}{u^4}\right)\rd u,\\
 \beta_{(1)}(\lambda) &= -4\lambda^2 \int_0^{\lambda}  \rd x\, \frac{\sin(\frac{\pi x}{2})}{\pi x}\frac{\Gamma(1+x)}{\Gamma(1+\frac{x}{2})^2}  \frac{x+1}{x+2} .
\ea
\end{equation}
 Because it is a convergent function, it holds no significant information about the asymptotic behavior of the series. 
 
The other three functions are more elaborate. The function $X$ is given by
 \be
\label{Xlambda}
\ba
X(\lambda)&= \int_0^1 {\rd z\over z^{2}} \CX_0\big(\lambda\sqrt{1-z},z\big) + \frac{1}{2}\int_0^1 {\rd z \over z (1-z) } \CX_1 (\lambda\sqrt{1-z},z), \\
\CX_0(y,z) &= \log \left[1-y \log \left(\frac{\sqrt{z}}{2}\right)\right] - \log \left[1 - y \log \left(\frac{\sqrt{z}}{1+\sqrt{1-z}}\right)\right]\\
&\qquad -\frac{y z}{4}\frac{1}{1 - y \log\left(\frac{\sqrt{z}}{2}\right)}, \\
\CX_1(y,z)&= \frac{1}{1- y \log\left(\frac{\sqrt{z}}{1+\sqrt{1-z}}\right)}-\frac{1}{1-y\log\left(\frac{\sqrt{z}}{2}\right)}. 
\ea
\end{equation}
This function contributes to the discontinuity, due to the logarithms and the poles in $\CX_{0,1}$ but only if $\lambda<0$. Similarly, the function $Z$ is given by
 \be
 \ba
Z(\lambda)&= \int_0^1 {\rd z\over  (1-z)^{2}}\left[ \frac{\CZ(z,\lambda)-1}{2} - 2 \log\left( \frac{1+\sqrt{\CZ(z,\lambda)}}{2} \right) \right],\\
\CZ(z,\lambda) &=  \frac{1+F(z)\lambda}{1+zF(z)\lambda}, \qquad F(z) \frac{\tanh^{-1}(\sqrt{z})}{\sqrt{z}},
\ea
\ee
which can only have a logarithmic discontinuity and a residue contribution when $\lambda<0$. These discontinuities at $\lambda<0$ are important to study the large order behavior of the perturbative series in general, but they contribute to alternating terms that do not obstruct resummation. They correspond to UV renormalons, rather than IR ones.

Lastly, the function $Y$ is given by
\be
\ba
\label{Ylambda}
Y(\lambda) &= \frac{1}{4}\int_0^1 \frac{\rd z}{z (1-z)} \CY (\lambda\sqrt{1-z}, z), \\
\CY(y,z) &=- z + 2z(1-z) + \frac{z+2}{1 - y\log \left(\frac{\sqrt{z}}{2}\right)}\\
&\qquad+ \frac{z\,\re^{-2/y}}{y}\, \text{E}_1\left[- \frac{2}{y} \left( 1 - y\log\left(\frac{\sqrt{z}}{2}\right)\right)\right] .
\ea
\ee
For $\lambda<0$, this function also has a complicated discontinuity coming from the exponential integral, which has a logarithm like branch cut $E_1\sim-\log(x)$, and its poles. However it is also discontinuous if $\lambda>0$. In this case, the argument of the exponential integral is always negative, which leads to a discontinuity of $2\pi\ri$. Integrating we find
\begin{equation}
\frac{1}{4} \int_0^1 \frac{\re^{-\frac{2}{\lambda\sqrt{1-z}}}}{\lambda (1-z)^{3/2}} (2\pi\ri)\rd z = \frac{\ri\pi}{2}\re^{-2/\lambda},
\label{Ydisc}
\end{equation}
which has the form of a trans-monomial.

The implication of \eqref{Ydisc} is that the resummed free energy is  discontinuous for positive $\lambda$ with discontinuity
\be
\disc \CF_{(1)}(h) = - \frac{h^2}{4\pi}\left(\ri\pi\re^{-2/\lambda}\right).
\ee
Using the resurgence relation \eqref{resurgence-relation}, we extract that the leading order behavior of the perturbative series is
\begin{equation}
v_m \sim 2^{-m-1} \Gamma(m) + (-1)^m 2^{-m}\big(\cdots\big),
\end{equation}
where we define $v_m$ such that
\begin{equation}
\CF_{(1)}(h) \approx - \frac{h^2}{4\pi} \left\{\sum_{m\geq 1} v_m \lambda^m+2W(\lambda)\right\},
\label{vm-series}
\end{equation}
excluding $W(\lambda)$ which can be removed with a redefinition of the coupling.
Thus, from ring diagrams we identify clearly, albeit not effortlessly, an IR renormalon effect.

\subsection{Discussion}

In this section, we outlined a calculation of the large $N$ free energy which is perturbative in $\lambda$. However, one does not have to make this additional assumption, as shown by \cite{dpmss}. There, they simply take the large $N$ limit and pick the saddle point with $\sigma = 0$ and $\chi\neq 0$. As we mentioned, we did not choose this vacuum because it does not connect evidently to perturbation theory when $\lambda\rightarrow 0$. The physics of these vacua is seemingly very distinct, the $\chi=0$ saddle is ungapped, often called the ``ordered phase'', while the other saddle is gapped and thus the ``disordered phase''. However, with different techniques, in \cite{dpmss}  the free energy is extracted from the large $N$ effective action without expanding in $\lambda$. The resulting integral formula for $\CF_1(h)$ is finite, real and unambiguous as a function of $\lambda$ and can be shown to equal the real part of the Borel resummation of \eqref{vm-series}. Roughly, results from the disordered phase seem to be already ``resummed''. Coherently, if this formula is then asymptotically expanded, the ring diagram series is recovered.

A similar picture was found in \cite{mr-2dren} for the ``linear sigma model'', which is a theory of an $N$-component boson with a quartic potential in $1+1$ dimension. In higher dimensions, this would be the paradigmatic example of the Goldstone mechanism. However, in $1+1$ the Coleman--Mermin--Wagner theorem shows that the $O(N)$ symmetry is unbroken in the quantum vacuum. In this model, one can also the tackle the problem at large $N$ in two ways: either finding the gapped vacuum with unbroken $O(N)$ symmetry from the large $N$ effective action or by taking the classical vacuum with broken symmetry and inspecting the large $N$ limit of perturbation theory. The second option is prima facie IR divergent, but a regulator can be introduced leading to a well defined asymptotic series for the free energy, as originally argued in \cite{Jevicki1977}. As shown in \cite{mr-2dren}, at leading order in large $N$, this series is given by ring diagrams. With a similar argument to the one used in this chapter, resummation of this series has a perturbative ambiguity corresponding to an IR renormalon, even though the theory is super-renormalizable. Since the analogous calculation in $2+1$ dimensions, done in \cite{Coleman1974}, leads to a Borel summable series, it seems that in this case the IR renormalon is the price to pay for expanding around a ``wrong'' vacuum. 

While the presence of the $h$ field in the NLSM does not allow us to carry over the discussion of symmetry breaking from the linear model, the idea of two vacua, or phases, seems to share a conceptual space. 
There is a further analogous case in the theory of superconductors, as we shall see in chapter \ref{cha_GY}, where the superconductor phase plays the role of the gapped ``true'' vacuum. 

We then start to see a subtle interplay between the ``trivial vacuum'' of perturbation theory and the ``true vacuum'', which in the case of the NLSM with a magnetic field would be the gapped disordered phase. On one hand, the perturbation theory around the trivial vacuum gives the correct asymptotic approximation of physical observables. This was shown for the NLSM case in \cite{dpmss} and had been previously argued in \cite{Jevicki1977,Elitzur1983}. On the other hand, the myopia of the trivial vacuum  to non-perturbative physics seems to be manifested as renormalon effects. But we find this same ``myopia'' even when we take a description of the gapped phase, like the exact large $N$ result of \cite{dpmss}, and re-expand it. As we explore in this thesis, the same happens with the perturbative expansion of the Bethe ansatz wave-function, which is an exact description of the ``true'', gapped, ground state. 

There are further subtleties coming from the large $N$ limit itself.
When we expand the exact large $N$ results we find perturbation theory as an asymptotic series within a trans-series. For example, at leading order, in \cite{dpmss} they find
\begin{equation}
\CF_{(0)}(h) = - \frac{h^2}{4\pi}\Bigg\{\frac{1}{\bar{\alpha}}-\frac{1}{2}+\frac{\re^{-2/\bar\alpha}}{2}\Bigg\},
\end{equation}
where $\bar\alpha\sim\lambda$ (we define this series properly in chapter \ref{cha_antrans}).
Since the perturbative part is a truncated series, not even resurgence techniques could unearth the non-perturbative term. It could be that this term is inaccessible from the trivial perturbative vacuum, but this turns out to be a pathology of the large $N$ limit. In fact, the interplay of large $N$ and trans-series is a difficult affair. Therefore, it is natural to wonder what happens with renormalons and non-perturbative effects at \textit{finite} $N$.

In the next chapter, we tackle the question of finite $N$ trans-series by using integrability. For integrable models, the exact ground state, the ``all orders'' true vacuum, can be captured by the Bethe ansatz, as we discussed in chapter \ref{cha_intro}.
Like with exact large $N$ results, when expanding at weak coupling we obtain not only the asympotic series but also the trans-series. This will in fact reveal that important physics is hidden from the large $N$ analysis.

\part{Renormalons in Integrable Field Theories}
\label{part-iqft}

\chapter[Analytic trans-series from the Bethe ansatz][Analytic trans-series from the Bethe ansatz]{Analytic trans-series\\ from the Bethe ansatz}
\label{cha_antrans}


In this chapter, we derive the central results of this part of the thesis: the analytic trans-series for a broad set of integrable models and the discovery of a new class of renormalon effects. These results were published in \cite{mmr-antrans} and the presentation in this chapter follows similar strokes. We start with the case of Gross--Neveu where we follow a thoroughly detailed route, filling in many of the aspects implied in \cite{mmr-antrans}. This should serve as a pedagogical introduction to the method and resulting structure. For the bosonic models, we explain the recipe but avoid many of the gritty calculations. The goal is that this section provides a complete reference to reproduce and generalize the method. We also recapitulate the model-specific results of \cite{mmr-antrans,mmr-theta} with additional details. Lastly, we review \cite{mmr-theta} in compact form, to tie the results therein to the physical lessons of this chapter. This chapter contains only the analytic results of \cite{mmr-antrans,mmr-theta}. We discuss the numeric aspects in the chapter \ref{cha_volin}.

\section{Summary of the results}
\label{sec-summary}

In this chapter, we show that from the kernel of the integral equation \eqref{iqft_geneq_hm}, we can build a function
\begin{equation}
\sigma(\omega) = \frac{G_-(\omega)}{G_+(\omega)}
\label{sigma_def}
\end{equation}
whose analytic structure determines the structure of the trans-series for the free energy. In particular, the position of its poles along the positive imaginary axis correlates with non-perturbative terms of the trans-series or, equivalently, with the Borel singularities
of perturbation theory. Precisely, if $\sigma(\omega)$ has a pole at $\omega=\ri\xi$, then the trans-series has a term
\begin{equation}
\re^{-\frac{\xi}{\beta_0 \bar{g}^2(h)}}
\end{equation}
which we can identify with a Borel singularity at
\begin{equation}
\zeta = \frac{\xi}{\beta_0}.
\end{equation}
From the residues at these poles, we can obtain the trans-series parameters, although we only inspect the leading terms.

The standard renormalon lore we reviewed in chapter \ref{sec_renormalons}  implies that the singularities lie at 
\begin{equation}
 \frac{\ell}{\beta_0}, \quad \ell \in\IN.
 \label{renormalon-guess-zeta}
\end{equation}
However, what we find is that in general there is an isolated singularity at
\begin{equation}
\zeta = \frac{1}{\beta_0},
\end{equation}
followed by singularities of the type
\begin{equation}
\zeta = \frac{\ell}{1-r(1/N)}\frac{1}{\beta_0},\quad \ell\in\IN,
\label{gen-new-renormalon}
\end{equation}
where $r$ is some model-dependent rational function such that $r(0)=0$. These singularities match \eqref{renormalon-guess-zeta} only in the large $N$ limit. We show that these ``new renormalons'' are at leading order tied to families of divergent diagrams. 

In some models, we will also find singularities
\begin{equation}
\zeta \sim \frac{N\ell}{\beta_0},\quad \ell\in\IN,
\label{gen-instanton}
\end{equation}
which are technically compatible with \eqref{renormalon-guess-zeta}, but since they disappear at large $N$ they do not correspond to renormalons. We speculate they are the signature of instanton effects, both stable and unstable. In a model with both \eqref{gen-new-renormalon} and \eqref{gen-instanton} singularities, one can also find ``mixed'' terms, associated with singularities which are sums of ``new renormalons'' and ``instanton-like'' singularities.

Finally, we find that these new renormalons have real contributions to the final results which are not merely related to the cancellation of ambiguities of the Borel summation of perturbation theory. For the coset sigma models, we show that these contributions change sign under a topological $\vartheta=\pi$ angle, much like what happens to instantons.

\section{Trans-series for the Gross--Neveu model}
\label{sec-gn-1}

In this section, we explore how one can obtain a trans-series from the Bethe ansatz integral equation. The key is to exploit the Wiener--Hopf formalism, combining the perturbative approach to Gross--Neveu of \cite{fnw1,fnw2} with the non-perturbative analysis of the Sine--Gordon model in \cite{zamo-mass,sb-book}. This section is meant to be more introductory, while the section \ref{sec-gn-2} presents more systematic aspects of the resulting trans-series.

\subsection{Wiener--Hopf analysis of the integral equations}
\label{gn-inteq}
In the introduction, we presented the Bethe ansatz integral equation which applies to the Gross--Neveu model, \eqref{iqft_geneq_hm}. We want to find a trans-series expansion of the solution of this equation as well as the free energy. As we will show later, the weak coupling limit corresponds to $B\rightarrow \infty$. In the current form of the equation, this limit is naively singular and does not allow for a straightforward expansion. In this section, we will show how to render this integral equation in a more workable form. Our first manipulations will be rather general and we will specify to Gross--Neveu later. The first part is the same manipulation done in \cite{hmn,hn}, adding some details to the accessible presentation of appendix A of \cite{fnw1}. The latter part follows the set up of \cite{zamo-mass}, which is reviewed in Chapter 26 of \cite{sb-book}.

Equation \eqref{iqft_geneq_hm} is not unlike a Fredholm equation. A common approach to this type of equation is to take them to Fourier space, where, for example, the convolution with the kernel becomes a simple product. We must first continue the equation from its original domain, the interval $[-B,B]$ to the full real line. There are two simple ways to do so. One way would be to define the distribution $\epsilon$ outside the original interval such that the equation remains valid. Another way, which turns out to be more convenient is to restrict the support of $\epsilon$ to the interval and then add a new function $Y$ which is defined such that the equation holds outside the interval. That is, we write
\begin{equation}
\epsilon(\theta) - \int_{-B}^B K(\theta-\theta')\epsilon(\theta')\rd \theta' = g(\theta) + Y(\theta-B)+Y(-\theta-B),
\label{iqft_R}
\end{equation}
where $Y$ is supported on the positive reals and $g$ is the driving term, i.e.
\begin{equation}
g(\theta) = h-m\cosh\theta,\quad \theta\in[-B,B].
\end{equation}
One naturally needs to specify $g$ outside the interval. But since we have added $Y$, the choice of the extension of $g$ is simply absorbed into the definition of $Y$. Nevertheless, even though any extension of $g$ is a priori correct, not all of them are equally convenient. For example, we want to keep $g$ integrable. The ideal choice turns out to be
\begin{equation}
g(\theta) = 
\begin{cases}
      - \displaystyle\frac{m\re^\theta}{2} & \text{if } \theta < -B, \\[1.5mm]
     h - m \cosh \theta & \text{if } \theta\in[-B,B], \\[1.5mm]
     - \displaystyle\frac{m\re^{-\theta}}{2} & \text{if } \theta > B.  
\end{cases}
\label{gt_def}
\end{equation}

We can now take a Fourier transform of \eqref{iqft_R}. We use $\tilde{f}(\omega)$ to designate the Fourier transform of $f(\theta)$, for example
\begin{equation}
\tilde K (\omega) = \int_\IR \re^{\ri\omega\theta}K(\theta)\rd\theta.
\end{equation}
Then, Fourier transform of \eqref{iqft_R} is
\begin{equation}
\big(1-\tilde{K}(\omega)\big)\tilde \epsilon(\omega)=\tilde{g}(\omega) + \re^{\ri B\omega} \tilde Y({\omega})+ \re^{-\ri B\omega} \tilde Y({-\omega}).
\end{equation}

The natural approach to this type of equation is the Wiener--Hopf formalism. The key tool in this formalism is the Wiener--Hopf decomposition of a function $\psi$, defined over the real line $\IR$, into  two functions $[\psi]_\pm$, or ``projections'', which are respectively analytic in the upper (lower) complex half plane $\mathbb{H}_\pm$.  
For $\omega$ in the upper/lower half plane, we define the decomposition as
\begin{equation}
[\psi(\omega)]_\pm = \pm \frac{1}{2\pi\ri}\int_\IR \frac{\psi(\omega')}{\omega'-(\omega\pm\ri 0)}\rd\omega',\quad \omega\in \mathbb{H}_\pm.
\label{WH-decomp}
\end{equation}
Each projection can then be extended beyond its original plane of definition such that
\begin{equation}
\psi(\omega) = [\psi(\omega)]_+ + [\psi(\omega)]_-,
\end{equation}
when both $[\psi]_\pm$ are defined, including $\IR$.

A convenient property of this decomposition is that if a function $\Psi_\pm(\omega)$ is analytic and bounded in the upper (lower) half plane, then 
\begin{equation}
[\Psi_\pm]_\mp = 0, \qquad [\re^{\pm \ri a \omega}\Psi_\pm]_\mp = 0,\quad a>0.
\label{WH_cancel}
\end{equation}
Another useful property is that if $\phi(\omega) = \varphi(-\omega)$ then
\begin{equation}
[\phi]_-(\omega) = [\varphi]_+(-\omega).
\end{equation}
The Wiener--Hopf decomposition can also be seen as a ``partial Fourier transform''.
If the function $\psi(\omega)$ is the Fourier transform of an integrable function $\phi(\theta)$, then we can equivalently define the Wiener--Hopf decomposition as
\begin{equation}
[\psi(\omega)]_+ = \int_0^\infty \re^{\ri \omega \theta}\phi(\theta)\rd\theta,\quad [\psi(\omega)]_- = \int^0_{-\infty} \re^{\ri \omega \theta}\phi(\theta)\rd\theta.
\end{equation}
A particularly important but subtle extension of this decomposition is
\begin{equation}
[2\pi \delta(\omega)]_\pm  = \pm \frac{\ri}{\omega\pm\ri 0} = \pm \int_0^{\pm \infty} \re^{\ri \omega \theta}\rd\theta.
\label{dirac_WH}
\end{equation}
One can easily extend this additive decomposition into a multiplicative decomposition by taking logarithms, i.e.
\begin{equation}
\psi(\omega) = \Psi_+(\omega)\Psi_-(\omega), \quad \Psi_\pm(\omega) = \re^{[\log \psi (\omega)]_\pm}.
\end{equation}
By analogy with the Wiener--Hopf decomposition, we will extend the subscript $(\pm)$ to label other functions which are also analytic in $\mathbb{H}_\pm$, unless otherwise specified. 

To harmonize with Wiener--Hopf analysis, we write
\begin{equation}
Y_\pm(\omega) = \tilde{Y}(\pm \omega),
\end{equation}
and we introduce
\begin{equation}
\epsilon_\pm (\theta) = \re^{\pm\ri B \omega}\tilde{\epsilon}(\omega), \quad g_\pm (\theta) = \re^{\pm\ri B \omega}\tilde{g}(\omega),
\end{equation}
where, despite the subscript, $g_+$ actually has a pole in the upper half plane at $\omega=\ri$ as it can be seen from its explicit form,
\be
 g_+(\omega) = \ri h \frac{  (1-\re^{2\ri B\omega})}{\omega} + \frac{\ri m\re^B}{2}\left(\frac{\re^{ 2\ri B \omega}}{\omega-\ri}-\frac{1}{\omega+\ri}\right).
 \label{gplus-def}
\ee
Crucially, we perform a multiplicative decomposition of the kernel 
\begin{equation}
1-\tilde{K}(\omega) = \frac{1}{G_+(\omega)G_-(\omega)}.
\label{WH-kernel}
\end{equation}
Since, in integrable models in general, we have have that $K(\theta)$ is both even and integrable, it follows respectively that $G_-(\omega)=G_+(-\omega)$ and
\begin{equation}
G_+( \ri x) = 1 + \CO\left(\frac{1}{x}\right) \quad  \text{when} \quad x\rightarrow +\infty.
\end{equation}
With these definitions at hand, we can write
\begin{equation}
\frac{\epsilon_-(\omega)}{G_-(\omega)}= G_+(\omega) g_-(\omega) + G_+(\omega) Y_+(\omega) + \re^{-2\ri B \omega}G_+(\omega) Y_-(\omega).
\label{fourier_WH}
\end{equation}
We can take the Wiener--Hopf decomposition of both sides of the equation \eqref{fourier_WH}. This provides an integral equation for $Y_+$ which depends only on known data and an equation which specifies $\epsilon_-$ from $Y_+$,
\begin{align}
 G_+(\omega) Y_+(\omega)&= - [\re^{-2\ri B \omega} G_+(\omega) Y_-(\omega)]_+  -\left[G_+(\omega) g_-(\omega)\right]_+  ,
\label{Yplus_og}\\
\frac{\epsilon_-(\omega)}{G_-(\omega)} &= \left[G_+(\omega) g_-(\omega)\right]_- + [\re^{-2\ri B \omega} G_+(\omega) Y_-(\omega)]_- .
\label{eminus_og}
\end{align}
Lastly, the original integral equation \eqref{iqft_geneq_hm} has to be solved in tandem with the boundary condition $\epsilon(\pm B) =0$.  In Fourier space, this becomes the condition
\begin{equation}
\lim_{\kappa\rightarrow +\infty} \kappa \epsilon_+(\ri\kappa) = 0.
\label{eps_bc_og}
\end{equation}
Once we solve for $Y_+$ and thus $\epsilon_+$ we can find the free energy from formula \eqref{fh-cosh}, which can be rewritten as
\begin{equation}
\CF(h)= - \frac{m\re^B}{2\pi}\epsilon_+(\ri).
\label{fh-eps}
\end{equation}

So far, we have been quite general in our assumptions. Let us now concentrate on the case where $G_+(0)$ is finite. In the dichotomy introduced in section \ref{sec_iqft}, this is true in the case of ``fermionic'' models, which includes the Gross--Neveu models, but not in the case of ``bosonic'' models, which will require a different treatment. We then introduce the function
\begin{equation}
y(\omega) = G_+(\omega) Y_+(\omega) - h \frac{\ri G_+(\omega)}{\omega+\ri0}+ \frac{m \re^B}{2}\frac{\ri G_+(\omega) }{\omega-\ri}.
\label{y_def}
\end{equation}
Let us begin by rewriting \eqref{Yplus_og} in terms of $y$. The first term on the l.h.s. becomes
\begin{multline}
[\re^{-2\ri B \omega} G_+(\omega) Y_-(\omega)]_+ = \left[\frac{G_+(\omega)}{G_-(\omega)} \re^{-2\ri B \omega} y(-\omega)\right]_+ 
\\
- \ri h  \left[\re^{-2\ri B \omega}\frac{G_+(\omega)}{\omega-\ri 0}\right]_+ + \frac{\ri m \re^B}{2}\left[\re^{-2\ri B \omega}\frac{ G_+(\omega)}{\omega+\ri}\right]_+.
\end{multline}
On the other hand we can write\footnote{Since the term proportional to $h$ is analytic at $0$, the sign of the regulator $- \ri 0$ in the denominator is arbitrary. The same derivation can also be done with $+ \ri 0$ instead, but the details differ.}
\begin{equation}
\ba 
\left[G_+(\omega) g_-(\omega)\right]_+ &=  - h \frac{\ri G_+(\omega)-\ri G_+(0)}{\omega-\ri 0} +\ri h \left[\re^{-2\ri B\omega}\frac{G_+(\omega)}{\omega-\ri 0}\right]_+\\
&-\frac{\ri m\re^B}{2}\left[\re^{ -2\ri B \omega}\frac{G_+(\omega)}{\omega+\ri}\right]_+
+\frac{\ri m\re^B}{2}\frac{G_+(\omega)-G_+(\ri)}{\omega-\ri}.
\ea
\end{equation}
Noting that
\begin{equation}
\frac{\ri G_+(\omega)}{\omega+\ri 0} - \frac{\ri G_+(\omega)-\ri G_+(0)}{\omega-\ri 0} = 2\pi \delta(\omega) G_+(\omega) + \frac{\ri G_+(0)}{\omega-\ri 0} = \frac{\ri G_+(0)}{\omega+\ri 0}
\end{equation}
where we used \eqref{dirac_WH} twice,
we can simplify \eqref{Yplus_og} to
\begin{equation}
y(\omega) + \left[\frac{G_+(\omega)}{G_-(\omega)} \re^{-2\ri B \omega} y(-\omega)\right]_+ = - \frac{\ri hG_+(0)}{\omega+\ri 0} + \frac{ m \re^B}{2}\frac{\ri G_+(\ri)}{\omega-\ri}.
\end{equation}
From similar manipulations in \eqref{eminus_og}, we obtain 
\begin{equation}
\frac{\epsilon_-(\omega)}{G_-(\omega)} = -\frac{\ri h G_+(0)}{\omega-\ri 0} + \frac{ m \re^B}{2}\frac{\ri G_+(\ri)}{\omega-\ri} + \left[\frac{G_+(\omega)}{G_-(\omega)} \re^{-2\ri B \omega} y(-\omega)\right]_-.
\end{equation}
In integral form, these equations are
\begin{align}
y(\omega) &=  - \frac{\ri h G_+(0)}{\omega+\ri 0} + \frac{ m \re^B}{2}\frac{\ri G_+(\ri)}{\omega-\ri}+\frac{1}{2\pi\ri}\int_\IR \frac{\re^{2\ri B \omega'}\sigma(\omega')y(\omega')}{\omega'+\omega+\ri 0} \rd \omega',
\label{inteq_smally}
\\
\frac{\epsilon_+(\omega)}{G_+(\omega)} &= \frac{\ri h G_+(0)}{\omega+\ri 0} - \frac{ m \re^B}{2}\frac{\ri G_+(\ri)}{\omega+\ri} + \frac{1}{2\pi\ri}\int_\IR \frac{\re^{2\ri B \omega'}\sigma(\omega')y(\omega')}{\omega'-\omega-\ri 0}\rd\omega',
\end{align}
where we have introduced $\sigma(\omega)$ as in \eqref{sigma_def}. These are the integral equations presented in \cite{fnw1} to study the perturbative expansion.
Finally, the boundary condition \eqref{eps_bc_og} becomes
\begin{equation}
\ri h G_+(0) - \frac{ m \re^B}{2}\ri G_+(\ri) - \frac{1}{2\pi\ri}\int_\IR \re^{2\ri B \omega'}\sigma(\omega')y(\omega')\rd\omega' = 0.
\label{bc_smally}
\end{equation}

One can still further simplify these equations by removing the term in $m$ from the integral equation \eqref{inteq_smally} using the boundary condition \eqref{bc_smally},
\begin{equation}
y(\omega) =  \ri h G_+(0)\frac{\ri}{\omega(\omega-\ri)} - \frac{1}{2\pi\ri}\int_\IR \re^{2\ri B \omega'} \frac{\sigma(\omega')y(\omega')(\omega'+\ri)}{(\omega'+\omega+\ri 0)(\omega-\ri)} \rd \omega'.
\label{inteq_preu}
\end{equation} 
This suggests the introduction of 
\begin{equation}
u(\omega) = \frac{\omega-\ri}{\ri h G_+(0)}y(\omega),
\label{u_def}
\end{equation}
which, unlike $y$, is analytic in the upper half plane.  It is  also useful to write
\begin{equation}
\rho(\omega) = - \frac{\omega+\ri}{\omega-\ri}\sigma(\omega),
\label{rhofunc_def}
\end{equation}
which has the same analyticity properties as $\sigma(\omega)$ in $\mathbb{H}_+$. In terms of these new functions, we reduce \eqref{inteq_preu} to
\begin{equation}
u(\omega) = \frac{\ri}{\omega}+ \frac{1}{2\pi\ri}\int_\IR \frac{\re^{2\ri B \omega'}\rho(\omega')u(\omega') }{\omega'+\omega+\ri 0} \rd \omega'.
\label{inteq_u}
\end{equation}
Identical manipulations allow us to write
\begin{equation}
\frac{\epsilon_+(\omega)}{G_+(\omega)}=\frac{\ri h G_+(0)}{\omega+\ri}\left(\frac{\ri}{\omega}- \frac{1}{2\pi\ri}\int_\IR \frac{\re^{2\ri B \omega'}\rho(\omega')u(\omega') }{\omega'-\omega+\ri 0} \rd \omega'\right).
\label{eps_u}
\end{equation}

From the definition of $y$ in \eqref{y_def}, we have that only the term proportional to $m$ is singular at $\omega=\ri$. Then the definition of $u$ in \eqref{u_def} implies
\begin{equation}
u(\ri) = \frac{m \re^B}{2h} \frac{G_+(\ri)}{G_+(0)}.
\label{u_bc}
\end{equation}
Since we have already incorporated the original boundary condition \eqref{eps_bc_og} into the integral equation \eqref{inteq_u}, we can instead use \eqref{u_bc} as the ``boundary condition'' for $u$, since it imposes the relation between $h/m$ and $B$. Combining this boundary condition \eqref{u_bc}, \eqref{fh-eps}, and \eqref{eps_u}, we can write the free energy as
\begin{equation}
\CF(h) = - \frac{h^2}{2\pi}u(\ri)G_+(0)^2\left(1- \frac{1}{2\pi\ri}\int_\IR \frac{\re^{2\ri B \omega'}\rho(\omega')u(\omega') }{\omega'-\ri} \rd \omega'\right).
\label{fh-u}
\end{equation}
This redefinition of the Bethe ansatz in terms of $u$ and $\rho$ was originally introduced by \cite{zamo-mass} to identify non-perturbative corrections in the Sine-Gordon model, see \cite{sb-book} for a review.

Equations \eqref{inteq_u}, \eqref{u_bc} and \eqref{fh-u} encapsulate our problem. We want to calculate the free energy as a function of $h$. For that, we need $u$, which is found by solving the integral equation as a function of $B$ and then by solving the boundary condition to find $B$ as a function of $h$. In practice, it will be easier to write our results in terms of an auxiliary coupling defined from $h$, but this is the roadmap we will follow in the next section.

\subsection{Trans-series in the complex integral}
\label{sec-integral}

In this section, we will introduce the key computational step which will allow us to solve \eqref{inteq_u} and \eqref{fh-u} in a trans-series form.
We turn our attention to equation \eqref{inteq_u} in particular. Ultimately, we want to find a solution in the $B\gg 1$ regime. So we must expand the integral
\be
\frac{1}{2\pi\ri}\int_\IR \frac{\re^{2\ri B \omega'}\rho(\omega')u(\omega') }{\omega'+\omega+\ri 0} \rd \omega'
\ee
for large $B$. From the exponential factor, we see that a natural procedure is to deform the integration contour from the real line into the upper half plane. This is what both \cite{fnw1,zamo-mass} did in analogous situations. One can bring the contour just around the positive imaginary axis. At this point it becomes important to look into the function $\rho(\omega)$ in more detail.

Focusing on the Gross--Neveu model, the kernel $K(\theta)$ for this model is known from \cite{fnw1,fnw2,zamo-zamo}, and its Fourier transform is given by
\begin{equation}
1-\tilde K(\omega)=\frac{1+\re^{-\pi  \Upsilon |\omega|}}{1+\re^{-\pi  |\omega|}}, 
\end{equation}
where
\begin{equation}
\Upsilon=1-2\Delta,
\end{equation}
and $\Delta$ was defined in \eqref{Delta2}. The decomposition of the kernel is given by
\begin{equation}
G_+(\omega)=\re^{-\frac{\Upsilon}{2}\ri\omega\left[1-\log\left(-\frac{\Upsilon}{2}\ri\omega\right)\right]+\frac{1}{2}\ri\omega[1-\log(-\frac{1}{2}\ri\omega)]} \frac{\Gamma\big(\frac{1}{2} - \frac{\Upsilon}{2}\ri\omega\big)}{\Gamma\big(\frac{1}{2} - \frac{1}{2}\ri\omega\big)}, 
\label{GN_Gplus_usec}
\end{equation} 
which is indeed finite at $\omega=0$. The function $G_+(\omega)$ is by construction analytic in the upper half plane, with zeros in the positive real axis due to the $\Gamma$-function in the denominator. Meanwhile, in the lower half plane, it is discontinuous along the negative imaginary axis, due to the logarithms, and has poles due to the $\Gamma$-function in the numerator.

\begin{figure}
\centering
\begin{subfigure}{\textwidth}
\begin{tabular}[c]{lr}
\includegraphics[width=0.5\textwidth]{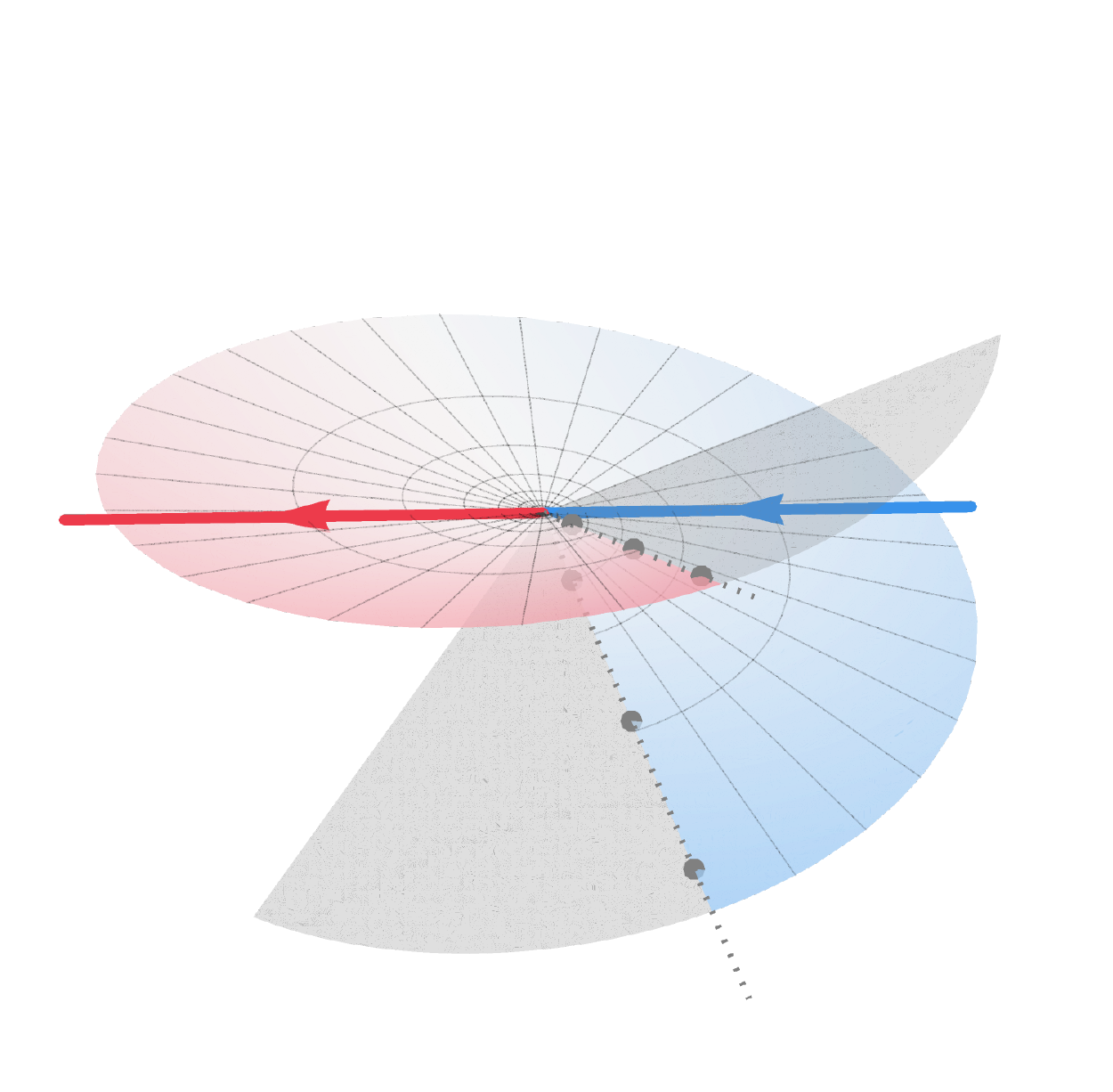}
&
\includegraphics[width=0.4\textwidth]{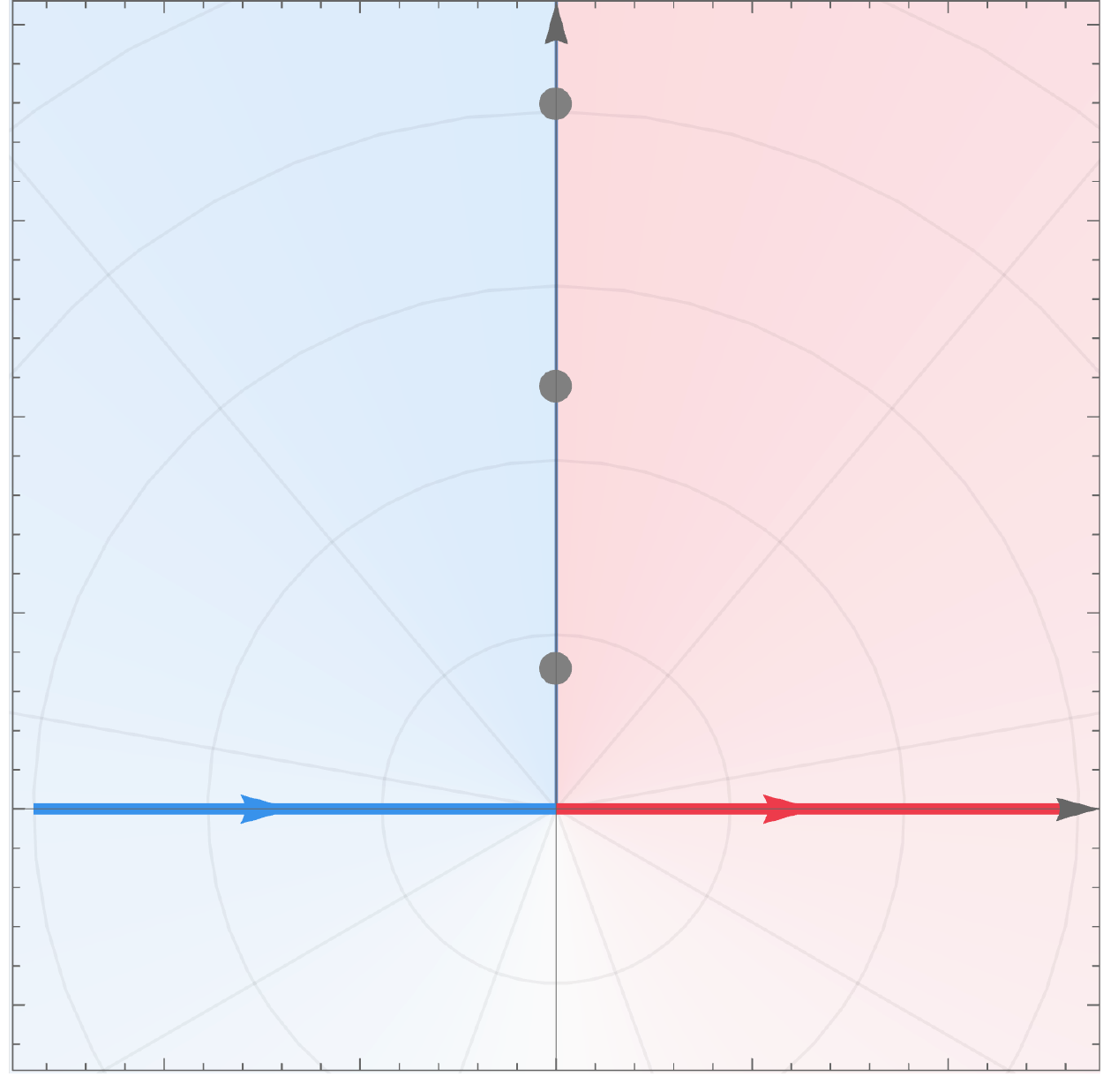}
\end{tabular}
\caption{Initial contour}
\end{subfigure}
\caption{The integration contour over $\IR$ drawn on part of a Riemann surface where $\log[\ri\omega]$ is continuous (left) and on the complex plane with a branch cut along the positive imaginary axis (right). One can think of the complex plane with a branch cut as the Riemann surface seen ``from above''. The grey parts of the Riemann surface are those which are not projected onto the complex plane on the right, while the dotted lines mark the two lines projected onto the positive imaginary axis. In $\mathbb{H}_+$, the discontinuity of $\rho$ comes from $\log[\ri\omega]$, so it is continuous on the displayed Riemann surface. On the other hand, it has poles (marked by grey dots) on both sheets, which are ambiguous from the perspective of the complex plane with a branch cut. For the solution to this ambiguity, see figure \ref{fig-deformed-integrals}.}
\label{fig-initial-integral}
\end{figure}

Through its definition \eqref{rhofunc_def}, $\rho$ inherits many of these properties. In particular, $\rho$ is both discontinuous and singular along the positive imaginary axis, the first due to a factor of $\log[\ri\omega]$, the second due to poles of the $\Gamma$-function. One has to be careful when handling these singularities. A rigorous way would be to think about them in the Riemann surface on which $\log[\ri\omega]$ is bijective\footnote{$\rho(\omega)$ inherits a factor $\log[\ri\omega]$ from $G_-$, see \eqref{GN_Gplus_usec}, which causes the discontinuity along $\ri \IR_+$. There is also a factor of $\log[-\ri\omega]$ from $G_+$, which is discontinuous along $\ri \IR_-$, but we will ignore it for now since it is does not impact this calculation. We inspect it later in section \ref{sec-uv}.}. The upper half plane is replaced by two sheets, the upper one where the $\log$ picks up a factor of $+\pi\ri$ and the lower one where it picks a factor of $-\pi\ri$, see figure \ref{fig-initial-integral}. There are then two immediate copies of the positive imaginary axis, one in each sheet, and there are poles in both, with different residues. When we deform the contour from the real axis, the half-piece which starts in the positive reals will end up in the upper sheet, and the one from negative reals will end up in the lower sheet. 

\begin{figure}
\begin{subfigure}{\textwidth}
\begin{tabular}[c]{lr}
\includegraphics[width=0.5\textwidth]{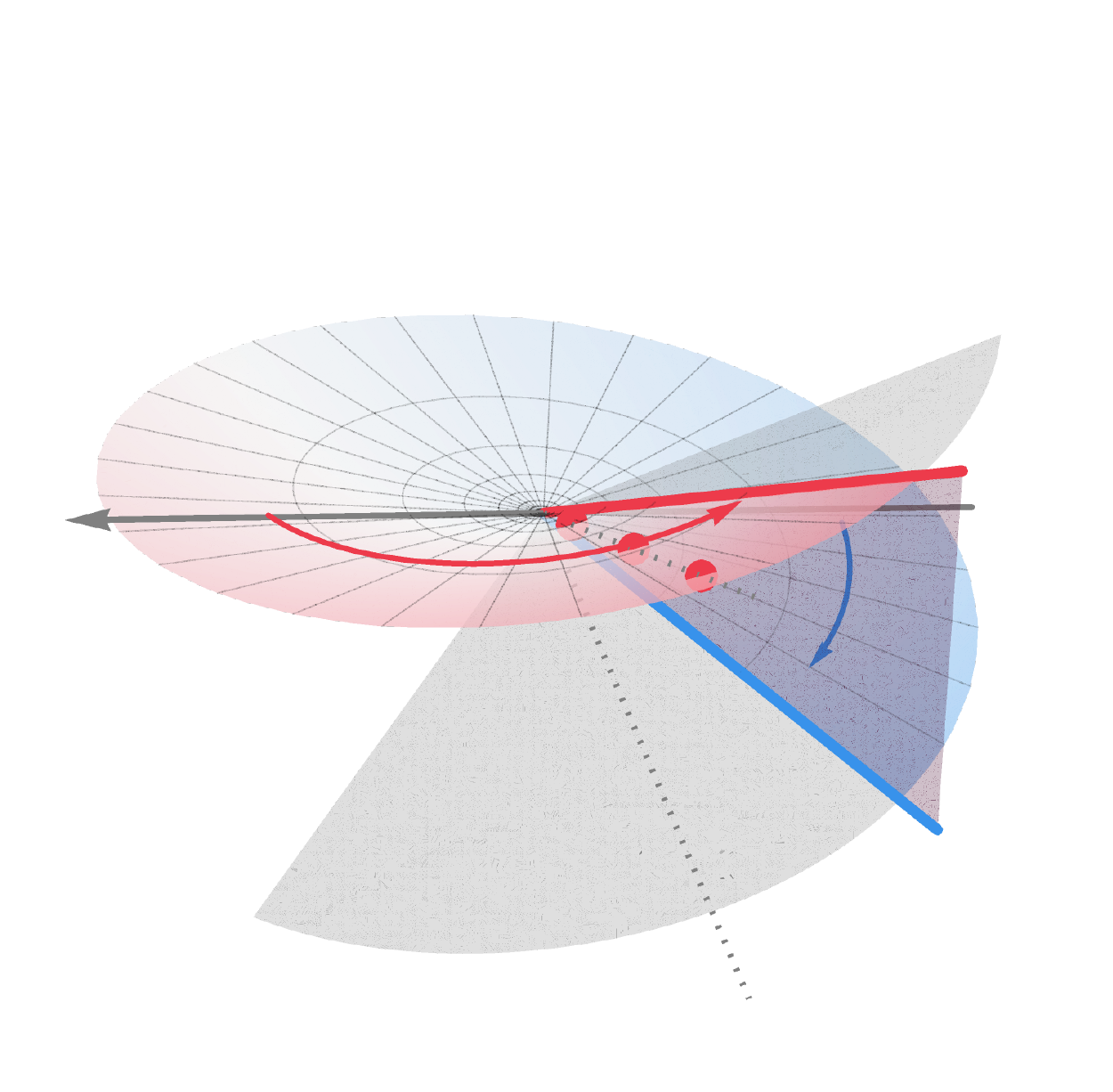}
&
\includegraphics[width=0.4\textwidth]{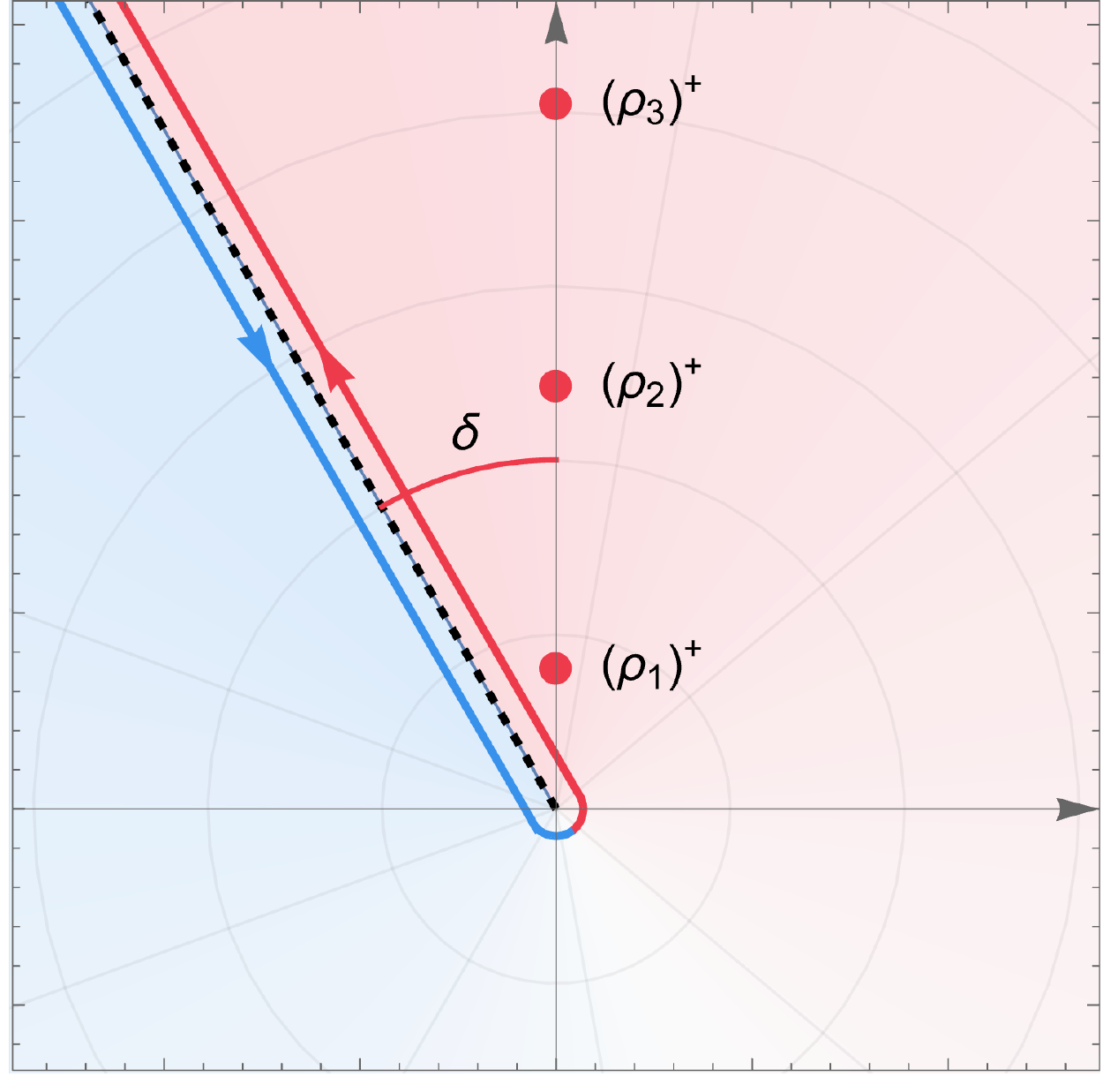}
\end{tabular}
\caption{Angle $\delta>0$ leading to integration contour $\CC_+$.}
\label{sub-fig-plus}
\vspace{-12.5pt}
\end{subfigure}
\begin{subfigure}{\textwidth}
\begin{tabular}[c]{lr}
\includegraphics[width=0.5\textwidth]{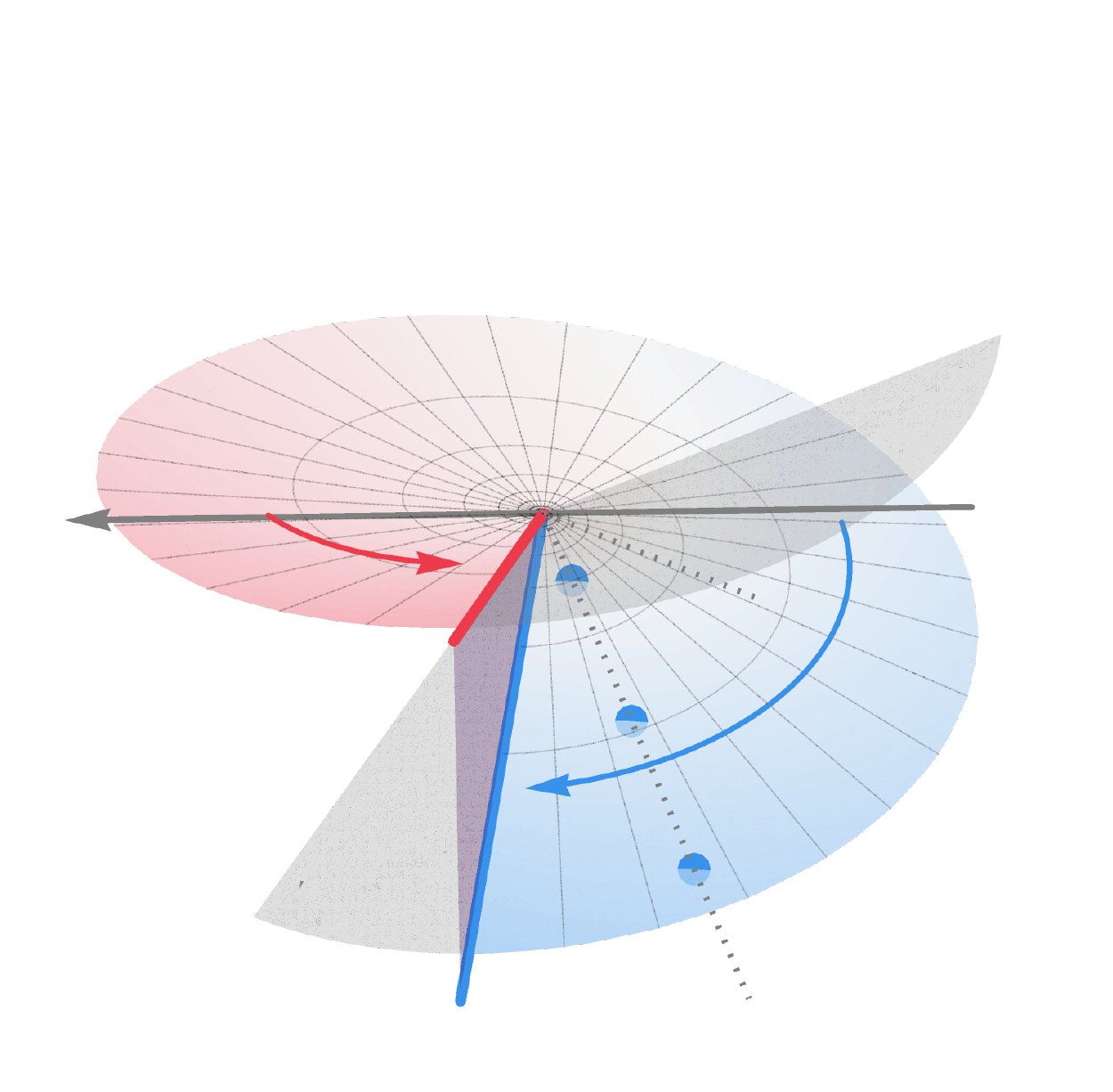}
&
\includegraphics[width=0.4\textwidth]{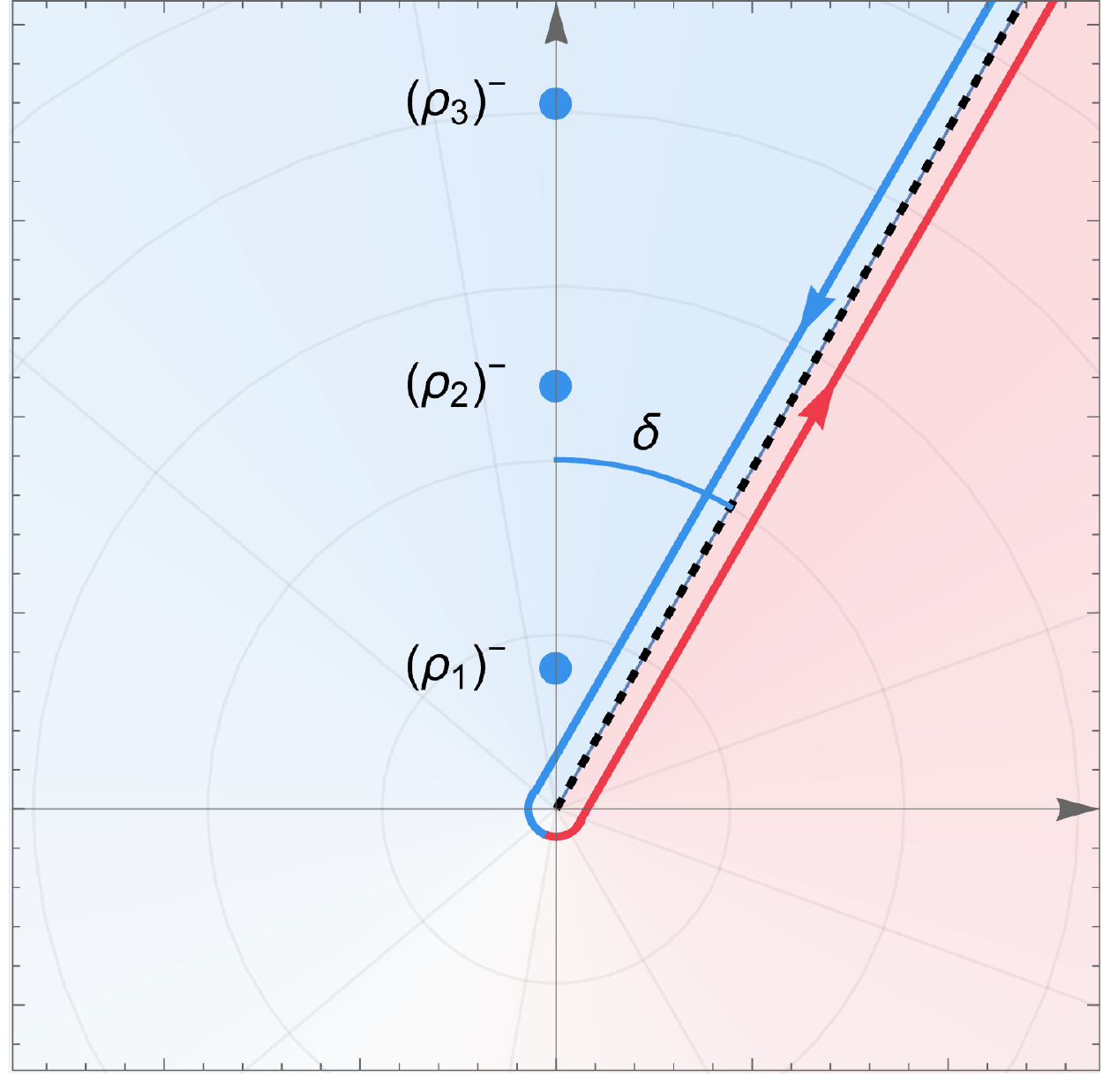}
\end{tabular}
\caption{Angle $\delta<0$ leading to integration contour $\CC_-$.}
\label{sub-fig-minus}
\end{subfigure}
\caption{The deformed integration contour drawn on part of a Riemann surface where $\log[\ri\omega]$ is continuous (left) and on the complex plane with a branch cut angled $\delta$ from the positive imaginary axis (right). The purple plane on the left indicates that the red integral is precisely above the blue integral, and also corresponds to the position of the branch on the right. The red, or blue, dots represent the poles whose residues we must consider when we deform the contour. The blue and red lines together form the Hankel contour $\mathcal{H}_+$ in figure \ref{sub-fig-plus}, and $\mathcal{H}_-$ in \ref{sub-fig-minus}. The dashed lines represent the discontinuity integrals $\CC_\pm$, respectively. See also figure \ref{fig-initial-integral}.}
\label{fig-deformed-integrals}
\end{figure}

If there were no poles, we could put one integral above the other and the final result would be given by the difference, the discontinuity. However, since they are singular along the axis itself, we must align them with some small angle $\delta$ from the positive imaginary axis in their respective sheet. If we choose a negative angle, such that both contours are to the right of the positive imaginary axis, then the half contour originating in the ``negative reals'' had to cross the poles in the lower sheet to get there, but the other half didn't cross any. Thus, we get a discontinuity integral along a ray with negative angle $\delta<0$ from the positive imaginary axis and a sum over the residues in the lower sheet. If we choose a positive angle $\delta>0$, we get a sum over the residues in the upper sheet instead. These two choices can be seen in the figures on the left in figure \ref{fig-deformed-integrals}.

Another, perhaps more straightforward, way of organizing this calculation is to think only of the starting sheet of the Riemann surface. The placement of the branch cut, which must connect the origin to infinity, is arbitrary. The conventional placement, due to the factor of $\log[\ri\omega]$, would be along the positive imaginary axis. However, this is ill-defined due to the fact that the poles also fall in this semi-axis. Then, it is better to place at angle $\delta$ from the positive imaginary axis. Let us say that we position the branch cut with $\delta > 0$, to the left of the semi-axis, and then we deform the integration contour around it. The half of the contour which starts in the positive real axis will pick up the residues when it passes by the positive imaginary axis. Because of the position of the branch cut, these poles are in the upper sheet of the Riemann surface. Had we chosen $\delta <0$, we would have picked up the poles in the lower sheet, but now from the half of the contour which starts in the negative real line.
In figures \ref{fig-initial-integral} and \ref{fig-deformed-integrals}, we illustrate these contour deformations both by drawing them in the Riemann surface and on the complex plane with a branch cut. 

In the end, we conclude that we can deform the integral along the real line into an Hankel contour integral around a ray angled $\delta\gtrless 0$ from the positive imaginary axis, which we write as $\CH_\pm$, as well as a sum over residues,
\begin{equation}
\ba
\frac{1}{2\pi\ri}\int_\IR \frac{\re^{2\ri B \omega'}\rho(\omega')u(\omega') }{\omega'+\omega+\ri 0} \rd \omega' = \frac{1}{2\pi\ri}\int_{\CH_\pm}\frac{\re^{2\ri B \omega'}\rho(\omega')u(\omega') }{\omega'+\omega+\ri 0} \rd \omega' 
+ \sum_{n\geq 1}
\frac{\re^{-2 B \xi_n}\rho_n^\pm u(\ri\xi_n)}{\omega+\ri\xi_n}.
\ea
\end{equation}
We label the poles of $\rho(\omega)$ as $\omega=\ri\xi_n$, where
\begin{equation}
\xi_n = \frac{2n-1}{\Upsilon},\quad n\in\IN,
\end{equation}
and their respective residues as $\rho_n^\pm$, 
\begin{equation}
\begin{aligned}
\rho^\pm_n 
&= \text{Res}_{\omega=\ri\xi_n\pm 0}\, \rho(\omega)\\
&= \ri\re^{\mp \ri \pi \Delta\frac{2n-1}{\Upsilon}}  \frac{2}{\Upsilon} \frac{(-1)^{n}}{\Gamma(n)^2} \left(\frac{2n-1}{2\re}\right)^{2n-1} \left( \frac{2n-1}{2\Upsilon\re} \right)^{-\frac{2n-1}{\Upsilon}} \frac{\Gamma\big(\frac{3}{2} + \frac{2n-1}{2\Upsilon}\big)}{\Gamma\big(\frac{3}{2} - \frac{2n-1}{2\Upsilon}\big)},
\end{aligned}
\label{polepm}
\end{equation}
where $\pm$ labels the upper (lower) sheet respectively and thus must be in sync with the choice of $\CH_\pm$. Noting the discontinuity of $\rho$ by\footnote{In this notation, the discontinuity of the conventional logarithm along the negative real axis is $\delta\log(\omega) = +2\pi\ri$.}
\begin{equation}
\delta \rho(\omega) = \rho\left(\re^{-\ri 0}\omega\right)-\rho\left(\re^{+\ri 0}\omega\right),
\label{disc_conv}
\end{equation}
which is given explicitly by
\begin{equation}
\delta \rho(\ri \xi) = - 2\ri \re^{[2\Delta (1+\log 2) +\Upsilon \log\Upsilon]\xi-2\Delta \xi \log \xi} \sin(\pi \Delta \xi)  \frac{\Gamma\big(\frac{1}{2}-\frac{1}{2}\Upsilon\xi\big)\Gamma\big(\frac{3}{2}+\frac{1}{2}\xi\big)}{\Gamma\big(\frac{1}{2}+\frac{1}{2}\Upsilon\xi\big)\Gamma\big(\frac{3}{2}-\frac{1}{2}\xi\big)},
\label{rhodisc}
\end{equation}
we can further simplify the previous integral, assuming $\omega=\ri \xi$ with $\xi>0$,
\begin{multline}
\frac{1}{2\pi\ri}\int_\IR \frac{\re^{2 \ri B \omega'}\rho(\omega')u(\omega') }{\omega'+\ri\xi+\ri 0} \rd \omega' = \frac{1}{2\pi\ri}\int_{\CC_\pm}\frac{\re^{-2 B \xi'}\delta\rho(\ri\xi')u(\ri\xi') }{\xi+\xi'} \rd \xi' 
\\
- \sum_{n\geq 1}
\frac{\re^{-2 B \xi_n}\ri\rho_n^\pm u(\ri\xi_n)}{\xi+\xi_n},
\label{ref-decomposition}
\end{multline}
where $\CC_\pm$ is the integration contour $\xi\in\re^{\pm\ri 0}\IR_+$.

The first check one should do of this construction is that both choices $\pm$ retrieve the same result. Let us subtract one choice from the other, finding
\be
\frac{1}{2\pi\ri}\left(\int_{\CC_+}-\int_{\CC_-}\right)\frac{\re^{-2 B \xi'}\delta\rho(\ri\xi')u(\ri\xi') }{\xi+\xi'} \rd \xi' - \sum_{n\geq 1}
\frac{\re^{-2 B \xi_n} u(\ri\xi_n)}{\xi+\xi_n} (\ri\rho_n^+-\ri\rho_n^-).
\ee
By construction, the discontinuity $\delta\rho$ is not itself discontinuous, as is manifest in \eqref{rhodisc}. However, it still has poles at $\ri\xi_n$. Thus, the difference can be put together as
\begin{equation}
\sum_{n\geq 0}\frac{\ri \re^{-2 B \xi_n} u(\ri\xi_n)}{\xi+\xi_n}\left[\text{Res}_{\omega=\ri\xi_n} \delta\rho(\omega) - (\rho_n^+-\rho_n^-)\right].
\label{sum_over_diff}
\end{equation}

We can show that the sum \eqref{sum_over_diff} is zero in fairly general terms. Let $r(\omega)$ be a meromorphic function such that
\begin{equation}
r(\omega) = \frac{r_n}{\omega-\ri \xi_n} + \text{holomorphic terms},
\end{equation}
and $f(\omega)$ be a discontinuous function such that, in a neighborhood of $\ri\xi_n$, can be written as 
\begin{equation}
f(\omega) = \re^{a \ri \omega \log[\ri b \omega]} r(\omega),\quad a,b>0.
\end{equation}
Then it follows, on one hand, that the residues have the form
\begin{equation}
f_n^\pm = \re^{- a\xi_n \log(b \xi_n) \mp \ri\pi a\xi_n}r_n,
\label{fnm_poles}
\end{equation}
and, on the other, that the discontinuity, using convention \eqref{disc_conv}, is
\begin{equation}
\delta f(\ri\xi) = -2\ri \sin(a\pi\xi) \re^{- a\xi \log(b \xi)} r(\ri\xi).
\end{equation}
Then its residues are simply
\begin{equation}
\text{Res}_{\omega=\ri\xi_n} \delta f(\omega)  = -2\ri \sin(a\pi\xi_n) \re^{- a\xi_n \log(b \xi_n)} r_n.
\label{delta_f_poles}
\end{equation}
Comparing the difference of the residues \eqref{fnm_poles} with the residues of the discontinuity \eqref{delta_f_poles}, we observe that
\begin{equation}
\text{Res}_{\omega=\ri\xi_n} \delta f(\omega) = (f_n^+-f_n^-),
\end{equation}
which for the case of $f=\rho$ can also be shown directly from manipulating \eqref{polepm} and \eqref{rhodisc}. We then conclude that \eqref{sum_over_diff} is zero and the two choices of sign are equivalent.

Let us inspect integral \eqref{ref-decomposition}. The sum over residues is very reminiscent of the trans-series introduced in chapter \ref{cha_resurgence}: the terms are exponentially suppressed in the $1/B$ expansion and the parameters are ambiguous. However, to take this analysis even further, we need to start solving for $u(\ri\xi)$.

\subsection{Finding the trans-series in the integral equations}

While ultimately our interest lies in the non-perturbative corrections, let us focus on the perturbative corrections for an instant. This will allow us to later organize both types of correction systematically. Thus, putting aside the residue contributions in \eqref{inteq_u} through \eqref{ref-decomposition}, we start from
\begin{equation}
u(\ri\xi) = \frac{1}{\xi} + \frac{1}{2\pi\ri}\int_{\CC_\pm}\frac{\re^{-2B \xi'}\delta\rho(\ri\xi')u(\ri\xi') }{\xi+\xi'} \rd \xi' + \CO(\re^{-2B\xi_0}).
\end{equation}
We want to expand in the weak coupling regime $B\gg 1$. These expansions are known to have terms both in $1/B$ and $\log B/B$, see for example \cite{fnw1,hn,volin}. However, this additional complication can be avoided by using the appropriate variables, as pointed out in \cite{fnw1}. We then introduce
\begin{equation}
\frac{1}{v}-2\Delta\log v = 2B, \quad \xi = v\eta.
\label{v-def}
\end{equation}
We write $u(\ri \xi(\eta))$ as $u(\eta)$ for simplicity. The discontinuity can also be recast in a more convenient form
\begin{equation}
\re^{-2B\xi}\delta \rho(\ri\xi)=-2\ri v \re^{-\eta}P(\eta),
\label{P-def}
\end{equation}
where $P(\eta)$ can be explicitly expanded as
\begin{equation}
 \quad P(\eta) \approx \sum_{n\geq 0} v^n \sum_{m=0}^n d_{n+1,m} \eta^{n+1} (\log\eta)^m\sim d_{1,0}\eta + \CO(v).
 \label{P-expansion}
\end{equation}
The $d_{n,m}$ can be computed from the Taylor expansion of the $\Gamma$-functions in $\rho$. Lastly we formally organize $u(\eta)$ as
\begin{equation}
u(\eta) = \sum_{n\geq 0} v^{n-1} u_{(n)}(\eta),
\end{equation}
where we admit that each $u_{(n)}(\eta)$ might have further corrections in $v$. Plugging everything together we see that
\begin{align}
u_{(0)}(\eta) &= \frac{1}{\eta}+\CO\left(\re^{-2B\xi_0}\right),\\
u_{(1)}(\eta) &= -\frac{1}{\pi}\int_0^{\re^{\pm\ri0}\infty}\frac{\re^{-\eta'} P(\eta')}{\eta+\eta'}u_{(0)}(\eta')\rd\eta' ,\\
u_{(n)}(\eta) &= -\frac{1}{\pi}\int_0^{\re^{\pm\ri0}\infty}\frac{\re^{-\eta'} P(\eta')}{\eta+\eta'}u_{(n-1)}(\eta')\rd\eta'.
\label{rec_un}
\end{align}
This allows us to recursively solve the perturbative part. 

We can then restore the non-perturbative contributions by reinstating them in $u_{(0)}$, such that
\be
u_{(0)}(\eta) = \frac{1}{\eta} -  \sum_{n\geq 1}\frac{\re^{-2B\xi_n}\ri\rho_n^\pm u_n}{\eta+v^{-1}\xi_n},
\ee
where we label $u(\ri\xi_n)$ as $u_n$. However, we now have a more elaborate recursive problem since we must recursively solve for the $u_n$ which will have perturbative and non-perturbative corrections themselves. In order to organize the two series in conjunction, it is useful to introduce the non-perturbative coupling $q$ such that
\begin{equation}
q= \re^{-1/(\Upsilon v)}v^{2\Delta/\Upsilon}\Rightarrow\re^{-2B\xi_n} = q^{2n-1}.
\end{equation}

Let us then take $u_1$ for example. Considering contributions up to $u_{(2)}$ and NLO in the $q$ expansion, we have 
\begin{equation}
\ba
u_1 = &\Upsilon
 - \frac{v}{\pi} 
\int_0^{\re^{\pm\ri0}\infty} 
\hspace{-5mm} \rd \eta \frac{ \re^{-\eta} P(\eta)}{\eta+(v \Upsilon)^{-1}} \left(\frac{1}{ v\eta}-  \frac{ q\ri\rho^\pm_1 u_1}{v\eta+\Upsilon^{-1}} - 
 \cdots\right)\\
& +   \frac{v^2}{\pi^2}
\int_0^{\re^{\pm\ri0}\infty} 
\hspace{-5mm} \rd \eta' \frac{ \re^{-\eta'} P(\eta')}{\eta'+(v\Upsilon)^{-1}}  \int_0^{\re^{\pm\ri0}\infty} 
\hspace{-5mm} \rd \eta \frac{ \re^{-\eta} P(\eta)}{\eta+\eta'} 
\left(\frac{1}{v\eta}- \frac{q\ri\rho^\pm_1 u_1}{v\eta+\Upsilon^{-1}}+\cdots\right)
\\
&-  \frac{q \Upsilon }{2}\ri \rho^\pm_1 u_1  -  \frac{q^3\Upsilon }{4}\ri\rho^\pm_2 u_2 +  \CO(v^3) + \CO(q^3)\,.
\ea
\end{equation} 
Expanding the integral kernel $P(\eta)$ to leading order, we have
\begin{equation}
\ba
u_1 &= \Upsilon- \frac{v}{\pi}\int_0^{\re^{\pm\ri0}\infty} \rd \eta \left( \Upsilon e^{-\eta } d_{1,0} + \CO(v^2)\right) - \frac{\Upsilon q}{2} \ri\rho^\pm_1u_1 + \CO(q^3)\\
&= \Upsilon -\frac{\Upsilon  d_{1,0}}{\pi }v+\CO(v^2)-   \ri\rho^\pm_1 q \Upsilon^2\left(\frac{1}{2} -\frac{  d_{1,0}}{2 \pi }v+\CO(v^2)\right)+\CO(q^3),
\ea
\end{equation}
where in the last line we plug in the perturbative solution $u_1\sim \Upsilon -\frac{\Upsilon  d_{1,0}}{\pi }v$ for the r.h.s. term in $u_1$.

Now that we understand how to get $u(\eta)$ and the $u_k$, we can start to move towards our goal of calculating the free energy. Applying the same decomposition in \eqref{ref-decomposition} to \eqref{fh-u}, and not neglecting to the additional pole at $\omega'=\ri$, we find
\begin{multline}
\mathcal{F}(h) = 
-\frac{h^2}{2\pi} u(\ri) G_+(0)^2 \biggl\{  1 - \frac{1}{2\pi \ri} \int_{\CC_\pm}  \frac{\re^{-2B\xi'}\delta \rho(\ri\xi')u(\ri\xi')}{\xi'-1} \rd \xi'\\
- \re^{-2B}\rho(\ri\pm 0)u(\ri) + \sum_{n\ge 1} \frac{\re^{-2B\xi_n}\ri\rho_n^{\pm} u_n}{\xi_n-1}\biggr\},
\label{eq_freeF}
\end{multline}
The discontinuity integral in \eqref{eq_freeF} can be expanded using the same logic used when solving for $u_0$. Meanwhile, the term originating from the extra pole can be explicitly simplified using the boundary condition \eqref{u_bc},
\be
\re^{-2B}\rho(\ri\pm 0)u(\ri)
=  \frac{m\re^{-B}}{2h}\tilde{\rho}^\pm,
\ee
where
\be
\label{tilderho}
\tilde{\rho}^\pm = \re^{\pm \ri \pi  \Delta  }(2\re)^{\Delta } (1-2 \Delta)^{\frac{1}{2}-\Delta } \Gamma (\Delta ) .
 \ee
We can then write
\begin{multline}
\CF(h)= {h^2 \over 2 \pi} u(\ri) G_+(0)^2
\biggl\{-1+\frac{m\re^{-B}}{2h}\tilde{\rho}^\pm +\frac{d_{1,0}}{\pi }v +\CO\left(v^2\right)
\\
- \ri\rho^\pm _1 q\left(\frac{\Upsilon^2  }{1-\Upsilon}-\frac{d_{1,0}\Upsilon ^2  }{\pi  (1-\Upsilon )}v +\CO\left(v^2\right)\right)+ \CO(q^2)\biggr\}.
\end{multline}

The coupling $v$ and $q$ are defined from $B$, which is indirectly related to $h/m$ (or, equivalently $h/\Lambda$). We would now like to make the dependency on $h$ more  clear. The relationship $B(h)$ is enforced by the boundary condition \eqref{u_bc}. We need to calculate $u(\ri)$, which is a similar calculation to the $u_k$ with the simplification that no recursion is necessary. We find 
\be
u(\ri)=1-\frac{d_{1,0}}{\pi }v +\CO\left(v^2\right)
-\ri \rho^\pm_1 q\left(\frac{ \Upsilon ^2}{\Upsilon +1}-\frac{ d_{1,0} \Upsilon ^2 }{\pi  (\Upsilon +1) } v +\CO\left(v^2\right)\right) + \CO(q^2). 
\ee
While we could proceed from here and write $v,q$ as trans-series in $h$, this generates unwieldy results with terms such as $\log\log h$. Instead, much like we introduced $v$ to account for logarithm dependencies on $B$, it is useful to introduce $\tilde\alpha$,
\begin{equation}
\frac{1}{\tilde{\alpha}}-\Delta\log\tilde{\alpha}=\log\frac{h}{\Lambda}.
\label{tilde-alpha-def}
\end{equation}
This auxiliary coupling, which is the first of many in this thesis, is aligned with previous studies of asymptotically free integrable theories \cite{bbbkp,volin,mr-ren}, and has parallels in similar definitions in QCD \cite{bly, jamin-mira}. Comparing to the original renormalized coupling we have that
\begin{equation}
\tilde \alpha \sim 2 \beta_0 \bar g^2 (h).
\end{equation}
 It is also useful to recall that the mass gap for the Gross--Neveu model is known exactly from \cite{fnw1,fnw2}\footnote{These references use a different definition of $\Lambda$}, and is given simply by
\begin{equation}
\frac{m}{\Lambda} = \frac{(2\re)^\Delta}{\Gamma(1-\Delta)}.
\end{equation}

Using the formulae of the mass gap and $G_+$, we can write the boundary condition as
\begin{equation}
 \left(\frac{1}{2v}-\Delta\log v\right) - \log u(\ri) = \left(\frac{1}{\tilde{\alpha}}-\Delta\log\tilde{\alpha}\right)+\frac{\Upsilon}{2}\log\Upsilon +\log 2.
\end{equation}
Since $u(\ri)\sim 1 + \CO(v)$, we see that $v\sim\tilde\alpha/2+\CO(\tilde{\alpha}^2)$. However, since $u(\ri)$ has non-perturbative corrections, so does $v(\tilde\alpha)$. To leading order we have that
\begin{multline}
v = \frac{\tilde\alpha }{2}+\frac{1}{4}  \big[ (\Delta-1) \log (4)-\Upsilon  \log (\Upsilon )\big]\tilde\alpha ^2 + \CO\left(\tilde\alpha ^3\right)\\
-\ri \rho^\pm_1 \re^{-\frac{2}{\Upsilon \tilde{\alpha}}}\left(\tilde{\alpha}\right)^{\frac{2\Delta}{\Upsilon}}  \left(-\frac{2^{-\frac{2}{\Upsilon }-1} \Upsilon }{\Upsilon +1}\tilde\alpha ^2 + \CO\left(\tilde\alpha ^3\right)\right)+ \CO\left(\re^{-\frac{4}{\Upsilon \tilde{\alpha}}}\right).
\end{multline}

For the term with $\tilde\rho^\pm$, the boundary condition \eqref{u_bc} can be imposed directly and, to all orders,
\begin{equation}
\frac{h^2}{2\pi} u(\ri) G^+(0)^2\times\left(\frac{m\re^{-B}}{2h}\tilde{\rho}^\pm\right) = \frac{m^2}{8\pi} G_+(\ri)G_+(0) \tilde{\rho}^\pm.
\end{equation}
We can then finally put together a result for the free energy,
\begin{equation}
\ba
\CF(h)\sim  
- \frac{h^2}{2\pi}\Biggl\{&1-\Delta  \tilde{\alpha }+\frac{1}{2} \Delta  \big[\Delta -2+2 \log (2)\big]\tilde{\alpha }^2 + \CO\left(\tilde{\alpha }^3\right)
\\
&-\ri \rho _1^\pm\re^{-\frac{2}{\Upsilon \tilde{\alpha}}}\left(\tilde{\alpha}\right)^{\frac{2\Delta}{\Upsilon}}\frac{2^{\frac{2}{2 \Delta -1}} (1-2 \Delta )^2 }{2\Delta (\Delta -1)}\left(
1+\frac{2\Delta^2}{1-2\Delta}\tilde{\alpha}
+\CO\left(\tilde{\alpha }^2\right)\right)
\\
&+\CO\left(\re^{-\frac{4}{\Upsilon \tilde{\alpha}}}\right)\Biggr\}
 \mp \frac{\ri m^2}{8}+\frac{m^2}{8} \cot (\pi  \Delta )
.
\ea
\label{fhWHZ-v0}
\end{equation}
This result is interesting on many fronts, so let us unwrap it. 

The first thing an attentive reader might have already noticed is that once we started to perturbatively expand the discontinuity integral, the dependency on the sign of the angle disappeared. Now, in the final result, the perturbative series is completely indifferent to whether we initially chose $\CC_+$ or $\CC_-$. However, a trace is left in the residues. One might suppose that such a result is then ambiguous, or that we should have had to ``fix'' the residues somehow, but this would be ill thought. In light of the theory of resurgence and the resummation of trans-series, as we presented in chapter \ref{cha_resurgence}, these ambiguities are not a bug but rather a feature, in fact, a necessity. Since the perturbative series is asymptotic, a fact which we will test extensively in chapter \ref{cha_volin}, it must be rendered finite through Borel summation. However, the Borel summation will itself be ambiguous, and a finite real result can only be found if the trans-series parameters are ambiguous in the precise opposite way. Therefore, the ambiguous non-perturbative terms in \eqref{fhWHZ-v0} seem to precisely fulfill this role, and in chapter \ref{cha_volin} we verify numerically that they cancel the ambiguities of the perturbative series. Because it is an exact solution, the Bethe ansatz provides all the necessary ingredients for a Borel summable trans-series.

We now focus on the non-perturbative terms we calculated. The outer term $\mp \ri m^2/8$ might seem at first elusive but we can restore the overall factor of $h$ and find $\propto h^2 (\re^{-2/\tilde{\alpha}})$. Since $\tilde\alpha\sim 2\beta_0 \bar g^2(h)$, this corresponds to \eqref{IR-ren} with $\ell=2$, and is thus a typical IR renormalon.
It is also a remarkably simple term, with no associated perturbative corrections. We shall henceforth refer to it as ``the IR renormalon''. The article ``the'' reveals that the other corrections are not of the IR renormalon form in \eqref{IR-ren}. To see why this is the case, note that the Borel space dual to $\bar g^2(h)$ the next exponential term corresponds a singularity at
\begin{equation}
\zeta = \frac{1}{\beta_0} \frac{N-2}{N-4},
~\label{first-renpole}
\end{equation}
which is a rational multiple of $1/\beta_0$ and hence incompatible with conventional renormalons. We call this and similar effects ``new renormalons'', although we will hold the discussion of why we maintain the term ``renormalon'' to section \ref{largeN_antrans}. Instead, we obtain the higher trans-series sectors first.

\section{Further results for the Gross--Neveu model}
\label{sec-gn-2}
In this section, we dive into finer details of the trans-series in Gross--Neveu. First we organize it in a more structured form, then we present our results in the canonical formalism and lastly we will discuss the UV phenomena also visible with these methods.

\subsection{Higher orders in perturbative and non-perturbative expansions}

To have a better grasp on the structure of the trans-series expression of $\CF(h)$ let us systematize the previous discussion. Let us define the ``seed solution''
\begin{equation}
\bar u(\eta) = v^{-1} u_{(0)}(\eta),
\end{equation}
and the integral operator
\begin{equation}
\left[\mathfrak{I} f\right](\eta) = - \frac{1}{\pi}\int_{\CC_\pm}\frac{\re^{-\eta'}P(\eta')}{\eta+\eta'} f(\eta')\rd\eta'.
\end{equation}
Then recursive relation \eqref{rec_un} can be written simply as 
\begin{equation}
u_{(n)}(\eta) = \left[\mathfrak{I} u_{(n-1)}\right](\eta),
\end{equation}
from which follows the exact solution of \eqref{inteq_u},
\begin{equation}
u(\eta) = \sum_{n\geq 0} v^n \left[\mathfrak{I}^n \bar u\right](\eta) = \frac{\bar u(\eta)}{1-v \mathfrak{I}},
\label{sum_un}
\end{equation}
where the last identity should be read simply as a formal power series of $\mathfrak{I}$. The integral operator, which is to leading order $\propto v^0$, crucially comes multiplied with a factor of $v$ in the infinite sum \eqref{sum_un}. Should one want to solve the equation only up to some fixed order $v^n$ it suffices to apply the operator only $n$-times.
Finally, it is also convenient to define the integrals
\begin{equation}
\mathfrak{I}_{k} f = \left[\mathfrak{I} f\right]\left(\tfrac{2k-1}{v\Upsilon}\right),\quad k\in\IN.
\end{equation}
so that the equation for the $u_k$ can be compactly written as
\begin{equation}
u_k=\frac{\Upsilon}{2k-1}+v \mathfrak{I}_k u - \frac{\Upsilon}{2}\sum_{n\geq 1}\frac{q^{2n-1}\ri\rho_{n}^\pm u_n}{k+n-1},\quad k\in\IN.
\label{eq_uk_v2}
\end{equation}
It is useful to note that $\mathfrak{I}_k u\sim 1+\CO(v)$.

One can also organize the solution of the $u_k$ in powers of $q$.
We start by factoring the residues of $\rho$ in \eqref{polepm} as
\begin{equation}
\rho_n^\pm =  \left(\re^{\mp\frac{\ri\pi\Delta}{\Upsilon}}\right)^{2n-1}\ri \bar\rho_n,\quad \bar\rho_n\in\IR.
\label{rho_factor}
\end{equation}
This way we can identify that the ambiguity for a term in $q^{k}$ is always the same. Then, let
\begin{equation}
u_k = \sum_{n\geq 0} q^n \left(\re^{\mp\frac{\ri\pi\Delta}{\Upsilon}}\right)^{n} u_k^{(n)},
\end{equation}
and similarly organize the ``seed solution'' as
\begin{equation}
\ba
\bar u^{(0)}(\eta) &= \frac{1}{v\eta},\\
 \bar u^{(s)}(\eta) &= \sum_{\tfrac{s+1}{2}\geq r \geq 1}\frac{\bar\rho_r u_r^{\left(s+1-2r\right)}}{\Upsilon v\eta + 2r-1},\\
 \bar u(\eta) &= \sum_{n\geq 0} q^n \left(\re^{\mp\frac{\ri\pi\Delta}{\Upsilon}}\right)^{n} \bar u^{(n)}(\eta).
\ea
\end{equation}
Then the equation for $u_k$ can be written order by order in $q$ as
\begin{equation}
\ba
u_k^{(0)}&=\frac{\Upsilon}{2k-1}+v\mathfrak{I}_k\frac{\bar u^{(0)}}{1-v\mathfrak{I}},\\
u_k^{(s)}&=v \mathfrak{I}_k\frac{\bar u^{(s)}}{1-v\mathfrak{I}} + \frac{\Upsilon}{2}\sum_{\tfrac{s+1}{2}\geq r\geq 1} \frac{\bar\rho_r u_r^{(s+1-2r)}}{k+r-1}.
\ea
\end{equation}
The r.h.s. does not contain $u^{(s)}_k$ itself, It only includes $u_{k'}^{(s')}$ with $s<s'$ and $k'\leq \frac{s+1}{2}$. Thus, we can solve the equations recursively. For example, we have
\begin{equation}
\bar u^{(1)}(\eta) = \frac{\bar\rho_1 u_1^{(0)}}{1+\Upsilon v \eta},
\end{equation}
and thus
\begin{equation}
\ba
u_1^{(0)} &= \Upsilon+v\mathfrak{I}_1\frac{\bar u^{(0)}}{1-v\mathfrak{I}} \sim \Upsilon - \frac{\Upsilon d_{1,0}}{\pi}v+\CO(v^2),\\
u_1^{(1)} &= \bar\rho_1 u_1^{(0)}\left(\frac{\Upsilon}{2} + v \mathfrak{I}_1\frac{1}{1-v\mathfrak{I}}\frac{1}{1+\Upsilon v \eta} \right) \sim \bar\rho_1\Upsilon^2\left(\frac{1}{2} + \frac{ d_{1,0}}{2\pi}v+\CO(v^2)\right),
\ea
\end{equation}
where what is left to do is expand the iterated integrals $\mathfrak{I}^k$ to obtain the perturbative series in $v$.

It should be noted that, regrettably, expanding the iterated integrals is not elementary. Up to second order, \texttt{Mathematica} can find exact values for the perturbative expansion, but not for terms with $\mathfrak{I}^3$. It is possible that, with some effort, more general formulae could be found. When it comes to the perturbative sector, it is better to avoid the iterated integrals and use Volin's method \cite{volin} and its generalizations \cite{mr-ren} instead, as we shall discuss in chapter \ref{cha_volin}. For the moment, this means that at each power in $q$ we can only calculate the first few terms in the expansion in $v$.

One advantage of this organization is that it makes evident that even though the ``seed solution'' only contains odd powers of $q$, they mix in the recursion relations leading to correction in even powers of $q$ to the $u_k$. For example, we have at order $q^2$ the term
\begin{equation}
u_1^{(2)} = \bar\rho_1 u_1^{(1)}\left(\frac{\Upsilon}{2} + v \mathfrak{I}_1\frac{1}{1-v\mathfrak{I}}\frac{1}{1+\Upsilon v \eta} \right) \sim (\bar\rho_1)^2\Upsilon^3\left(\frac{1}{4} - \frac{d_{1,0}}{4\pi}v+\CO(v^2)\right).
\end{equation}

Since
\begin{equation}
q \sim \re^{-\frac{2}{\Upsilon\tilde{\alpha}}}\alpha^\frac{2\Delta}{\Upsilon}\left(2^{-\frac{2\Delta}{\Upsilon}}+\CO(\tilde{\alpha})\right),
\end{equation}
we are lead to the following trans-series structure for the free energy
\begin{multline}
\tilde\Phi^\pm(\tilde{\alpha}) = \tilde\varphi_0(\tilde{\alpha}) \pm \ri C_0 \re^{-\frac{2}{\tilde{\alpha}}} \tilde{\alpha}^{2\Delta}
\\
+ \sum_{\ell\geq 1} \left\{\cos\left(\tfrac{\ell\pi\Delta}{1-2\Delta}\right)\mp\ri \sin \left(\tfrac{\ell\pi\Delta}{1-2\Delta}\right)\right\}\re^{-\frac{2}{1-2\Delta}\frac{\ell}{\tilde{\alpha}}}\tilde\alpha^{\frac{2\Delta}{1-2\Delta}\ell}\,C_\ell\,\tilde\varphi_\ell(\tilde{\alpha}) ,
\label{free-energy-ts}
\end{multline}
where $\varphi_{k\geq 0}$ are real asymptotic formal power series and the $C_{\ell\geq 0}$ are real constants. From \eqref{fhWHZ-v0}, we have in particular
\begin{equation}
\ba 
\tilde\varphi_0(\tilde\alpha) &=1-\Delta  \tilde{\alpha }+\frac{1}{2} \Delta  \big[\Delta -2+2 \log (2)\big]\tilde{\alpha }^2 + \CO\left(\tilde{\alpha }^3\right),\\
\tilde\varphi_1(\tilde\alpha) &= 1+\frac{2\Delta^2}{1-2\Delta}\tilde{\alpha}
+\CO\left(\tilde{\alpha }^2\right),\\
C_0 &= \frac{\pi(2\re)^{2\Delta}}{4\Gamma(1-\Delta)^2},\quad C_1 = \frac{2^{\frac{2}{2 \Delta -1}-1} (1-2 \Delta )^2 }{\Delta (\Delta -1)} \bar\rho_1.
\ea
\end{equation}
The decomposition \eqref{rho_factor} allows us to know the form of the ambiguities of all trans-series parameters.
The free energy itself is given by the unambiguous expression
\begin{equation}
\CF(h) = - \frac{h^2}{2\pi}s_\pm[\tilde\Phi^\pm]\big(\tilde{\alpha}(h)\big) + \frac{m^2}{8}\cot(\pi\Delta),
\label{free-energy-resum}
\end{equation}
where $s_\pm$ are lateral Borel summations above (below) the positive real axis. 

The most important thing that we learn from \eqref{free-energy-ts} are the positions of the Borel poles,
\begin{equation}
\zeta_\ell = \frac{\ell}{\beta_0} \frac{N-2}{N-4},\quad \ell\in\IN.
\end{equation}
Much like we observed from $\ell=1$, this goes against the canonical prediction \eqref{IR-ren}. Furthermore, beyond the IR renormalon effect, the family of poles corresponding to \eqref{IR-ren} seems to be absent except for the coincidence $\ell = k (N-4),\,k\in\IN$. In the large $N$ limit, however, we recover the traditional renormalons. The fact that these poles contribute when $N\rightarrow\infty$ is part of the reason we identify them as renormalons. Instanton effects should vanish at large $N$.

As we mentioned in the introduction, the Bethe ansatz description breaks down at $N=4$ and this is visible in the trans-series structure. The $N=5$ case is also special because in this case the $\rho_n^\pm$ vanish, and consequently so do all $C_{\ell\geq 1}$. The only IR effect in this case is the standard IR renormalon.

A more technical observation is that, since the imaginary ambiguity of each trans-series sector factors out, the series $\varphi_\ell$ are likely all encoded in the large order behavior of $\varphi_0$. This would make this an example of strong resurgence, as discussed in chapter \ref{cha_resurgence}.

Lastly, the constant term in \eqref{free-energy-resum} has a physical interpretation. Recall that, by definition,
\begin{equation}
\CF(h) = F(h)-F(0).
\end{equation}
One could then expect $\CF(0) =0$ but the definition hides some subtleties. As discussed in \cite{dpmss}, the free energy $F(h)$ and $F(0)$ refer to the different states which are energetically favoured when $h=0$ and $h\neq 0$. One could write $F_h(h)-F_0(0)$, instead. Around $h\sim m$, they are equal since this is when $h$ is critically large enough to ``create'' particles, which makes the populated state more favourable than the original vacuum, and thus $\CF(\sim m) =0$. We then identify the $h$-independent term as $-F(0)$ which matches the known prediction of \cite{saleur},
\begin{equation}
F(0)=-\frac{m^2}{8}\cot(\pi\Delta).
\end{equation}
This follows a similar analysis in \cite{zamo-mass,foz}.

\subsection{Energy in the canonical formalism}

The result \eqref{free-energy-resum} can in principle be tested, however it is not in its most convenient form. The reason is twofold: on one hand, the integral equation \eqref{iqft_geneq_hm} is not ideal for numerical implementation since its boundary condition $\epsilon(\pm B)$ requires several iterations of the equation to find $B(h)$. On the other hand, the large order behavior of the perturbative series is easier to obtain directly in the canonical formalism
. Thus, although it would certainly be possible to test the free energy directly, it is easier to test instead the energy density $e(\rho)$. Fortunately, it is easy to transform the trans-series solution of one into the other.

We recall that the internal energy $e$ and the particle density $\rho$ can be obtained from $\CF$ and $h$ through the Legendre transform
\begin{equation}
\rho = - \CF'(h), \quad e = \CF(h) + h\rho.
\end{equation}
In terms of the trans-series \eqref{free-energy-ts},
\begin{equation}
\rho \approx  \frac{h}{2\pi} \left(2 \tilde\Phi^\pm(\tilde{\alpha}) + \left(\frac{\rd \tilde{\alpha}}{\rd\log h}\right)\frac{\rd }{\rd\tilde{\alpha}}\tilde\Phi^\pm(\tilde{\alpha})\right).
\label{rho_trans}
\end{equation}
Using the definition of $\tilde{\alpha}$ in \eqref{tilde-alpha-def}, the derivative factor can easily be written as a function power series in $\tilde{\alpha}$,
\begin{equation}
\frac{\rd \tilde{\alpha}(h)}{\rd\log h} = - \frac{\tilde\alpha^2}{1+\Delta\tilde\alpha} \sim -\tilde\alpha^2+\tilde\alpha^3 \Delta +O\left(\tilde\alpha^4\right),
\end{equation}
and so we naturally obtain a representation of $\rho/h$ as a trans-series in $\tilde{\alpha}$. Then, we can write a trans-series for the normalized energy density using
\begin{equation}
\frac{e}{2\pi\rho^2}\approx   \frac{\tilde\Phi^\pm(\tilde{\alpha}) +  \frac{\rd \tilde{\alpha}}{\rd\log h}\frac{\rd \tilde\Phi^\pm(\tilde{\alpha})}{\rd\tilde{\alpha}}}{4\left( \tilde\Phi^\pm(\tilde{\alpha}) + \frac{1}{2}\frac{\rd \tilde{\alpha}}{\rd\log h}\frac{\rd \tilde\Phi^\pm(\tilde{\alpha})}{\rd\tilde{\alpha}}\right)^2} - \frac{F(0)}{2\pi\rho^2}.
\label{trans-ratio}
\end{equation}

The coupling $\tilde{\alpha}$ is defined in terms of $h$, and is thus inadequate for the canonical formalism. It is more useful to define
\begin{equation}
\frac{1}{\alpha}-\Delta\log\alpha = \log\left(\frac{2\pi\rho}{\Lambda}\right).
\label{alpha-gn}
\end{equation}
This coupling is related to the previous on through
\begin{equation}
\frac{1}{\alpha}-\frac{1}{\tilde\alpha}= \Delta\log\left(\frac{\alpha}{\tilde{\alpha}}\right)+\log\left(\frac{2\pi\rho}{h}\right).
\end{equation}
We can see that $\alpha\sim\tilde{\alpha}+\CO(\tilde{\alpha}^2)$, however this relation is a trans-series itself due to the term $\rho/h$. Once this trans-series is inverted, which can be done order by order, we find
\begin{equation}
\ba
\frac{e}{2\pi\rho^2} \sim \frac{1}{4} & +\frac{\Delta}{4}\alpha +\frac{1}{8} \Delta  (\Delta +2) \alpha ^2+\CO\left(\alpha ^3\right)
\\ &+\re^{-\frac{2}{\alpha}}\alpha^{2\Delta} 
  \left\{\re^{\mp\ri\pi\Delta}\frac{(2 \re)^{2 \Delta } \Gamma (\Delta )}{4 \Gamma (1-\Delta )}\right\}
 \\
&- \re^{-\frac{2}{\Upsilon\alpha}}\alpha^\frac{2\Delta}{\Upsilon} \left\{\re^{\mp\frac{\ri\pi\Delta}{1-2\Delta}} \frac{(1-2 \Delta )^2 \bar \rho_1}{8 \Delta (\Delta -1)}\right\} \Bigg(1
- \frac{2\Delta^2  }{(1-2 \Delta ) }\alpha+\CO\left(\alpha ^2\right)\Bigg)
\\&+ \re^{-\frac{4}{\Upsilon\alpha}}\alpha^\frac{4\Delta}{\Upsilon} \left\{\re^{\mp\frac{2\ri\pi\Delta}{1-2\Delta}}\frac{(1-2 \Delta )^2 (\bar \rho_1)^2}{8 (\Delta -1)^2}\right\}\Bigg(1+\CO(\alpha)\Bigg)
+\CO\left(\re^{-\frac{6}{\Upsilon\alpha}}\right).
\label{efromWHZ}
\ea
\end{equation}
The exact answer should then be given by
\be 
\label{exacterho}
{e \over 2\pi\rho^2} = s_\pm (\Phi^\pm)(\alpha), 
\ee
where
\be
\label{transerho}
\Phi^\pm (\alpha)= \varphi(\alpha) + \CC_0^\pm  \re^{-2/\alpha} \alpha^{2 \Delta} + \sum_{\ell=1}^\infty \CC_\ell ^\pm \re^{-\frac{2 \ell}{\Upsilon\alpha}}\alpha^\frac{2 \ell \Delta}{\Upsilon} \varphi_\ell(\alpha), 
\ee
which includes the perturbative series 
\be
\label{var-p}
\varphi(\alpha)\equiv \sum_{k \ge 0} e_k \alpha^k=\frac{1}{4}+\frac{\Delta}{4}\alpha +\frac{1}{8} \Delta  (\Delta +2) \alpha ^2+\CO\left(\alpha ^3\right).
\ee
as  well as the higher series associated with non-perturbative corrections. For example, we can read the leading non-perturbative series
\begin{equation}
\varphi_1(\alpha) = 1-\frac{2\Delta^2}{1-2\Delta}\alpha
+\CO\left(\alpha^2\right).
\end{equation}

The trans-series parameters follow as well, with
\begin{align}
\label{coefC0}
\CC_0^\pm &= 
  \left\{\cos(\pi\Delta)\mp\ri \sin(\pi\Delta)\right\}\frac{(2 \re)^{2 \Delta } \Gamma (\Delta )}{4 \Gamma (1-\Delta )}, \\
 \label{coef_C10}
\CC_1^\pm &=- \left\{\cos\left(\tfrac{\pi\Delta}{1-2\Delta}\right)\mp\ri \sin\left(\tfrac{\pi\Delta}{1-2\Delta}\right)\right\}(1-2 \Delta )^{\frac{2 \Delta }{1-2 \Delta }}\frac{(2 \re)^{\frac{2 \Delta }{1-2 \Delta }}\Gamma \left(\frac{\Delta }{1-2 \Delta }\right)}{4\Gamma \left(1-\frac{\Delta }{1-2 \Delta }\right)}.
\end{align}
Naturally, they inherit from \eqref{free-energy-ts} the structure
\begin{equation}
\CC_\ell^\pm = \re^{\mp\frac{\ri\ell\pi\Delta}{1-2\Delta}} \CC_\ell, \quad \CC_\ell\in\IR,\quad\ell\in\IN.
\end{equation}
More generally one can see from \eqref{trans-ratio}, that the structure of the trans-series for $\CF(h)/h^2$ or $e/\rho^2$ is identical. We will thus discuss them interchangeably.

\subsection{UV renormalons}
\label{sec-uv}

As we discussed in section \ref{sec_renormalons}, there can also be UV renormalons.  These correspond to poles in the negative real axis. While they do not obstruct resummation, they are also connected to the large order behavior of the perturbative series. In this section, we will briefly review how to obtain UV renormalons, which turns out to be a simple modification of the procedure developed so far.

In order to make the UV poles appear in the calculation, we need to ``analytic continue'' to negative $B$. 
It is also important that we analytically continue in such a way that the perturbative part remains the same, except for the sign of the coupling, since our ultimate goal is to find the UV large order behavior of the perturbative theory in the $B>0$ case. If we just take negative $B$ in \eqref{inteq_u}, perturbation theory will be off. So let us construct an auxiliary model from equations \eqref{inteq_u}, \eqref{fh-u} with $B<0$, $\rho\rightarrow-\rho$ and $u$ analytic in $\mathbb{H}_-$. This model will have the same perturbative expansion but with a negative coupling. By assuming the cancellation of the ambiguities of Borel summation with those of trans-series, we will find results for the large order behavior of the asymptotic series for $\CF(h)$, which is identical to the physical case\footnote{This auxiliary model is also schematically similar to what happens with the repulsive Gaudin--Yang model which we discuss in chapter \ref{cha_GY}.}.

\begin{figure}
\centering
\begin{subfigure}[t]{0.4\textwidth}
\centering
\begin{tabular}[t]{c}
\includegraphics[width=0.9\textwidth]{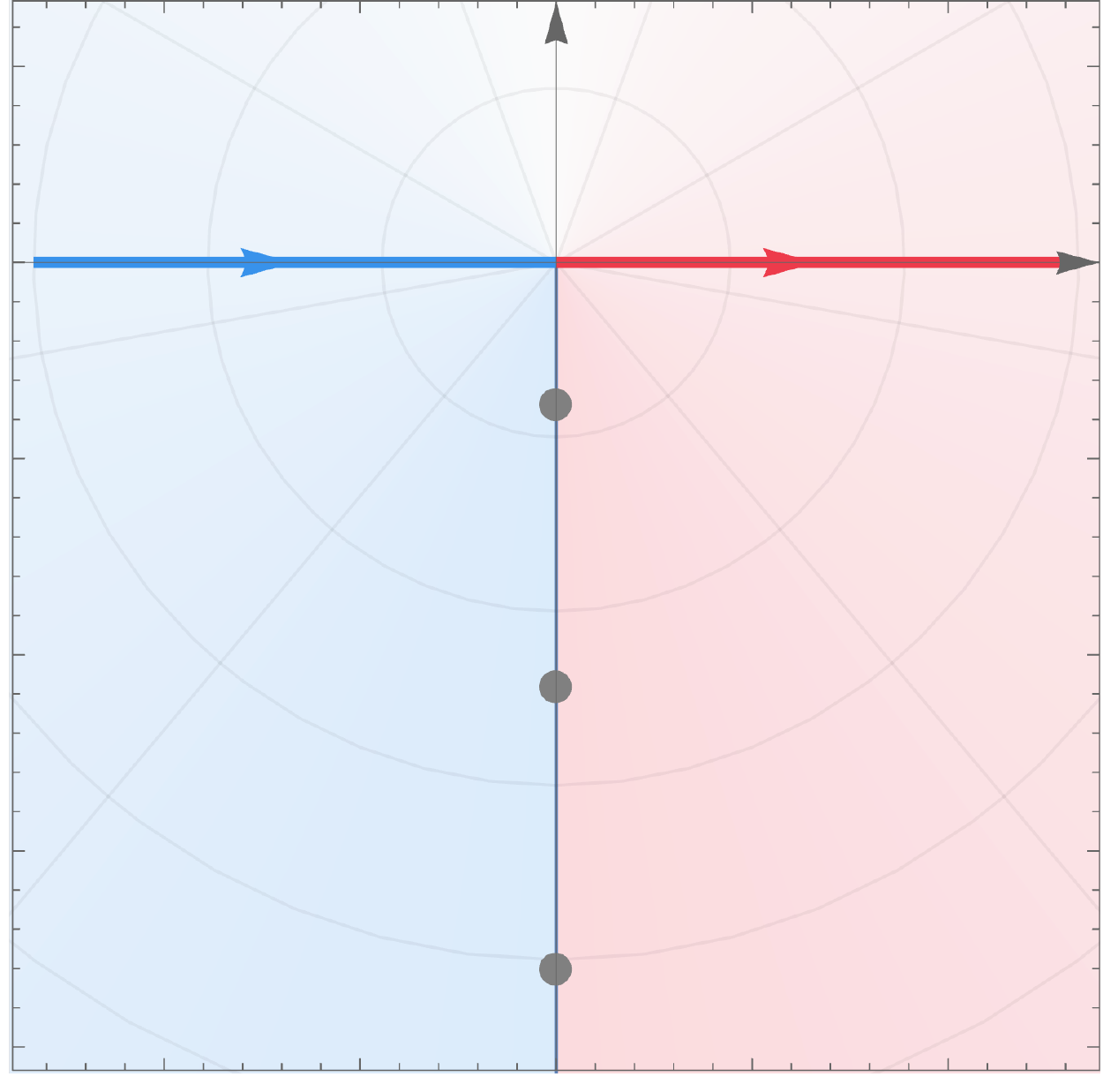}
\end{tabular}
\caption{Initial contour}
\end{subfigure}
\vspace{15pt}
\\
\begin{subfigure}[t]{0.4\textwidth}
\centering
\begin{tabular}[t]{c}
\includegraphics[width=0.9\textwidth]{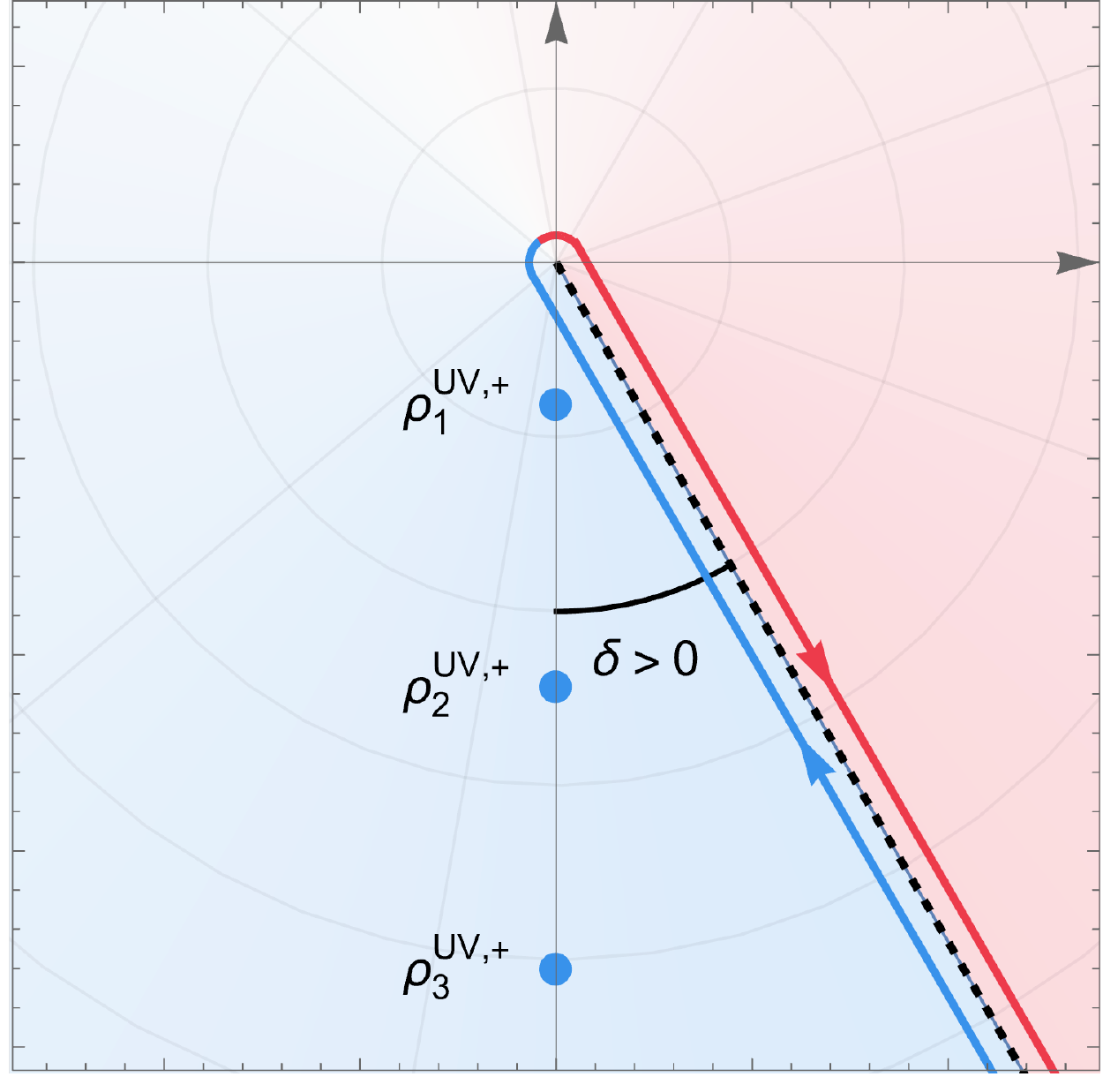}
\end{tabular}
\caption{$\delta>0$, with $\CC_+,\rho_n^{UV,+}$}
\end{subfigure}
\hspace{0.00001\textwidth}
\begin{subfigure}[t]{0.4\textwidth}
\centering
\begin{tabular}[t]{c}
\includegraphics[width=0.9\textwidth]{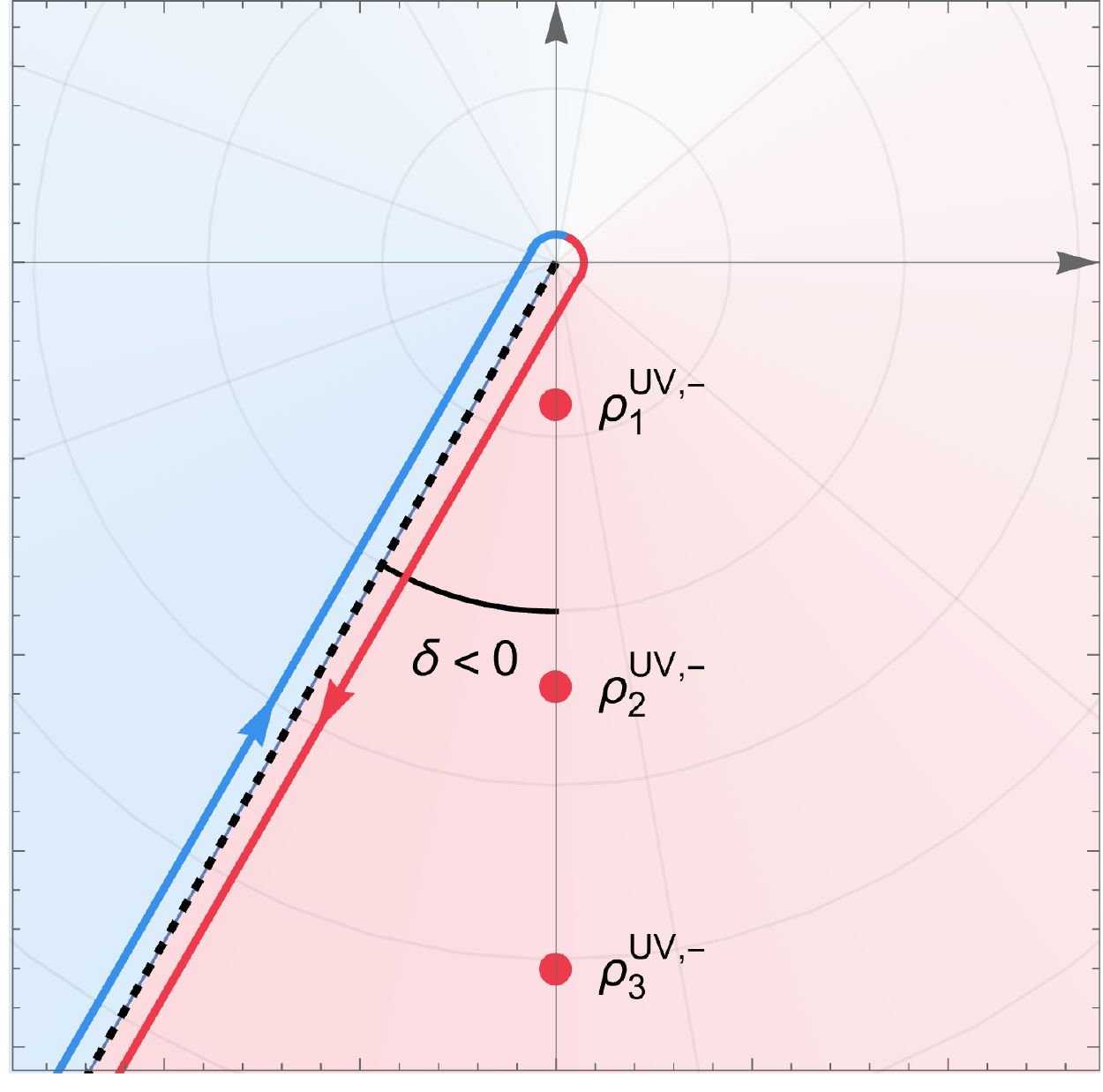}
\end{tabular}
\caption{$\delta<0$, with $\CC_-,\rho_n^{UV,-}$}
\end{subfigure}
\caption{The integration contour drawn on the complex plane with a branch cut for $\log[-\ri\omega]$. The grey dots represent the positions of the poles $-\ri\xi_n$. The branch cut is at an angle from the negative imaginary axis to avoid ambiguities with the poles. From the perspective of the Riemann surface, this is similar to figure \ref{fig-deformed-integrals}. In order to match Borel summation, we maintain the convention that $\CC_+$ corresponds to an anti-clockwise angle ($\delta>0$), and $\CC_-$ to a clockwise angle ($\delta<0$). The sheet where the residues are evaluated follows.}
\label{fig-uv}
\end{figure}

The  major difference is that change of sign in $B$ forces us to deform the contour into the lower half plane, rather than the upper. Through the factor of $\sigma(\omega)$ in \eqref{rhofunc_def}, which is defined in \eqref{sigma_def}, we see that $\rho$ is also singular and discontinuous in the negative imaginary axis, but due to $G_+(\omega)^{-1}$ rather than due to $G_-(\omega)$. The poles in the lower half plane are positioned at $\omega=-\ri\xi_n^{UV}$ such that
\begin{equation}
\xi_n^{UV} = 2n-1, \quad n\in \IN_{\geq 2},
\end{equation}
and we label their residues as $\rho^{UV,\pm}_n$, defined as the residues of $\rho$ at $-\ri\xi\mp 0$.
As we will see later, it will be useful to label $\xi_1 = 1$ as well, even though there is no pole of $\rho$ there. We maintain the convention \eqref{disc_conv} for the discontinuity and that $\CC_\pm$ corresponds to integrating $\xi$ over $\re^{\pm\ri 0}\IR_+$ (now, in the $\omega$-plane, $\CC_+$ is to the right of the negative imaginary axis). The analogue of \eqref{ref-decomposition} is then
\begin{multline}
-\frac{1}{2\pi\ri}\int_\IR \frac{\re^{2 \ri B \omega'}\rho(\omega')u(\omega') }{\omega'-\ri\xi+\ri 0} \rd \omega' = \frac{1}{2\pi\ri}\int_{ \CC_\pm}\frac{\re^{2 B \xi'}\delta\rho(-\ri\xi')u(-\ri\xi') }{\xi+\xi'} \rd \xi' 
\\
+ \sum_{n\geq 2}
\frac{\re^{2 B (2n-1)}\ri\rho^{UV,\pm}_n u_n^{UV}}{\xi+2n-1},
\label{UV-decomposition}
\end{multline}
where $u_n^{UV}=u(-\ri(2n-1))$. See figure \ref{fig-uv} for an illustration of the placement of the cuts and poles.

One would expect that the analytic continuation would keep the perturbative expansion intact. To see this, we show that the discontinuity in the lower half plane turns out to be related to the discontinuity in the upper half plane. Since $G_+(\omega)$ is of the form
\begin{equation}
G_+(\omega) = \re^{-\ri \Delta \omega \log[-\ri\omega]}\mathfrak{g}_+(\omega), \quad a>0,
\end{equation}
where $\mathfrak{g}_+(\omega)$ is a meromorphic function analytic in the $\mathbb{H}_+$. Then $\rho$ is of the form
\begin{equation}
\rho(\omega) = \re^{\ri \Delta \omega \log[\ri\omega]} \re^{\ri \Delta \omega \log[-\ri\omega]} \mathfrak{r}(\omega),
\end{equation}
where $\mathfrak{r}$ is some meromorphic function. Then the discontinuities in $\mathbb{H}_\pm$ are
\begin{equation}
\ba 
\delta\rho (\ri\xi) &= -2\ri\sin(\pi\Delta\xi) \re^{-2\Delta\xi\log\xi} \mathfrak{r}(\ri\xi),\\
\delta\rho (-\ri\xi) &= +2\ri\sin(\pi\Delta\xi) \re^{2\Delta\xi\log\xi} \mathfrak{r}(-\ri\xi),
\ea
\end{equation}
where we assume $\xi>0$. Much like in the usual case, it is useful to change from $B,\xi$ to $v,\eta$. In order to account for negative $v$, we define this change of variables as
\begin{equation}
\frac{1}{v}-2\Delta\log|v| = 2B,\quad \xi = |v| \eta, \quad \eta>0.
\end{equation}
In terms of these variables, we can write
\begin{equation}
\re^{2B\xi}\delta\rho(-\ri\xi) = - 2\ri v \re^{-\eta} P(\eta),
\end{equation}
where $P$ is the same function of $\eta,v$ as in \eqref{P-def} and \eqref{P-expansion}.
Note also that we define $u(\eta)$ as
\begin{equation}
u(\eta) = u\big(-\ri\xi(\eta)\big) = u(\ri v\eta),
\end{equation}
which in terms of the $\omega$ plane is the same definition of $u(\eta)$ as before, only with a different sign of $v$. 
This implies that $u(\eta)$ is $u$ evaluated along the negative imaginary axis.
Finally, we have the equation
\begin{equation}
u(\eta) = \frac{1}{v\eta} - \frac{v}{\pi}\int_0^{\re^{\pm\ri 0}\infty} \frac{\re^{-\eta'}P(\eta')u(\eta')}{\eta+\eta'}\rd\eta'+ \sum_{n\geq 2}
\frac{\re^{2 B (2n-1)}\ri\rho^{UV,\pm}_n u_n^{UV}}{(2n-1)-v\eta}, \quad \eta >0, v<0.
\end{equation}
If we drop the exponentially suppressed terms, this is the exact same perturbative problem as before, as desired.
 The techniques for finding $u(\eta)$ and $u_n^{UV}$ are precisely the same as in the case of $B,v>0$ so we will not dwell on them.

Curiously, for the free energy and the boundary condition the leading non-perturbative effects are rearranged. The free energy is given by
\begin{multline}
\CF(h) = - \frac{h^2}{2\pi} u(\ri)G_+(0)^2\Bigg\{1+\frac{v}{\pi}\int_0^{\re^{\pm\ri 0}\infty}\frac{\re^{-\eta'}P(\eta')u(\eta')}{\eta'-v^{-1}}\rd\eta'
\\
-\frac{1}{2}\sum_{n\geq 2}\frac{\re^{2B(2n-1)}\ri\rho^{UV,\pm}_nu_n^{UV}}{n-1}\Bigg\},
\end{multline}
which while perturbatively identical has a different non-perturbative structure, lacking the IR renormalon pole as expected. Meanwhile, when deforming the integral in \eqref{inteq_u} downwards to calculate $u(\ri)$, there is a pole at $\omega'=-\ri$.
\begin{multline}
u(\ri) = 1 - \frac{v}{\pi}\int_0^{\re^{\pm\ri 0}\infty} \frac{\re^{-\eta'}P(\eta')u(\eta')}{\eta'+ v^{-1}}\rd\eta'\\ 
+\re^{2B}\rho^{\rm{UV},\pm}_1  u(-\ri)
+\frac{1}{2}\sum_{n\geq 2}\frac{\re^{2B(2n-1)}\ri\rho_n^{\rm{UV},\pm}u_n^{\rm{UV}}}{n-1}
,
\end{multline}
where we define
\be
\rho^{\rm{UV},\pm}_1=\rho(-\ri\mp 0)= \re^{\pm \ri \pi  \Delta  } \frac{(1-2 \Delta )^{2 \Delta -1} }{\pi  (2 \re)^{2 \Delta }}\sin (\pi  \Delta ) \Gamma (1-\Delta )^2.
 \ee
We can find $u(-\ri)$ from the original equation, without any need for recursion. This term has the important role of, through the boundary condition, changing the leading term exponential correction from $\re^{6B}$ to $\re^{2B}$.

Introducing the negative coupling $\tilde{\alpha}$ and $\alpha$ as 
\begin{equation}
\frac{1}{\tilde{\alpha}}-\Delta\log|\tilde{\alpha}| = \log\frac{h}{\Lambda},\quad \frac{1}{{\alpha}}-\Delta\log|{\alpha}| = \log\left(\frac{2\pi\rho}{\Lambda}\right),
\end{equation}
we can write the free energy,
\begin{multline}
\CF(h)=- \frac{h^2}{2\pi}\bigg\{ 1- \Delta \tilde{\alpha} + \CO\left(\tilde{\alpha}^2\right)
\\
-  \re^{\frac{2}{\tilde{\alpha}}}\left|\tilde{\alpha}\right|^{-2\Delta}\rho_1^{\rm{UV},\pm} 4\Upsilon^\Upsilon\left(1 -2 \Delta  \tilde{\alpha}+\CO\left(\tilde{\alpha}^2\right)\right)+\CO\left(\re^{\frac{4}{\tilde{\alpha}}}\right)\bigg\}.
\end{multline}
or, equivalently, the internal energy,
\begin{multline}
\frac{e}{2\pi\rho^2}=\frac{1}{4}+\frac{\Delta }{4}\alpha +\CO\left(\alpha^2\right)
+\re^{\frac{2}{\alpha}}|\alpha|^{-2\Delta}
\rho_1^{\rm{UV},\pm}\frac{\Upsilon^\Upsilon}{4} \left(1 +2\Delta \alpha +\CO\big(\alpha^2\big)\right)+\CO\left(\re^{\frac{4}{\alpha}}\right).
\label{eUV}
\end{multline}
Looking at the position of the poles of $\rho$ in the lower half plane we see that the energy density, for example, will have a trans-series form 
\begin{equation}
\frac{e}{2\pi\rho^2} \approx \varphi_0(\alpha) + \sum_{\ell=1}^\infty \left\{\cos(\ell\pi\Delta)\pm\ri\sin(\ell\pi\Delta)\right\} C^{UV}_\ell \re^{\frac{2\ell}{\alpha}}|\alpha|^{-2\Delta} \varphi^{UV}_{\ell}(\alpha).
\label{int-energy-uv}
\end{equation}
Here, we use the structure of the discontinuity of $\rho$ to factor the ambiguity in the residues.
 \eqref{int-energy-uv} actually satisfies the canonical prediction for UV renormalons \eqref{UV-ren}. In the UV expansion, the constant term $-F(0)$ term does not appear. However, it is important to keep in mind that this is an unphysical auxiliary model whose purpose is merely to illuminate the asymptotic growth of the perturbative series. We will test the large order behavior of the perturbative series implied in \eqref{eUV} in chapter \ref{cha_volin}.

\section{Trans-series for bosonic models}
\label{sec_bosonic}

We generalize the analysis of the Gross--Neveu model to the broader class of ``bosonic models''.  From the perspective of Wiener--Hopf analysis, bosonic models are characterized by a function $G_+$ that is singular at zero, $G_+(\sim\xi)\propto \xi^{-1/2}$. This contrasts with the Gross--Neveu case where we had $G_+(\sim\xi)\propto 1$. Unfortunately this causes several computational complications, even if the conceptual picture remains the same. The first is that simplifying the integral equations in terms of $y$ such as in \eqref{y_def} is not possible, since $G_+(0)=\infty$. We will thus have to work with a more complicated driving term. Furthermore, the integral kernel is no longer $\sim B^{-1}$ but rather $\sim B^{0}$.  This means that we can no longer generate perturbative corrections by iterating the kernel. If we split the solution into a power series in $1/B$, 
we need to solve a non-trivial integral equation at each other. However, the non-perturbative corrections follow the same structure and there is no additional complication coming from there.

In this section, we review the calculation as studied in \cite{mmr-antrans} for the bosonic models. We will not be as exhaustive as we were for the Gross Neveu case, in part because the details are much more involved. 

\subsection{The integral equations}

As we have mentioned too many times, the key distinction of bosonic models under Wiener--Hopf analysis is the behavior of $G_+(\ri\xi)$ for $\xi\ll 1$. More specifically, the kernel decomposition obeys
\begin{equation}
G_+(\ri\xi) = \frac{k}{\sqrt{\xi}}\re^{-a\xi\log\xi}\left\{1-b\xi+\CO(\xi^2)\right\},
\label{Gplus-bosonic}
\end{equation}
for some real constants $a,b,k$ which will be useful later. 
We restart from equations \eqref{Yplus_og} and \eqref{eminus_og}, which were obtained without any constraints on $G_+(0)$, and introduce
\begin{equation}
Q(\omega) = G_+(\omega) Y_+(\omega).
\end{equation}
Using definition \eqref{sigma_def}, the integral versions of the aforementioned equations are, for $Q(\omega)$,
\be
\label{Q-eq}
Q(\omega)-{1\over 2 \pi \ri} \int_\IR {\re^{2 \ri B\omega'} \sigma (\omega') Q(\omega') \over \omega+ \omega'+ \ri 0} \rd \omega'= {1\over 2 \pi \ri} \int_\IR {G_-(\omega') g_+(\omega') \over \omega+ \omega'+ \ri 0} \rd \omega'. 
\ee
and, for $\epsilon_+(\omega)$,
\be
\label{epsom}
{\epsilon_+(\omega) \over G_+(\omega)} = {1\over 2 \pi \ri} \int_\IR {G_-(\omega') g_+(\omega') \over  \omega'-\omega- \ri 0}\rd \omega'+ 
{1\over 2 \pi \ri} \int_\IR {\re^{2 \ri B\omega'} \sigma (\omega')Q(\omega') \over\omega'- \omega- \ri 0} \rd \omega'. 
\ee

In order to explore the limit $B\gg 1$, we need once again to split the integral into an integral over a discontinuity and a sum over residues. In upper half plane, $\sigma$ has the same analytic structure as $\rho$, so we can apply the discussion of section \ref{sec-integral} to the integrals with $\sigma$ or $G_-$ (to make this evident we can write $G_-=\sigma G_+$). For example, after deforming the integrals with $Q$, we find:
\begin{multline}
\label{Q-eq-0-to-1}
{1\over 2 \pi \ri} \int_\IR {\re^{2 \ri B\omega'} \sigma (\omega') Q(\omega') \over \ri\xi+ \omega'+ \ri 0} \rd \omega'=
\\
{1\over 2 \pi \ri} \int_{\CC_\pm} \frac{\re^{-2  B\xi'} \delta\sigma (\xi') Q(\ri\xi') }{ \xi+ \xi'} \rd \xi' 
- \sum_{n\geq 1}\frac{\re^{-2B\xi_n}\ri\sigma_n^\pm Q_n}{\xi+\xi_n},
\end{multline}
where $\delta\sigma$ is the discontinuity of $\sigma$ according to convention \eqref{disc_conv}, the $\ri\xi_n$ are the positions of the poles of $\sigma$ in the positive imaginary axis, $Q_n=Q(\ri\xi_n)$ and
\begin{equation}
\sigma_n^\pm=\text{Res}_{\omega=\ri\xi_n\pm0} \sigma(\omega), \quad n\in\IN.
\label{sigma-poles}
\end{equation}

We can also decompose the integral on the r.h.s. of \eqref{Q-eq}. The terms proportional to $h$ in \eqref{gplus-def} are arranged as
\begin{multline}
\frac{\ri h}{2\pi\ri}\int_\IR \frac{G_-(\omega')}{\ri\xi+\omega'+\ri 0} \frac{1-\re^{2\ri B\omega'}}{\omega'}\rd\omega' 
= \frac{\ri h}{2\pi\ri}\int_{M_{-\epsilon}} \frac{G_-(\omega')}{\ri\xi+\omega'+\ri 0} \left(\frac{1}{\omega'}-\frac{\re^{2\ri B\omega'}}{\omega'}\right)\rd\omega'\\
= \frac{h G_+(\ri\xi)}{\xi}-\frac{h}{2\pi\ri}\int_{\mathcal{H}_{\pm}} \frac{\re^{-2B\xi'}G_-(\ri\xi')}{\xi'(\xi+\xi')}\rd\xi'
+h\sum_{n\geq 1}\frac{\re^{-2B\xi_n}\ri\sigma_n^\pm G_+(\ri\xi_n)}{\xi_n(\xi+\xi_n)}.
\label{h-integrals}
\end{multline}
In the first line, we deform the contour into a mousehole contour $M_{-\epsilon}$ which goes slightly under the origin. We are allowed to do this since the integrand is $\sim |\omega|^{-1/2}$ around the origin, which is sufficiently regular. Then, we deform the integral with the first term downwards, picking up the pole at $-\ri\xi-\ri 0$, while we deform the second upwards. For this second integral, we need to decompose into an angled Hankel contour and residues as before. 

In order to turn the angled Hankel contour in the last line of \eqref{h-integrals} into an integral of a discontinuity, we must proceed carefully. The integrand behaves as $\xi^{-3/2}$ when $\xi\rightarrow 0$ and thus turning it directly into an integral over $\CC_\pm$ would lead to a divergent integral. We must then regulate at the origin before we ``squeeze'' the positive imaginary axis (or rather the positive real axis in the $\xi$ plane). Guided by \eqref{Gplus-bosonic}, we factor out the leading behavior of $G_-(\ri\xi)$,
\begin{multline}
-\frac{h}{2\pi\ri}\int_{\mathcal{H}_{\pm}} \frac{\re^{-2B\xi'}G_-(\ri\xi')}{\xi'(\xi+\xi')}\rd\xi' = 
- \frac{h}{2\pi\ri}\int_{\mathcal{H}_{\pm}} \frac{1}{\xi'(\xi+\xi')}\left(\frac{k}{\sqrt{-\xi'}}\right)\rd\xi'
\\
-\frac{h}{2\pi\ri}\int_{\mathcal{H}_{\pm}} \frac{1}{\xi'(\xi+\xi')}\left(\re^{-2B\xi'} G_-(\ri\xi')-\frac{k}{\sqrt{-\xi'}}\right)\rd\xi' 
.
\label{h-pre-hankel}
\end{multline}
Because the Hankel contours $\mathcal{H}_{\pm}$ avoid the origin, we can deform the new contour, in the first line of \eqref{h-pre-hankel}, into the negative real axis (in the $\xi'$ plane) where it only has a pole at $\xi'=-\xi$. Meanwhile, the second integral is now sufficiently regular at the origin, so we can turn the Hankel contour into an integral of discontinuity (of both $G_-(\ri\xi')$ and $\sqrt{-\xi'}$)  along the ray $\CC_\pm$. We find
\begin{multline}
-\frac{h}{2\pi\ri}\int_{\mathcal{H}_{\pm}} \frac{\re^{-2B\xi'}G_-(\ri\xi')}{\xi'(\xi+\xi')}\rd\xi' = 
 - \frac{k h}{\xi\sqrt{\xi}}\\
-\frac{h}{2\pi\ri}\int_{\mathcal{C}_{\pm}} \frac{\re^{-2B\xi'}}{\xi'(\xi+\xi')}\left(\re^{-2B\xi'} \delta \sigma(\ri\xi') G_+(\ri\xi')+\frac{2\ri k}{\sqrt{\xi'}}\right)\rd\xi'.
\label{h-hankel}
\end{multline}

As for the terms proportional to $m$ we obtain
\begin{multline}
\frac{\ri m\re^B}{4\pi\ri}\int_\IR \frac{G_-(\omega')}{\ri\xi+\omega'+\ri 0} \left(-\frac{1}{\omega'+\ri}+\frac{\re^{2\ri B\omega'}}{\omega'-\ri}\right)\rd\omega' 
= \frac{m\re^B}{2}\frac{G_+(\ri\xi)-G_+(\ri)}{1-\xi} \\
+ \frac{ m \re^B}{2}\frac{1}{2\pi\ri}\int_{\CC_\pm} \frac{\re^{-2B\xi'}\delta\sigma(\ri\xi')G_+(\ri\xi')}{(\xi+\xi')(\xi'-1)}\rd\xi' - \frac{m\re^B}{2} \sum_{n\geq 1}\frac{\re^{-2B\xi_n}\ri\sigma_n^\pm G_+(\ri\xi_n)}{(\xi+\xi_n)(\xi_n-1)},
\label{int-m-Q}
\end{multline}
where simply need to deform the contour for the first term downwards, picking up residues at $-\ri\xi$ and $-\ri$, and for the second term upwards, leading to the decomposition into discontinuity integral and sum over poles. 

All together we have an equation for $Q(\ri\xi)$ factored into calculable pieces
\begin{multline}
Q(\ri\xi)-{1\over 2 \pi \ri} \int_{\CC_\pm} 
\frac{\re^{-2  B\xi'} \delta\sigma (\ri\xi') \left\{Q(\ri\xi') +\mathfrak{g}(\xi')\right\}-2\ri kh (\xi')^{-\frac{3}{2}}}{ \xi+ \xi'} 
\rd \xi' 
=\\
\left(\frac{m\re^B}{2}\frac{G_+(\ri)}{\xi-1}-\frac{k h}{\xi\sqrt{\xi}}-\mathfrak{g}(\xi)\right)
-\sum_{n\geq 1}\frac{\re^{-2B\xi_n}\ri\sigma_n^\pm \left\{Q_n+\mathfrak{g}(\xi_n)\right\}}{\xi+\xi_n},
\label{Q-eq-0}
\end{multline}
where
\begin{equation}
\mathfrak{g}(\xi) =  G_+(\ri\xi)\left(\frac{m\re^B}{2(\xi-1)}-\frac{h}{\xi}\right).
\end{equation}

The manipulation of \eqref{epsom} for $\omega=\ri\kappa$ is nearly identical up to two distinctions coming from the integral with $g_+$. One is a technicality, while the other has important physical consequences. The former comes from the regulation of the integral with $h$. Now we have
\begin{equation}
\ba
-\frac{h}{2\pi\ri}&\int_{\mathcal{H}_{\pm}} \frac{\re^{-2B\xi'}G_-(\ri\xi')}{\xi'(\xi'-\kappa)}\rd\xi' =\\
&= -\frac{h}{2\pi\ri}
\left\{
\int_{\mathcal{H}_{\pm}} \left(\frac{\re^{-2B\xi'} G_-(\ri\xi')}{\xi'(\xi'-\kappa)}+\frac{k}{\kappa \xi'\sqrt{-\xi'}}\right)\rd\xi' 
-
\int_{\mathcal{H}_{\pm}} \frac{k\, \rd\xi'}{\kappa \xi'\sqrt{-\xi'}}
\right\},\\
&= -\frac{h}{2\pi\ri}\int_{\mathcal{C}_{\pm}} \left(\frac{\re^{-2B\xi'} \delta \sigma(\ri\xi') G_+(\ri\xi')}{\xi'(\xi'-\kappa)}-\frac{2\ri k}{\kappa (\xi')^{3/2}}\right)\rd\xi',
\ea
\label{h-hankel-2}
\end{equation}
where the ``extra'' integral evaluates to zero due to the absence of poles outside of the Hankel contour. This leads to the equation for $\kappa\neq 1$,
\begin{multline}
    \frac{\epsilon_+(\ri\kappa)}{G_+(\ri\kappa)}= 
      \frac{1}{2\pi\ri}\int_{\CC_\pm} \left(\frac{\re^{-2B\xi}\delta\sigma(\ri\xi)
       \left\{Q(\ri\xi)+\mathfrak{g}(\xi)\right\}}{\xi-\kappa}-\frac{2\ri kh}{\kappa \xi^{3/2}}\right)\rd\xi 
   \\
-\frac{m \re^ B}{2}\frac{G_+(\ri)}{\kappa+1}
- \sum_{n\ge 1} \frac{\re^{-2B\xi_n} \ri\sigma_n^\pm \left\{Q_n+\mathfrak{g}(\xi_n)\right\}}{\xi_n-\kappa}+\CO(\re^{-2B\kappa})
    \label{epsilon_kappa}.
\end{multline}
The suppressed terms proportional to $\re^{-2B\kappa}$ come from the pole at $\omega'=\ri\kappa$ and are of little interest. 

The more important change comes from the special case $\kappa=1$. One could expect poles at $\omega'=\ri$ in all of the integrals in \eqref{epsom}, but for almost all models studied\footnote{One notable model that elides this property is the $O(3)$ sigma model, which is so exceptional we dedicate section \ref{sec-O3} to it.} we have that $G_-(-\ri)=0$. This regularizes any would be pole except from the term
\begin{equation}
\textrm{Res}_{\omega=\ri} \frac{G_-(\omega')}{\omega'-\ri-\ri 0}\left(\frac{\ri m \re^B }{2}\frac{\re^{2\ri B \omega'}}{\omega'-\ri}\right)  = \frac{m\re^{-B}}{2} \ri G'_-(\ri\pm 0).
\label{ir-pole-bos}
\end{equation}
We find
\begin{multline}
    \frac{\epsilon_+(\ri)}{G_+(\ri)}= 
       -\frac{m \re^ B}{4}G_+(\ri)+\frac{1}{2\pi\ri}\int_{\CC_\pm} \left(\frac{\re^{-2B\xi}\delta\sigma(\ri\xi)
       \left\{Q(\ri\xi)+\mathfrak{g}(\xi)\right\}}{\xi-1}+\frac{2\ri kh}{\xi^{3/2}}\right)\rd\xi 
   \\
    + \frac{m\re^{-B}}{2} \ri G'_-(\ri\pm 0)    - \sum_{n\ge 1} \frac{\re^{-2B\xi_n} \ri\sigma_n^\pm \left\{Q_n+\mathfrak{g}(\xi_n)\right\}}{\xi_n-1}
    \label{epsilon_0}.
\end{multline}
As for the boundary condition, obtained by imposing $\kappa\epsilon(\ri\kappa)\rightarrow\infty=0$ in \eqref{epsilon_kappa} with $G_+(\ri\infty)=1$, it is given by 
\begin{multline}
0= \frac{m \re^ B}{2}G_+(\ri) 
    +\frac{1}{2\pi\ri}\int_{\CC_\pm}\left(\re^{-2B\xi}\delta\sigma(\ri\xi)\left\{Q(\ri\xi)+\mathfrak{g}(\xi)\right\}-2\ri kh\xi^{-3/2}\right) \rd\xi 
  \\
   -\sum_{n\ge 1} \re^{-2B\xi_n}\ri\sigma_n^\pm \left\{Q_n+\mathfrak{g}(\xi_n)\right\}.
    \label{epsilon_bc}
\end{multline}

At this stage, and informed by our study of Gross--Neveu, one can already identify the trans-series structure we found before. There are regular discontinuity integrals which are amenable to a perturbative expansion, trans-monomials with exponential weights determined by the poles of $\sigma(\omega)$ and a special residue contribution from $\omega=\ri$ which we will later identify with the standard IR renormalon. But to write an actual trans-series, we must first solve these integral equations.

\subsection{Perturbative expansion}
\label{sec-pert-bos}

In this section, we will inspect some of the details of the solution of equations \eqref{Q-eq-0}, \eqref{epsilon_0} and \eqref{epsilon_bc} to leading order in $1/B$ and then present the NLO results, which were originally derived in \cite{hn,hmn,pcf}. The technical details of the NLO calculation using the techniques presented in this section can be found in the appendices of \cite{mmr-antrans}.

In order to proceed with the perturbative expansion, it is useful to introduce a rescaled variable $\xi = x/(2B)$, which is analogous to the variable $\eta$ used in the Gross--Neveu analysis. Since the integral equations are linear, it is also useful to separate the problem into terms proportional to $h$ and proportional to $m$. With both of these ideas in mind we introduce
\begin{equation}
Q\left(\frac{\ri x}{2B}\right) = h \left\{ q^h_0(x) + \cdots \right\} + \frac{m\re^B}{2} \left\{ q^m_0(x) + \cdots \right\},
\label{qdef}
\end{equation}
where the $\cdots$ denote the expansion in $1/B$ (we will be more precise once we ascertain the overall power of $B$ of each expansion). In the terms of the variable $x$, \eqref{Gplus-bosonic} implies
\begin{equation}
\ba
G_+\!\left(\frac{\ri x}{2B}\right) &= \sqrt{2B} \frac{k}{\sqrt{x}} \left\{1 + \CO \left(\frac{1}{B}\right) \right\},
\\
 \delta\sigma\left(\frac{\ri x}{2B}\right)  &= -2\ri  \left\{1 + \CO \left(\frac{1}{B}\right) \right\}.
\ea
\label{expand-kernels}
\end{equation}

Collecting the terms proportional to $m$ in \eqref{Q-eq-0} and expanding with \eqref{qdef} and \eqref{expand-kernels}, we find
\begin{equation}
q_0^m(x) + \frac{1}{\pi}\int_0^\infty \frac{\re^{-y} q_0^m(y)}{x+y} \rd y =  \frac{k\sqrt{2B}}{\sqrt{x}} + \frac{k\sqrt{2B}}{\pi}\int_0^\infty \frac{\re^{-y}}{(x+y)\sqrt{y}}\rd y. 
\end{equation}
On the right hand side, we dropped terms of order $\CO(B^0)$\footnote{One could fear this implies a series in $1/\sqrt{B}$ rather than $1/B$, however such terms turn out to be spurious. Other approaches manifestly avoid this.}. We  now introduce the Airy operator $\mathsf{A}$, which is given by the integral
\begin{equation}
\left[\mathsf{A}f\right] (x) = \frac{1}{\pi}\int_0^\infty \frac{\re^{-y}}{x+y} f(y)\rd y.
\label{airy-def}
\end{equation}
and we quickly solve the integral equation
\begin{equation}
(1+\mathsf{A})  q_0^m(x)  =  (1+\mathsf{A}) \frac{k\sqrt{2B}}{\sqrt{x}}  \Leftrightarrow q_0^m(x)  = \frac{k\sqrt{2B}}{\sqrt{x}}.
\label{q0m-solution}
\end{equation}

Not all integral equations in the perturbative expansion are as simple as the last one. However, due to the expansion of $\delta \sigma$, we will in general find integral equations of the form
\begin{equation}
\left(1+\mathsf{A}\right) q(x) = r(x).
\label{gen-pert-eq}
\end{equation}
These can be tackled by using a spectral decomposition of the Airy operator \eqref{airy-def}, as shown in appendix B of \cite{mmr-antrans} using the results of \cite{mtw,cfz}. In particular, the solution can be written in integral from by considering the continuous family of eigenfunctions,
\begin{equation}
\ba
  \label{chiK}
  \chi_p(x) &={  {\sqrt{2 p \sinh(\pi p)}} \over \pi} {\re^{x/2} \over {\sqrt{x}}} K_{\ri p}\left({x \over 2} \right)
  \\
   \left[\mathsf{A}\chi_p\right](x) &= \textrm{sech}(\pi p) \chi_p(x),\quad p\in \IR_+,
  \ea
\end{equation}
defined with  the modified Bessel function $K_\nu(x)$.
Using the appropriate measure for the eigenbasis \eqref{chiK}, an inverse of the Airy kernel is given by
\begin{equation}
\left(1+\mathsf{A}\right)^{-1}r(x) = \int_0^\infty\frac{\chi_p(x)}{1+\textrm{sech}(\pi p)}\left\{\int_0^\infty  \re^{-y}\chi_p(y) r(y)\rd y\right\}\rd p
\label{airy-inverse}
\end{equation}
which solves equation \eqref{gen-pert-eq}.

For example, for $q_0^h$ we find the integral equation
\begin{equation}
(1+\mathsf{A})q_0^h(x) =    \frac{k h (2B)^{3/2}}{\pi}\int_0^\infty \frac{\re^{-y}-1}{(x+y)y^{3/2}}\rd y = - k h (2B)^{3/2}\,\mathsf{A}\,\frac{\re^x-1}{x^{3/2}}.
\end{equation}
Using \eqref{airy-inverse}, after some involved integration, one can find
\begin{equation}
q_0^h(x) =  kh (2B)^{3/2} \left(\frac{1}{x^{3/2}}-\frac{\re^{x/2}}{\sqrt{x}} K_1\left(\frac{x}{2}\right)\right)
\label{q0h-solution}
\end{equation}
In some cases, the integrals \eqref{airy-inverse} cannot be solved explicitly as a function of $x$. Nevertheless, they can suffice to obtain integrals involving the solution, which are the required ingredients.

Now that we have the leading behavior of $Q$ one can enforce the boundary condition. Expanded to LO in $1/B$, \eqref{epsilon_bc} reduces to
\begin{multline}
0= \frac{m\re^B}{2} -\frac{1}{2\pi B}\int_0^\infty \re^{-x} \left(h q_0^h(x) +\frac{m\re^B}{2} q_0^m\right)\rd x \\
+ \frac{km\re^B}{2\pi \sqrt{2B}} \int_0^\infty \frac{\re^{-x}}{\sqrt{x}}\rd x - \frac{\sqrt{2B}kh}{\pi} \int_0^\infty \frac{1-\re^{-x}}{x^{3/2}}\rd x+\cdots.
\label{bc-lo}
\end{multline}
From \eqref{q0h-solution}, we can obtain that
\begin{equation}
\int_0^\infty \re^{-x}q_0^h(x) = -kh (2B)^{3/2} \frac{\sqrt{\pi}}{2}(4-\pi),
\end{equation}
and, plugging in \eqref{q0m-solution}, \eqref{bc-lo} becomes
\begin{equation}
 (2B)^{1/2}\frac{k h \sqrt{\pi}}{2} +\cdots = \frac{m\re^B G_+(\ri)}{2}+\cdots,
 \label{bc-mebh}
\end{equation}
which can be solved for $B$,
\begin{equation}
B = \log\left(\frac{h}{m}\right) + \frac{1}{2}\log\log\left(\frac{h}{m}\right) + \log\left(\frac{k\sqrt{2\pi}}{G_+(\ri)}\right)+\cdots,
\label{B-hm}
\end{equation}
where we assumed that corrections in $1/\sqrt{B}$ cancel, which can be checked explicitly.

Finally, we can tackle the free-energy itself. To leading order, putting the integral equation \eqref{epsom} in the definition of the free energy \eqref{fh-eps} and expanding in $1/B$ leads to
\begin{multline}
\CF(h) = -\frac{m\re^B}{2\pi}G_+(\ri)\Bigg\{-\frac{m\re^B}{4}G_+(\ri)+\frac{1}{2\pi B}\int_0^\infty \re^{-x} \left(h q_0^h(x) +\frac{m\re^B}{2} q_0^m\right)\rd x 
\\
- \frac{km\re^B}{2\pi \sqrt{2B}} \int_0^\infty \frac{\re^{-x}}{\sqrt{x}}\rd x 
+ \frac{\sqrt{2B}kh}{\pi} \int_0^\infty \frac{1-\re^{-x}}{x^{3/2}}\rd x+\cdots\Bigg\}.
\end{multline}
Calculating the integrals, and removing factors of $m$ with \eqref{bc-mebh}, we can simplify the expression to
\begin{equation}
\CF(h) = - \frac{k^2h^2}{4}B+\cdots.
\end{equation}

At higher orders, the procedure is similar but more computationally intensive. For example, one starts to also have terms in $\log B/B$. Carrying out this procedure to one higher order in $1/B$, which implies solving for the higher corrections to $Q$ and to the boundary condition, one can obtain the generic result
\begin{multline}
\mathcal{F}(h) = -{k^2 h^2 \over 4} \biggl\{ \log\left( {h \over m} \right)+\left( a+{1\over 2} \right) \log \log\left( {h \over m} \right)+ \log \left( {{k\sqrt{2 \pi}}  \over G_+(\ri)} \right)\\
- a \big(1- 3 \log 2- \gamma_E \big) - b -1 + \mathcal{O}\big(\log^{-1/2}(h/m)\big) \biggr\}.
\end{multline}
which was known from \cite{pcf}, generalizing the formulae from \cite{hn,hmn}.
The constants $a,b,k$ are the constants in \eqref{Gplus-bosonic}.

Since the expansion in $B$, as well as in $\log(h/m)$, contains $\log$ terms, it quickly becomes unwieldy. Much like in Gross--Neveu, it is useful to introduce a more natural coupling.
Following \cite{volin,mr-ren,bbbkp}, we define
\begin{equation}
\frac{1}{\tilde{\alpha}}+\xi \log\tilde\alpha = \log\left( \frac{h}{\Lambda} \right),
\label{tilde-alpha-bosonic}
\end{equation}
where
\begin{equation}
\xi= \frac{\beta_1}{2\beta_0^2} = a+\frac{1}{2}. 
\end{equation}
The second equality can be seen as a consistency check of $G_+$, as argued in \cite{pcf}.
Then we have
\begin{equation}
\CF(h) = -\frac{h^2 k^2}{4 \tilde\alpha } \left\{1-\tilde{\alpha}\left[1+b
-a\gamma_E
+\log \left(\frac{\re^a G_+(\ri)}{2^{3a}k\sqrt{2 \pi }  }\frac{m}{\Lambda}\right)
\right]+O\left(\tilde\alpha^2\right)\right\},
\label{cFa-pert}
\end{equation}
which does away with $\log\tilde\alpha$ terms. 

\subsection{Non-perturbative corrections}

Our main interest in these integral equations lies in extracting the non-perturbative corrections. In order to find the leading exponentially-suppressed to the free-energy, we must focus on the terms coming from the residues at the poles of $\sigma$ in \eqref{Q-eq-0}, \eqref{epsilon_0} and \eqref{epsilon_bc}. The new element we need to solve for is $Q_1$ which can be found by setting $\xi=1$. We focus only on the leading order term in $1/B$, and neglect further exponential corrections.
The equation reads 
\begin{multline}
\label{Q1eq}
Q_1-\frac{1}{2\pi\ri}\int _0^\infty \frac{\re^{-2B \xi} \delta\sigma(\ri\xi)Q(\ri\xi)}{\xi_1+\xi} \rd\xi  =
\left(\frac{m\re^B}{2}\frac{G_+(\ri)}{\xi_1-1}-\frac{k h}{\xi_1\sqrt{\xi_1}}-\mathfrak{g}(\xi_1)\right)\\
+\frac{1}{2\pi\ri}\int_{\CC_\pm} \frac{\re^{-2B\xi'}\delta\sigma(\ri\xi')\mathfrak{g}(\xi')-2\ri kh (\xi')^{-3/2}}{\xi_1+\xi'}\rd\xi'
 +\CO\big(\re^{-2B\xi_1}\big)
.
\end{multline}
The integral on the left can be computed to leading order using the $q_0^{h,m}$ computed in \eqref{q0m-solution} and \eqref{q0h-solution}, while the terms on the right follow straightforwardly. We find then
\begin{equation}
Q_1 =  -kh(2B)^{1/2}\frac{\sqrt{\pi}}{2\xi_1} -\frac{m\re^B}{2}\frac{G_+(\ri\xi_1)-G_+(\ri)}{\xi_1-1}+\CO\big(B^0\big).
\end{equation}
We can then use this directly in the free energy, which is given by
\begin{multline}
\CF(h) = - \frac{m\re^B G_+(\ri)}{2\pi}\Bigg\{
h k \sqrt{\frac{\pi B}{2}} 
-\frac{ m\re^B}{4} G_+(\ri)
+
\frac{m \re^{-B}}{2}  \ri G'_-(\ri\pm 0)
\\+\ri \sigma^\pm_1 \re^{-2 B \xi _1} \left(\frac{h G_+\left(\ri \xi _1\right)}{\left(\xi _1-1\right) \xi _1}-\frac{m \re^B G_+\left(\ri \xi _1\right)}{2 \left(\xi _1-1\right){}^2}-\frac{Q_1}{\xi _1-1}\right)
+\cdots
\Bigg\}.
\end{multline}
Here we neglected exponentially suppressed contribution to $Q(\ri\xi)$, that would manifest through the discontinuity integral in \eqref{epsilon_0}. Let us quickly see that these corrections would be sub-leading in $1/B$. The terms with $\re^{-2\ri\xi_1}$ in \eqref{Q-eq-0} that are not $Q(\omega)$ itself are of order $\CO(B^0)$. Thus, one could expect at leading order a correction of the form $Q(\ri x/2B) = \cdots +  \re^{-2B\xi_1} q_{(1)}(x) +\cdots$. The leading contribution of this term to the free energy would be
\begin{equation}
\frac{\re^{-2B\xi_1}}{\pi}\int_0^\infty \frac{\re^{-x}q_{(1)}(x)}{2B-x}\rd x +\cdots \sim \CO\left(\frac{\re^{-2B\xi_1}}{B}\right),
\end{equation}
and hence sub-leading.

We need to impose the boundary condition. A priori, \eqref{epsilon_bc} has itself non-perturbative corrections which would contribute to the final result. However, one can check that non-perturbative corrections to the boundary condition lead to sub-leading corrections in $1/B$ to the first non-perturbative correction of the free-energy. Since we are focusing only on the leading term in $1/B$, we can use the perturbative boundary conditions. We then find
\begin{equation}
\CF(h)= - \frac{k^2 h^2}{4} B \left\{1- 
\frac{2\ri \sigma^\pm_1 \re^{-2 B \xi _1}}{\left(\xi _1-1\right)^2 \xi _1}+ \cdots
\right\} - \frac{m^2}{4\pi} \ri G_+(\ri) G_-'(\ri\pm 0). 
\end{equation}
where the $\cdots$ include both perturbative and non-perturbative corrections.
In terms of the coupling ${\tilde{\alpha}}$, we have 
\begin{multline}
\label{fh-bos}
\CF (h) = - \frac{k^2 h^2}{4\tilde\alpha}\left\{ 1
-
2\ri \sigma^\pm_1\left(\re^{-\frac{2}{\tilde{\alpha}}}\tilde{\alpha}^{1-2 \xi }\right)^{\xi _1}\! 
\frac{1}{\left(\xi _1-1\right)^2 \xi _1}
\left(\frac{G_+(\ri)^2}{2 \pi  k^2}
\left(\frac{m}{\Lambda}\right)^{2}\right)^{\xi_1}+ \cdots \right\}\\
 - \frac{m^2}{4\pi} \ri G_+(\ri) G_-'(\ri\pm 0).
\end{multline}

As we discussed in the Gross--Neveu model, the term independent of $h$ has particular significance. Its imaginary part plays the role of the IR renormalon, while its real part should correspond to $-F(0)$, i.e. the ground state energy in the absence of the $h$ field. We can write the generic expression, 
\be
\label{F0-gen}
F(0)=  \frac{m^2}{4\pi} {\rm Re}\left(\ri G_+(\ri) G'_-(\ri\pm 0)\right). 
\ee
%
%

Let us also write our results in the canonical formalism, in terms of $e, \rho$. If we write the transseries for $\CF(h)$ as
\begin{equation}
\CF(h) \approx - \frac{k^2 h^2}{4\tilde{\alpha}}\tilde{\Phi}^\pm(\tilde{\alpha}) - F(0),
\end{equation}
then
\begin{equation}
\frac{e}{\rho^2}\approx   \frac{\tilde{\alpha}}{k^2}\frac{\tilde\Phi^\pm(\tilde{\alpha}) +  \frac{\rd \log\tilde{\alpha}}{\rd\log h}\left(\tilde{\alpha}\frac{\rd \tilde\Phi^\pm(\tilde{\alpha})}{\rd\tilde{\alpha}}-\tilde\Phi^\pm(\tilde{\alpha})\right)}{\left( \tilde\Phi^\pm(\tilde{\alpha}) + \frac{1}{2}\frac{\rd \log\tilde{\alpha}}{\rd\log h}\left(\tilde{\alpha}\frac{\rd \tilde\Phi^\pm(\tilde{\alpha})}{\rd\tilde{\alpha}}-\tilde\Phi^\pm(\tilde{\alpha})\right)\right)^2} - \frac{F(0)}{\rho^2},
\label{trans-ratio-bosonic}
\end{equation}
where we use
\begin{equation}
\frac{\rd \log\tilde{\alpha}}{\rd\log h} = - \frac{\tilde{\alpha}}{1-\xi\tilde{\alpha}} \sim - \tilde{\alpha}+\CO(\tilde{\alpha}^2), 
\label{frakC-choice}
\end{equation}
which follows from \eqref{tilde-alpha-bosonic}. However, it is better for the canonical formalism to use a coupling defined with respect to $\rho$. We introduce
\begin{equation}
\frac{1}{\alpha }+(\xi -1) \log (\alpha )=\log \left(\frac{\mathfrak{C} \rho }{m}\right),
\label{alpha-def-bos}
\end{equation}
which incorporates a choice of an arbitrary constant $\mathfrak{C}$\footnote{In the notation of \cite{mr-ren,mmr-antrans,mmr-theta}, we write the choice of $\mathfrak{C}$ through $\mathfrak{c} = 2\beta_0\mathfrak{C}\Lambda/m$.}. This new coupling $\alpha$ is related to the previous one by a trans-series, which at very leading order is $\alpha\sim\tilde{\alpha}$.

One way of choosing the constant $\mathfrak{C}$ is by fixing the NLO coefficient of the  energy density. This coefficient is the first to depend on $\mathfrak{C}$ as can been seen by using \eqref{cFa-pert} in conjunction with \eqref{trans-ratio-bosonic} and \eqref{alpha-def-bos}, leading to
\be
\frac{e}{\rho^2} \sim \frac{\alpha}{k^2}\left\{1+  \alpha\left(1+b+a ( 1-3\log 2-\gamma_E)+\log \left[\frac{\mathfrak{C}G_+(\ri) k}{2 \sqrt{2 \pi }}\right]\right) +\CO(\alpha^2)\right\}.
\ee
A conventional choice is $\frac{e k^2}{\rho^2}\sim \alpha+\alpha^2/2+\CO(\alpha^3)$, which requires
\begin{equation}
\mathfrak{C} = \frac{2^{1+3a}\sqrt{2 \pi }}{ \re^{\frac{1}{2}+b+a(1-\gamma_E )} k G_+(\ri)} .
\label{Cfrak-onehalf}
\end{equation}
This choice is such that it keeps the form of the coefficients in the $\alpha$ expansion simple, canceling terms with $\log 2$, $\log\Delta$, $\gamma_E$, etc. . Another possible convenient choice would be $\frac{e k^2}{\rho^2}\sim \alpha+\CO(\alpha^3)$.

With the non-perturbative corrections, \eqref{trans-ratio-bosonic} leads to
\begin{equation}
\label{bos-ts}
\frac{e}{\rho^2}= \frac{\alpha}{k^2}\left\{\varphi(\alpha)
+\CC_0^\pm \re^{-\frac{2}{\alpha}}  \alpha ^{1-2 \xi }
+\CC_1^\pm \left(\re^{-\frac{2}{\alpha}}  \alpha ^{1-2 \xi }\right)
^{\xi_1}(1+\cdots)+\cdots \right\}. 
\end{equation}
The series $\varphi(\alpha)$ is the full perturbative series
\begin{equation}
\varphi(\alpha) = \sum_{n\geq 0} e_n \alpha^n.
\label{pert-gs}
\end{equation}
 The first trans-series parameters are given by
\begin{align}
\label{C0-generic-gs}
\CC_0^\pm &= -\frac{\mathfrak{C}^2 G_+(\ri) k^2 }{4 \pi}\ri G'_-(\ri\pm 0),\\
\CC_1^\pm &=\frac{ 2\ri \sigma^\pm _1 }{\left(\xi _1-1\right)^2 \xi _1}\left(\frac{\mathfrak{C}^2 G^2_+(\ri) k^2}{8\pi}\right)^{\xi _1}.
\label{C1-generic-gs}
\end{align}
These constants will allow us to numerically test our results in chapter \ref{cha_volin}.

The overall structure that we find is then quite similar to the Gross--Neveu case, with a leading IR renormalon associated with an isolated Borel singularity at $\zeta=2$ and a more complicated series coming from a Borel singularity at $2\xi_1$. We will discuss the possible physical interpretation of this singularity in specific models, since it is not universal. Further corrections will come from the other poles of $\sigma$ and their linear combinations, as discussed in the beginning of this chapter. Unlike in Gross--Neveu, we cannot in general factor out the ambiguous part of the coefficients, and thus the form of the trans-series parameters might vary more between models.

To study the UV behavior one could, in principle, proceed as in section \eqref{sec-uv} and analytically continue the integral equations for $B<0$ to find a trans-series with terms $\propto \re^{2B}$. Naturally, these would come from poles in the negative imaginary axis of $\sigma$. As we discussed in the case of Gross--Neveu, these contributions are relevant for the asymptotic analysis of the perturbative series $\varphi(\alpha)$, even if this analytic continuation in $B$ or $\alpha$ is unphysical. While we will not carry out the calculations for this case, we will keep in mind the result that the poles of $\sigma$ along the negative imaginary axis determine the UV structure of the theory. In particular, for all models studied, these will turn out to be positioned at negative integers, and thus correspond to the standard UV renormalons \eqref{UV-ren}.

Finally, an improved approach to the calculation of the trans-series in bosonic models was recently developed by \cite{bbhv} and worked out for the case of the $O(N)$ sigma model. In that approach, the driving term is simplified to a term without an integral kernel. They achieve this by working in the canonical formalism, starting from \eqref{iqft_geneq}, and introducing a function similar to \eqref{y_def}, without requiring $G_+(0)$ to be finite since there is no term proportional to $h$. One then has to obtain the density $\rho\propto \epsilon_+(0)$ indirectly from \eqref{fourier_WH} to avoid the divergence coming from $G_+(0)$. The perturbative expansion still requires more sophisticated methods than the Gross Neveu case, since it still cannot be solved by iteration. A further simplification done in \cite{bbhv} is to reexpress the perturbative series in an auxiliary coupling $v$, like we did for Gross--Neveu with \eqref{v-def}, avoiding the complication of $\log B$ and $\sqrt B$ in intermediate calculations. 

\section{Results for bosonic models}
\label{sec-bos}

In this section, we will analyze the implications of our analysis in the specific bosonic models we introduced in section \ref{sec_iqft}. This includes discussing the structure of the trans-series, its physical implications and the leading order terms in the trans-series for the energy density \eqref{bos-ts}. For completeness, we will also include some higher order terms of the perturbative series $\varphi(\alpha)$. We will discuss in depth how to obtain such a series in chapter \ref{cha_volin}.

\subsection{\texorpdfstring{$O(N)$}{O(N)} non-linear sigma model}

We first consider the $O(N)$ non-linear sigma model with $N\geq 4$. This is perhaps the best studied of the bosonic models here presented, and the most simple. We start by recalling  the key data 
\be 
\beta_0 = \frac{1}{4\pi\Delta} ,\quad 
\xi = \Delta, \quad \Delta=\frac{1}{N-2}.
\ee
The perturbative analysis for bosonic models generalizes the work of \cite{hn,hmn}, where the mass gap was derived,
\begin{equation}
\frac{m}{\Lambda}=\left(\frac{8}{\re}\right)^{\Delta }\frac{1}{\Gamma (\Delta +1)}.
\end{equation}
The Wiener--Hopf decomposed kernel was also derived therein, finding
\begin{equation}
G_+(\omega)=\frac{\exp\left\{-{1 \over 2} \ri \omega [(1-2\Delta)(\log(-\frac{1}{2}\ri \omega)-1)-2\Delta\log(2\Delta)]\right\}}{\sqrt{-\ri \Delta \omega}}\frac{\Gamma(1-\ri \Delta \omega)}{\Gamma\big(\tfrac{1}{2}-\tfrac{1}{2}\ri\omega\big)}. 
\label{GplusON}
\end{equation}
For our methods, the following constants are determined by \eqref{GplusON}
\be
k=\frac{1}{\sqrt{\pi\Delta}}, \quad \mathfrak{C} = \frac{m}{2\beta_0\Lambda},
\ee
where $\mathfrak{C}$ was chosen according to \eqref{frakC-choice}, which is consistent with the choice of \cite{volin}.

The perturbative series of this model was first computed to high order in \cite{volin} and expanded on analytically in \cite{mr-ren} and \cite{bbv}. Numerically, it was calculated to extremely high order for $N=4$ in \cite{abbh1,abbh2}. The first few orders of \eqref{pert-gs} are as follows
\begin{equation}
\label{on-pert}
\ba
\varphi(\alpha) &=1 +\frac{\alpha}{2}+\frac{\alpha ^2 \Delta }{2}+\frac{\alpha ^4 \Delta}{32}   \left(-8 \Delta ^2 (3 \zeta (3)+1)+14 \Delta  (3 \zeta (3)+2)-21 \zeta (3)+8\right)\\
&+\frac{\alpha ^4 \Delta }{96}  \big(-24 \Delta ^3 (19 \zeta (3)+1)+\Delta ^2 (918 \zeta (3)+60)\\
&\quad-7 \Delta  (87 \zeta (3)-20)+3 (35 \zeta (3)+8)\big)\\
&+\frac{\alpha ^5 \Delta}{6144}  \left(-96 \Delta ^4 (1024 \zeta (3)+405 \zeta (5)+10)+24 \Delta ^3 (8544 \zeta (3)+4185 \zeta (5)+4)\right.\\
&\left.\quad-8 \Delta ^2 (19236 \zeta (3)+12555 \zeta (5)-2200)+12 \Delta  (3878 \zeta (3)+93 (45 \zeta (5)+16)\big)\right.\\
&\left.\quad-9 (980 \zeta (3)+1395 \zeta (5)-256)\right)
+ \CO(\alpha^6).
\ea
\end{equation}

\begin{table}[t]
\begin{center}
\renewcommand{\arraystretch}{1.5}
\begin{tabular}[\textwidth]{||c|c||c|c||}
\hline
\multicolumn{2}{||c||}
{ \textbf{IR Renormalon}}
& Real part/$F(0)$ & Imaginary part\\
\hline 
\multicolumn{2}{|| c||}{$\xi=1$} &  Yes, if $N>4$. & Yes.\\
\hline
\hline
\multicolumn{2}{||c||}
{\textbf{Poles of} $\sigma(\omega)$}
&
\multicolumn{2}{c||}
{\textbf{Residue} $\ri\sigma^\pm$}\\
\hline
$\xi$ Position & Type  & Real part & Imaginary part  \\
\hline 
$(N-2)\{2\ell-1\}$ & Instanton & No. & 
 \begin{tabular}{@{}c@{}}Yes, if $N$ even,\\
No, if $N$ odd.\end{tabular}
 \\
\hline 
$(N-2)\{2\ell\}$ & Instanton & No. & Yes.\\
\hline
$- \{2\ell-1\}$ & UV Renormalon & Yes, if $N>4$. & Yes.\\
\hline
\hline
$\widehat\varphi$ poles & 
\multicolumn{3}{c||}
{$-2\ell,\,2,\,2(N-2)\ell,\quad \ell\in\IN$}
\\
\hline
\textbf{Trans-series terms} & 
\multicolumn{3}{c||}
{
 \begin{tabular}{c}$\re^{-2/\alpha},\,\re^{-2(N-2)\ell/\alpha},\quad \ell\in\IN$, if $N$ even,\\
$\re^{-2/\alpha},\,\re^{-4(N-2)\ell/\alpha},\quad \ell\in\IN$, if $N$ odd.\end{tabular}}
\\
\hline
\multicolumn{1}{||c|}
{\textbf{Leading poles}} & 
\multicolumn{3}{c||}{
 \begin{tabular}{@{}c@{}}
 $\xi_{\text{IR}}=1,\quad \xi_1=\Delta^{-1}\quad N$ even,\\
$\xi_{\text{IR}}=1,\quad \xi_1=2\Delta^{-1}\quad N$ odd.\end{tabular}
 }\\
\hline
\end{tabular}
\end{center}
\caption{Trans-series structure for the $O(N)$ non-linear sigma model with $N\geq 4$.}
\label{table-on}
\end{table}

The trans-series structure that follows from \eqref{GplusON} is summarized in table \ref{table-on}. Before we comment on the results, some notes on how to read the tables in this section:
\begin{itemize}
\item Unless otherwise specified, $\ell$ takes values in $\IN$.
\item The column ``Type'' should be seen as a schematic idea and read in tandem with the accompanying discussion. It is not a definite classification.
\item Residues which are labeled as ``Yes.'' for their real/imaginary part might still have vanishing values for specifics values of $\ell$ and $N$.
\item ``$\widehat{\varphi}$ poles'' refers to the poles of the Borel transform of the perturbative series $\varphi(\alpha)$ extrapolated from the trans-series structure. This includes poles in the negative real axis.
\item ``Trans-series terms'' lists the exponentially suppressed factors one must include in sum \eqref{bos-ts}. Each of these factors would come multiplied by a tran-series parameter, an overall rational power of $\alpha$ and a non-trivial asymptotic series.
\item ``Leading poles'' lists the contributions considered in equations \eqref{bos-ts}, \eqref{C0-generic-gs} and \eqref{C1-generic-gs}.
\end{itemize}

At first non-perturbative order, we find the IR renormalon contribution, which matches the analysis of \cite{volin, mr-ren, abbh2, mmr}. It comes with the trans-series parameter
\be
\CC_0^\pm =-\frac{ \re^{\pm \ri \pi  \Delta  }}{2}\left(\frac{64}{\re^2}\right)^{\Delta }\frac{ \Gamma (1-\Delta )}{ \Gamma (1+\Delta)}.
\label{C0-on}
\ee
The real part of this constant is related to the free energy when $h=0$, which can be read from \eqref{F0-gen},
\begin{equation}
F(0) = \frac{m^2}{8}  \cot (\pi  \Delta ).
\end{equation}
This agrees with the same formula in \cite{saleur}, and extends the result of \cite{bcr}.

After the simple term of the IR renormalons, we find a series of exponentially suppressed effects associated with singularities at
\begin{equation}
\zeta = 2\ell (N-2)
\label{on-sings}
\end{equation}
in the Borel plane. As per \eqref{bos-ts}, these multiply non-trivial asymptotic series. The trans-series parameter of the leading contribution, for $N$ even, comes from the pole at $\xi = (N-2)$
\begin{equation}
\CC_1^\pm = \re^{\mp\frac{\ri \pi }{2 }\left(1+\frac{1}{\Delta}\right)}
 \, 16 \left(2^{\Delta -1} \Delta \right)^{1/\Delta }   \frac{\Gamma \left(\frac{1}{2\Delta }-\frac{1}{2}\right)}{ \re^2\Delta ^2 \Gamma \left(\frac{3}{2}-\frac{1}{2 \Delta }\right)}.
 \label{C1-on}
\end{equation}
For $N$ odd, this constant vanishes and one must look at residue when $\xi = 2(N-2)$.
For $N=4$, using $\Delta=1/2$ in formulae \eqref{C0-on} and \eqref{C1-on} in \eqref{bos-ts} reproduces the leading results of \cite{abbh1,abbh2}.
 
The singularities \eqref{on-sings} disappear when $N\rightarrow \infty$, which makes us speculate that they correspond to unstable instantons. Their position seems to be compatible with the known action of such effects \cite{volin-thesis} and the fact that their residues are purely imaginary seems consistent with their unstable nature. The UV renormalons appear at the conventional positions \eqref{UV-ren}. 

\subsection{\texorpdfstring{$\CN=1$}{N=1} supersymmetric \texorpdfstring{$O(N)$}{O(N)} non-linear sigma model}

The relevant parameters for the \texorpdfstring{$\CN=1$}{N=1} supersymmetric \texorpdfstring{$O(N)$}{O(N)} non-linear sigma model are
\begin{equation}
\begin{gathered}
\Delta=\frac{1}{N-2},\quad \beta_0 = \frac{1}{4\pi\Delta},\quad \xi = 0,
\\
 \quad k = \frac{1}{\sqrt{\pi \Delta}},\quad \frac{m}{\Lambda}=\frac{2^{2 \Delta } \sin (\pi  \Delta )}{\pi  \Delta },\quad \mathfrak{C} = \frac{m}{2\beta_0\Lambda}. 
\end{gathered}
\end{equation}
where we consider only $N>4$. From \cite{eh-ssm}, we read the Wiener--Hopf decomposition of the kernel,
\begin{multline}
G_+(\omega) = \frac{\re^{-\frac{1}{2}\ri(1-2\Delta)\omega[1-\log(-\frac{1}{2}\ri(1-2\Delta)\omega)]} \re^{-\ri \Delta \omega[1-\log(-\ri\Delta\omega)]}}{ \re^{-\ri\omega[1-\log(-\frac{1}{2}\ri\omega)]} \sqrt{-\ri\Delta\omega}}\\
\times
\frac{\Gamma \big(\frac{1}{2}-\frac{1}{2} \ri (1-2 \Delta ) \omega \big) \Gamma (1-\ri \Delta  \omega )}{\Gamma \left(\frac{1}{2}-\frac{1}{2}\ri \omega\right)^2}.
\end{multline}
The perturbative series was found in \cite{mr-ren} to order 42, starting with
\begin{equation}
\label{susyon-pert}
\ba
\varphi(\alpha)&=1 +\frac{\alpha }{2}\\
&-\frac{3}{32} \alpha ^3 \left(\Delta  \left(8 \Delta ^2-14 \Delta +7\right) \zeta (3)\right)+\frac{5}{32} \alpha ^4 \Delta  \left(8 \Delta ^2-14 \Delta +7\right) \zeta (3)\\
&-\frac{15\Delta\alpha ^5 }{2048} 
\big(864 \Delta ^4 \zeta (5)-2232 \Delta ^3 \zeta (5)+8 \Delta ^2 (28 \zeta (3)+279 \zeta (5))
\\
&
\quad-4 \Delta  (98 \zeta (3)+279 \zeta (5))+196 \zeta (3)+279 \zeta (5)\big)
+\CO(\alpha^6).
\ea
\end{equation}

\begin{table}[t]
\begin{center}
\renewcommand{\arraystretch}{1.5}
\begin{tabular}[\textwidth]{||c|c|c||c|c||}
\hline
\multicolumn{3}{||c||}
{\textbf{IR Renormalon}}
& Real part/$F(0)$ & Imaginary part\\
\hline 
\multicolumn{3}{|| c||}{Absent.} & No. & No.\\
\hline
\hline
\multicolumn{3}{||c||}
{\textbf{Poles of} $\sigma(\omega)$}
&
\multicolumn{2}{c||}
{\textbf{Residue} $\ri\sigma^\pm$}\\
\hline
$\xi$ & Position & Type  & Real part & Imaginary part  \\
\hline 
$\xi_{2\ell-1}$ &
$
\begin{aligned}
 &(N-2)\\
 &\times\{2\ell-1\}
\end{aligned}
$
 & Instanton & No. & 
 \begin{tabular}{@{}c@{}}Yes, if $N$ even,\\
No, if $N$ odd.\end{tabular}
 \\
\hline 
$\xi_{2\ell}$ & $(N-2)\{2\ell\}$ & Instanton & No. & Yes.\\
\hline
$\xi'_\ell $ & $  \frac{N-2}{N-4} \{\ell\}$ & New Renormalon &  Yes, if $N>6$. & Yes\footnotemark.\\
\hline
$\xi^{UV}_\ell $ & $- \{2\ell-1\}$ & UV Renormalon & Yes. & Yes.\\
\hline
\hline
\multicolumn{2}{||c|}
{$\widehat\varphi$ poles} & 
\multicolumn{3}{c||}
{$-2\ell,2\xi_\ell, 2\xi'_{\ell'}, 2(\xi_\ell+\xi'_{\ell'})\quad \ell,\ell'\in\IN$}
\\
\hline
\multicolumn{2}{||c|}
{\textbf{Trans-series terms}} & 
\multicolumn{3}{c||}{
 $\re^{-2\xi_\ell/\alpha},\re^{-2\xi'_{\ell'}/\alpha},\re^{-2(\xi_\ell+\xi'_{\ell'})/\alpha},\quad \ell,\ell'\in\IN$
 }
\\
\hline
\multicolumn{2}{||c|}
{\textbf{Leading pole}} & 
\multicolumn{3}{c||}{
$\xi'_1=\frac{N-2}{N-4}$.
 }
\\
\hline
\end{tabular}
\end{center}
\caption{Trans-series structure for the  $\mathcal{N}=1$ supersymmetric $O(N)$ non-linear sigma model with $N> 4$.}
\label{table-susy}
\end{table}
\footnotetext{If $N=5$, the residues with $\ell$ odd vanish.}

The position of the singularities read from $\sigma(\ri \xi)$ are summarized in table \ref{table-susy}.
A first important result is that, at finite $N$, the IR renormalon is entirely absent in this model, both its real and imaginary contributions. Consequently
$\CC_0^\pm =0$ and $F(0)=0$.
This means that the leading asymptotics of the perturbative series are, while divergent, exponentially suppressed when compared with the usual $2^{-k}k!$. This caused some confusion with the original numerical analysis in \cite{mr-ren}, since one could discern the series was asymptotic but it seemed to be off from the expected behavior. 

Beyond the non-existing IR singularity, this model has an interesting non-perturbative structure. 
Its kernel $\sigma(\omega)$ contains two families of poles:
\be
\label{12sson}
\xi_\ell= {\ell \over \Delta}= (N-2)\ell ,\quad \xi'_\ell= {\ell \over 1-2\Delta}= {N-2 \over N-4}\ell,  \qquad \ell \in \IN.
\ee
The former are identical to those found in the $O(N)$ non-linear sigma model, and thus we identify them as ``instanton-like''. The latter are identical to the new family of renormalons we identified in the Gross--Neveu model, and also survive large $N$. Thus, we identify them as ``renormalon-like''. However, much like in the Gross Neveu model, they defy standard renormalon expectations, and thus we dub them ``New Renormalons'' in table \ref{table-susy}. As one extracts the free energy from the integral equations, these sequences will be multiplied and mixed. Hence we expect the final result to have singularities in the Borel plane at
\be
\label{mix-seq}
{2\ell \over \Delta}+{2\ell' \over 1-2\Delta}, \qquad (\ell, \ell') \in \IN^2\setminus\{0\}.
\ee 
From the residue at the first ``renormalon''-like singularity, we find, 
\begin{equation}
\label{susy-cs}
\begin{aligned}
\CC_1^\pm &=  \re^{\mp\frac{\ri \pi}{2}\left(1+\frac{1}{1-2\Delta}\right)}\frac{   \pi  2^{\frac{1}{1-2 \Delta }} (1-2 \Delta )^{\frac{2-2 \Delta }{1-2 \Delta }} }{4 \Delta  \Gamma \big(\frac{\Delta }{2 \Delta -1}\big)^2\sin \big(\frac{\pi -\pi  \Delta }{2 \Delta -1}\big)}.
\end{aligned}
\end{equation}

\subsection{Principal chiral field}

We can apply this analysis to the PCF field in both the charge choice of \cite{pcf}, which we call the ``standard charge choice'', and the setting of \cite{fkw1,fkw2}, which we call FKW charges. From the perspective of the integral equation, this amounts to different choices of $G_+$, as well as some normalisation factors due to the multiplicity of particles. Irrespective of the choice of charge, the following data holds for the PCF model
\begin{equation}
\Delta=\frac{1}{N},\quad \beta_0 = \frac{1}{16\pi\Delta},\quad 
\beta_1 = \frac{1}{256\pi^2\Delta^2},\quad 
\xi = \frac{1}{2},\quad \frac{m}{\Lambda}=\sqrt{\frac{8 \pi }{\re}}\frac{ \sin (\pi  \Delta )}{\pi  \Delta }.
\label{PCFpars}
\end{equation}

\subsubsection{Standard choice of charge}

We consider the PCF in the setting discussed in \cite{pcf}. One has the Wiener--Hopf decomposition of the kernel 
\begin{equation}
G_+(\omega)=\frac{\re^{- \ri \omega  [-(1-\Delta) \log (1-\Delta)-\Delta  \log (\Delta )]}}{\sqrt{-2 \pi\ri (1-\Delta ) \Delta \omega } }\frac{\Gamma (1-\ri (1-\Delta ) \omega ) \Gamma (1-\ri \Delta  \omega ) }{\Gamma (1-\ri \omega )}.
\label{GplusPCF}
\end{equation}
From which follows, with the choice implied by \eqref{frakC-choice},
\begin{equation}
k = \frac{1}{\sqrt{2 \pi  (1-\Delta ) \Delta }},\quad \mathfrak{C} = \frac{m}{8\beta_0\Lambda}.
\label{sPCFpars}
\end{equation}
The long perturbative series was first derived in \cite{mr-ren}, and starts as
\begin{equation}
\ba
\varphi(\alpha)&=1 +\frac{\alpha}{2}+\frac{\alpha ^2}{4}+\frac{1}{16} \alpha ^3 \left(6 {\Delta} ^2 \zeta (3)-6 {\Delta}  \zeta (3)+5\right)\\
&+\frac{1}{96} \alpha ^4 \left(54 {\Delta} ^2 \zeta (3)-54 {\Delta}  \zeta (3)+53\right)\\
&+\frac{1}{384} \alpha ^5 \left(405 {\Delta} ^4 \zeta (5)-810 {\Delta} ^3 \zeta (5)+81 {\Delta} ^2 (7 \zeta (3)+10 \zeta (5))
\right.
\\
&\left.
-81 {\Delta}  (7 \zeta (3)+5 \zeta (5))+487\right)
+\CO\left(\alpha ^6\right).
\ea
\label{pcf-ps}
\end{equation}
As in the case of $O(N)$ sigma model, this series had been sufficient to identify the existence of the IR singularity.

Using our results, we confirm the IR renormalon effect, which is purely a formally ambiguous imaginary value. In other words, $F(0)=0$. The further structure of the trans-series is summarized in table \ref{table-pcf}. Much like in the supersymmetric $O(N)$ sigma model we find two sequences of poles, one ``instanton-like'', which disappears at infinity when $N\rightarrow\infty$, and one ``renormalon-like'', which becomes the ``standard'' renormalons in the same limit. They are given by
\be
\label{xiz-pcf}
\xi_\ell= \frac{\ell}{\Delta}, \qquad \xi'_\ell= \frac{ \ell}{1-\Delta}, \qquad \ell \in \IN.
\ee
However, the residues associated to the ``instanton'' poles $\xi_\ell$ vanish at integer $N$\footnote{We thank Janos Balog for pointing out this correction.}. Thus, in physical realisations, we only have the ``renormalon-like'' non-perturbative effects,
\be
\label{true-pcf}
\xi'_\ell= \frac{N \ell}{N-1}, \qquad \ell \in \IN.
\ee
This shows that the phenomenon of ``new renormalons'' is not always exactly that of Gross--Neveu, nor is it confined to models with fermions.

\begin{table}[t]
\begin{center}
\renewcommand{\arraystretch}{1.5}
\begin{tabular}[\textwidth]{||c|c|c||c|c||}
\hline
\multicolumn{3}{||c||}
{\textbf{IR Renormalon}}
& Real part/$F(0)$ & Imaginary part\\
\hline 
\multicolumn{3}{|| c||}{$\xi=1$} & No. & Yes.\\
\hline
\hline
\multicolumn{3}{||c||}
{\textbf{Poles of} $\sigma(\omega)$}
&
\multicolumn{2}{c||}
{\textbf{Residue} $\ri\sigma^\pm$}\\
\hline
$\xi$ & Position & Type  & Real part & Imaginary part  \\
\hline 
$\xi_{\ell}$ & $N\{\ell\}$ & Instanton & No. & 
 \begin{tabular}{@{}c@{}}No.\\
(Yes, if $N\notin\IN$.)\end{tabular}
 \\
\hline
$\xi'_\ell $ & $  \frac{N}{N-1} \{\ell\}$ & New Renormalon & No. & Yes.\\
\hline
$\xi^{UV}_\ell $ & $- \{\ell\}$ & UV Renormalon & No. & Yes.\\
\hline
\hline
\multicolumn{2}{||c|}
{$\widehat\varphi$ poles} & 
\multicolumn{3}{c||}
{$-2\ell,2,2\xi_\ell, 2\xi'_{\ell'}, 2(\xi_\ell+\xi'_{\ell'}) \quad \ell,\ell'\in\IN$}
\\
\hline
\multicolumn{2}{||c|}
{\textbf{Trans-series terms}} & 
\multicolumn{3}{c||}{
 $\re^{-2/\alpha},\re^{-2\xi_\ell/\alpha},\re^{-2\xi'_{\ell'}/\alpha},\re^{-2(\xi_\ell+\xi'_{\ell'})/\alpha},\quad \ell,\ell'\in\IN$
 }
\\
\hline
\multicolumn{2}{||c|}
{\textbf{Leading poles}} & 
\multicolumn{3}{c||}{
$\xi_{\text{IR}}=1,\quad \xi'_1=\frac{N}{N-1}$,
 }
\\
\hline
\end{tabular}
\end{center}
\caption{Trans-series structure for the  $SU(N)$ principal chiral field with $N\geq 2$.}
\label{table-pcf}
\end{table}
%

The leading trans-series parameters for this model are
\begin{equation}
\label{pcf-cs}
\ba
\CC_0^\pm&= \mp\frac{ 2\ri}{\re (1-\Delta ) \Delta },\\
\CC_1^\pm &= \pm  \frac{2\ri \,\Gamma \big(\frac{\Delta }{1-\Delta }\big)}{\re^{\frac{1}{1-\Delta }} (1-\Delta ) \Gamma \big(\frac{1}{1-\Delta }\big)}.
\ea
\end{equation}
A curious feature of the PCF model with this choice of charges is that all IR singularities derive from purely imaginary residues. Naturally, in the final trans-series we find arbitrary products of the residues with varying reality. However, if the intuition we will develop later that the imaginary part of residues forms the necessary and sufficient trans-series to cancel all ambiguities from the Borel summation of the perturbative series, including higher Alien derivatives, then this would suggest that the PCF model is amenable to median resummation.
In fact, median resummation is known to work in this model at large $N$ \cite{dpmss}. 
From the point of view of the integral equation, this feature comes from the ``simplicity'' of \eqref{GplusPCF} which has no $\log\omega$ dependency ($a=0$ in terms of \eqref{Gplus-bosonic}\footnote{From the point of view of Volin's method, which we shall discuss in the next chapter, this also leads to the disappearance of $\log s$, $\log z$ and $\log B$ terms.}), and thus the ambiguity of the residues come simply from the discontinuity of the square root. The same simplification happens in the $O(4)$ non-linear sigma model and in \cite{abbh1,abbh2} it was shown that median resummation is successful in that model to highly exponentially suppressed order.

\subsubsection{FKW choice of charge}

As for the choice of charges of \cite{fkw1,fkw2}, the kernel introduced in \eqref{kernelfkw}
leads to
\begin{equation}
G_+(\omega) = 2^{\ri \Delta  \omega }\frac{(1-\ri \omega )}{\sqrt{-\ri \omega }} \frac{  \Gamma (1-\ri \Delta  \omega )}{ \Gamma \left(1-\frac{\Delta}{2}-\frac{\ri   \Delta \omega}{2}\right) \Gamma \left(1+\frac{\Delta}{2}-\frac{\ri\Delta\omega}{2} \right)},\quad k= \frac{2 \sin \left(\frac{\pi  \Delta }{2}\right)}{\pi  \Delta },
\end{equation}
and
\begin{equation}
k=\frac{\sqrt{2} \sin \left(\frac{\pi  \Delta }{2}\right)}{\pi  \Delta  }, \quad \mathfrak{C} = \frac{m}{2  \beta_0 \Lambda \,c_{\text{FKW}}},
\end{equation}
with
\begin{equation}
c_{\text{FKW}}= \frac{16}{\Delta\sqrt{\re}}\sin \left(\frac{\pi  \Delta }{2}\right) \sin (\pi  \Delta ) \re^{\frac{\Delta}{2}   \left(\psi^{(0)}\left(1+\frac{\Delta }{2}\right)+\psi^{(0)}\left(1-\frac{\Delta }{2}\right)+2\gamma_E \right)},
\end{equation}
 where $\psi^{(0)}$ is the digamma function. This is a slightly different choice from \eqref{Cfrak-onehalf}, but keeps an already unwieldy perturbative series simpler. We consider $e,\rho, \CF(h)$ as normalized in section \ref{sec_iqft}, see \eqref{rhoe-fkw}, due to the multiplicity of particles.

The perturbative series was studied at large $N$ in \cite{fkw1,fkw2,ksz} and at finite $N$ in \cite{mr-ren}. It is given by
 \begin{equation}
 \ba
\varphi(\alpha)&= 1 +\frac{\alpha ^2}{4}+\frac{1}{8} \alpha ^3 \left(\Delta ^3 Z_{\Delta}(3)-1\right)+\frac{1}{48} \alpha ^4 \left(20-3 \Delta ^3 Z_{\Delta}(3)\right)
 \\
 &+\frac{1}{384} \alpha ^5 \left(81 \Delta ^5 Z_{\Delta}(5)+177 \Delta ^3 Z_{\Delta}(3)-110\right)
+\CO\left(\alpha ^{6}\right).
 \ea
 \label{efkw}
 \end{equation}
We have defined a modified $\zeta$-function
\begin{equation}
Z_{\Delta}(n)=\zeta(n)+\frac{(-1)^{n+1}}{2^n\Gamma(n)}\left(\psi^{(n-1)}\left(1+\frac{\Delta}{2}\right)+\psi^{(n-1)}\left(1-\frac{\Delta}{2}\right)\right).
\label{FKWzeta}
\end{equation}
In higher terms, all terms with a polygamma function or $\zeta$-function can be rearranged into terms with \eqref{FKWzeta}.

The trans-series structure deducted from $\sigma(\omega)$ is written in table \ref{table-fkw}.
A remarkable feature is that the contributions are different from the standard choice of charges. The renormalon-like family $\xi'_\ell$ in \eqref{xiz-pcf} is absent. Instead, we have only the ``instantons'' $\xi_\ell$, which do contribute at integer $N$, as well as the conventional IR singularity. 
While it naively has some different UV structure, this is of little interest since these poles end up corresponding to standard UV renormalons.

\begin{table}[t]
\begin{center}
\renewcommand{\arraystretch}{1.5}
\begin{tabular}[\textwidth]{||c|c|c||c|c||}
\hline
\multicolumn{3}{||c||}
{\textbf{IR Renormalon}}
& Real part/$F(0)$ & Imaginary part\\
\hline 
\multicolumn{3}{|| c||}{$\xi=1$} & No. & Yes.\\
\hline
\hline
\multicolumn{3}{||c||}
{\textbf{Poles of} $\sigma(\omega)$}
&
\multicolumn{2}{c||}
{\textbf{Residue} $\ri\sigma^\pm$}\\
\hline
$\xi$ & Position & Type  & Real part & Imaginary part  \\
\hline 
$\xi_{\ell}$ & $N\{\ell\}$ & Instanton & No. & Yes.
 \\
\hline
$\xi^{UV}_\ell $ & $- 1$ & UV Renormalon & No. & Yes.\\
\hline
$\xi^{UV}_\ell $ & $- N\{2\ell\}+1$ & UV Renormalon & No. & Yes.\\
\hline
$\xi^{UV}_\ell $ & $- N\{2\ell\}-1$ & UV Renormalon & No. & Yes.\\
\hline
\hline
\multicolumn{2}{||c|}
{$\widehat\varphi$ poles} & 
\multicolumn{3}{c||}
{$-2\ell,2,2\xi_\ell,\quad \ell\in\IN$}
\\
\hline
\multicolumn{2}{||c|}
{\textbf{Trans-series terms}} & 
\multicolumn{3}{c||}{
 $\re^{-2/\alpha},\re^{-2\xi_\ell/\alpha},\quad \ell,\in\IN$
 }
  \\
\hline
\multicolumn{2}{||c|}
{\textbf{Leading poles}} & 
\multicolumn{3}{c||}{
$\xi_{\text{IR}}=1,\quad \xi_1=N$.
 }
\\
\hline
\end{tabular}
\end{center}
\caption{Trans-series structure for the  $SU(N)$ principal chiral field with $N\geq 2$, when $h$ couples to the FKW charge choice.}
\label{table-fkw}
\end{table}

\subsection{Fendley's coset models}

Lastly we have the two families of coset sigma models introduced by Fendley in \cite{fendley}, $SU(N)/SO(N)$ and $O(2P)/O(P)\times O(P)$. For both of them, we have
\begin{equation}
\beta_0= \frac{1}{ 16 \pi \Delta},\quad\xi=\frac{1}{2}+\Delta.
\end{equation}
And we recall that for $SU(N)/SO(N)$ we have
\be
\label{SUN-data}
\Delta={1\over N},\quad {m \over\Lambda}= {\sqrt { \pi}}  {2^{3\Delta+2 } \re^{-{1\over 2}-\Delta } \over \Gamma(1- \Delta) \Gamma(1+ 2 \Delta)}. 
\ee
while for $O(2P)/O(P)\times O(P)$,
\be
\Delta={1\over 2(P-1)},\quad {m \over\Lambda}= {\sqrt { \pi}}  {2^{5 \Delta+2 } \re^{-{1\over 2}-\Delta } \over \Gamma(1- \Delta) \Gamma(1+ 2 \Delta)}. 
\label{O2P-data}
\ee
With the appropriate choice of charges, as discussed in section \ref{sec_iqft}, both models are described by the same kernel \eqref{eq_kernel_fend}. This leads to the dual purpose decomposition,
\begin{multline}
G_+(\omega) =\frac{ \re^{ \left(\ri \omega  \left(-\Delta +2 \Delta  \log (2 \Delta )+\left(\frac{1}{2}-\Delta \right) \log \left(\frac{1}{2}-\Delta \right)+\frac{\log (2)}{2}\right)+\ri \Delta  \omega  \log (-\ri \omega )\right)}}{2 \sqrt{-\pi \ri \Delta\omega }}\\
\times\frac{\Gamma (1-2 \ri \Delta  \omega ) \Gamma \left(\frac{1}{2}-\ri \left(\frac{1}{2}-\Delta \right) \omega \right) }{ \Gamma \left(\frac{1}{2}-\frac{\ri \omega }{2}\right)},
\label{Gplus-fendley}
\end{multline}
which can be read in \cite{fendley, mmr-theta}.
We then choose
\begin{equation}
\mathfrak{C}=\frac{\pi ^{3/2} 2^{5 \Delta +\frac{7}{2}} \re^{-\Delta -1/2} \Delta }{\Gamma (1-\Delta ) \Gamma (2 \Delta +1)},
\end{equation}
which respects \eqref{frakC-choice} for both models.

In these models, we consider the observables as defined per excited particle species. 
This is the standard definition for the case $O(2P)/O(P)\times O(P)$, but incorporates a factor of $2$ in $SU(N)/SO(N)$ coming from the two identical particle types in the ground state. The total ground state energy would be twice the ground state per particle, which we present, and so forth. Furthermore, we consider only the case $\Delta<1/2$, since $\Delta=1/2$ reduces to the $O(3)$ non-linear sigma model which requires special treatment.

The perturbative series of the energy density was presented in \cite{mmr-theta},
\begin{multline}
\varphi(\alpha) = 
1+\frac{\alpha }{2}+\alpha ^2 \left(\frac{1}{4}+\frac{\Delta }{2}\right)
+\alpha ^3 \bigg(\frac{5}{16}-\frac{21 \zeta (3)-30}{32} \Delta
+\frac{21 \zeta (3)+8}{16}  \Delta ^2\\
-\frac{1+3 \zeta (3)}{4}  \Delta ^3\bigg)
+\alpha ^4 \bigg(
\frac{53}{96}
+\frac{394-189 \zeta (3)}{192} \Delta 
+\frac{97-105 \zeta (3)}{48} \Delta ^2\\
+\frac{115 \zeta (3)+2}{16} \Delta ^3
-\frac{19 \zeta (3)+1}{4} \Delta ^4
\bigg)
+\CO\left(\alpha ^5\right).
\label{fendley-gs}
\end{multline}

The non-perturbative corrections are summarized in table \ref{table-fendley}. Like in previous cases, there is a family of ``instanton-like'' singularities which disappear at large $N$, and a family of ``renormalon-like'' singularities which not only survives large $N$ but at finite $N$ is associated with unconventional Borel poles. The residues of the instanton singularities suggest that in this case some might be ``stable effects'' (the purely unambiguous real contributions) while others would be ``unstable'' (the purely ambiguous imaginary ones). However, this would require further study before any conclusion. In particular, it would be useful to clarify the relationship of such would-be saddles with the $\mathbb{Z}_2$ topological invariant introduced in section \ref{sec_iqft}, which we will revisit later in this chapter.

Finally, the energy at $h=0$, found through \eqref{F0-gen}, is 
\begin{equation}
F(0) = - \frac{m^2}{16\sin(2\pi\Delta)}.
\label{F0_fendley}
\end{equation}
For $\Delta\leq 1/4$, the leading trans-series parameters read
\begin{equation}
\begin{aligned}
\CC_0^\pm &= 
\bigl[\sin (\pi  \Delta )\mp\ri  \cos (\pi  \Delta )\bigr]
\left(\frac{64}{\re^2}\right)^{\Delta }\frac{ \Gamma \left(\frac{1}{2}-\Delta \right) }{\left(\re \Delta \right) \Gamma \left(\Delta +\frac{1}{2}\right)},\\
\CC_1^\pm &= 
\left[-\sin \biggl(\frac{\pi  \Delta }{1-2 \Delta }\biggr)\pm\ri \cos \biggl(\frac{\pi  \Delta }{1-2 \Delta }\biggr)\right]\\
&\qquad\times
 \left(\frac{1024^{\Delta }}{4 \re^2 (1-2 \Delta )}\right)^{\frac{1}{1-2 \Delta }}
\frac{2 \re \left(1-2\Delta\right) \Gamma \bigl(\frac{1-4 \Delta }{2-4 \Delta }\bigr) }{ \Delta \Gamma \bigl(\frac{3-4 \Delta }{2-4 \Delta }\bigr)}.
\end{aligned}
\label{fenCpm_0}
\end{equation}
For $\Delta=1/3$ one has instead
\begin{equation}
\CC^\pm_1 = -\frac{12 }{\re^4 \pi ^2}\Gamma \left(\frac{1}{4}\right)^4,
\end{equation}
which is invisible to perturbation theory. The leading coefficient after the IR singularity seen by perturbation theory in this case is \eqref{fenCpm_0} with $\Delta =1/3$.

\begin{table}[t]
\begin{center}
\renewcommand{\arraystretch}{1.5}
\begin{tabular}[\textwidth]{||c|c|c||c|c||}
\hline
\multicolumn{3}{||c||}
{\textbf{IR Renormalon}}
& Real part/$F(0)$ & Imaginary part\\
\hline 
\multicolumn{3}{|| c||}{$\xi=1$} & Yes. & Yes.\\
\hline
\hline
\multicolumn{3}{||c||}
{\textbf{Poles of} $\sigma(\omega)$}
&
\multicolumn{2}{c||}
{\textbf{Residue} $\ri\sigma^\pm$}\\
\hline
$\xi$ & Position & Type  & Real part & Imaginary part  \\
\hline 
$\xi_{2\ell-1}$ & $ \frac{1}{2\Delta}\{2\ell-1\}$ & Instanton & 
\begin{tabular}{@{}c@{}}
Yes, except if\\
$1/\Delta-2 \in 4\IN$.
\end{tabular}
 &  
 \begin{tabular}{@{}c@{}}
No, except if\\
$1/\Delta \in 4\IN$.
\end{tabular}
\\
\hline 
$\xi_{2\ell}$ & $\frac{1}{2\Delta}\{2\ell\}$ & Instanton & No. & Yes.\\
\hline
$\xi'_{2\ell-1} $ & $  \frac{1}{1-2\Delta} \{2\ell-1\}$ & New Renormalon & Yes. & Yes.\\
\hline
$\xi'_{2\ell} $ & $  \frac{1}{1-2\Delta} \{2\ell\}$ & New Renormalon & No. & Yes.\\
\hline
$\xi^{UV}_\ell $ & $- \{2\ell-1\}$ & UV Renormalon & Yes. & Yes.\\
\hline
\hline
\multicolumn{2}{||c|}
{$\widehat\varphi$ poles} & 
\multicolumn{3}{c||}
{$-2\ell,2, 2\xi_\ell, 2\xi'_{\ell'},2(\xi_\ell+2\xi'_{\ell'}),\quad \ell,\ell'\in\IN$}
\\
\hline
\multicolumn{2}{||c|}
{\textbf{Trans-series terms}} & 
\multicolumn{3}{c||}{
 $\re^{-2/\alpha},\re^{-2\xi_\ell/\alpha},\re^{-2\xi'_{\ell'}/\alpha},\re^{-2(\xi_\ell+\xi'_{\ell'})/\alpha},\quad \ell,\ell'\in\IN$
 }
 \\
\hline
\multicolumn{2}{||c|}
{\textbf{Leading poles}} & 
\multicolumn{3}{c||}{
 \begin{tabular}{@{}c@{}}
 $\xi_{\text{IR}}=1,\quad \xi_1=3/2\quad \Delta=\frac{1}{3}$,\\
$\xi_{\text{IR}}=1,\quad \xi'_1=\frac{1}{1-2\Delta}\quad \Delta<\frac{1}{3}$.\end{tabular}
 }
\\
\hline
\end{tabular}
\end{center}
\caption{Trans-series structure for Fendley's coset sigma models.}
\label{table-fendley}
\end{table}

\subsection{Brief summary}

In this section, we presented many different models with different physical aspects. Here we summarize some of the main lessons from the ensemble of models.
\begin{itemize}
\item We generally find non-perturbative effects associated with
\begin{itemize}
\item the isolated IR renormalon singularity $\zeta=2$,
\item singularities $\zeta\sim 2N\ell$ that disappear at large $N$, we call them ``instanton-like'',
\item singularities  $\zeta\sim \frac{2\ell}{1-\CO(1/N)}$ that become $\zeta= 2\ell$ at large $N$, which we call ``renormalon-like'',
\end{itemize} 
From the point of view of the integral equation, they appear in equal footing.
\item Instanton-like singularities seem to be associated with stable and unstable instantons, hence their name. However, this connection, while consistent, is speculative.
\item Renormalon-like singularities, with the exception of the IR singularity, always appear as ``new renormalons'', i.e. at non-integer positions. They appear in all models except the $O(N)$ sigma models, so new renormalons are not an uncommon feature.
\item Generically, all these contributions are multiplied by trans-series parameters that include contributions which are priori invisible to perturbation theory. This is a priori in line with ``strong resurgence'' but not the strongest version using median resummation.
\item Conventional IR renormalons other than the leading IR renormalon, as predicted in \eqref{IR-ren}, only appear coincidentally (instanton-like singularities are at integers, and new renormalons can fall on the integer position for particular values of $\ell$). 
\item UV renormalons always appear at negative integers, as predicted by \eqref{UV-ren}.
\end{itemize}
In the next section, we comment on the large $N$ limit with more detail. In the next chapter, we will review some methods to test the predictions of this section.

\section{Trans-series at large \titleN}
\label{largeN_antrans}

When distinguishing ``instanton-like'' and ``renormalon-like'', we invoked the common test that renormalons are believed to contribute in the 't Hooft limit while instantons do not. However, a natural question is what happens to our analytic trans-series as a whole when we take this limit? This is an interesting question since at large $N$ there are many known non-perturbative results for the inspected models. Furthermore, at large $N$ one has much better diagrammatic control of the perturbative series, as we saw in section \ref{cha_largeN}. This elucidates why we call the unconventional poles ``renormalons'' rather than a new name like ``abnormalons''.

One important general remark for this analysis is that
\begin{equation}
\alpha \sim \tilde\alpha \sim 2 \beta_0 \bar g(h)^2,
\end{equation}
with $\bar g$ the renormalized coupling. And, in all the models we study, the convention is such that
\begin{equation}
\beta_0\propto \frac{1}{\Delta} \sim N,\quad N\gg 1.
\end{equation}
Thus, the 't Hooft limit,
\begin{equation}
N\rightarrow\infty,\qquad \lambda = N g\quad \text{finite},
\end{equation}
is equivalent to 
\begin{equation}
N\rightarrow\infty,\qquad \alpha\quad \text{finite}.
\end{equation}
So $\alpha$ and $\tilde\alpha$ are appropriate couplings to use when taking the large $N$ limit.

\subsection{Gross--Neveu model}

We start with the Gross--Neveu model. The large $N$ limit of this theory has been solved in \cite{fnw1,fnw2}. In this limit, the Bethe ansatz integral equation reduces to a chain of integral equations which can be solved in closed form for the first few orders. It can also be calculated from field theoretical tools. Further subleading corrections in $1/N$ were then calculated numerically in \cite{dpmss}.
From these results, we know that the free energy can be expanded as
\begin{equation}
\CF(h) = - \frac{h^2}{2\pi}\sum_{k\geq 0}\Delta^k f_k (h).
\end{equation}

While we did just state that $\tilde\alpha$ is an appropriate coupling for large $N$ expansion, series in $\tilde\alpha$ at large $N$ produce ugly, albeit correct, coefficients. So in the interest of aesthetics and harmonising with the notation \cite{dpmss}, we introduce one more cousin to our family of couplings, 
\begin{equation}
\frac{1}{\bar{\alpha}}-\Delta\log\bar{\alpha}=\log \left(\frac{2h}{m} \right),
\end{equation}
which is related to our previous coupling, $\tilde \alpha \sim \bar{\alpha}$.
In terms of this coupling, the free energy is can be written as
\begin{equation}
\ba
\CF(h) \sim -\frac{h^2}{2\pi}\Bigg\{&
\left(1-\Delta  \bar{\alpha }+\CO(\bar{\alpha }^2)\right)
- \re^{-\frac{2}{\bar{\alpha }}} \bar{\alpha }^{2 \Delta }  \big(\pi\cot (\pi  \Delta )\mp\ri \pi\big)\\
&+
\left(\re^{-\frac{2}{\bar{\alpha }}}\bar{\alpha }^{2 \Delta }\right)^{\frac{1}{1-2 \Delta}}
\left(\pi  \cot \left(\tfrac{\pi  \Delta }{1-2 \Delta }\right)\mp \ri \pi   \right)\\
&\qquad\times
\frac{\big[(1-2 \Delta )^\Delta \Gamma (1-\Delta )\big]^{\frac{2}{1-2 \Delta }}}{\Gamma \left(1-\frac{\Delta }{1-2 \Delta }\right)^2}
\left(1+\frac{2 \Delta ^2 \bar{\alpha }}{1-2 \Delta }+\Delta \CO(\bar\alpha^2)\right)
 \\
&+
\left(\re^{-\frac{2}{\bar{\alpha }}}\bar{\alpha }^{2 \Delta }\right)^{\frac{2}{1-2 \Delta}}
\left(\pi  \cot \left(\tfrac{\pi  \Delta }{1-2 \Delta }\right)\mp \ri \pi   \right)^2\\
&\qquad\times
\frac{\big[(1-2 \Delta )^\Delta \Gamma (1-\Delta )\big]^{\frac{4}{1-2 \Delta }}}{\Gamma \left(1-\tfrac{\Delta }{1-2 \Delta }\right)^4}
\left(\frac{2 \Delta ^2}{(1-2 \Delta )^2}+\CO(\bar\alpha)\right)
\\
&+\CO\left(\re^{-\frac{6}{(1-2 \Delta)\bar\alpha}}\right)
\Bigg\},
\ea
\label{f-alphabar}
\end{equation}
where we incorporate the $m^2$ term inside the curly bracket and we have organized the coefficients to make their leading order behavior in the $\Delta\rightarrow 0$ limit more evident.

Let us inspect how these terms behave in the large $N$ limit.
The trans-monomials themselves change
\begin{equation}
\left(\re^{-\frac{2}{\bar{\alpha }}}\bar{\alpha }^{2 \Delta }\right)^{\frac{\ell}{1-2 \Delta}} = \re^{-\frac{2\ell}{\bar{\alpha }}}\left\{1+2 \Delta  \ell \left(\log \bar{\alpha }-\frac{2}{\bar{\alpha }}\right)+\CO\left(\Delta ^2\right)\right\}.
\end{equation}
We see that the non-perturbative scale becomes $\re^{-\frac{2\ell}{\bar{\alpha }}}$ which was the prediction from standard renormalon lore. This is why the unconventional position of the renormalon poles in the Gross--Neveu model eluded previous large $N$ studies. Immediately, we see that the IR renormalon and the leading new renormalon now contribute at the same order, $\re^{-\frac{2}{\bar{\alpha }}}$. The IR renormalon expands as
\begin{equation}
- \re^{-\frac{2}{\bar{\alpha }}} \bar{\alpha }^{2 \Delta }  \big(\pi\cot (\pi  \Delta )\mp\ri \pi\big) = \re^{-\frac{2}{\bar{\alpha }}}\left\{
-\frac{1}{\Delta }
-2 \log \bar{\alpha }\pm \ri \pi  
+\CO\left(\Delta\right)\right\},
\end{equation}
while the new renormalon contribution is of the form
\begin{multline}
\left(\re^{-\frac{2}{\bar{\alpha }}}\bar{\alpha }^{2 \Delta }\right)^{\frac{1}{1-2 \Delta}}
\frac{\big[(1-2 \Delta )^\Delta \Gamma (1-\Delta )\big]^{\frac{2}{1-2 \Delta }}}{\Gamma \left(1-\frac{\Delta }{1-2 \Delta }\right)^2}
\left(1+\frac{2 \Delta ^2 \bar{\alpha }}{1-2 \Delta }+\CO(\bar\alpha^2)\right)\\
=\re^{-\frac{2}{\bar{\alpha }}}
\Bigg\{
\frac{1}{\Delta}
+
\left(
-\frac{4}{\bar{\alpha }}
+
\big(2 \log \bar{\alpha }\mp \ri \pi   -2 \big)+\CO\left(\bar{\alpha }\right)\right)\\
+
\Delta  
\left(
\frac{8}{\bar{\alpha }^2}
+
\frac{- 8 \log \bar{\alpha }\pm \ri \pi}{\bar{\alpha }}
+\CO(\bar\alpha^0)\right)+\CO\left(\Delta ^2\right)\Bigg\}.
\label{abnormalon-largeN}
\end{multline}
The two renormalons reducing to the same non-perturbative scale is crucial, so that the term in $1/\Delta$ is canceled and the correct overall scaling with $N$ is found. This shows that even the leading order behavior in the large $N$ limit comes from a non-trivial rearrangement of the finite $N$ trans-series. 

Furthermore, notice that the ambiguous terms at order $\Delta^0 \bar\alpha^0$ cancel between the two contributions. The leading imaginary term is at order $\Delta \bar\alpha^{-1}$ and comes from the second line of \eqref{abnormalon-largeN}. This is important because this term is responsible for canceling the leading ambiguity from Borel summation of perturbation theory. In fact, it is known from \cite{fkw1,fkw2,dpmss} that perturbative series for the free energy large $N$ is exactly
\begin{equation}
\tilde\varphi(\bar\alpha) = 1- 2\Delta \sum_{n\geq 1} 2^{-n}\Gamma(n+1) \bar\alpha^n +\CO(\Delta^2).
\end{equation}
The large order behavior is matched by the aforementioned leading imaginary term. Since at large $N$ the perturbative series should be dominated by ring-like diagrams, we conclude that the Borel pole at $\zeta = \frac{2}{1-2\Delta}$ is a renormalon effect. 
From a finite $N$ perspective, ring diagrams sense not only the standard IR renormalon but also the new renormalon behind it. 
 This highlights that the large $N$ analysis can be quite deceiving: the finite $N$ case is not given merely by corrections to the coefficients of the large $N$ trans-series but by a more subtle non-perturbative structure which degenerates into the large $N$ one.

As a further test, one can put together more terms from \eqref{f-alphabar},
\begin{equation}
\ba
f_0(h) &=
\begin{multlined}[t]
1
+\re^{-\frac{2}{\bar{\alpha}}} 
\left(-\frac{4}{\bar{\alpha}}-2+\CO\left(\bar{\alpha} ^2\right)
\right)
+\re^{-\frac{4}{\bar{\alpha}}} 
\big(2+\CO(\bar{\alpha})\big)
+\CO\left(\re^{-\frac{6}{\bar{\alpha}}}\right),
\end{multlined}
\\
f_1(h) &=
\begin{multlined}[t]
-\bar{\alpha} -\bar{\alpha}^2+\CO\left(\bar{\alpha}^3\right)
+\re^{-\frac{2}{\bar{\alpha}}} 
\left(\frac{8}{\bar{\alpha}^2}+\frac{-8 \log (\bar{\alpha} )\pm 4 \ri \pi}{\bar{\alpha} }-4+\CO\left(\bar{\alpha}\right)\right)
\\
+\re^{-\frac{4}{\bar{\alpha}}} 
\left(-\frac{16}{\bar{\alpha}}+\CO\left(\bar{\alpha} ^0\right)\right)
+\CO\left(\re^{-\frac{6}{\bar{\alpha}}}\right),
\end{multlined}
\ea
\label{largeNFGN}
\end{equation}
This corresponds to the leading terms of the trans-series found in \cite{dpmss}. As remarked there, looking at each order in large $N$ one cannot deduce the non-perturbative effects from the large order behavior of the perturbative part. Strikingly, at leading order the perturbative series is trivial but there is a series in $\re^{-2\ell/\bar\alpha}$. This contrasts with the finite $N$ case, where we remarked that strong resurgence likely holds. The fact that ``strong resurgence breaks at large $N$'' is a recurring pattern in this section.

\subsection{\texorpdfstring{$O(N)$}{O(N)} non-linear sigma model}

In the $O(N)$ sigma model, the large $N$ trans-series is very simple. Since the instanton-like poles are exponentially suppressed in $N$, only the expansion in $\Delta$ of the contribution of the IR renormalon matter. Expanding our trans-series result for $e$ in $\Delta$ we find
\begin{equation}
\ba
\frac{e}{\pi\Delta\rho^2} &\approx \alpha \varphi(\alpha) 
-\frac{1}{2} \re^{-\frac{2}{\alpha} } \alpha ^2
+\Delta   \re^{-\frac{2}{\alpha} } \alpha ^2 \left(\mp\frac{\ri \pi}{2}+\log \alpha -\gamma_E +1-3 \log 2\right) \\
&\quad+\re^{-\frac{2}{\alpha} } \sum_{k\geq 2} \Delta^k p^\pm_k(\log\alpha)+\CO\left(\re^{-\frac{2}{\Delta\alpha}}\right).
\ea
\label{ON-largeN}
\end{equation}
The ambiguous imaginary term in this expansion agrees with the prediction from ring diagrams discussed in chapter \ref{cha_largeN} and found in \cite{mmr}. Further terms in $\Delta^k$ are given by degree $k$ polynomials of $\log\alpha$, whose coefficients include both real and ambiguous imaginary components. However, these terms are not very meaningful, they merely result from a poor choice of $\alpha$ at large $N$. The problem with our current definitions of $\alpha$, and $\tilde\alpha$, is that when we express $\rho/m$, or $h/m$, as a function of $\alpha$ there are corrections in $\Delta$ from both the $\log$ term and the mass gap in \eqref{alpha-def-bos} and \eqref{tilde-alpha-bosonic}.

In order to better study the large $N$ limit of the trans-series, let us focus instead on the free energy,
\begin{equation}
\CF(h) = \sum_{k\geq 0} \Delta^{k-1} \CF_{(k)}(h).
\end{equation}
These $\CF_{(k)}(h)$ were computed in \cite{dpmss} to NLO both from a pure large $N$ expansion and numerically from the Bethe ansatz, so they serve as a further non-trivial test. As we discussed in section \ref{cha_largeN}, the results from these techniques are already unambiguously resummed from the perspective of perturbation theory. 

To better handle the large $N$ limit, we introduce one final coupling
\begin{equation}
\frac{1}{\bar \alpha} = \log\frac{h}{m}.
\end{equation}
With this definition, once we express $h/m$ in terms of $\bar\alpha$ there is no mix between the different $\CF_{(k)}(h)$.
Since the large $N$ trans-series terms come entirely from the $m^2$ term,
\begin{equation}
\CF(h) = - \left\{\text{perturbation theory}\right\} -\frac{h^2}{4\pi} \re^{-2/\bar\alpha}\Bigg\{\frac{\pi}{2}\cot(\pi \Delta) \pm \frac{\ri\pi}{2} \Bigg\}+\CO\big(\re^{-N}\big).
\label{mlargeN}
\end{equation}
We expand in $\Delta$, and add the perturbative results invoked in section \ref{cha_largeN}, finding
\begin{multline}
\CF(h) = - \frac{h^2}{4\pi\Delta}\Bigg\{\frac{1}{\bar{\alpha}}-\frac{1}{2}+\frac{\re^{-2/\bar\alpha}}{2}\Bigg\}
 -\frac{h^2}{4\pi}\Bigg\{\varepsilon(\bar{\alpha})+ \re^{-2/\bar\alpha}\left(\pm \frac{\ri \pi}{2}\right) \Bigg\} + \CO(\Delta),
\end{multline}
where $\varepsilon(\bar{\alpha})$ is the asymptotic series found with ring diagrams\footnote{$\varepsilon(\bar{\alpha})$  also includes some additional constants and a $\log\bar\alpha$ term that come from $\Delta$ corrections to the change $\tilde\alpha\rightarrow\bar\alpha$ in the $1/(\Delta\tilde\alpha)$ term.} re-expressed in terms of $\bar\alpha$, which can be found from the results of \cite{mmr}.
This is precisely what was derived in \cite{dpmss} to NLO in $\Delta$. We can see from \eqref{mlargeN} that at each higher order in $\Delta$ the perturbative series must be Borel summable, since there are no ambiguous terms, but there will always be unambiguous ``invisible'' terms proportional to $\re^{-2/\bar\alpha}$ coming from $F(0)$. 

In this model at least, ring diagrams do correspond to a standard renormalon prediction, and this pole does not ``move'' when compared to finite $N$. However, we have once again that at large $N$ strong resurgence visibly does not hold. Even at leading order the perturbative series is trivial but there is a non-perturbative effect. 

\subsection{\texorpdfstring{$\CN=1$}{N=1} supersymmetric \texorpdfstring{$O(N)$}{O(N)} non-linear sigma model}

The large $N$ limit of the \texorpdfstring{$\CN=1$}{N=1} supersymmetric \texorpdfstring{$O(N)$}{O(N)} non-linear sigma model is quite illustrative of the nature of the new renormalons. In this model, the IR renormalon is absent, but the leading new renormalons ``move'' like in Gross--Neveu, $\exp\left(-\frac{2}{(1-2\Delta)\alpha}\right)\rightarrow \exp(-2/\alpha)$. So at large $N$ we expect singularities at $\zeta=2\ell$. Expanding the trans-series we have 
\begin{equation}
\ba
\frac{e}{\pi\Delta\rho^2}&\approx\alpha\varphi(\alpha)+\re^{-\frac{2}{\alpha} } \left(-\frac{\alpha ^2}{2}+\CO\left(\alpha ^3\right)\right)
\\
&+\re^{-\frac{2}{\alpha} }
 \Delta  \left(2 \big(\alpha +\CO\big(\alpha^2\big)\big)\pm\frac{\ri\pi}{2}\big(\alpha^2 +\CO\big(\alpha^3\big)\big)\right)+\cdots,
\label{SUSY-largeN}
\ea
\end{equation}
where $\cdots$ includes both higher corrections in $\Delta$ as well as possibly higher real corrections in $\re^{-4/\alpha}$ even at order $\Delta$. From the results of \cite{mmr}, we know there are no further imaginary terms at order $\Delta$.
Once again we see that at least at order $\Delta^0$ strong resurgence fails.

\begin{figure}
\centering
\includegraphics[width=0.9\textwidth]{figures/ringdiagON.pdf}
\\
\includegraphics[width=0.9\textwidth]{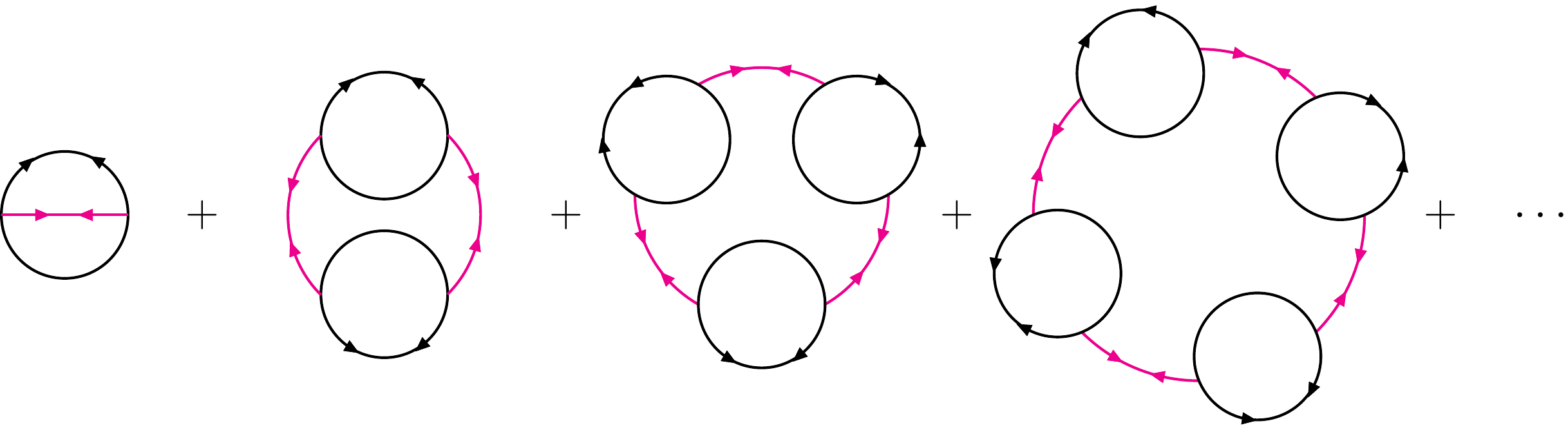}
\caption{The two types of ring diagrams that contribute at leading order in $N$ in the \texorpdfstring{$\CN=1$}{N=1} supersymmetric \texorpdfstring{$O(N)$}{O(N)} non-linear sigma model. At the top, ring diagrams with bosons which also appear in the $O(N)$ NLSM. At the bottom, ring diagrams with both bosons and Majorana fermions. Fermion-only ring diagrams do not contribute at large $N$. See \cite{mmr} for details.}
\label{fig-susy}
\end{figure}

The most enlightening aspect of this model is that it was shown explicitly in \cite{mmr} that the imaginary ambiguity in \eqref{SUSY-largeN} comes from two families of ring diagrams, see figure \ref{fig-susy}. In this case, the only non-perturbative effect that survives at large $N$ are the new renormalons. This model shows clearly that the new renormalons are associated with ring diagrams and thus satisfy the ``perturbative definition'' of a renormalon effect.

\subsection{Principal chiral field}

For the principal chiral field with the standard charge choice, the results of \cite{dpmss} indicate a rich trans-series already at leading order, which is of order $\Delta$. These IR renormalons, which correspond to singularities at $2\ell$, are the large $N$ limit of the new renormalons we identified for the PCF model. We have,
\begin{multline}
\frac{e}{2\pi\rho^2}= \Delta  \Big(\alpha +\CO(\alpha^2)\Big) 
\mp \frac{2\ri\alpha}{ \re}\re^{-\frac{2}{\alpha }} 
\\
\pm\left\{ \frac{2\ri\alpha}{ \re} - \frac{4\ri\Delta}{\re} \big(1 + \CO(\alpha)\big) \right\}\re^{-\frac{2}{\alpha }}  + \CO\left(\Delta\re^{-\frac{4}{\alpha }} \right)+\CO\left(\Delta ^2\right),
\end{multline}
where the non-perturbative term in the first line is the full contribution of the IR renormalon and the terms in curly brackets are the leading contributions of the new renormalon sector. Much like in Gross--Neveu, the IR renormalon cancels with new renormalon at the ``would-be leading order'', leading to the first non-exponential order of the result \cite{dpmss}. Meanwhile, the leading ambiguous term, which dictactes the leading large order behavior of perturbation theory, purely comes from the unconventional renormalon.

It can be checked explicitly that for Fendley models the perturbative series \eqref{fendley-gs} agrees at leading order in large $N$ with that of the PCF model \eqref{pcf-ps}. Coherently the large $N$ limit of the Fendley model leading trans-monomials is also identical to  that of the PCF model, in the appropriate normalisation. However, the identity is only found after the coalescence of the IR renormalon and the leading unconventional renormalon, since the contribution from each term differs from the PCF case.

Lastly, we can also compare the FKW charge of the PCF model setting with known results from the literature. We focus on the large $N$ limit of the free energy, rather than the energy density. Recall that in the FKW setting there are many species of particle  coupled to $h$, so the free energy must be defined as
\begin{equation}
\CF(h) = - \frac{m}{8\sin(\pi\Delta)^2}\int_{-B}^B \epsilon(\theta)\cosh\theta\rd\theta = - \frac{m\re^B }{8\sin(\pi\Delta)^2} \epsilon_+(\ri).
\end{equation}
With this in consideration we find at finite $N$,
\begin{multline}
\CF(h) = - \frac{h^2 \log\frac{h}{m}
}{16\pi\Delta^2\cos\left(\frac{\pi\Delta}{2}\right)^2}
\left\{
1
\mp
\ri
\frac{h^{-2/\Delta}}{m^{-2/\Delta}}\left(
\log\frac{h}{m}\right)^{-\frac{1}{\Delta }}
\frac{\left(\pi\Delta^2/2\right)^{1+\frac{1}{\Delta}}}{\sin^{2/\Delta} \left(\frac{\pi  \Delta }{2}\right)}
+ \cdots\right\} 
\\
\mp  \frac{\ri m^2}{8\sin ^2(\pi  \Delta )} .
\label{FKWfiniteN}
\end{multline}
In the large $N$ limit, the instanton-like terms proportional to $h^{-2/\Delta}$ do not contribute and only the IR renormalon survives,
\begin{equation}
\Delta^2\CF(h) = -\frac{h^2}{16\pi}\left\{\log\frac{h}{m}+\cdots\right\}\mp \ri \frac{m^2 }{8\pi^2}\,.
\label{FKWlargeN}
\end{equation}
In this last equation, the $\cdots$ include the full asymptotic perturbation theory. We know that the leading terms of the long perturbative series \eqref{efkw} agree  with the large $N$ results of \cite{fkw1,fkw2,ksz}, see \cite{mr-ren}.

Let us compare \eqref{FKWlargeN} with the exact large $N$ results of \cite{fkw2}. There they derive that the Borel summation of the perturbative part of the free energy can be written as an integral,
\begin{equation}
 \frac{h^2}{16\pi}s_\pm\left(\log\tfrac{h}{m}+\cdots\right) =   \frac{h^2 B^2}{16\pi}\int_{\CC_\pm} \rd t\, \re^{-t B} {}_2F_1\left(-\frac{1}{2},\frac{3}{2},1,\frac{t^2}{4}\right),
\end{equation}
and they observe that taking the real part reproduces their exact result for the free energy. Recalling that
\begin{equation}
\text{Im} \int_0^{\re^{\pm\ri\epsilon}\infty}\rd t\,\re^{-t B} {}_2F_1\left(-\frac{1}{2},\frac{3}{2},1,\frac{t^2}{4}\right) =  \mp  \frac{2}{\pi }K_1(B)^2,
\end{equation}
and using the large $N$ exact ``boundary condition'' from \cite{fkw2},
\be 
\frac{m}{h} = B K_1(B),
\ee
we find that the imaginary ambiguity from the perturbative series is
\begin{multline}
-\frac{h^2 B^2}{16\pi} \int_{\CC_\pm} \rd t\, \re^{-t B} {}_2F_1\left(-\frac{1}{2},\frac{3}{2},1,\frac{t^2}{4}\right) = \\ -\frac{ B^2}{16\pi} \text{Re}\int_0^{\infty} \rd t\, \re^{-t B} {}_2F_1\left(-\frac{1}{2},\frac{3}{2},1,\frac{t^2}{4}\right) \pm \ri \frac{m^2}{8\pi^2}.
\end{multline}
This ambiguity cancels precisely with the term in \eqref{FKWlargeN}, as desired.

\section{\texorpdfstring{$O(3)$}{O(3)} non-linear sigma model}
\label{sec-O3}

The $O(3)$ non-linear sigma model is a model of particular interest in physics. It is also an integrable ``bosonic'' model but the preceding analysis does not apply. The difference lies in the Wiener--Hopf decomposition of the kernel, which is given by
\begin{equation}
G_+(\omega) = \frac{\re^{-\ri\omega\bigl(\frac{1}{2}-\frac{\log(2)}{2}\bigr) + \frac{1}{2}\ri \omega \log(-\ri\omega)}}{\sqrt{-\ri \omega}} \frac{\Gamma (1-\ri\omega) }{\Gamma \left(\tfrac{1}{2}-\tfrac{1}{2}\ri\omega\right)}. 
\label{eq_G+_O3}
\end{equation}
and does not satisfy $G_-(\ri)=0$. This somewhat general property was used in deriving \eqref{Q-eq-0} and \eqref{epsilon_0}. Without it, new non-perturbative terms appear, changing substantially the structure of the trans-series. As we shall see, they result in non-trivial exponentially suppressed trans-series which are seemingly invisible to perturbation theory. We will further explore this model by adding a $\vartheta$ angle to its Lagrangian in section \ref{sec-theta}.

\subsection{The integral equations}

The additional non-perturbative terms due to $G_-(\ri) \neq 0$ show up in all of the integral equations. First, in \eqref{Q-eq-0} one has an additional residue coming from the integral discussed in \eqref{int-m-Q}, which is the contribution from the terms with $m$ in $g_+$, in particular the term singular at $\omega=\ri$. Accounting for this additional contribution, we have 
\begin{multline}
Q(\ri\xi)-{1\over 2 \pi \ri} \int_{\CC_\pm} \frac{\re^{-2  B\xi'} \delta\sigma (\ri\xi') Q(\ri\xi') }{ \xi+ \xi'} \rd \xi' 
+\sum_{n\geq 1}\frac{\re^{-2B\xi_n}\ri\sigma_n^\pm Q_n}{\xi+\xi_n}
=
\\
\left(\frac{m\re^B}{2}\frac{G_+(\ri)}{\xi-1}-\frac{k h}{\xi\sqrt{\xi}}-\mathfrak{g}(\xi)\right)
+\frac{1}{2\pi\ri}\int_{\CC_\pm} \frac{\re^{-2B\xi'}\delta\sigma(\ri\xi')\mathfrak{g}(\xi')-2\ri kh (\xi')^{-3/2}}{\xi+\xi'}\rd\xi'
\\+ \frac{m\re^B}{2} \frac{\re^{-2B}G_-(\ri)}{\xi+1}-\sum_{n\geq 1}\frac{\re^{-2B\xi_n}\ri\sigma_n^\pm\mathfrak{g}(\xi_n)}{\xi+\xi_n},
\label{Q-eq-O3}
\end{multline}
where the new term is the leftmost term in the third line. Then one can inspect what happens to the boundary condition. Similarly, the sole change is a residue coming from the simple pole of $g_+$ at $\omega=\ri$. We find
\begin{multline}
0= \frac{m \re^ B}{2}G_+(\ri) 
    +\frac{1}{2\pi\ri}\int_{\CC_\pm}\left(\re^{-2B\xi}\delta\sigma(\ri\xi)\left\{Q(\ri\xi)+\mathfrak{g}(\xi)\right\}-2\ri kh\xi^{-3/2}\right) \rd\xi 
  \\
   +\frac{m\re^{B}}{2}\re^{-2B}G_-(\ri)-\sum_{n\ge 1} \re^{-2B\xi_n}\ri\sigma_n^\pm \left\{Q_n+\mathfrak{g}(\xi_n)\right\}.
    \label{epsilon-bc-O3}
\end{multline}

For the free energy, there is a more subtle term coming from
\begin{equation}
\frac{\ri m \re^B}{2}\textrm{Res}_{\omega=\ri\pm 0} \left(\frac{\re^{2\ri B\omega}G_-(\omega)}{(\omega-\ri)^2}\right) = \frac{m \re^{-B}}{2} \left(-2 B G_-(\ri)+\ri G_-'(\ri\pm 0)\right)
\label{m-double-pole}
\end{equation}
as well as residues from the other simple poles at $\omega'=\ri$. In the end, we find
\begin{multline}
    \frac{\epsilon_+(\ri)}{G_+(\ri)}= 
       -\frac{m \re^ B}{4}G_+(\ri)+\frac{1}{2\pi\ri}\int_{\CC_\pm} \left(\frac{\re^{-2B\xi}\delta\sigma(\ri\xi)
       \left\{Q(\ri\xi)+\mathfrak{g}(\xi)\right\}}{\xi-1}+\frac{2\ri kh}{\xi^{3/2}}\right)\rd\xi 
   \\
+   \re^{-2B} \left(\frac{G_-(\ri)}{G_+(\ri)}Q(\ri)-h G_-(\ri) + \frac{m \re^{B}}{2} \left(-2 B G_-(\ri)+\ri G_-'(\ri\pm 0)\right)\right)
   \\ - \sum_{n\ge 1} \frac{\re^{-2B\xi_n} \ri\sigma_n^\pm \left\{Q_n+\mathfrak{g}(\xi_n)\right\}}{\xi_n-1}
    \label{epsilon-0-O3}.
\end{multline}

\subsection{The free energy}

The perturbative part is the same as before, so we can conserve those results using the appropriate $G_+$. 
The non-perturbative corrections organize in a similar way.
Since the singularities of $\sigma(\ri\xi)$, with $G_\pm$ as in \eqref{eq_G+_O3}, are at
\begin{equation}
\xi_n = n, \quad n\in \IN_{\geq 2},
\end{equation}  
the most natural convention for the non perturbative corrections is to label $\xi_n = n$. This way we have $\sigma_1^\pm =0$ but the new terms multiplied by $\re^{-2B}$ are associated with $\xi_1$. The residues $\ri\sigma^\pm_{n\geq 2}$ are purely imaginary ambiguous terms, like the unstable instantons in the $O(N\geq 4)$ NLSM .

Since $Q_1 = Q(\ri)$ no longer contributes to the equation for $Q$, \eqref{Q-eq-O3}, we need only to set $\omega=\ri$ to find it, no recursion necessary. Using the results from perturbation theory, we find that
\begin{equation}
Q(\ri) = - \frac{1}{2}\left(kh \sqrt{2\pi B}+m\re^B \ri G_+'(\ri)\right)+\cdots.
\end{equation}
Much like in the regular bosonic case, exponential corrections to the boundary condition contribute only to sub-leading effects in $1/B$ in the free energy (in fact, sub-sub-leading), and thus we will skip them. 

Plugging the perturbative results in \eqref{epsilon-0-O3}, we finally find 
\begin{multline}
\mathcal{F}(h) = -\frac{k^2 h^2}{4} \Biggl[ \left(B + \frac{\log(B)- 2 + 4\log(2)}{2} + \mathcal{O}\big(B^{-1}\big)\right) 
\\
+  \re^{-2B} \left(-\frac{4 B^2}{\re}+\frac{-2 \log(B) - 3 + 2 \gamma_E -6 \log (2)}{\re}B + \mathcal{O}\big(B^0\big)\right) 
\\
+ \mathcal{O}\big(\re^{-4B}\big)\Biggr] 
\mp \ri \frac{m^2}{16}.
\label{eq_free_energy_exp4B}
\end{multline}
It is more useful to re-express it in terms of $\tilde{\alpha}$, using \eqref{tilde-alpha-bosonic} with $\xi=1$. Incorporating the calculation of higher orders done in \cite{mmr-theta}, we write
\begin{equation}
\label{fh-ts}
\mathcal{F}(h) = -\frac{ h^2}{4 \pi} \biggl[ \mathcal{F}_{(0)}(h) +  \mathcal{F}_{(1)}(h)\re^{-2/\tilde\alpha} + \mathcal{F}_{(2)}(h)\re^{-4/\tilde\alpha} + \mathcal{O}\big(\re^{-6/\tilde\alpha}\big)\biggr] \mp \ri \frac{m^2}{16},
\end{equation}
where
\be
\label{f1h}
\ba
\mathcal{F}_{(0)}(h) &= \frac{1}{\tilde\alpha} - \frac{1}{2} + \mathcal{O}\big(\tilde\alpha\big),\\
\mathcal{F}_{(1)}(h) &= -\frac{64}{\re^2 \tilde\alpha^3} + \frac{32 (-\log\tilde\alpha -3 + \gamma_E +5\log 2)}{\re^2 \tilde\alpha^2}+O\big(\tilde\alpha^{-1}\big),\\
\mathcal{F}_{(2)}(h) &= \frac{512(1 \mp \ri)}{\re^4 \tilde\alpha^3} + \mathcal{O}\big( \tilde\alpha^{-2} \big).
\ea
\ee
This trans-series has several peculiarities. A particularly striking one is the $\log\tilde{\alpha}$ term, which one can track down to extra term proportional to $B$ in the residue of the double pole \eqref{m-double-pole}. More importantly, the entire series $\CF_{(1)}(h)$ is a non-trivial formal power series which is entirely unambiguous. There is an ambiguity of the same exponentially suppressed magnitude, which is given by the single imaginary ambiguous $m^2$ term on the right, but it is much different from the unambiguous contribution of the same order. This imaginary term seems to fulfill the role of the IR renormalon and cancels any ambiguity from the pole of the Borel transform at $\zeta=2$, as we shall test in chapter \ref{cha_volin}. Thus, from the perturbative series alone it seems impossible to deduce the contribution of $\CF_{(1)}(h)$. 

A physical interpretation can be conjectured for this phenomenon, since the $O(3)$ non-linear sigma model, unlike the $O(N\geq4) $ model, has stable instanton solutions. Hence, one might expect that this sector results from the contribution of a stable instanton, with the higher sector coming from many-instanton contributions. In principle, one should be able to work out such a contribution from a path integral calculation, but such a computation is riddled with  divergences. We will gather more evidence for such an interpretation in section \ref{sec-theta} when we introduce the topological $\vartheta$ term.

All these features can also be seen in the canonical formalism,
\begin{multline}
\frac{e}{\pi\rho^2} = \alpha+\frac{\alpha^2}{2} + \mathcal{O}\big(\alpha^3\big)
\\
+ \frac{32}{\re^2}\left\{ \frac{2}{\alpha } + \log\alpha  +3  - \gamma_E - 5\log 2
+\frac{\alpha}{2}
+\mathcal{O}\big(\alpha^2\big) \right\}  \re^{-2/\alpha}\\
\mp  \frac{16\ri\pi}{\re^2}  \re^{-2/\alpha}
+ \frac{512}{\re^4}\left( \frac{1 \pm \ri}{\alpha} + \mathcal{O}\big(\alpha^0\big) \right) \re^{-4/\alpha} + \mathcal{O}\big( \re^{-6/\alpha} \big).
\label{eq_normalized_energy}
\end{multline}
Here the leading amibiguous term is grouped into the $\re^{-2/\alpha}$ sector, as usual. We also include the order $\alpha \re^{-2/\alpha}$ term, which was calculated in \cite{bbh}. Further subleading perturbative corrections to the first sectors have been computed in \cite{bbhv}, analytically to order $\alpha^3$, and  numerically to order $\alpha^5$.

\section{Models with \texorpdfstring{$\vartheta=\pi$}{theta=pi}}
\label{sec-theta}

As we discussed in section \ref{sec-theta-intro}, one can add a topological $\vartheta$ term to the Lagrangian without breaking integrability. This can be done in the $O(3)$ non-linear sigma model and Fendley's coset models, both with $\vartheta=\pi$.  The topological $\vartheta$ term is a purely non-perturbative deformation, and it is interesting to see how it interacts with the trans-series. For the $O(3)$ sigma model, there is a clear prediction \eqref{instanton-action-O3}, that real $n$-instantons effects should pick up a sign $(-1)^n$ according to their action. Furthermore, both models have an IR renormalon and the coset models have new renormalons, so it also provides a clear test of what happens to renormalons under a topological term. As we shall see, the change is identical to what happens to instantons, picking up a sign in their real parts.

In \cite{zz-theta,foz,fendley}, the S-matrices and Bethe ansatze for these models were derived, providing all the necessary tools for our analysis. In these references, a very similar analysis to ours was done but only after deforming the models with an additional parameter. In the $O(3)$ NLSM case, this deformation is called the ``sausage model'' due to the effect of ``stretching'' the curvature of the target sphere\footnote{There is some discussion about which is the best version of string theory, but this one is definitively the wurst.}. However, in both cases, the limits were the deformation is undone and the standard sigma models are recovered is singular, and the structure of trans-series is transformed in a non-trivial way.

Reviewing our results from \cite{mmr-theta}, we will study these systems at $\vartheta=\pi$ without any further deformation.
At first glance, the physics of the system with $\vartheta=\pi$ is radically different, in particular the theory becomes gapless, as shown by Haldane in \cite{haldane}. But, since the term is purely non-perturbative, perturbation theory should remain the same.  Using our techniques, we extract a transseries description of this ground state, and in particular of the free energy
\begin{equation}
\CF(h,\vartheta) = F(h,\vartheta)-F(0,\vartheta).
\end{equation}
In order to compare with the $\vartheta=0$ cases, we start by showing how the Bethe ansatze for the massless theories might be put into a familiar form. Then we derive the pattern of how the trans-series changes when comparing the two angles. Lastly, we detail the physical implications in each model.

\subsection{Wiener--Hopf analysis at \texorpdfstring{$\vartheta=\pi$}{theta=pi}}

In order to connect with our analysis of the $\vartheta=0$ case, we need to render the integral equations \eqref{orig_elr_inteq} into a more workable form using the Wiener--Hopf language developed in this chapter. First, we extend them to the full real line, by setting $\epsilon_{1,2}$ to zero outside their original semi-infinite domains and introducing an unknown function $Y(\theta)$, which is supported in the positive half-line. We also define the extended driving term
\begin{equation}
g(\theta) = \begin{cases}
      th- \dfrac{M\re^\theta}{2} & \theta < B, \\[2mm]
      (t-1)h  &  \theta>B.
\end{cases}
\label{gextend}
\end{equation}
Once again, we remind the reader that the extension of $g$ for $\theta>B$ is arbitrary and different choices amount to redefinitions of $Y$ by subtracting a function with support on $\IR^+$.

To start the Wiener--Hopf analysis of the integral equations (\ref{orig_elr_inteq}), we have to extend their
domain of validity to the full real line. With this choice, we have
\begin{equation}
\begin{aligned}
\epsilon_1(\theta) - \int_{-\infty}^B \varphi_1(\theta-\theta')\epsilon_1(\theta')\rd \theta' - \int^{\infty}_{-B} \varphi_2(\theta-\theta')\epsilon_2(\theta')\rd \theta' &= g(\theta) + Y(\theta-B) , \\
\epsilon_2(\theta) - \int_{-\infty}^B \varphi_2(\theta-\theta')\epsilon_1(\theta')\rd \theta' - \int^{\infty}_{-B} \varphi_1(\theta-\theta')\epsilon_2(\theta')\rd \theta'&= g(-\theta)+ Y(-\theta-B) ,
\end{aligned}
\label{extended_elr_inteq}
\end{equation}
and we can take their Fourier transform,
\begin{align}
\tilde{\epsilon}_1-\phi_1 \tilde{\epsilon}_1- \phi_2 \tilde{\epsilon}_2&=
 \re^{\ri  B \omega}Y_++\re^{\ri  B \omega} g^m_- - th \frac{\ri\re^{\ri B \omega}}{\omega-\ri 0} + (t-1) h\frac{\ri\re^{\ri B \omega}}{\omega+\ri 0},\label{epsilon1_fourier}\\
 \tilde{\epsilon}_2-\phi_2 \tilde{\epsilon}_1- \phi_1 \tilde{\epsilon}_2&=\re^{-\ri  B \omega}Y_-+\re^{-\ri  B \omega} g^m_+  
+ th \frac{\ri\re^{-\ri B \omega}}{\omega+\ri 0} - (t-1) h\frac{\ri\re^{-\ri B \omega}}{\omega-\ri 0}
  .
\label{epsilon2_fourier}
\end{align}
The Fourier transforms of $\epsilon_i$, $\varphi_i$ and $Y$ are noted as $\tilde\epsilon_i$, $\phi_i$ and $Y_+$, respectively. We also  define $Y_-(\omega) = Y_+(-\omega)$, as well as
\begin{equation}
g_-^{m}(\omega) = g_+^{m}(-\omega) = 
\frac{m \re^B}{2} \frac{\ri}{\omega - \ri},
\end{equation}
to better organize the Fourier transform of the driving term $g(\omega)$.

Since we can obtain the free energy \eqref{free-en-massless} from $\epsilon_1$ alone, it is useful to remove $\epsilon_2$ from the problem. \eqref{epsilon2_fourier} leads to
\begin{equation}
\tilde{\epsilon}_2 = \frac{1}{1-\phi_1}\left(\phi_2 \tilde{\epsilon}_1+\re^{-\ri  B \omega} Y_- + \re^{-\ri  B \omega} g^m_+ 
+ th \frac{\ri\re^{-\ri B \omega}}{\omega+\ri 0} - (t-1) h\frac{\ri\re^{-\ri B \omega}}{\omega-\ri 0}
\right).
\end{equation}
And thus \eqref{epsilon1_fourier} can be reduced to
\begin{multline}
\label{yint}
\left(1-\phi_1 - \frac{\phi_2^2}{1-\phi_1}\right) \epsilon_- = Y_+ + g^m_- + \re^{-2\ri B \omega}\frac{\phi_2}{1-\phi_1}\left(Y_-+g^m_+\right)\\
+ h 2\pi \delta(\omega) \left((t-1)+ \frac{t\, \phi_2(0)}{1-\phi_1(0)}\right) 
- h \left(\frac{\ri}{\omega-\ri 0} - \frac{\phi_2}{1-\phi_1}\frac{\ri\re^{-2\ri B \omega}}{\omega-\ri 0}\right),
\end{multline}
with
\be
\epsilon_-(\omega) = \re^{-\ri  B \omega} \tilde{\epsilon}_1(\omega).
\ee
In \eqref{yint}, we made use of \eqref{dirac_WH} to simplify the terms proportional to $h$.
We want to choose the normalization $t$ such that the perturbative theory of $\vartheta=0$ and $\vartheta=0$ are identical when expressed in terms of $h$. As will become evident shortly, this requires us to cancel the Dirac-$\delta$ term with the choice
\begin{equation}
t = \frac{H}{h} = \left(1+\frac{\phi_2(0)}{1-\phi_1(0)}\right)^{-1}.
\end{equation}
This reproduces the choice of $H$ in \cite{foz,fendley}.

As we inch closer to more familiar equations let us introduce a Wiener--Hopf decomposition of the kernel,
\begin{equation}
\label{eff-kernel}
1-\phi_1(\omega) - \frac{\phi_2^2(\omega)}{1-\phi_1(\omega)}  
= \frac{1}{K_+(\omega)K_-(\omega)}.
\end{equation}
In analogy with the $\vartheta=0$ case, define 
\begin{equation}
\sigma(\omega) = \frac{K_-(\omega)}{K_+(\omega)}.
\end{equation}
and
\begin{equation}
Q= K_+ Y_+.
\end{equation}
We will also define a new function $\tilde\sigma$, such that
\begin{equation}
\tilde\sigma(\omega) = \left(\frac{\phi_2(\omega)}{1-\phi_1(\omega)}\right)\sigma(\omega),
\end{equation}
and it is useful to keep in mind that the $\phi_i$ are even.
Using these functions, \eqref{yint} becomes%
\begin{equation}
\ba
\frac{\epsilon_-(\omega)}{K_-(\omega)} &= Q(\omega) +
\re^{-2\ri B \omega}\tilde\sigma(-\omega) K_-(\omega)g^m_+(\omega)  + K_+(\omega)g^m_-(\omega) \\
&\quad+\re^{-2\ri B \omega}\tilde\sigma(-\omega) Q(-\omega)
- \frac{\ri h K_-(\omega)}{\omega-\ri 0}  \left(\sigma(-\omega) - \re^{-2\ri B \omega}\tilde{\sigma}(-\omega)\right).
\label{completeeq_genform}
\ea
\end{equation}
If $\tilde\sigma\rightarrow \sigma$, we recover the $\vartheta=0$ equation \eqref{fourier_WH} with $K$ instead of $G$. To see this, we must use
\begin{equation}
g_-(\omega) = -\frac{\ri h}{\omega-\ri 0}\left(1-\re^{-2\ri B\omega}\right) + \re^{-2\ri B\omega} g_+^m(\omega)+g_-^m(\omega),
\end{equation}
which is a trivial rewriting of \eqref{gplus-def}. In fact, both for the $O(3)$ sigma model and for the coset sigma models, it was derived from explicit calculation of the S-matrices in \cite{foz,fendley} that
\begin{equation}
K_+(\omega) = G_+(\omega),
\end{equation}
where $ G_+(\omega)$ is the Wiener--Hopf decomposition of the kernel used in section \ref{sec_bosonic} for $\vartheta=0$. 

We can write the equations for $Q$ and $\epsilon_+$ by taking the Wiener--Hopf decomposition of \eqref{completeeq_genform},
\begin{equation}
\ba 
Q(\omega)  &= \frac{\ri m \re^B}{2}\frac{G_+(\omega)-G_+(\ri)}{\omega-\ri} + \big[(\re^{2\ri B \omega}\tilde\sigma \{Q+ G_+ g_-^m\})\big]_-(-\omega)
\\
&\quad + \left[\frac{\ri h\left( G_- - \re^{2\ri B \omega} \tilde{\sigma} G_+ \right)}{\omega+\ri 0}  \right]_-(-\omega),\\
\frac{\epsilon_+(\ri\omega)}{G_+(\ri\omega)}&= - \frac{\ri m \re^B}{2}\frac{G_+(\ri)}{\omega+\ri} + \big[(\re^{2\ri B \omega}\tilde\sigma \{Q+ G_+ g_-^m\})\big]_+(\omega)
\\
&\quad+ \left[\frac{\ri h \left( G_- - \re^{2\ri B \omega}  \tilde{\sigma} G_+ \right)}{\omega+\ri 0}  \right]_+(\omega).
\ea
\label{theta-WH}
\end{equation}

We want to show the perturbative expansion of $\CF(h,\vartheta)$ and then derive how the non-perturbative corrections change. Recall from section \ref{sec_bosonic}, that there are three type of contributions in these equations:
\begin{itemize}
\item driving terms not multiplied by $\re^{2\ri B\omega}$ contribute only to the perturbative expansion,
\item terms multiplied by $\re^{2\ri B\omega}$ contribute to perturbation theory through their discontinuity along the positive imaginary axis,
\item non-perturbative terms contributions come from the poles of terms multiplied by $\re^{2\ri B\omega}$.
\end{itemize}
We can see that the first of these three is already identical between $\vartheta=\pi$ and $\vartheta=0$.
Furthermore, we can quickly change between the $\vartheta=0$ and $\vartheta=\pi$ cases by swapping $\sigma\leftrightarrow\tilde\sigma$ in \eqref{theta-WH}. 

We now show that in both the $O(3)$ sigma model and the Fendley coset models, we have that $\tilde\sigma$ is the reflection of the conjugate of $\sigma$ along the positive imaginary axis, that is
\begin{equation}
\tilde \sigma (\ri \xi \pm 0) = - \overline{ \sigma (\ri \xi \pm 0)}=  -\Re\big\{\sigma (\ri \xi)\big\}+\ri\Im\big\{ \sigma (\ri \xi \pm 0)\big\},\quad \xi\in \IR_+.
\label{tildesigma-sigma}
\end{equation}
Since it is the imaginary part of $\sigma(\omega)$ which is discontinuous in the models of interest, this implies
\begin{equation}
\delta \tilde \sigma (\ri \xi) =  \delta \sigma (\ri \xi).
\label{equal-disc}
\end{equation}
It also follows from \eqref{tildesigma-sigma} that the poles of $\tilde\sigma$ lie in the same positions as those of $\sigma$.
Their residues are related using the definition,
\begin{equation}
 \tilde{\sigma}_n^\pm = \lim_{\omega\rightarrow\ri\xi_n\pm 0}  (\omega-\ri\xi_n\mp 0) \tilde\sigma(\omega)
= \lim_{\xi \rightarrow \xi_n} \ri(\xi - \xi_n)\left\{- \overline{\sigma(\ri\xi\pm 0)}\right\},
\end{equation}
which leads to
\begin{equation}
\ri \tilde \sigma_n^\pm =  -\Re\big\{\ri \sigma _n^\pm\big\}+\ri\Im\big\{\ri \sigma _n^\pm\big\},
\label{equalres}
\end{equation}
in the form that residues appear in \eqref{Q-eq-0} and \eqref{epsilon_0}. 
Because $G_+$ is analytic along the positive real axis, it also follows from \eqref{tildesigma-sigma} that
\begin{equation}
\textrm{Res}_{\omega=\ri\pm 0} \left(\frac{\re^{2\ri B\omega}\ri\tilde\sigma(\omega) G_+(\omega)}{(\omega-\ri)^2}\right) = 
-\overline
{\left\{\textrm{Res}_{\omega=\ri\pm 0} \left(\frac{\re^{2\ri B\omega}\ri G_-(\omega)}{(\omega-\ri)^2}\right)\right\}},
\label{equalIRpole}
\end{equation}
which up to overall real factors corresponds to the ``IR singularity'', analogous to \eqref{ir-pole-bos} (or \eqref{m-double-pole} for the $O(3)$ model). Similarly, this also applies to the other residues involving $\tilde\sigma$ (or $\sigma$) at $\omega=\ri$ which contribute to the integral equations in the $O(3)$ non-linear sigma model.

To show \eqref{tildesigma-sigma} in the coset models, including the $O(3)$ sigma model as the special case $\Delta=1/2$, we start by using the known $\phi_1,\phi_2$ from \cite{fendley} to write 
\begin{equation}
\frac{\phi_2 (\omega)}{1-\phi_1(\omega) } = \re^{-2\pi \Delta |\omega|}.
\label{phifraction}
\end{equation}
To continue \eqref{phifraction} from the real line to complex plane we introduce $d_\pm(\omega)$,
\begin{equation}
d_+(\omega) =d_-(-\omega)= \frac{\re^{-\ri \Delta\omega[1-\log(-\ri \Delta \omega)]}}{\sqrt{-2\pi\ri \Delta\omega}} \Gamma(1-\ri \Delta\omega),
\label{eq_dformula}
\end{equation}
which satisfy
\begin{equation}
\frac{\phi_2(\omega)}{1-\phi_1(\omega)}= 1-\frac{1}{d_+(\omega)d_-(\omega)},
\label{eq_ddef}
\end{equation}
 such that
\begin{equation}
\tilde{\sigma}(\omega) = \sigma(\omega) - \frac{\sigma(\omega) }{d_+(\omega)d_-(\omega)}.
\label{sigma-dd}
\end{equation}
Close to the positive imaginary axis, in particular, we obtain
\begin{equation}
d_+(\ri \xi \pm  0) d_-(\ri \xi \pm  0) = \frac{\re^{\pm\ri\pi\big(\Delta\xi-\tfrac{1}{2}\big)}}{2\sin\big(\pi \Delta\xi \big)}, \qquad \xi>0.
\end{equation}
At the same time, from the form of the respective $G_+$ in \eqref{Gplus-fendley}, we can write $\sigma$ as
\begin{equation}
\sigma(\ri\xi\pm 0) = \re^{\pm\ri\pi\big(\Delta\xi-\tfrac{1}{2}\big)}s(\xi) ,
\end{equation}
where $s(\xi)$ is a real function. Then it follows that
\begin{equation}
\frac{\sigma(\ri\xi)}{d_+(\ri\xi)d_-(\ri\xi)} = 2\sin\big(\pi \Delta\xi\big) s(\xi)  = 2\Re\big\{\sigma(\ri\xi)\big\}, \qquad \xi>0,
\end{equation}
which realizes \eqref{tildesigma-sigma} through \eqref{sigma-dd}. 

Because the discontinuities of $\tilde\sigma$ and $\sigma$ are identical, perturbation theory is identical for both values of $\vartheta$, as expected. As for non-perturbative terms in the integral equation, which come from residues such as  \eqref{equalres} and \eqref{equalIRpole},  the real part changes sign while the imaginary part stays the same. If we focus on the leading non-perturbative  contribution, this means that if in the absence of the topological $\vartheta$ term, we have an effect of the form
\begin{equation}
(\Re \CC_1 \pm \ri \Im \CC_1 ) \re^{-2B\xi_1}, \quad \vartheta=0,
\end{equation}
when we turn the $\vartheta$ term to $\pi$, we will have 
\begin{equation}
(-\Re \CC_1 \pm \ri \Im \CC_1 ) \re^{-2B\xi_1}, \quad \vartheta=\pi.
\end{equation}
That the ambiguous imaginary part does not change is quite important. As we have discussed, these formally ambiguous imaginary contributions cancel the imaginary ambiguities from Borel summation of the divergent asymptotic series coming from perturbation theory. Since perturbation theory remains the same, the cancelation of ambiguities must as well. It is the real part, which is invisible to perturbation theory, which changes sign. 

For higher orders, the idea is schematically the same but more complicated. The final trans-series mixes the non-perturbative contributions of the integral equation, and one gets arbitrary products of the residues. On the other hand, the real unambiguous exponentially suppressed terms might themselves be asymptotic and thus require doubly exponentially suppressed imaginary ambiguous terms to cancel the ambiguity of their Borel resummation. Naturally, since these real terms change signs, those higher order imaginary ambiguities must also change sign. This is precisely compatible with the structure of the final trans-series, where for example the product of two residues $\sigma_1$ will lead to effects such as
\begin{equation}
\ba
(\Re \sigma_1 \pm \ri \Im \sigma_1)^2 \re^{-4B\xi_1} = \big(\Re (\sigma_1)^2 \pm 2\ri  \Im \sigma_1 \Re \sigma_1\big) \re^{-4B\xi_1} ,\quad \vartheta =0,\\
(-\Re \sigma_1 \pm \ri \Im \sigma_1)^2 \re^{-4B\xi_1} = \big(\Re (\sigma_1)^2 \mp 2\ri  \Im \sigma_1 \Re \sigma_1\big) \re^{-4B\xi_1} ,\quad \vartheta =\pi,
\ea
\end{equation}
as desired. 

One could presumably formulate the intricacies of this pattern with the language of Alien calculus. However, the emerging philosophy is simpler: terms which are related to the cancellation of ambiguities of perturbation theory remain the same, while terms invisible to the perturbation theory have their sign changed in a predictable pattern. Furthermore, using this real part/imaginary part dichotomy in the integral equation, one can track which terms are ``seen'' by perturbation theory and which are not. The real part of the residues is naively invisible to perturbation theory while the imaginary part is necessary and sufficient to cancel all ambiguities of Borel summation. One could naturally speculate whether this pattern still holds in models which do not admit a $\vartheta$ angle. It is certainly trivially true at leading exponential order. We find it likely to be the case, but present techniques do not allow us to test it.

The physical consequences of the trans-series at $\vartheta=\pi$ are distinct for the $O(3)$ non-linear sigma model  when compared to the other coset sigma models, so we will discuss them separately.

\subsection{\texorpdfstring{$O(3)$}{O(3)} non-linear sigma model at \texorpdfstring{$\vartheta=\pi$}{theta=pi}}

As we saw in section \ref{sec-O3}, in the $O(3)$ non-linear sigma model there are two distinct type of non-perturbative contributions in the integral equations. There are real unambiguous terms coming from the pole at $\omega=\ri$, which are proportional to $\re^{-2B}$, and there are the residues $\ri\sigma_n^\pm$, which are purely imaginary and contribute at order $\re^{-2Bn}$. There is also the imaginary ambiguous part of the pole at $\omega=\ri$ which plays the role of the IR renormalon. Because the real terms in the final trans-series all come from powers of the contributions of $\omega=\ri$ residues, and since they change their sign according to \eqref{equalIRpole}, we can find the trans-series at $\vartheta=\pi$ by tracking the effects of $\re^{-2B}\rightarrow(-\re^{-2B})$ in \eqref{Q-eq-O3}, \eqref{epsilon-bc-O3} and \eqref{epsilon-0-O3}.
With this prescription, we can take the $\vartheta=0$ result for energy density \eqref{eq_normalized_energy} and obtain for $\vartheta=\pi$:
\begin{multline}
\frac{e}{\pi\rho^2} = \alpha+\frac{\alpha^2}{2} + \mathcal{O}\big(\alpha^3\big)
- \frac{32}{\re^2}\left( \frac{2}{\alpha } + \log(\alpha ) +3  - \gamma_E - 5\log(2) 
+\frac{\alpha}{2}
+\mathcal{O}\big(\alpha^2\big) \right)  \re^{-2/\alpha}\\
\mp \frac{16\ri\pi}{\re^2}\re^{-2/\alpha}
+ \frac{512}{\re^4}\left( \frac{1 \pm \ri}{\alpha} + \mathcal{O}\big(\alpha^0\big) \right) \re^{-4/\alpha} + \mathcal{O}\big( \re^{-6/\alpha} \big).
\label{eq_normalized_energy_theta_pi}
\end{multline}
This result was numerically tested in \cite{mmr-theta}, validating both the sign change at order $\re^{-2/\alpha}$ and the lack of it at leading order in $\re^{-4/\alpha}$.

The change is precisely what we would expect if the real exponentially suppressed terms come from stable instantons whose action obeys \eqref{instanton-action-O3}. Specifically, the instantons with charge $\CQ=1$ should pick up a minus sign from $\re^{\ri\pi\CQ}$, but the two instantons term with charge $\CQ=2$ does not.

One can conjecture that for intermediate values of $\vartheta$ the transseries follows from applying the rule $\re^{-2B}\rightarrow \cos(\vartheta) \re^{-2B}$ in \eqref{Q-eq-O3}, \eqref{epsilon-bc-O3} and \eqref{epsilon-0-O3}. This keeps the final result real and does not seem to disturb the sensitive cancellation of ambiguities from Borel summation. We would find
\be
\CF(h, \vartheta)=  -\frac{ h^2}{4 \pi}  \mathcal{F}_{(0)}(h) \mp \ri \frac{m^2}{16}-\cos(\vartheta) \frac{ h^2}{4 \pi}  \mathcal{F}_{(1)}(h)\re^{-2/\tilde\alpha} + \mathcal{O}\big(\re^{-4/\tilde\alpha}\big), 
\label{cos-trans-series}
\end{equation}
where $\mathcal{F}_{(0)}(h)$ and $\mathcal{F}_{(1)}(h)$ were defined in \eqref{f1h}.
 This cannot be tested against the solution of integral equations because integrability is lost, though one could test it with lattice methods, like in \cite{bruckmann}. A natural observable that codifies this behavior is the  topological susceptibility,
 \be
\chi_t(h)= \left( {\rd^2 \CF(h, \vartheta) \over \rd \vartheta^2} \right)_{\vartheta=0},
\ee
  which \eqref{cos-trans-series} predicts to be 
\be
\label{chith}
\chi_t(h)\approx  \frac{ h^2}{4 \pi}  \left(\frac{64}{\re^2 \tilde\alpha^3} - \frac{32 (-\log(\tilde\alpha) -3 + \gamma_E +5\log (2))}{\re^2 \tilde\alpha^2}+O\big(\tilde\alpha^{-1}\big) \right)\re^{-2/\tilde\alpha}. 
\ee
The advantage of studying \eqref{chith} is that it is purely non-perturbative, evading contributions from perturbation theory.

\subsection{Fendley's coset models at at \texorpdfstring{$\vartheta=\pi$}{theta=pi}}

For Fendley's coset models, we don't have the purely real contributions from the poles at $\omega=\ri$. Instead, we have the more standard bosonic model, where non-perturbative effects come from either the IR renormalon or the residues $\ri \sigma_n^\pm$, following the rules \eqref{equalres} and \eqref{equalIRpole}. As we previously discussed, the real part of the IR renormalon is tied to the free energy when $h=0$. Therefore, the change of sign in the real part implies that the free energy, given by \eqref{F0_fendley} at $\vartheta=0$, becomes
\begin{equation}
F(0, \pi) = \frac{m^2}{16\sin(2\pi\Delta)}.
\end{equation}
The impact of the change in the residues can be seen in trans-series parameters. Instead of \eqref{fenCpm_0} we find, for $\Delta<1/4$, 
\begin{equation}
\begin{aligned}
\CC_0^{\pm}(\pi) &= 
\bigl[-\sin (\pi  \Delta )\mp\ri  \cos (\pi  \Delta )\bigr]
\left(\frac{64}{\re^2}\right)^{\Delta }\frac{ \Gamma \left(\frac{1}{2}-\Delta \right) }{\left(\re \Delta \right) \Gamma \left(\Delta +\frac{1}{2}\right)},\\
\CC_1^{\pm}(\pi) &= 
\left[\sin \biggl(\frac{\pi  \Delta }{1-2 \Delta }\biggr)\pm\ri \cos \biggl(\frac{\pi  \Delta }{1-2 \Delta }\biggr)\right]\\
&\qquad\times
 \left(\frac{1024^{\Delta }}{4 \re^2 (1-2 \Delta )}\right)^{\frac{1}{1-2 \Delta }}
\frac{2 \re \left(1-2\Delta\right) \Gamma \bigl(\frac{1-4 \Delta }{2-4 \Delta }\bigr) }{ \Delta \Gamma \bigl(\frac{3-4 \Delta }{2-4 \Delta }\bigr)}.
\end{aligned}
\label{fenCpm_pi}
\end{equation}

It is important to note that the coefficient $\CC^\pm_1$ in \eqref{fenCpm_pi} comes from the effect of the leading new renormalon, see table \ref{table-fendley}. Furthermore, these renormalon effects have in general real parts, which will be affected by the topological $\vartheta$ term. This shows that, in these models, the real non-perturbative effects from renormalons are changed by the introduction of the topological term, in much the same way that instantons are. This had been anticipated by \cite{grunberg,david-cpn}.

In these coset models, there are also singularities which we speculatively associated with instanton effects, both stable and unstable, listed in table \ref{table-fendley}. Those which have a real part should also change sign under \eqref{equalres}, as expected. In fact, once again, admitting that these singularities are indeed instantons, our Bethe ansatze methods are completely agnostic to the semi-classic status of the singularities. Instantons and renormalons appear identically as poles of $\sigma$ and transform identically under the $\vartheta$ topological term.

\chapter[Testing renormalons with exact series][Testing renormalons with exact series]{Testing renormalons \\ with exact series}
\label{cha_volin}

In this chapter, we review how to obtain the large order behavior of perturbation theory and how to use it to test the analytic trans-series.
In the first part, sections \ref{sec_volin} and \ref{sec_iqft_pert}, we review Volin's method as used in \cite{mr-ren, mr-long}, with additional theoretical clarifications. Although not entirely rigorous, this presentation hopes to more precisely motivate the elements of Volin's method, which can otherwise appear somewhat miraculous. It should also serve as an introduction to the method, for those seeking to apply it or generalize it. In the second part of the chapter, section \ref{sec-tests}, we review the methods used to numerically test the results of chapter \ref{cha_antrans}, which are based on the series calculated with Volin's method. This contains the numerical analysis of \cite{mr-ren, mmr-antrans}. We are not comprehensive in presenting all the tests that have been done in the different models, but rather focus on laying down the tools of each type of test paired with the presentation of some examples.

\section{Motivation}

In chapter \ref{cha_antrans}, we found a series with perturbative and non-perturbative terms which we associated with the trans-series expansion of the observable. This step was not fully rigorous: we never showed unequivocally that the energy is given by the Borel summation of the trans-series we identified. For example, it is a priori possible we missed some subtle non-perturbative contribution. It is important to check through other methods that the trans-series we wrote are indeed correct and, at least to testable order, their Borel summation is consistent and matches the exact result.

That Borel summation is the correct procedure to sum the trans-series found  chapter \ref{cha_antrans} is a well motivated assumption. As we discussed in chapter \ref{cha_resurgence}, it is the natural method for summing trans-series, as has been rigorously shown in the study of non-linear ODEs (see e.g. \cite{costin-ode}). Thus, we would expect that Borel summation retrieves the exact result from its trans-series expansion, even though this does not immediately follow from our calculations. As we show in this chapter, numeric tests validate that we have the correct trans-series and that Borel summation reproduces the correct result to available precision.

In the previous chapter, we performed some analytic tests of the trans-series at large $N$.  However, our results also allow for numeric tests at finite $N$. 
There are two types of test that follow from the assumption that the Borel summation of the trans-series we found gives the exact unambiguous result. First, ambiguities coming from the Borel summation of the perturbative series must cancel the ambiguities of the trans-series parameters. That is, the large order behavior of the perturbative series must be the one dictated by our trans-series. Second, once we subtract from the exact result the unambiguous part of the Borel summation of the perturbative series, what is left should be the 
the Borel summation of the non-perturbative terms. As we only know the first few terms, we can but test the behavior for very small coupling. Since numeric values of the exact result can be obtained with great precision from the integral equation \eqref{iqft_geneq}, the key requirement to test our trans-series is to calculate the perturbative series to very high order.

The task of finding the perturbative series to very high order in an integrable field theory at finite $N$ is a difficult one. From field theoretic methods alone, it is borderline impossible, and one can rarely make it past $11$ or so coefficients in a generic field theory \cite{dunne-phi4}. Lattice methods are another option, but they are extremely costly. Naturally, one would like to use the mighty power of integrability to extract these series, but the integral equations at weak coupling are not easy to expand. While these problems have a long history, only the first few orders were known until very recently. In \cite{volin, volin-thesis}, Volin developed an astoundingly useful method that transforms the unwieldy integral equations into recursive algebraic equations, allowing for the determination of long perturbative series, both exactly and numerically. Volin's original formulation was targeted at the $O(N)$ non-linear sigma model. In \cite{mr-ren, mr-long}, we reformulated part of the method to make it easier to generalize to the models we studied in chapter \ref{cha_antrans}, and to non-relativistic systems, which we review in  chapter \ref{cha_GY}. We review this method in the next section.

To conclude this introduction, we note that, in good scientific tradition, the position of this chapter in this thesis is completely ahistorical. The history of the perturbative expansion of this family of integral equations starts arguably in the 1870's with \cite{kirchhoff} where it is implicit in the problem of the disk capacitor, which we leave to the end of appendix \ref{Lieb--Liniger}. In
integrable models, it is first explored with non-relativistic systems like \cite{popov} in the 1960's, which we only discuss in the beginning of chapter \ref{cha_GY}.  In integrable \textit{relativistic} field theories, a significant part of the literature ranges from the late 1970's, e.g. \cite{zamo-zamo}, to the 1990's, with the mass gaps calculations of \cite{hn, hmn, pcf, fnw1,fnw2,eh-ssm} among others. In 2009, Volin discovered his method in \cite{volin}, after which we used it to study numerically the leading renormalon in integrable theories in 2019 \cite{mr-ren}, which is the object of study of this chapter. Only then, motivated by our results with large order behavior of perturbation theory did we attack the non-perturbative effects hidden in Bethe ansatz in \cite{mmr-antrans}, which we already presented in chapter \ref{cha_antrans}. Because the results of \cite{mmr-antrans} completely supersede those of \cite{mr-ren}, it is not useful to discuss the latter before the former. The end result is that 
if a tachyon were to read this thesis, they would find the concepts to be in chronological order.

\section{Generalized Volin's method}
\label{sec_volin}

The method we call ``Volin's method'' is the procedure presented in \cite{volin,volin-thesis} to systematically solve the equation for the ground state rapidity distribution of the $O(N)$ non-linear sigma model as an expansion in $1/B$ to arbitrary order. In this section, we will present a slightly generalized framework based on this method which is applicable to integral equations of the form
\begin{equation}
f(\theta) - \int_{-B}^B K(\theta-\theta') f(\theta')\rd \theta' = m \cosh(p \theta),\, -B<\theta<B,
\label{volin_mostgeneral}
\end{equation}
for some $m$, $p\in \IR_0$. $K(\theta)$ is an even continuous integrable function and $B$ is the parameter which we want to expand in the $B\gg 1$ limit. Form \eqref{volin_mostgeneral} includes, for example, equation \eqref{volin_eq_TBA_GY} for $p=0$ and $m=2$ and equations \eqref{iqft_geneq} for $p=1$ and $m$ the mass of the excited particle. The case of integral equations with a polynomial function instead of a hyperbolic cosine on the r.h.s. is treated in appendix A of \cite{mr-hubbard}.

The weak coupling limit (i.e. $1/B\ll 1$) of these equations is very challenging and traditional techniques cannot go beyond the first few orders (unlike, curiously, the strong coupling limit). Iterative methods, for example, are difficult to apply. Volin's method tackles this by introducing the resolvent and exploring two distinct limits of the equation when $B\gg 1$, the \textit{bulk limit} and the \textit{edge limit}. The coherence between these two limits imposes sufficient algebraic constraints to fix the appropriate ansatze, which are also provided by the procedure.
 
\subsection{The resolvent}

One of the crucial tools of Volin's method is that it reorganizes the problem around the resolvent, rather than the function $f(\theta)$ itself. We define the resolvent as
\begin{equation}
R(\theta) = \int_{-B}^B  \frac{f(\theta')}{\theta-\theta'}\rd \theta'\,,
\label{volin_res_def}
\end{equation}
which we extend over the complex plane.
The resolvent has two key properties. Firstly, it encodes the function $f$ in its discontinuity
\begin{equation}
R(\theta+\ri 0) - R(\theta-\ri 0) = 
- 2\pi \ri f(\theta) 
,
\quad \theta \in [-B,B].
\label{volin_res_disc}
\end{equation}
Secondly, it encodes the momenta of $f$ in its expansion at infinity
\begin{equation}
R(\theta) \approx \sum_{k=1}^\infty \frac{1}{\theta^{k+1}} \int_{-B}^B \rd \theta'\left(\theta'\right)^k f(\theta')\,. 
\label{volin_mom_from_res}
\end{equation}
One important complex-analytic feature of the resolvent is that, by construction, it must vanish at infinity $R(\infty)\rightarrow0$. This kills any holomorphic part that could be added without modifying the moments or the discontinuity, making it unique.
For the edge limit, it will be useful to introduce the inverse Laplace transform of the resolvent as well,
\begin{equation}
\hat{R}(s)=\int_{-\ri\infty+0}^{\ri\infty+0}\frac{\re^{s z}  }{2\pi \ri}R(B+z/2) \rd z \Leftrightarrow R(B+z/2) = \int_0^\infty \re^{-s z} \hat R(s) \rd s.
\label{volin_lapres}
\end{equation}

\subsection{Bulk limit}

For the bulk limit, we want to study the integral equation in the limit
\begin{equation}
B\rightarrow \infty,\quad \frac{\theta}{B} \quad \text{finite}.
\end{equation}
It is elucidating to work out a specific example before presenting the general framework. Let us start with the integral equation for Gaudin--Yang model, \eqref{volin_eq_TBA_GY}.
Recall that for this equation it is conventional to write $x$ instead of $\theta$. 
Following \cite{volin,mr-long}, we introduce 
the shift operator
\begin{equation}
D R(x) = R(x+\ri)\,,\quad D = \re^{\ri \partial_x}\,.
\end{equation}
We recall that the resolvent has a discontinuity in the real interval $[-B,B]$ so some care should be taken on whether we are working above or below the real axis. We define $R^\pm(x)\equiv R(x\pm \ri 0)$, or equivalently as $R(x)$ restricted to the upper(lower) half complex plane. We can then rewrite  \eqref{volin_eq_TBA_GY} as
\begin{equation}
(1+D)R^+(x)-(1+D^{-1})R^-(x) = - 4 \pi \ri\,.
\label{volin_gy_Rshifts}
\end{equation}

For the bulk analysis, we want to inspect the behavior of the equation far away from the edges in the limit that $B$ is large. Thus, we introduce the ``dimensionless'' coordinate $u=x/B$ and expand in $1/B$. We have
\begin{equation}
\left(2+\sum_{n=1}^\infty \frac{\ri^n}{B^n}\partial_u^{(n)}\right)R^+(u)-\left(2+\sum_{n=1}^\infty \frac{(-\ri)^n}{B^n}\partial_u^{(n)}\right)R^-(u) = -4 \pi \ri\,,
\end{equation}
Let us posit that 
\begin{equation}
R(u) = R_0(u)+\frac{1}{B} R_1(u) + \frac{1}{B^2}R_2(u) +\cdots
\label{Rbulk_powers}
\end{equation}
and denote the discontinuity $\delta R_n(u)\equiv R_n^+(u)-R_n^-(u)$. This is the opposite convention of $\delta$ used in the previous chapter. To first order
\begin{equation}
2(\delta R_0(u)+2\pi \ri)+\frac{1}{B}\left(\delta R_1(u)+\ri\partial R_0^+(u)+i\partial R_0^-(u)\right)+\mathcal{O}\left(\frac{1}{B^2}\right)=0\,.
\end{equation}

Constraining that $R(u)$, from its definition, must be an odd function possibly discontinuous at $(-1,1)$ with at most singularities at $\pm 1$ and analytic elsewhere. This is sufficient to solve that
\begin{equation}
\ba
\delta R_0(u) = -2\pi \ri &\Rightarrow R_0(u) = -\log\left(\frac{u-1}{u+1}\right)\\
&\Rightarrow R_1(u)=\frac{1}{\pi}\frac{1}{u^2-1}\log\left(\frac{u-1}{u+1}\right)+r_1(u)
\ea
\label{R0R1sol}
\end{equation}
where $r_1(u)$ is a function with no discontinuity. With the constraints imposed by symmetry, analyticity and asymptotics, we assume it to be of the form $r_1(u)= r_1 u/(u^2-1)+\cdots$. Since the next order introduces $R_2(u)$ which will have a non discontinuous part as well, the coefficient $r_1$ will never be fixed by this method alone.

In order to see a pattern for higher order, we will rewrite \eqref{volin_gy_Rshifts} in a way which highlights the recursive perturbative relations. For this, it is useful to notice that, since in the bulk limit we treat the operators perturbatively, we can put aside problems of analytic continuation and formally treat them as power series of derivatives. After all, at finite $u$ and large $B$, a shift $D u = u + \ri/B$ is too small to cross the discontinuity. So we can apply, for example, $D^+$ to $R_-$ without worrying about branch cuts.

 Start then by applying $(D^{1/2}+D^{-1/2})^{-2}$ to \eqref{volin_gy_Rshifts}
\begin{equation}
\frac{D^{1/2}}{D^{1/2}+D^{-1/2}}R^+(x)-\frac{D^{-1/2}}{D^{1/2}+D^{-1/2}}R^-(x) = - \pi \ri\,.
\end{equation}
Then reorganize
\begin{equation}
R^+(x)-R^-(x)=-\frac{D^{1/2}-D^{-1/2}}{D^{1/2}+D^{-1/2}}(R^+(x) +R^-(x))-2\pi\ri.
\label{volin_gy_eq_bulk_comp}
\end{equation}
where the rational functions of operators should be understood purely in the sense of power series of derivatives. We can see that the operator on the r.h.s. is proportional to $1/B$, which allows us to define each order from the previous ones.  Thus, at each step in the expansion of the bulk equation we will have something of the form
\begin{equation}
\delta R_m(u) = \sum_{j=0}^{m-1} O_{m,j} \,\partial^{(m-j)}(R^+_j(u)-(-1)^{m-j}R^-_j(u))\,,
\label{volin_perteq}
\end{equation}
for some coefficients $O_{m,j}$.

Equation \eqref{volin_perteq} is solved by a combination of terms of type
\begin{equation}
R_m(u)=\sum_{j=1}^m\sum_{k=0}^m c_{j,m-j,k} \frac{u^{1-k\bmod2}}{(u^2-1)^j}\log\left(\frac{u-1}{u+1}\right)^k\,.
\label{volin_eq_term_m}
\end{equation}
It can be inductively verified that the terms  \eqref{volin_eq_term_m} are closed under  \eqref{volin_perteq}, in the sense that taking up to $m$ derivatives of $R_{j<m}$ does not generate terms which cannot be captured by some combination of $c_{j,m-j,k}$.
For example, the upper bound on $k$ is fixed by comparing discontinuity of $R_m(u)$ will have to account for logarithmic terms in the previous terms (e.g. the discontinuity of $R_2(u)$ will require a logarithmic term from the derivative of $R_1$, which means that $R_2(u)$ must have a square of the logarithm and so forth). 
The upper bound on $j$ for terms with logarithms is similarly inductively fixed. 
Meanwhile, the form of the homogeneous part is also constrained by the matching limit. All of this is independent of the actual coefficients $O_{m,j}$.

Assembling the full ansatz, and restoring the variable $x=B u$, we have
\begin{multline}
R(x)=-c_{0,0,1}\log\left(\frac{x-B}{x+B}\right)\\+\sum_{m=1}^\infty\frac{1}{B^m}\sum_{n=1}^m\sum_{k=0}^m c_{n,m-n,k}\frac{(x/B)^{1-k\bmod2}}{(x^2/B^2-1)^n}\log\left(\frac{x-B}{x+B}\right)^k\,.
\label{volin_eq_bulkGY}
\end{multline}
We managed to determine that $c_{0,0,1}=1$ for the solution of \eqref{volin_gy_Rshifts}, but the bulk equations are not enough to specify the $c_{n,m-n,k}$.

Let us try to generalize this picture for a generic equation of the form \eqref{volin_mostgeneral}.
The difficult step is to write the integral in terms of derivatives, or shifts, of the resolvent. Inspired by what is done \cite{volin,volin-thesis} for the specific case of the sigma model, we can establish it at a formal level. We can write the kernel as an inverse Fourier transform split into two integrals
\begin{equation}
K(\theta) = \frac{1}{2\pi}\int_{-\infty}^0 \re^{\ri \omega \theta} \tilde K(\omega)\rd\omega + \frac{1}{2\pi}\int_{0}^\infty \re^{\ri \omega \theta} \tilde K(\omega)\rd\omega.
\end{equation}
In the examples studied, the kernel is of the form
\begin{equation}
\tilde K(\omega) = \sum_{n\geq 0} \tilde{K}_n |\omega|^n = \hat{K}(|\omega|),
\label{analytic-K}
\end{equation}
where $\hat{K}$ is an analytic function. 
The even-ness of $\tilde{K}$ can be traced back to the unitarity of the S-matrix.
Formulation \eqref{analytic-K} allows us to expand the Fourier transforms as power series, finding the following formal identity
\begin{equation}
\ba
\frac{1}{2\pi}\int_{0}^\infty \re^{\ri \omega \theta} \tilde K(\omega)\rd\omega &= \sum_{n\geq 0}  \tilde K_n \left(\frac{1}{2\pi}\int_{0}^\infty \re^{\ri \omega \theta} (\omega)^n \rd\omega\right)
\\
& = \sum_{n\geq 0}  \tilde K_n \frac{(-\ri\partial_\theta)^n}{2\pi\ri} \frac{\ri}{\theta+\ri \epsilon}\\
& = - \frac{1}{2\pi\ri} \hat{K}(-\ri\partial) \frac{1}{\theta+\ri\epsilon},
\ea
\end{equation}
and conversely
\begin{equation}
\frac{1}{2\pi}\int_{-\infty}^0 \re^{\ri \omega \theta} \tilde K(\omega)\rd\omega =  \frac{1}{2\pi\ri} \hat{K}(\ri\partial) \frac{1}{\theta-\ri\epsilon}.
\end{equation}
Thus, combining with \eqref{volin_res_disc}, we can write
\begin{equation}
\left[1-\hat{K}(-\ri\partial)\right] R(\theta+\ri \epsilon) - \left[1-\hat{K}(\ri\partial)\right] R(\theta-\ri \epsilon) = - 2 \pi \ri m \cosh(p\theta).
\label{volin_genbulk_eq}
\end{equation}

When $p=0$ and $1-\hat{K}(0) \neq 0$, we can see that \eqref{volin_genbulk_eq} reduces to a problem of the form \eqref{volin_perteq}. Thus, the ansatz \eqref{volin_eq_bulkGY} is a good ansatz for such cases as well. If $p\neq 0$ and $1-\hat{K}(0) \neq 0$, we must do one additional trick, suggested by \cite{volin,volin-thesis}, which is applying the operator $(D^{p \pi/2} + D^{-p \pi/2})$ to both sides of the equation \eqref{volin_genbulk_eq}, annihilating the right hand side. This leads to
\begin{equation}
\left[D^{p \pi/2} + D^{-p \pi/2}\right]\left(\left[1-\hat{K}(-\ri\partial)\right] R^+(\theta) - \left[1-\hat{K}(\ri\partial)\right] R^-(\theta) \right) = 0.
\end{equation}
Since this operator is $(D^{p \pi/2} + D^{-p \pi/2}) \sim 2 + \CO(1/B)$ in the bulk limit and is symmetric as a power series in $\ri \partial$, we can perturbatively organize the equation as
\begin{equation}
R^+(\theta)-R^-(\theta) + \frac{1}{B} \left( \mathsf{O}(-\ri\partial) R^+(\theta) - \mathsf{O}(\ri\partial) R^-(\theta)\right) = 0,
\label{volin_eq_gen_fermionic}
\end{equation}
where $\mathsf{O}$ is an operator that can be written as a real power series in $\ri\partial$.  In the $1/B$ expansion of the bulk limit, higher powers in $\ri\partial$ correspond to higher powers in $1/B$ and equation \eqref{volin_eq_gen_fermionic} is of the form \eqref{volin_genbulk_eq}. I.e. the discontinuity at order $m$ is specified by the lower orders while the homogeneous part solves $\delta R=0$. Hence, the subsequent discussion that specifies the ansatz from closure under differentiation and discontinuities holds. However since the r.h.s. was set to zero the first term never appears and we can use as ansatz simply 
\begin{equation}
R(\theta)=\sum_{m=1}^\infty\frac{1}{B^m}\sum_{n=1}^m\sum_{k=0}^m c_{n,m-n,k}\frac{(\theta/B)^{1-k\bmod2}}{(\theta^2/B^2-1)^n}\log\left(\frac{\theta-B}{\theta+B}\right)^k\,.
\label{pre_eq_bulkGN}
\end{equation}
We will use this for the Gross--Neveu case. 

On the other hand, if $p\neq 0$ and $1-\hat K(0) = 0$, as is the case of the bosonic models, one cannot write \eqref{volin_eq_gen_fermionic}. Instead, we have something of the form 
\begin{equation}
\left[D^{p \pi/2} + D^{-p \pi/2}\right]\left(\left[- \ri \partial + \cdots \right] R^+(\theta) - \left[- \ri \partial - \cdots\right] R^-(\theta) \right) = 0
\end{equation}
Once we expand in $1/B$, the overall derivative $\partial$ acting on $R^\pm$ commutes with $D$ so it can be integrated out. In principle, this could create a constant term, but, since $R$ is odd, the constant must be zero. In the end, we find an equation of the form
\begin{equation}
R^+(\theta)+R^-(\theta) + \frac{1}{B} \left( \mathsf{O}(-\ri\partial) R^+(\theta) +  \mathsf{O}(\ri\partial) R^-(\theta)\right) = 0,
\label{volin_eq_gen_bosonic}
\end{equation}
for some $ \mathsf{O}$.
In this case, we have that 
\begin{equation}
\ba
R^+_m(u)+R^-_m(u) &= \sum_{j=0}^{m-1} O_{m,j} \,\partial^{(m-j)}(R^+_j(u)+(-1)^{m-j}R^-_j(u))
\ea
\label{volin_perteq_bos}
\end{equation}
and the homogeneous part satisfies $R^++R^- =0$. Clearly, this requires a different ansatz, but the constraints on analyticity remain similarly rigid. Using a similar reasoning to the Gaudin--Yang case but with $R^++R^- =0$ interchanged with $R^+-R^- =0$, one arrives at the ansatz of Volin in \cite{volin},
\begin{equation}
R(\theta)=\frac{1}{\sqrt{\theta^2/B^2-1}}\left(\sum_{m=0}^\infty\frac{1}{B^m}\sum_{n=0}^m\sum_{k=0}^{m} c_{n,m-n,k}\frac{(\theta/B)^{k\bmod 2}}{(\theta^2/B^2-1)^n} \log\left(\frac{\theta-B}{\theta+B}\right)^k\right).
\label{volin_bulk}
\end{equation}
The homogeneous part is still given by the $k=0$ terms, where now the square root enforces $R^++R^- =0$.
One last case of the general set up \eqref{volin_mostgeneral} is the case where $p=0$ and $1-\hat K(0)=0$. This case is subtle and we will solve it only for the specific case of Lieb--Liniger in appendix \ref{Lieb--Liniger}.

Before we proceed to the edge limit, let us note that the problem behaves rather differently depending on whether $1-\tilde K(0)$ is, or not, zero. Models where $1-\tilde K(0) \neq 0$ correspond to the models we introduced as fermionic, while models where $1-\tilde K(0) = 0$ are the bosonic models. It is easy to see that this is equivalent to the classification in section \ref{sec_iqft} and the criterion based on whether $G_+(0)$ is finite or infinite, which we used in the previous chapter.

\subsection{Edge limit}
\label{edge_limit}
The idea of the edge regime is to approximate the equation in the limit that
\begin{equation}
B\rightarrow \infty, \quad \theta \rightarrow B, \quad \theta-B \quad \text{finite}
\end{equation}
In Fourier (and Laplace) space, this corresponds to taking the large $B$ limit with a finite frequency, since the shift by $B$ is accounted for by an overall exponential. In this way, it becomes very similar to the analysis done in the previous chapter and in the mass gap literature.
Here we follow a more rigorous version of the presentation done in \cite{mr-long,mr-long}. It is equivalent to the approach of \cite{volin}, but it has the advantage of using the machinery of Wiener--Hopf decomposition which makes it easier to apply to a broad range of cases.

We start with an integral equation in the form \eqref{volin_mostgeneral}.
 The first step is to extend the integral equation to the real line. We define $f$ to be zero outside the $[-B,B]$ interval. Let
\begin{equation}
g(\theta) = \frac{m}{2}\re^{p \theta} \Theta(B-\theta),
\end{equation}
where $\Theta$ is the Heavyside step function, then we can write
\begin{equation}
f(\theta) - \int_{-B}^B K(\theta-\theta') f(\theta')\rd \theta' = g(\theta)+g(-\theta) + \xi(\theta-B) + \xi(-\theta-B)
\label{volin_lineeq}
\end{equation}
for some unknown function $\xi$ such that $\xi(\theta<0)=0$.

We take the Fourier transform of \eqref{volin_lineeq}, introducing the following functions
\begin{equation}
\ba
F_-(\omega) &= \int_{-2B}^0 \re^{\ri\omega\theta} f(\theta+B)\rd\theta,\\
\tilde K(\omega) &= \int_\IR \re^{\ri\omega\theta} K(\theta)\rd\theta,\\
X_+(\omega) &= \int_0^\infty \re^{\ri\omega\theta} \xi(\theta)\rd\theta.
\ea
\end{equation}
We find
\begin{multline}
\left(1-\tilde K(\omega)\right) \re^{\ri B \omega} F_-(\omega) = - \frac{\ri m \re^{pB}}{2}\left(\frac{\re^{\ri B \omega }}{\omega-\ri p}-\frac{\re^{-\ri B \omega }}{\omega+\ri p}\right)\\ + \re^{\ri B \omega} X_+(\omega) + \re^{-\ri B \omega} X_+(-\omega)\,.
\label{volin_preWH}
\end{multline}

This is the stage for the Wiener--Hopf methods introduced in section \ref{gn-inteq}, and we decompose the kernel as in \eqref{WH-kernel}.  
Equation \eqref{volin_preWH} can be recast as 
\begin{equation}
\frac{F_-(\omega)}{G_+(\omega)G_-(\omega)} = - \frac{\ri m \re^{pB}}{2}\left(\frac{1}{\omega-\ri p}-\frac{\re^{-2\ri B \omega }}{\omega+\ri p}\right) + X_+(\omega) + \re^{-2\ri B \omega} X_+(-\omega).
\end{equation}
The Wiener--Hopf decomposition of the equations gives us two familiar equations
\begin{align}
\frac{F_-(\omega)}{G_-(\omega)} &= 
\begin{multlined}[t]
- \frac{\ri m \re^{pB}}{2} \frac{G_+(\ri p)}{\omega-\ri p} + \frac{\ri m \re^{pB}}{2}\left[\frac{\re^{-2\ri B\omega} G_+(\omega)}{\omega+\ri p}\right]_-\\ 
\qquad+ \left[\re^{-2\ri B\omega} G_+(\omega)X_+(-\omega)\right]_- \hfill
\end{multlined}
\label{volineq_Fminus}
\\
- G_+(\omega) X_+(\omega) &=
\begin{multlined}[t]
 - \frac{\ri m \re^{pB}}{2} \frac{G_+(\omega)-G_+(\ri p)}{\omega-\ri p}  + \frac{\ri m \re^{pB}}{2}\left[\frac{\re^{-2\ri B\omega} G_+(\omega)}{\omega+\ri p}\right]_+\\ 
\qquad+ \left[\re^{-2\ri B\omega} G_+(\omega)X_+(-\omega)\right]_+ \hfill
\end{multlined}
\end{align}

So far we have not done any approximation. In chapter \ref{cha_antrans}, we solved analogous equations while trying to keep all perturbative and non-perturbative contributions. For Volin's method, however, it suffices to take the leading contribution in $1/B$ and constrain the form of further corrections. By using the leading behavior of $G_+(\omega)$, bosonic or otherwise, one can estimate that the projected terms in \eqref{volineq_Fminus} are subleading in $1/B$, and we get
\begin{equation}
F_-(\omega) \sim - \frac{\ri m \re^{pB}}{2} \frac{ G_+(\ri p)G_-(\omega)}{\omega-\ri p}\,.
\end{equation}
Comparing it with the bulk limit contraints the form of the correction in $1/B$, as can be seen explicitly in the example of appendix section \ref{app-volin-gy}. The minimal ansatz turns out to be 
\begin{equation}
F_-(\omega) = - \frac{\ri m \re^{pB}}{2} G_+(\ri p)G_-(\omega)\left(\frac{1}{\omega-\ri p} + \frac{1}{B \omega}\sum_{m=0}^\infty \sum_{n=0}^m \frac{M_{n,m-n}}{B^{m}\omega^n} \right).
\end{equation}

While we could in principle calculate $f(\theta)$ from the bulk and then Fourier transform it, this is not an efficient way of connecting the edge limit with the bulk limit. We can instead directly relate the Fourier transform to the Laplace transform  of resolvent \eqref{volin_lapres}. We take the countour along the imaginary axis and deform it around the negative real axis, picking up the discontinuity,
\begin{equation}
\ba
\hat{R}(s)&=\int_{-\infty+\ri\epsilon}^{0+\ri 0}\frac{\re^{ s z}  }{2\pi \ri}R(B+z/2) \rd z - \int_{-\infty-\ri 0}^{0-\ri 0}\frac{\re^{ s z}  }{2\pi \ri}R(B+z/2) \rd z\\
&=\int_{-\infty}^{0} \re^{ s z} \left(\frac{R(B+z/2-\ri\epsilon)-R(B+z/2+\ri\epsilon)}{2\pi \ri}\right)\rd z\\
&=\int_{-2B}^{0}\re^{ s z} f(B+z/2) \rd z = 2 F_-(-2\ri s) \,.
\label{res_to_Fou}
\ea
\end{equation}
We can thus write an almost general ansatz for the edge limit of the resolvent as
\begin{equation}
\hat R(s) =   \frac{m \re^{p B} G_+(\ri p)}{2} G_+(2 \ri s)\left(\frac{1}{ s +  \frac{p}{2}} + \frac{1}{B s}\sum_{m=0}^\infty \sum_{n=0}^m \frac{Q_{n,m-n}}{B^{m}s^n} \right).
\label{volin_Rhatansatz}
\end{equation}
The dichotomy of fermionic and bosonic models appears once again. At the level of the edge limit, this is manifested as in behavior of $G_+(0)$, just as in chapter \ref{cha_antrans}. For fermionic models, $G_+(0)$ is finite, while for bosonic models $G_+(2\ri s) \sim 1/\sqrt{s}$.

\subsection{Matching procedure}
\label{subsec_match}
Having solved the edge and bulk limits, we have two ansatze with coefficients $c_{n,m,k}$ and $Q_{n,m}$ left to specify. In both limits, at each order in $1/B$ there is a finite number of coefficients that need to be found. The wondrous insight of Volin's method is to combine the two limits, which produces a series of recursive linear equations that fix all $c_{n,m,k}$ and $Q_{n,m}$ order by order. 

The matching procedure requires a precise double limit.
Recall that the existing limits were
\begin{itemize}
\item edge limit: $B\rightarrow \infty$, $\theta-B = z/2$ finite, $\omega$ and $s$ finite
\item bulk limit: $B\rightarrow \infty$, $\theta/B = x$ finite.
\end{itemize}
Naively, one could expect that a possible matching would be to take $\theta-B = z$ small in the bulk and $\omega\rightarrow\infty$ large in the edge. However, this does not provide good matching conditions, we get terms in $\log(z)$ from the bulk but no $\log(s)$ in the edge, which prevents matching. The issue is that $z\rightarrow 0$ is outside the domain of validity of the bulk ansatz. On the other hand if we simply take $z\rightarrow \infty$ while $\omega\rightarrow 0$ we have terms with $\log(s)$ but no terms with $\log(z)$. In order to get a coherent double limit, we need to be careful with the $z/B$ ratio. 

The correct limit is then
\begin{itemize}
\item matching limit: $B\rightarrow \infty$, $z\rightarrow \infty$ but $z/B\rightarrow 0$ , $s\rightarrow 0$ (and $\omega\rightarrow 0$).
\end{itemize}
While conceptually we take $z\rightarrow \infty$, in practice the bulk ansatz depends on z only through $z/B$ so it is formally equivalent to taking $z\rightarrow 0$. The reason why we need $z\rightarrow \infty$ while $z/B\rightarrow 0$ is that the Laplace (Fourier) variable $s$ ($\omega$) scales inversely to $z$ and we want to expand it around 0. In practice, we carry out this limit by taking $z\rightarrow 0$ in the bulk and $s\rightarrow 0$ in the edge.

In this limit, the bulk ansatz for $R(z)$ is expanded order by order in $1/B$ into a power series in $(z/B)^n \log(z/B)^m$. Meanwhile, the edge ansatz $\hat R(s)$ is expanded in powers of $s^n \log(s)^m$. The idea is to then take the inverse Laplace transform of the bulk expansion and compare it to the edge expansion term by term. One can also take the Laplace transform of the edge limit expansion, which turns out to be computationally more favourable. 
An important feature of the ansatze which only becomes visible while matching is that the coefficients $c_{n,m,k}$ and $Q_{m,n}$ can depend on $B$ through $\log B$.
To see why, we refer to the concrete example of the matching procedure done in appendix \ref{app_volin}.

It is important to note that the number of coefficients necessary to calculate an observable up to some order in $1/B$ is finite. Roughly, to calculate something to order $1/B^M$ one needs to calculate all $c_{n,m-n,k}$ and all $Q_{n,m-n}$ with $n,m,k\leq M$. Thus, one can iterate the procedure until all such coefficients are determined. This routine can be implemented in \texttt{Mathematica}, or similar software. The procedure can in principle be carried out to arbitrary high order but computations become increasingly difficult and there are increasingly more coefficients to compute. If one is calculating the coefficients numerically, the time required to calculate order $1/B^M$ is $\propto M^3$ for large $M$, as can be simply estimated from the number of coefficients $c_{n,m-n,k}$. However to obtain the exact coefficients, computation time is significantly amplified by the complicated form of each coefficient and in total grows exponentially in $M$.

\subsection{Extracting the solution}

Carrying out the matching procedure to sufficiently high order we build a good approximation of the resolvent both in the edge and bulk limits. We can use this to extract different aspects of the solutions, which we can also use to test the method.

For example, we saw in \eqref{volin_mom_from_res} we can extract the moments of the solution by taking the large $\theta$ limit of the resolvent. We can do this explicitly order by order in $1/B$ by using the bulk ansatz. In the case of \eqref{volin_eq_bulkGY}, we find
\begin{equation}
\int_{-B}^B f(x)\rd x = \frac{2B}{\pi}+\frac{1}{\pi} \sum_{m\geq 0}\frac{c_{1,m,0}}{B^m}\,.
\end{equation}
However, we can even approximate the function itself in the bulk limit using \eqref{volin_res_disc}. For the same example, we find 
\begin{multline}
f(x)=1 - \sum_{m=0}^\infty \frac{1}{B^m} \sum_{n=1}^m  \sum_{k=0}^{m} \frac{(-1)^n c_{n,m-n,k}(x/B)^{1-k\bmod 2} }{ \left(1-x^2/B^2\right)^{n}}\times\\
\times \sum_{r=0}^{[\frac{k-1 }{ 2}]} \binom{k }{ 2r+1} (-1)^r \pi^{2r} \left(\log \left|  \frac{x-B }{ x+B} \right| \right)^{k-2r-1}.
\label{fitselfGY}
\end{multline}
We can test this approximation numerically. Naturally, it breaks down as $x$ moves closer to $\pm B$, since the ansatz is divergent but the exact solution is not. There we can use the information from the edge limit instead.

If we expand $\hat R(s)$ as
\begin{equation}
\hat R(s) = \frac{R_0}{s} + \frac{R_1}{s^2} +\frac{R_2}{s^3} + \cdots\,,
\label{Rhat_infinity}
\end{equation}
this implies, using \eqref{res_to_Fou}, that the Fourier transform in the lower half plane behaves asymptotically  as 
\begin{equation}
F_-(\omega) = \frac{R_0}{\ri\omega} + \frac{2R_1}{(\ri\omega)^2} + \frac{4R_2}{(\ri\omega)^3} + \cdots,\quad \omega\in\mathbb{H}_-
\end{equation}
which translates into
\begin{equation}
f(x) = R_0 - 2 R_1 (x-B)+ 2R_2 (x-B)^2 +\cdots\,.
\end{equation}
The latter expansion characterizes the edge limit of the solution itself and can also be numerically tested. The edge limit can further be used to calculate some integrals, such as
\begin{equation}
\int_{-B}^B f(\theta)\cosh(p\theta)\rd \theta = 
\frac{\re^{pB}}{2}\hat{R}\left(\frac{p}{2}\right).
\label{edge-energy}
\end{equation}
With \eqref{edge-energy} and \eqref{volin_mom_from_res} we can extract the desired observables from the ground state.

This concludes our presentation of the generalized Volin's method. In appendix \ref{app_volin}, we add some additional technical results and considerations. These mostly concern the calculation of the matching procedure to high order. As such they might be of interest to a reader who wishes to implement Volin's method. 
As we see in the following sections, this method is extraordinarily handy to perform numerical analysis of the large order behavior.
It is also possible to do an analytic analysis of the resulting perturbative series through these recurrence equations, as was done by \cite{bbv} for the $O(4)$ sigma model. There they relied on the particularity that $c_{n,m,k\neq 0}=0$ in that model (which also happens for the PCF model), but their method should hold in principle for other models. 

\section{Exact series in integrable field theories}
\label{sec_iqft_pert}
As we saw in chapter \ref{cha_intro}, the ground state of integrable field theories when excited by a chemical potential $h$ coupled to a charge is described by an equation of the form \eqref{iqft_geneq}. Volin's method was originally developed for the specific case of the $O(N)$ sigma model \cite{volin}. In \cite{mr-ren}, we extended the application of the method by doing the edge limit analysis in the language of Wiener--Hopf, as we reviewed in section \ref{sec_volin}, which allowed it to be applied to a broader class of integral models. In this section, we review how to put the data from the integrable models into Volin's method.

\subsection{The case of Gross--Neveu}

The kernel $K$ and its decomposition $G_+$ are the same as in the previous chapter, thus
$G_+$ is given by \eqref{GN_Gplus_usec},
which results in the following ansatz for the edge limit
 \be
\hat{R}(s)= m \, \re^B A \Phi(s)\left( \frac{1}{s+1/2}+\frac{1}{B s}\sum_{m=0}^\infty\sum_{n=0}^m\frac{Q_{n,m-n}(\log(B)) }{B^m s^n}\right)\,,
\label{R_GN}
\ee
where
\be
\ba
\Phi(s)&=G_+(2\ri s),\quad 
A&=\frac{\Gamma (1-\Delta )}{2 (2 \re)^{\Delta } (1-2 \Delta )^{\frac{1}{2}-\Delta }}.
\ea
\ee
In the opposite limit, the adequate bulk ansatz is \eqref{pre_eq_bulkGN}, which we normalise as
\begin{equation}
R(\theta)=\sum_{m=1}^\infty\frac{A}{B^m}\sum_{n=1}^m\sum_{k=0}^m c_{n,m-n,k}\frac{(\theta/B)^{1-k\bmod 2}}{(\theta^2/B^2-1)^n}\log\left(\frac{\theta-B}{\theta+B}\right)^k.
\label{iqft_bulkGN}
\end{equation}
Using these ansatze, we can write our key observables as
\begin{align}
\rho&=\frac{m \, \re^{B}A} {2\pi}\sum_m \frac{c_{1,m,0}}{B^m},\\
e &=
\frac{m^2 \, \re^{2B} A^2}{2\pi}\left(1+\sum_{m=1}^\infty\frac{1}{B^m}\sum_{s=0}^{m-1}
2^{s+1}Q_{s,m-1-s}\right).
\end{align}
Using the series for $\rho$ we can use \eqref{alpha-gn} to obtain $\alpha$ as a series in $B$ and then invert to find $B$ as a series in $\alpha$.

Using Volin's method, we are then able to calculate the perturbative expansion of the normalised energy density,
\begin{equation}
\label{GN-series}
\ba
4\varphi(\alpha)&=
1+\alpha  \Delta +\frac{1}{2} \alpha ^2 \Delta  (\Delta +2)
-\frac{1}{2} \alpha ^3 ((\Delta -3) \Delta )\\
&+\frac{1}{8} \alpha ^4 \Delta  \left(\Delta ^3 (1-24 \zeta (3))+42 \Delta ^2 \zeta (3)-\Delta  (21 \zeta (3)+25)+24\right)
\\
&+\frac{1}{12} \alpha ^5 \Delta  \left(120 \Delta ^4 \zeta (3)+\Delta ^3 (7-354 \zeta (3))\right.
\\
&\quad\left.+\Delta ^2 (357 \zeta (3)+43)-18 \Delta  (7 \zeta (3)+8)+90\right)
\\
& + \CO(\alpha^6). 
\ea
\end{equation}
We have calculated 45 exact coefficients in this expansion and 230 numeric coefficients for the integer values of $N$ between 5 and 12. As was tested in \cite{mr-ren}, the large $N$ behavior of this series is compatible with the results of \cite{fkw1,fkw2,dpmss}.

\subsection{The case of bosonic models}

For bosonic models we write the edge limit ansatz, based on \eqref{volin_Rhatansatz}, as
\begin{equation}
\hat{R}(s)=m \, \re^B \, A \Phi(s)\left(\frac{1}{s+\frac{1}{2}}+Q(s)\right).
\label{bosonic_edge}
\end{equation}
Where we define
\begin{equation}
\Phi(s) = \frac{G_+(2\ri s)}{k\sqrt{\pi}},\quad A= \frac{k G_+(\ri)}{2}\sqrt{\frac{\pi}{2}},
\end{equation}
such that
\begin{equation}
\Phi(s) = \frac{1}{\sqrt{\pi s}}\left(1+\CO(s)\right).
\end{equation}
From this ansatz, we can conveniently normalise the energy and density as
\begin{equation}
e= m^2 \e^{2B} \frac{A^2}{\pi^2k^2}\tilde{e}, \quad 
\rho=  \frac{m \re^B\sqrt{B}A}{\pi}\tilde{\rho}.
\end{equation}
Using \eqref{edge-energy}, $\tilde{e}$ is obtained directly from the ansatz coefficients,
\begin{equation}
\tilde{e}=\left(1+\sum_{m=0}^\infty\frac{1}{B^m}\sum_{s=0}^{m-1}
2^{s+1}Q_{s,n-1-s}\right).
\label{epst}
\end{equation}

As for the bulk, the appropriate ansatz is \eqref{volin_bulk}, which we now normalise as
\begin{equation}
R(\theta)=\sum_{m=0}^\infty\sum_{n=0}^\infty\sum_{k=0}^{m+n}2A\sqrt{B} c_{n,m,k}\frac{(\theta/B)^{k\bmod 2}}{B^{m-n}(\theta^2-B^2)^{n+1/2}}\log\left(\frac{\theta-B}{\theta+B}\right)^k.
\label{bulk}
\end{equation}
Using the formula for the moments \eqref{volin_mom_from_res}, we have
\begin{equation}
\tilde{\rho}= 1+\sum_{m=1}^\infty\frac{c_{0,m,0}-2 c_{0,m,1}}{B^m} 
\rho=  \frac{m \re^B\sqrt{B}A}{\pi}\tilde{\rho}.
\label{rhot}
\end{equation}
In the end, we can write the perturbative series \eqref{pert-gs} from the normalised observables,
\begin{equation}
\alpha \varphi(\alpha)=
\frac{\tilde{e}}{B \tilde{\rho}^2}.
\end{equation}

The perturbative series \eqref{on-pert}, \eqref{susyon-pert}, \eqref{pcf-ps}, \eqref{efkw} and \eqref{fendley-gs} were derived with this method.
For the $O(N)$ non-linear sigma model, we have calculated analytically the first $44$ terms of this expansion as functions of $\Delta$, and 165 numerically for the integer values of $N$ between 3 and 12. These were sufficient to test the leading non-perturbative effects. For the specific values of $N=3$, \cite{bbv,bbh} calculated 336 coefficients using Volin's method, while for $N=4$ the impressive numerical calculation in \cite{abbh1,abbh2} obtained beyond 2000 coefficients using Volin's method, which permitted them to study the exponential effects beyond the leading singularities. 

As for the supersymmetric  $O(N)$ non-linear sigma model we have calculated analytically the first $42$ terms in this expansion, and numerically 70 coefficients for $N=4,5,6,7$.
The PCF model is of particular simplicity since there is no $\log(s)$ in $\Phi(s)$, in both charge choices (this feature is a consequence of $\xi=1/2$ which is independent of $h$). Thus, all coefficients other than $c_{n,m,0}$ in \eqref{bulk} vanish. 
We have calculated analytically the first $54$ terms of the expansion with standard charges, and 200 numerical coefficients for all integers $N$ between 2 and 12.
For FKW charges, we have obtained the first $50$ exact coefficients of the series \eqref{efkw}. In higher terms, all terms with a polygamma function or $\zeta$-function can be rearranged into polynomials of the modified $\zeta$-functions \eqref{FKWzeta} evaluated at odd integers. For Fendley's coset models, we obtained between 75 numerical coefficients for $\Delta=1/3$ and 81 coefficients for all integer values of $\Delta^{-1}$ between $4$ and $8$. 

\section{Putting renormalons to the test}
\label{sec-tests}

In this section, we explain how to use the long perturbative series generated with Volin's method to test the trans-series results from chapter \ref{cha_antrans}. We sort the methods into three different groups: direct extrapolation from the large order behavior of perturbation theory, identification and calculation of ambiguities in the Borel plane, and comparison with the numeric solution of the integral equation. The first two are sensitive to the imaginary part of the trans-series parameters while the third tests the real part. We also use the first to test UV renormalons.

\subsection{Auxiliary series and acceleration}

Before we present the tests using the large order behavior of perturbative series, it is useful to review some basic techniques.
We expect the asymptotic behavior of a general class of perturbative series to be of the form
\begin{equation}
e_m \approx \sum_i (A_i)^{-m} C_i \Gamma(m+b_i) \sum_{k\geq 0} \frac{c_{i,k}}{m^k}\,,
\label{generic-asym}
\end{equation}
with the leading behavior being set by the smallest weight $|A_i|$. Combining the resurgence relation \eqref{resurgence-relation} and our knowledge of the leading UV and IR renormalons, we expect the leading IR and UV weights to be the same, i.e.
\begin{multline}
e_m \sim \left(A_0\right)^{-m} (-1)^m D_0 \Gamma(m+b_-)\left\{1+\CO\left(\frac{1}{m}\right)\right\} \\
+ \left(A_0\right)^{-m}S_0 \Gamma(m+b_+)\left\{1+\CO\left(\frac{1}{m}\right)\right\}.
\label{gen_IRUVseq}
\end{multline}

An elementary test one can perform is to verify the value of $A_0$, for example by taking
\begin{equation}
\frac{e_{m+1}}{m e_m} = \frac{1}{A_0}+\CO\left(\frac{1}{m}\right).
\label{A0_test}
\end{equation}
Once $A_0$ is settled, we can construct the normalised coefficients
\begin{equation}
c^\pm_m = \frac{A_0^{2m-1} e_{2m}}{\Gamma(2m+1)}\pm\frac{A_0^{2m} e_{2m+1}}{\Gamma(2m+2)}.
\label{auxseries}
\end{equation}
These series can be used to extract the coefficients $b_\pm$ as
\begin{equation}
b^\pm_m =m \log\left(\frac{c^\pm_{m+1}}{c^\pm_{m}}\right)+1 \sim b_\pm, \qquad m\gg 1.
\label{bpm-limit}
\end{equation}
With a good grasp of $A_0$ and $b_\pm$ we can extract the overall constants. We introduce
\begin{equation}
\ba
s_m &= \frac{A_0^{2m} e_{2m+1}}{\Gamma(2m+1+b_+)}+\frac{A_0^{2m-1} e_{2m}}{\Gamma(2m+b_+)},\\
d_m &= \frac{A_0^{2m} e_{2m+1}}{\Gamma(2m+1+b_-)}-\frac{A_0^{2m-1} e_{2m}}{\Gamma(2m+b_-)},
\label{sm-dm}
\ea
\end{equation} 
which converge faster than the $c_m^\pm$. Let, for example, $b_--b_+ < 1$ (otherwise there are some manageable subtleties), $S_0$ can then be found from the limit
\begin{equation}
s_m \sim S_0 +\CO\left(\frac{1}{m}\right),\quad  m\gg 1.
\label{s_m-limit}
\end{equation}
And similarly we can obtain $D_0$ from $d_m$.

Finally, a fundamental tool we use is Richardson transforms. Let $g(x)$ be a function such that
\begin{equation}
g(x) \sim g_0 + g_1 x^k + \cdots,\quad x\ll 1.
\end{equation}
The most generic form of a Richardson transform is 
\begin{equation}
R(g,s) = \frac{s^k g\left(\frac{x}{s}\right)-g(x)}{s^k-1},
\label{Rich-cont}
\end{equation}
where introduce an arbitrary parameter $s$.
As expected, the  transform converges faster,
\begin{equation}
R(g,s) \sim g_0 + \CO\left(x^{k+1}\right).
\label{Rich-fast}
\end{equation}
We can apply this transform to a series that behaves as 
\begin{equation}
f_m \sim f_0 + \frac{f_1}{m^{k_1}}+ \frac{f_2}{m^{k_2}}+\cdots,\qquad m\gg 1,
\end{equation}
with $k_2>k_1$. For simplicity, we construct the sequence $k_n$ such that for every $n$ there is an $m\geq n+1$ that satisfies $k_n+1=k_m$ (thus for every non-integer $k_q$ we include  $k_q+\IN$).
Then
 by defining
\begin{equation}
f^{(1|k_1)}_m = \frac{ \left(1+\frac{1}{m}\right)^{k_1}f_{m+1}-f_m}{\left(1+\frac{1}{m}\right)^{k_1}-1},
\end{equation}
\eqref{Rich-fast} guarantees 
\begin{equation}
f^{(1|k_1)}_m - f_0 \sim \CO\left( \frac{1}{m^{k_2}}\right)+\cdots,\qquad m\gg 1.
\end{equation}
One can take a subsequent Richardson transform  $\left[f^{(1|k_1)}\right]^{(1|k_2)}$, further improving convergence.  We would call this the second Richardson. For the $m$-th coefficient of this second transform one uses $f_m$, $f_{m+1}$ and $f_{m+2}$. 

Knowing the sequence of $k_1,\dots,k_n$ to some order $n$, which can be derived by knowing the structure of the $1/m$ corrections in \eqref{generic-asym}, one can then iterate until the $n$-th Richardson transform, recursively defined as
\begin{equation}
f^{(n|k_1,\dots,k_n)}_m = \frac{ \left(1+\frac{1}{m}\right)^{k_n}f^{(n-1|k_1,\dots,k_{n-1})}_{m+1}-f^{(n-1|k_1,\dots,k_{n-1})}_m}{\left(1+\frac{1}{m}\right)^{k_n}-1}\,.
\label{RT_genseq}
\end{equation}
For the particular case where the $k_i$ are simply the integers, we can simplify \eqref{RT_genseq} into a closed formula \cite{bender-book,msw},
\begin{equation}
f^{(n)}_m = \sum _{k=0}^n (-1)^{k+n}\frac{  (k+m)^n}{k! (n-k)!}f_{k+m},
\label{int_RTN}
\end{equation}
where $f^{(n)}_m$ is $f^{(n|1,\dots,n)}_m$ in the notation of \eqref{RT_genseq}. While we will not specify when we use which, to obtain the results presented in this section we often needed the more generic Richardson transform \eqref{RT_genseq} for rational $k_n$.

\subsection{Testing IR and UV renormalons with large order behavior}

In order to write formulas applicable to all models studied,  it is useful to define
\begin{equation}
\eta = \begin{cases}
1, & \text{bosonic models},\\
0, & \text{Gross--Neveu}.
\end{cases}
\label{eta-def}
\end{equation}
Our results in chapter \ref{cha_antrans}, particularly \eqref{efromWHZ} and \eqref{bos-ts}, using \eqref{resurgence-relation}, fix $A_0$ and $b_\pm$ in  \eqref{gen_IRUVseq} to be 
\begin{equation}
A_0 = 2,\quad  b_\pm = \pm (2\xi-\eta).
\label{A-bpm}
\end{equation}
We can test both at once by assuming $A_0=2$ and calculating the limit \eqref{bpm-limit}. This was the main numerical test in \cite{mr-ren}. We find that \eqref{A-bpm} works for all models studied, except for the super-symmetric $O(N)$ NLSM at finite $N$, since in that model there is no IR renormalon and the $b_+$ test breaks down. As an example, for the $SU(5)$ PCF model we predict $b_\pm=0$. From the numeric sequences \eqref{bpm-limit}, we have
\begin{equation}
\ba
b^+_{95} &=-0.0027575025316\dots\,,\\
b^{+,(5)}_{90} &=-0.0000000000085\dots\,,\\
b^-_{95} &=-0.0550439232385\dots\,,\\
b^{-,(5)}_{90} &=-0.0000000474430\dots\,,\\
\ea
\end{equation}
using the notation of \eqref{int_RTN} for Richardson  transforms. All our following tests assume \eqref{A-bpm} holds, so it is tested beyond reasonable doubt.
This test only senses that the IR renormalon  is a term proportional to $\Lambda^2$, which, through \eqref{Lambda-def} and $\alpha\sim 2\beta_0 \bar g ^2(\mu) $, specifies $A_0$ and $b_\pm$. It is not sensitive to the specific predictions of \eqref{efromWHZ} and \eqref{bos-ts}. 

\begin{figure}
\center
\begin{subfigure}{\figsize}
\centering
\includegraphics[width=\textwidth]{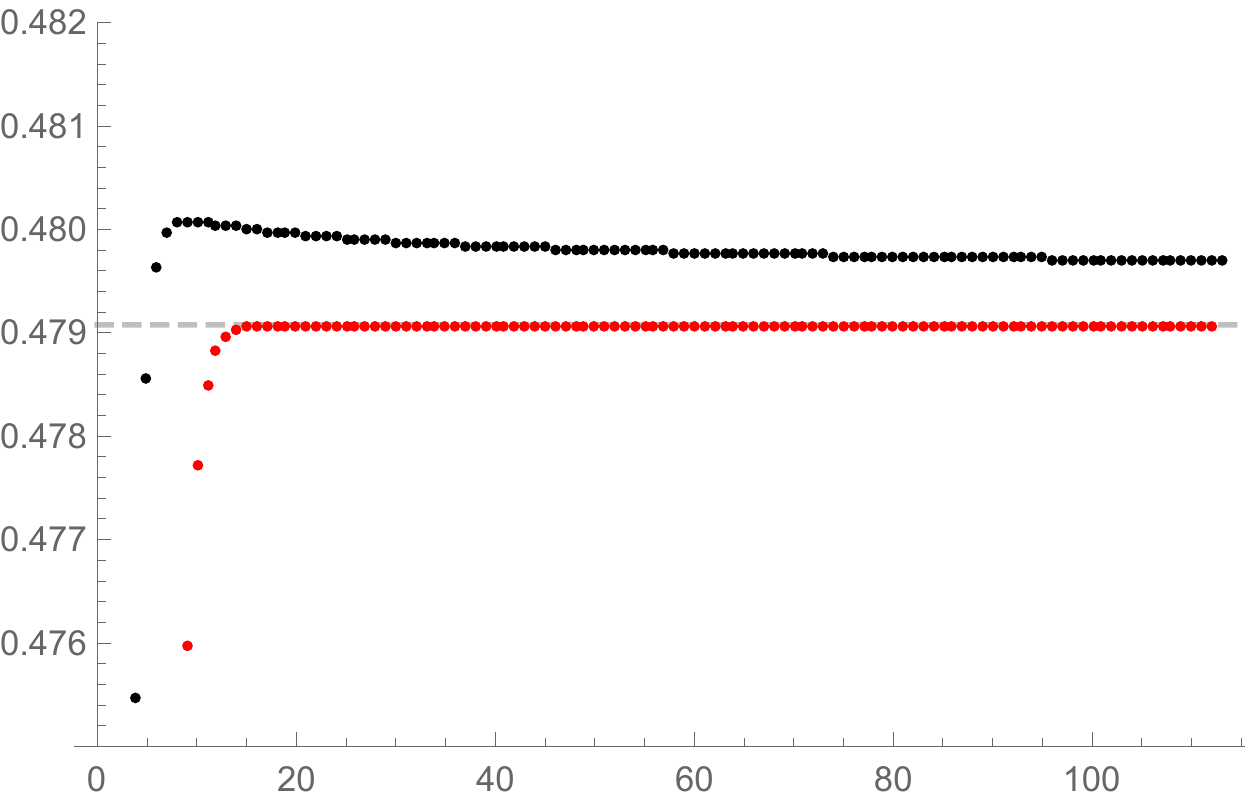}
\caption{$O(7)$ GN model}
\end{subfigure}
\\
\begin{subfigure}{\figsize}
\centering
\includegraphics[width=\textwidth]{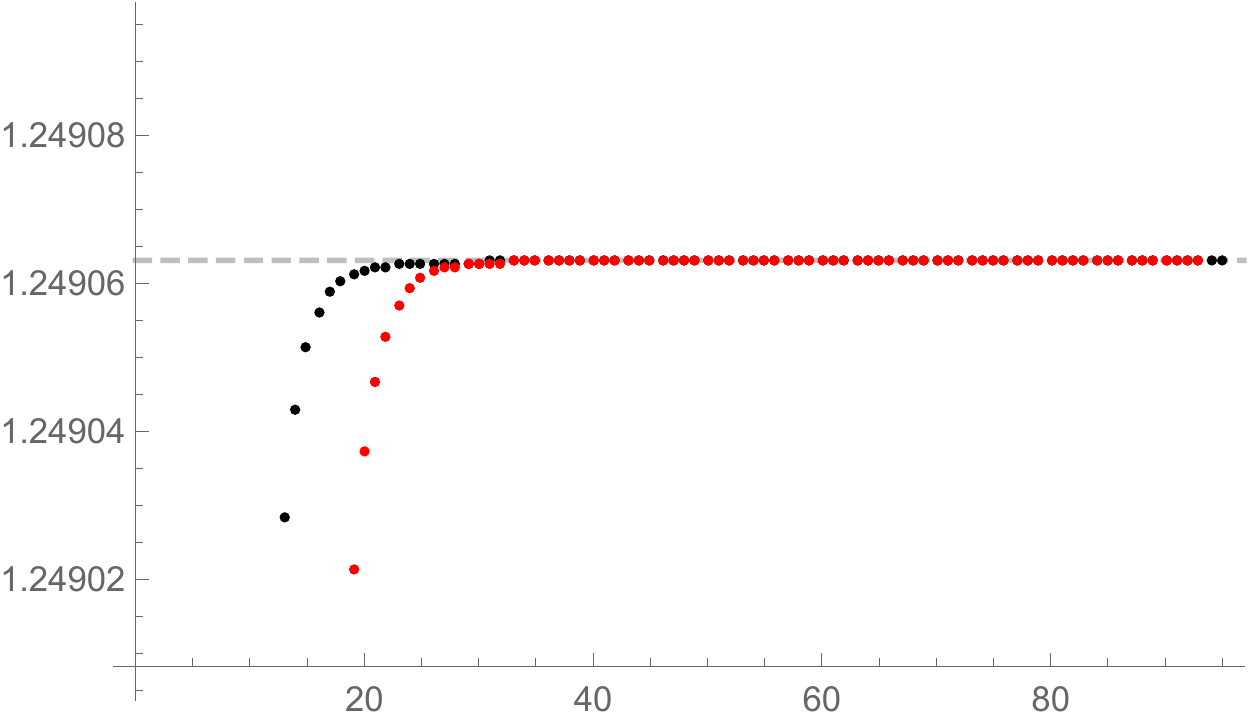}
\caption{$SU(4)$ PCF model}
\end{subfigure}
\caption{Figure adapted from \cite{mmr-antrans}. Plot of the sequence $s_m$ 
(black) and its respective second Richardson transform (red). The dashed line is the theoretical prediction.
}
\label{fig-asym-beh-GN}
\end{figure}

A more specific test is to determine the overall constants $S_0$ and $D_0$ in \eqref{gen_IRUVseq}. This directly tests the leading trans-series parameters. In general, we have that
\begin{equation}
e_m \sim A_0^{-b-m} \frac{\mathsf{S}_0}{2\pi}\Gamma(m+b) + \text{UV contribution}
\end{equation}
where $\mathsf{S}_0$ is the Stokes constant associated with the singularity at $A_0$, which is obtained from the trans-series parameters as
\begin{equation}
\mathsf{S}_0=-\ri(\CC_0^--\CC_0^+).
\end{equation}
Thus, we write the generally applicable formula
\begin{equation}
S_0 = \frac{2^{-2\xi+\eta}}{2\pi\ri} \big(\CC_0^--\CC_0^+\big).
\label{S0-generic}
\end{equation}
In figure \ref{fig-asym-beh-GN}, we test the limit \eqref{s_m-limit} using the sequence \eqref{sm-dm} for Gross--Neveu and the PCF model. We compare the prediction from combining \eqref{coefC0} and \eqref{pcf-cs} with \eqref{S0-generic}. We can see that the convergence is very good. Because the IR singularity is given by a single term, the leading IR asymptotics are fully captured by \eqref{S0-generic} and the $1/m$ corrections come fully from UV renormalons. One can obtain $S_0$ with even more precision by using methods which probe only IR effects, as we do in the next section.

\begin{figure}
\center
\begin{subfigure}{\figsize}
\centering
\includegraphics[width=\textwidth]{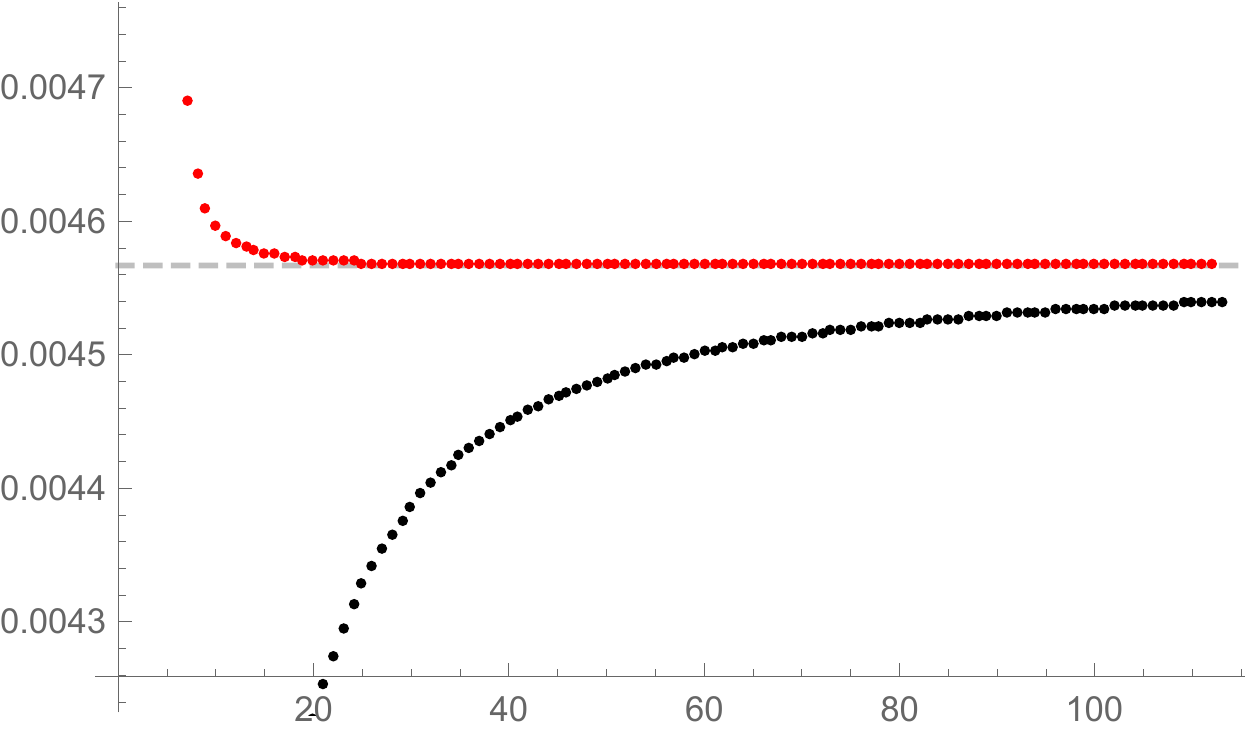}
\caption{Sequence $d_m$}
\end{subfigure}
\\
\begin{subfigure}{\figsize}
\centering
\includegraphics[width=\textwidth]{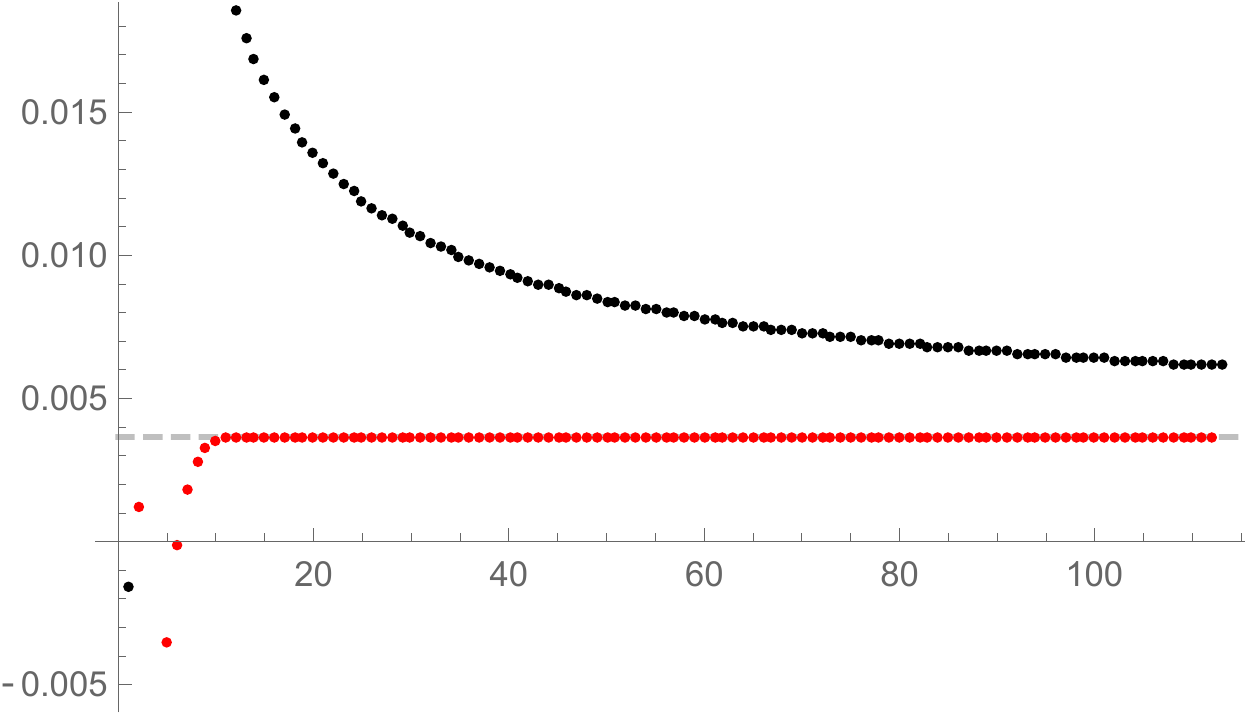}
\caption{Sequence $-2m\left(d_m+D_0\right)$}
\end{subfigure}
\caption{Figure adapted from \cite{mmr-antrans}. Analysis of the UV renormalons in the $O(7)$ GN model. Plot of two sequences (black) and their respective second Richardson transforms (red). The dashed lines are the predicted values $-D_0$ (top) and $-4\Delta D_0$ (bottom).
}
\label{fig-UV-GN}
\end{figure}

In section \ref{sec-uv}, we derived a trans-series with the leading UV renormalon singularity for the Gross Neveu model using an auxiliary ``analytically continued'' model. From the analogue of \eqref{S0-generic}, it follows that
\begin{equation}
D_0 = -\frac{(4 \re)^{-2 \Delta }}{4 \Gamma (\Delta )^2}.
\label{D0-generic}
\end{equation}
In fact, one can go one step further and also test the leading $1/m$ correction from the leading $\alpha$ correction in the UV renormalon sector of \eqref{eUV}. We should have
\begin{equation}
\ba
d_m &\sim \frac{(4 \re)^{-2 \Delta }}{4 \Gamma (\Delta )^2}\left(1+\frac{2\Delta}{m}+\cdots\right),\quad m\gg 1,
\ea
\end{equation}
which is tested in figure \ref{fig-UV-GN}.

We cannot use this type of strategies to find the next exponential sector with $A_1>A_0$ because the leading UV renormalon exponentially dominates in \eqref{gen_IRUVseq}. Thus, we must go to the Borel plane where we can more easily isolate IR effects from UV ones.

\subsection{Renormalon singularities from the Padé approximant}

In order to study the non-perturbative effects smaller than the IR singularity, it is convenient to subtract the IR singularity contribution to large order behavior. We introduce the new coefficients,
\begin{equation}
\bar e_k= e_k - 2^{-k}\left(\frac{2^{-2\xi+\eta}}{2\pi\ri} \big(\CC_0^--\CC_0^+\big)\right) \Gamma(k+2\xi-\eta),
\end{equation}
where $\eta$ is defined in \eqref{eta-def}.
As we just mentioned, it is impossible to perform the analysis of the previous section with these coefficients because the UV effects overwhelm exponentially subleading IR contributions. One could hope to subtract also the leading UV renormalon. However, unlike the IR renormalon, there is an infinite series of $1/m$ corrections which would need to be calculated from the UV trans-series. Instead we need to pick up the tools from chapter \ref{cha_resurgence} and travel to the Borel plane.

We introduce the new power series $\phi$ and its Borel transform,
\begin{equation}
\phi(\alpha) = \sum_{k\geq 0} \bar e_k \alpha^k \Rightarrow \widehat{\phi}(\alpha) =  \sum_{k\geq 0} \frac{ \bar e_k}{k!} \,\zeta^k.
\label{modphi-series}
\end{equation}
The leading singularity of $\widehat\phi$ is not $\zeta=2$ but rather $\zeta = 2\xi_1$ where $\xi_1$ is given by the leading pole of $\sigma$ in the respective model. This is a non-trivial prediction of our trans-series results. For models with new renormalons, it shows that the next contribution is indeed the new renormalon at its unconventional position. For models without new renormalons, $\xi_1$ is ``instanton-like'' and thus proportional to $N$, which puts it more distant than the expectation of the standard renormalons. So in those models the position of this singularity is also important to show that there are no other effects between the IR singularity and the instanton singularity.

Since we only have access to a finite truncation of the series \eqref{modphi-series}, one has to extrapolate the position of this pole by approximating the analytic structure of $\widehat{\phi}$. The ideal mathematical tool to capture this aspect are Padé approximants. For two integers $m$ and $n$, the Padé approximant $[m/n]_{\widehat{\phi}}$ is the defined as the rational function
\begin{equation}
[m/n]_{\widehat{\phi}}(\zeta) = \frac{\sum_{i=0}^m p_i \zeta^i }{1+\sum_{i=1}^n q_i \zeta^i},
\end{equation}
with $p_i$ and $q_i$ chosen such that
\begin{equation}
[m/n]_{\widehat{\phi}}(\zeta) = \sum_{k= 0}^{m+n} \frac{ \bar e_k}{k!} \,\zeta^k+ \CO(\zeta^{m+n+1}).
\end{equation}
Clearly the larger $m$ and $n$, the better approximation of $\widehat{\phi}$ is the Padé approximant. Also well known but less obvious is that the Padé approximant is a very good   approximation $\widehat{\phi}$ outside of the radius of convergence of its Taylor series. This fits our application well, because we need to identify the position of the singularities. As for the choice of $m$ and $n$, for a fixed maximum number of known terms $M$, the optimal choice of $n$ and $m$ is $m=n=M/2$, see for example \cite{mmbook}. We label the diagonal Padé approximant as
\begin{equation}
\mathsf{P}^M_{\widehat{\phi}}(\zeta) = \left[\frac{M}{2}\Big/\frac{M}{2}\right]_{\widehat{\phi}}(\zeta) = \sum_{k= 0}^{M} \frac{ \bar e_k}{k!} \,\zeta^k+ \CO(\zeta^{M+1}).
\end{equation}

The singularity predicted by our trans-series is in general a branch cut due to the rational power of $\alpha$ that accompanies the exponential. Clearly, the rational function $\mathsf{P}^M_{\widehat{\phi}}(\zeta)$ cannot have a branch cut. But it can have as many as $n$ poles. In practice, what happens is that the poles accumulate along the branch cut starting at the singularity. With enough coefficients and thus sufficiently high $m$ and $n$ this suffices to identify the position of the leading singularity with some precision.

\begin{figure}
\centering
\begin{subfigure}{0.925\textwidth}
\centering
\picdims[width=\textwidth]{\textwidth}{0.2\textwidth}{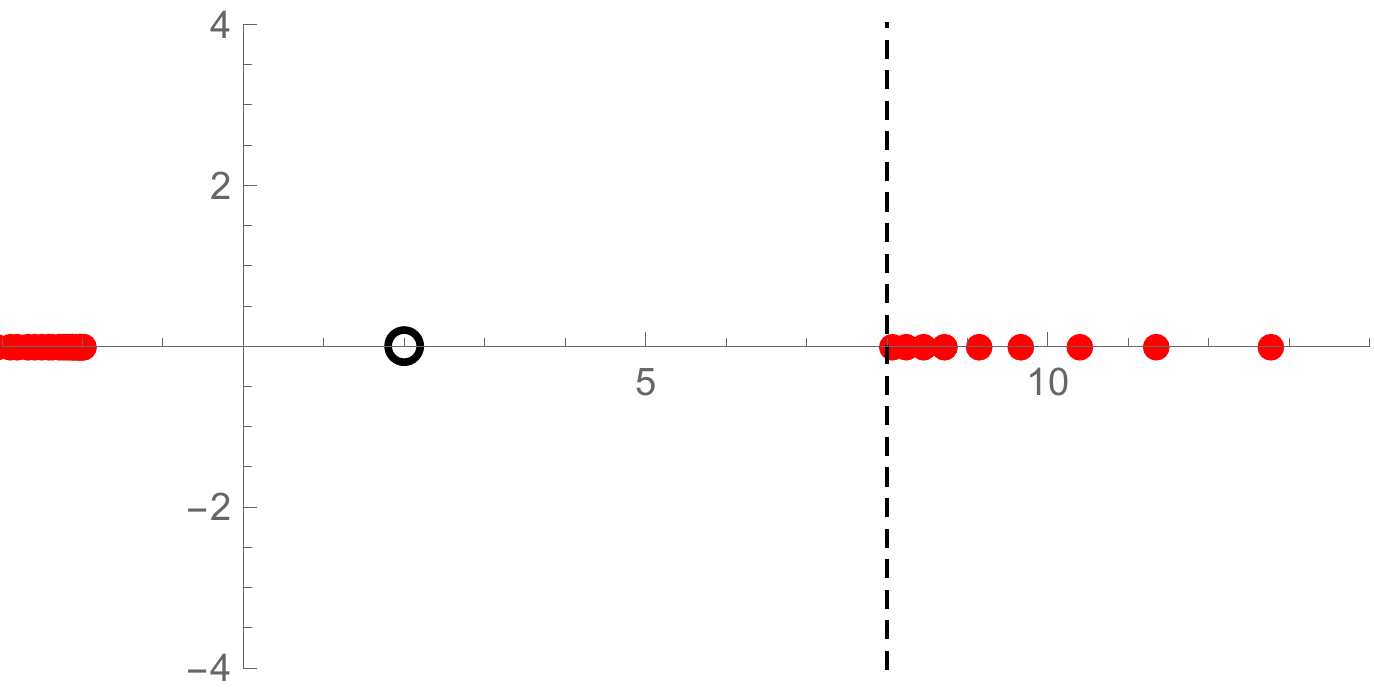}
\caption{$O(6)$ NLSM}
\end{subfigure}
\\
\vspace{\baselineskip}
\centering
\begin{subfigure}{0.925\textwidth}
\centering
\picdims[width=\textwidth]{\textwidth}{0.2\textwidth}{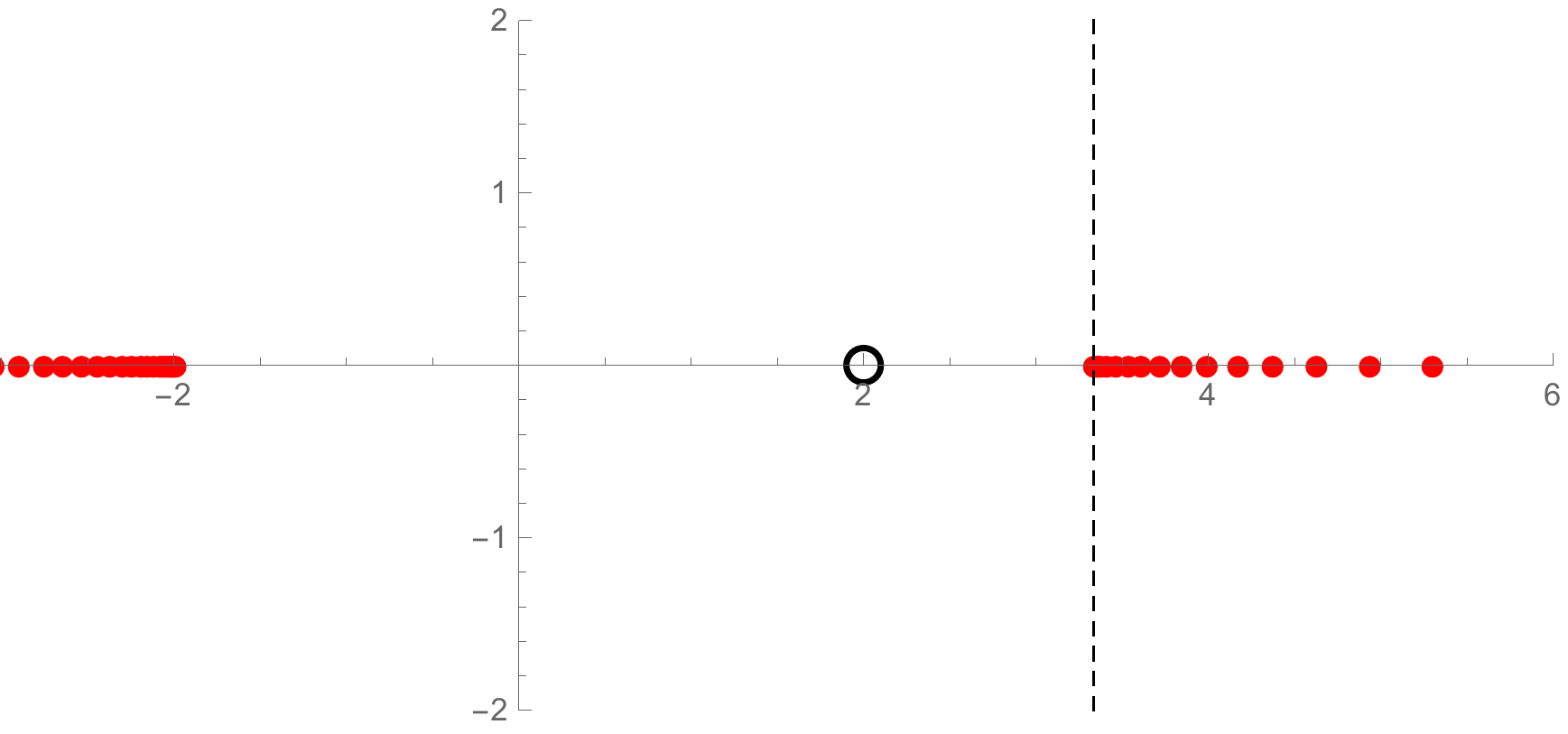}
\caption{$O(7)$ GN model}
\end{subfigure}
\\
\vspace{\baselineskip}
\begin{subfigure}{0.925\textwidth}
\centering
\picdims[width=\textwidth]{\textwidth}{0.2\textwidth}{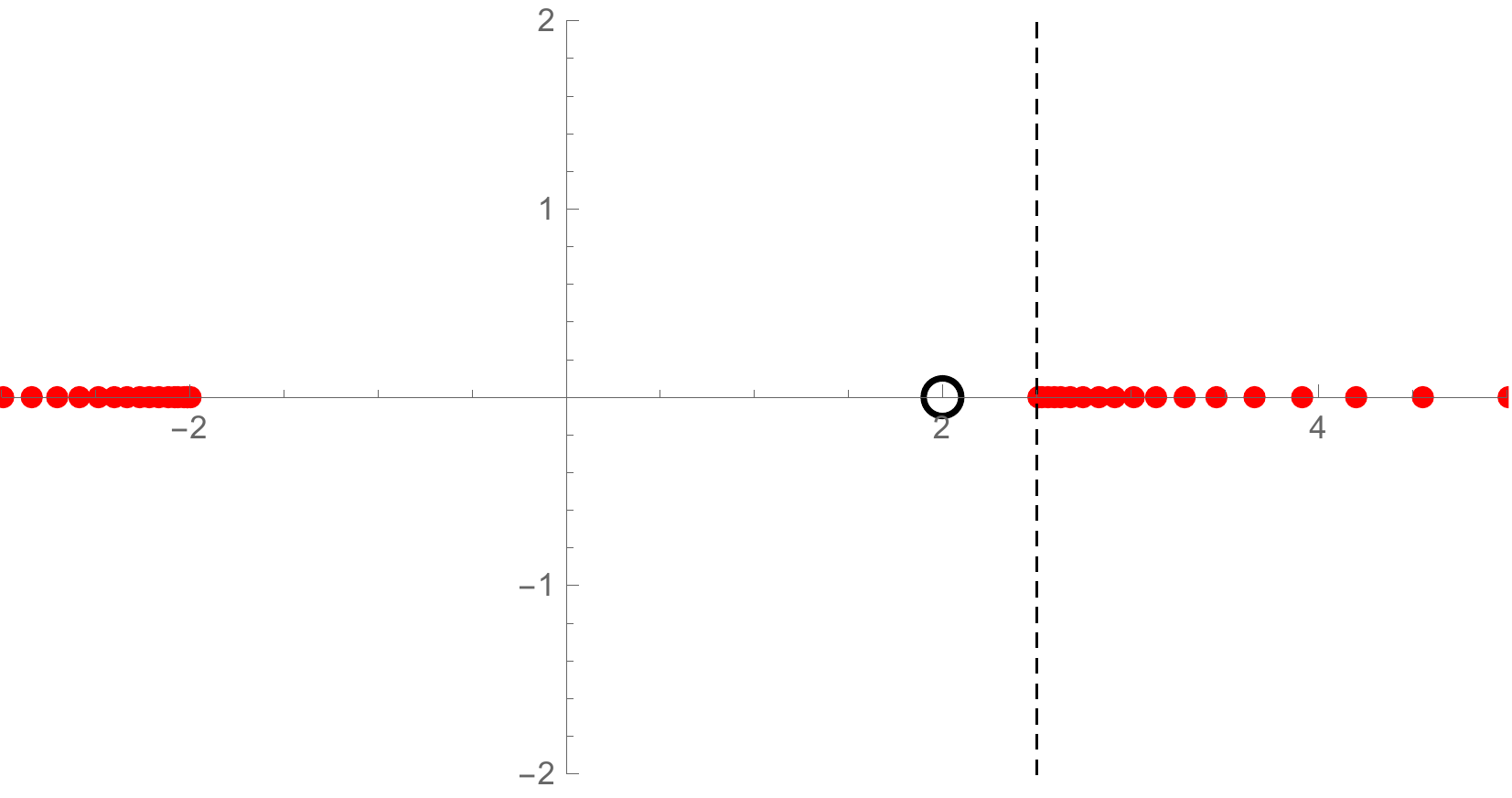}
\caption{$SU(5)$ PCF model}
\end{subfigure}
\\
\vspace{\baselineskip}
\caption{Figures adapted from \cite{mmr-antrans}. The poles of the Borel-Padé approximant of the series $\bar{e}_m$ in the Borel plane, for different models. All plots use 120 numeric terms of the perturbative series.
The dashed line indicates the predicted position from chapter \ref{cha_antrans} and the black circle indicates the position of the removed IR singularity at $\zeta=2$. 
}
\label{fig-borel-poles}
\end{figure}

In \ref{fig-borel-poles}, we plot this test for some of the models. They clearly show the position of the new renormalon, or, in its absence, of the leading instanton-like singularity. We can also identify the leading UV singularity appearing at $\zeta=-2$.

\subsection{Discontinuity of the Borel sum}

One can also use the Padé approximation of the Borel transform to extract the Stokes constants. Recall that we assume $s_\pm(\Phi^\pm)$, where $\Phi^\pm$ is the full trans-series, is the unambiguous exact result. This means that
\begin{equation}
\disc s(\varphi)(\alpha) = \ri\mathsf{S}_0\,\re^{-\frac{2}{\alpha}}\alpha^{\eta-2\xi} + \CO\left(\re^{-\frac{2\xi_1}{\alpha}}\right),\quad \alpha\ll 1,
\end{equation}
and, for the subtracted series,
\begin{equation}
\disc s(\phi)(\alpha) = \ri\mathsf{S}_1\,\re^{-\frac{2\xi_1}{\alpha}}\alpha^{(\eta-2\xi)\xi_1}\left(1+c_1^{(1)}\alpha\right) + \CO\left(\alpha^2\re^{-\frac{2\xi_1}{\alpha}}\right),
\end{equation}
where the $\mathsf{S}_{0,1}$ are specified by the $\CC^\pm_{0,1}$.

In principle, one could approximate $s_\pm$ by taking
\begin{equation}
s_\pm(\varphi)(\alpha) \sim \frac{1}{\alpha}\int_0^{\re^{\pm\ri\theta}\infty} \re^{-\zeta/\alpha} \mathsf{P}^M_{\widehat{\varphi}}(\zeta)\rd\zeta,\quad 1\gg\theta>0.
\label{s-numeric-disc}
\end{equation}
However, for the calculation of discontinuity we found this to be unnecessarily unstable when $\alpha\rightarrow 0$.
We can calculate instead a sum over the residues of the Padé approximant,
\begin{equation}
\disc s(\varphi) = -\frac{2\pi\ri}{\alpha}\sum_{j=0}^{M/2} \re^{-r_j/\alpha}\textrm{Res}_{\zeta = r_j} \mathsf{P}^M_{\widehat{\varphi}}(\zeta).
\label{disc-res-num}
\end{equation}
This requires first the calculation of the poles $r_j$ of the Padé approximant, which amounts to finding numerically all the solutions of $1/\mathsf{P}^M_{\widehat\varphi}(\zeta)=0$. The calculation of the residues can also be easily done numerically, for example by integrating in a small circle around the numerical approximation of the pole. Naturally, the radius must be significantly larger than the numerical precision but substantially smaller than the distance between neighboring poles. 

The advantage of this method is that the calculation of the poles and residues is done independently of $\alpha$. These values can be calculated first with high precision and saved. Then, the discontinuity at different values of $\alpha$, \eqref{disc-res-num}, simply amounts to summing the residues with different $\alpha$ dependent weights, which is a fast and simple calculation. This avoids a lot of the instabilities that plague the numeric integration of \eqref{s-numeric-disc} when $\alpha$ is small, and makes it easy to extrapolate since evaluating at new values of $\alpha$ is cheap. Nevertheless, there is no such thing as a free lunch and for very small $\alpha$ this approximation also breaks down, although at $\alpha$ much smaller than what we require to extrapolate with great precision. 

\begin{figure}
\begin{center}
\begin{subfigure}{\figsize}
\centering
\includegraphics[width=\textwidth]{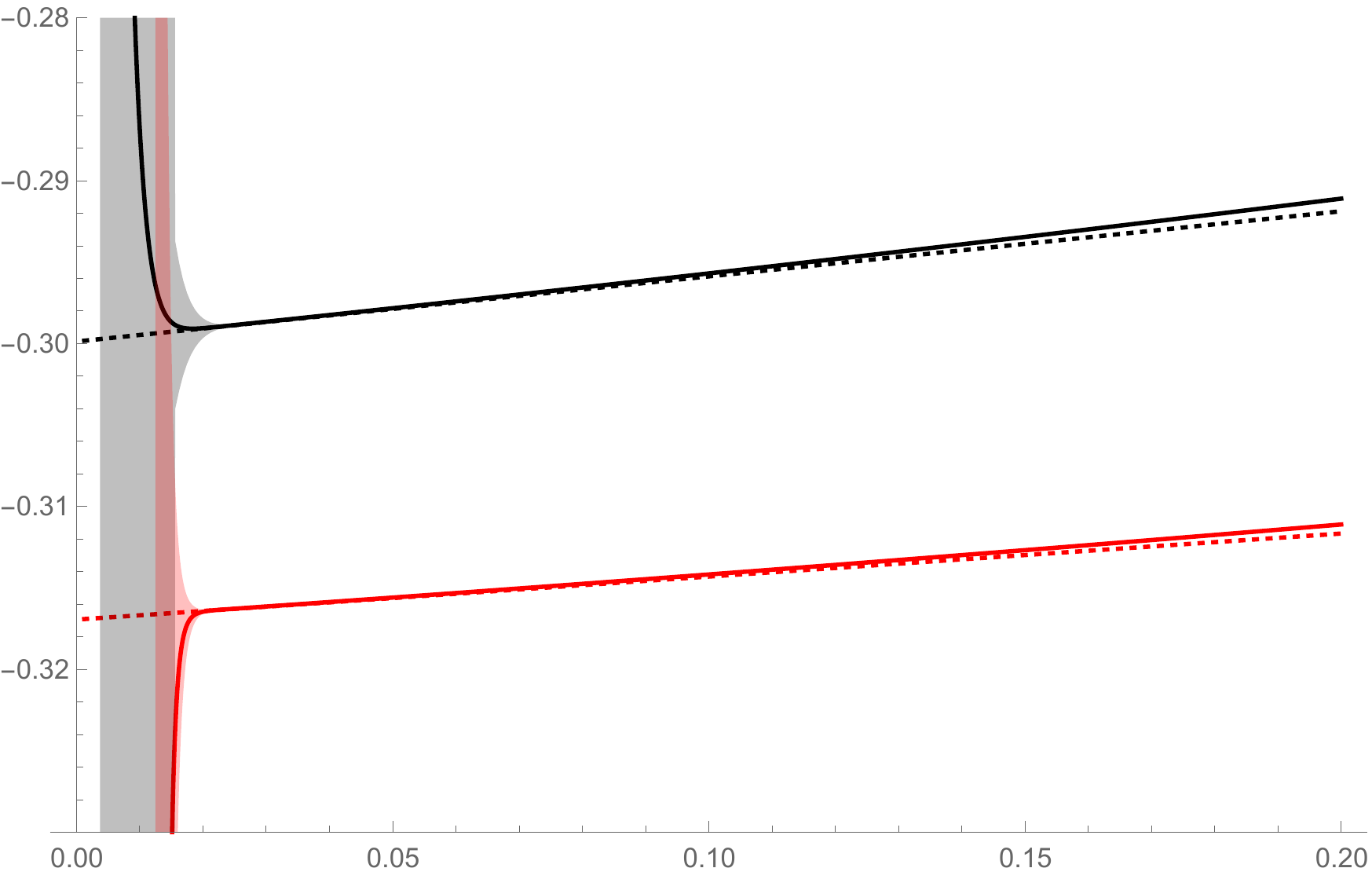}
\caption{$O(7)$ (black) and $O(8)$ (red) GN models}
\end{subfigure}
\\
\begin{subfigure}{\figsize}
\centering
\includegraphics[width=\textwidth]{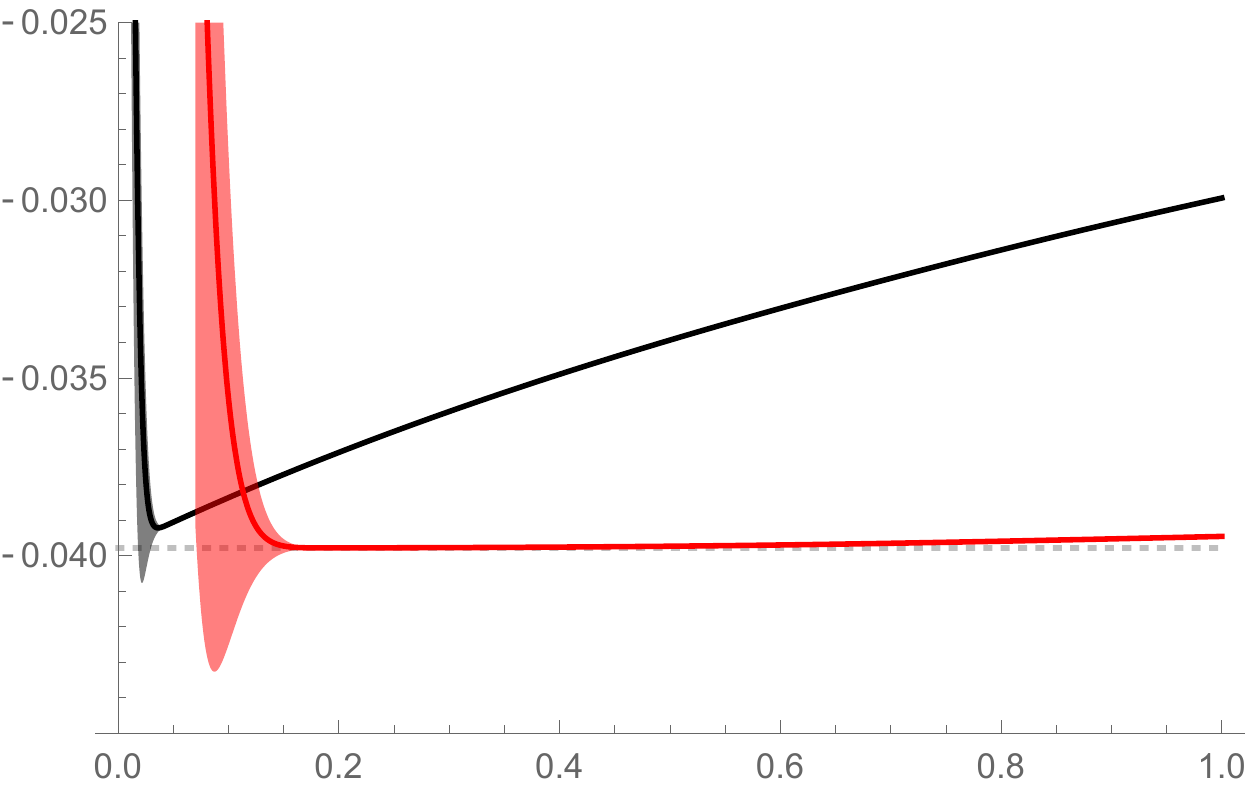}
\caption{$O(6)$ SUSY NLSM}
\end{subfigure}
\end{center}
\caption{Figures adapted from \cite{mmr-antrans}. We plot an approximation to $f_\phi(\alpha)$ defined in \eqref{f-disc-def}. At the top, the $O(7)$ GN model (black) and $O(8)$ GN model (red), using 71 coefficients in both cases. 
At the bottom, for the SUSY $O(6)$ NLSM we plot $f_\phi(\alpha)$ (black) and its second Richardson transform (red), using 68 coefficients.
The shaded areas represent the error of the corresponding color and the dashed lines represent the predicted asymptotic behavior, which includes the first linear correction for GN. The $x$-axis is the value of $\alpha$.}
\label{fig-disc}
\end{figure}

To find the limiting behavior, it is convenient to define the normalised discontinuities,
\begin{equation}
\ba
f_\varphi(\alpha) &= \frac{\disc s(\varphi)(\alpha)}{2\pi\ri \left(\re^{-\frac{2}{\alpha}}\alpha^{\eta-2\xi}\right)} \sim \frac{\mathsf{S}_0}{2\pi} + \CO\left(\re^{-\frac{2(\xi_1-1)}{\alpha}}\right),\\
f_\phi(\alpha) &= \frac{\disc s(\phi)(\alpha)}{2\pi\ri \left(\re^{-\frac{2\xi_1}{\alpha}}\alpha^{(\eta-2\xi)\xi_1}\right)} \sim \frac{\mathsf{S}_1}{2\pi} + \CO(\alpha).
\ea
\label{f-disc-def}
\end{equation}
For the case of $f_\phi$, we know that the convergence for small $\alpha$ is only linear. In the Gross--Neveu model, we have calculated the linear coefficient, so we can incorporate it in our prediction. But for bosonic models we have not. There, it is thus useful to accelerate its convergence with the Richardson transform \eqref{Rich-cont} to minimize the effect of this term. We often used $s=2$. See figure \ref{fig-disc} for two examples.

\subsection{Comparison with the integral equation}

\begin{figure}
\begin{center}
\begin{subfigure}{\figsize}
\centering
\includegraphics[width=\textwidth]{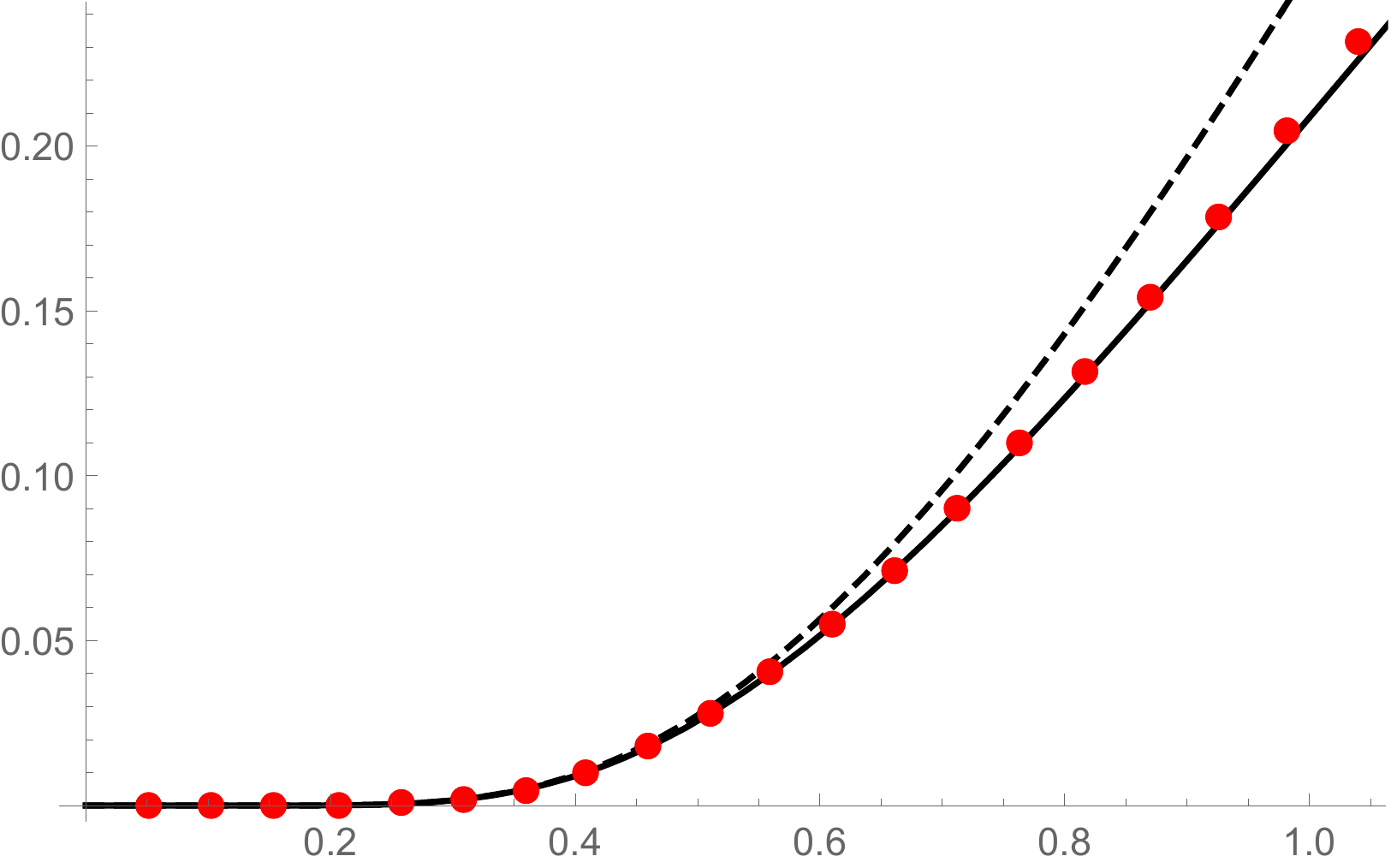}
\caption{$O(8)$ GN model}
\end{subfigure}
\\
\begin{subfigure}{\figsize}
\centering
\includegraphics[width=\textwidth]{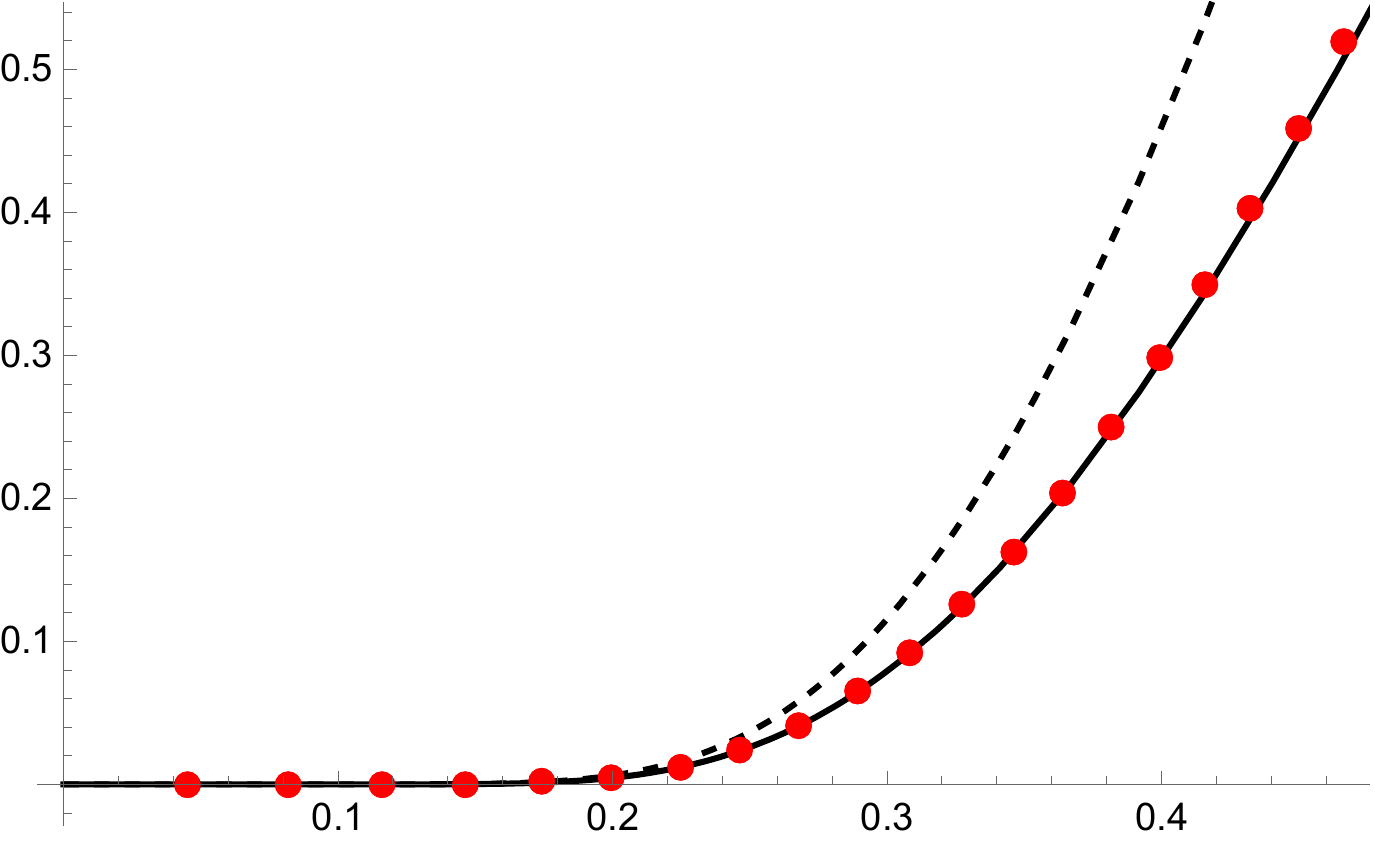}
\caption{$O(3)$ NLSM}
\end{subfigure}
\end{center}
\caption{
Plot of $r_\varphi(\alpha)$ defined in \eqref{r-rem-def}.
At the top, the $O(8)$ GN model, at the bottom the $O(3)$ NLSM.
The $x$-axis is the value of $\alpha$. 
The dots (red) are the numerical calculations of $r_\varphi$ evaluated at $B=20/k$ for $k=1,\dots,20$. The dashed line (black) is the leading theoretical prediction and the full line (black) includes further corrections.
For the $O(8)$ GN model, it includes up to the linear correction to the first new renormalon in \eqref{efromWHZ}.
For the $O(3)$ NLSM, it includes up to the linear term in the $\re^{-2/\alpha}$ sector of \eqref{eq_normalized_energy}, thus including the $\log\alpha$ term.
Figure on top adapted from \cite{mmr-antrans}. 
}
\label{fig-real}
\end{figure}

So far, the methods introduced only test the difference between the trans-series parameters $\CC^\pm_{0,1}$, that is to say their imaginary part. We can also test their real part. Let $\mathsf{R}_i$ be
\begin{equation}
\mathsf{R}_i = \frac{\CC_i^++\CC_i^-}{2}.
\end{equation}
Because we know that the imaginary part of the trans-series parameters cancels the imaginary ambiguity of Borel summation, the real part must contribute to the actual exact value. We do caution that this is only true at leading exponential order.
Let
\begin{equation}
r_\varphi(\alpha)=\frac{e}{\rho^2} - \frac{\alpha^\eta}{k^2} \Re\left\{s_\pm(\varphi)(\alpha)\right\},
\label{r-rem-def}
\end{equation}
where $k$ is defined in \eqref{Gplus-bosonic} for bosonic models and is $(2\pi)^{-1/2}$ for Gross--Neveu. Then we must have
\begin{equation}
r_\varphi(\alpha) \sim \mathsf{R}_0  \re^{-\frac{2}{\alpha}}\alpha^{\eta-2\xi} + \mathsf{R}_1\re^{-\frac{2\xi_1}{\alpha}}\alpha^{(\eta-2\xi)\xi_1}\left(1+c_1^{(1)}\alpha+\cdots\right)+\cdots.
\end{equation}
In order to perform this test, we need to have an independent calculation of $e/\rho^2$, which we obtain from solving \eqref{iqft_geneq} numerically.

The numeric solution of the integral equation can be achieved by discretizing the interval $[-B,B]$ according to a quadrature. We used the Gauss–Kronrod quadrature, for example. This reduces the integral equation to a linear system of equations for $\chi(\theta_i)$ where $\theta_i$ are the quadrature points. The observables $e$ and $\rho$ can be obtained by calculating their respective integrals with the same quadrature. The convergence of the results is exponential with the number of discretisation points, so it suffices to pick a sufficiently large but not huge number, around $20$--$200$ depending on desired precision. It should be noted that in this method we specify $B$ first, then we obtain $\rho$ from which we calculate $\alpha$.

As for the Borel summation, in this method we cannot avoid doing the integral \eqref{s-numeric-disc}. In order to stabilise the integral at small $\alpha$, the ideal method is the ``Padé-Conformal-Borel'' method of \cite{costin-dunne-conformal}. We start by observing that $\widehat{\varphi}$ should be analytic in $\mathbb{C}$ except for the branch cuts at $(-\infty,-2]\cup[2,\infty)$. Then we introduce the conformal map,
\begin{equation}
u = \frac{\zeta/2}{\sqrt{1-(\zeta/2)^2}+1} \leftrightarrow \zeta = \frac{4 u}{u^2+1} = \sum _{n=1}^{\infty } 4 (-1)^{n+1} u^{2 n-1},
\end{equation}
which maps the plane except the cuts to the interior of the unit disk $\mathbb{D}_1$ and the cuts to the boundary of the disk. In the disk, the mapped Borel transform 
\be 
\widehat\varphi = \sum_{k\geq 0} \frac{e_k}{k!} \zeta(u)^k = \sum_{k\geq 0} b_k u^k,
\ee
is convergent everywhere as a series in $u$. So in the integral of the Borel sum, it is better to map each point $\zeta$ to a point $u$ and sum the Borel transform there as a convergent series in $u$. This can be further improved if instead of just summing the Borel transform we calculate the Padé approximant as a rational function of $u$.
That is, we find $\mathsf{P}^M_{\widehat\varphi\circ\zeta}$ such that
\begin{equation}
\varphi(\alpha) 
\xrightarrow{\text{Borel}} 
\sum_{k\geq 0} \frac{e_k}{k!} \zeta^k
\xrightarrow{\text{Conformal}} 
\sum_{k\geq 0} b_k u^k 
\xrightarrow{\text{Padé}} 
\mathsf{P}^M_{\widehat\varphi\circ\zeta}(u)
 = \sum_{k= 0}^M b_k u^k+\CO(u^{M+1}),
\end{equation}
hence the name ``Padé $\circ$ Conformal $\circ$ Borel $\circ\,\varphi$''.

In the end, we calculate
\begin{equation}
s_\pm(\varphi)(\alpha) \sim \frac{1}{\alpha}\int_0^{\re^{\pm\ri\theta}\infty} \re^{-\zeta/\alpha}\mathsf{P}^M_{\widehat\varphi\circ\zeta}\big(u(\zeta)\big)\rd\zeta,\quad 1\gg\theta>0.
\label{s-pcb}
\end{equation}
While we do not care about the imaginary part, it is important to keep a small angle $\theta$ in the integral to avoid singularities in the integrand.
We found $\pi/20\gtrsim\theta\gtrsim\pi/50$ to be a good range for our purposes. As suggested in \cite{dpmss}, the best estimate of the numerical error is the convergence of the approximant. That is, we compare the calculation using $\left[\frac{M}{2}/\frac{M}{2}\right]$ with the calculation using $\left[\frac{M}{2}-1/\frac{M}{2}-1\right]$. This also applies to the computation of \eqref{disc-res-num}. See two examples of the numerical calculation of $r_\varphi$ in figure \ref{fig-real}.

We can also use this strategy to test the change in the real part of the parameters when $\vartheta=\pi$. However, the integral equations when $\vartheta=\pi$ are defined on a semi-infinite interval. This makes the convergence of the numeric solution much
worse and more sophisticated numeric solutions of the integral equation are required. These tests were done successfully in \cite{mmr-theta} and we refer to appendix B therein for details on the numeric solution of the integral equations.

The tests described in this chapter were applied to all models of section \ref{sec_iqft} for different values of $\Delta$ with uniform success. This confirms that the trans-series we find is the correct one and that we can trust our results about the structure of singularities in the Borel plane, at the very least up to the second singularity after the IR renormalon. 

\part{Renormalons in Condensed Matter Theory}
\label{part-qmb}

\chapter[Superconductivity in the Borel plane][Superconductivity in the Borel plane]{Superconductivity\\ in the Borel plane}
\label{cha_GY}


Quantum many-body systems were the original setting for the Bethe ansatz. The Lieb--Liniger model \cite{ll}, an integrable gas of bosons, is one the simplest integrable system, and the Gaudin--Yang model \cite{gaudin,yang}, an integrable gas of fermions, was the first system solved using the nested Bethe ansatz. These systems can also be solved perturbatively at weak coupling by calculating Feynman diagrams. And, in the case of the attractive Gaudin--Yang, it can also be described as a superconductor in light of BCS theory. 
As we study in this chapter, these three approaches are related. Through Volin's method, one can extract perturbative series from the Bethe ansatz skipping diagrammatic calculations. From resurgence, one can find superconductivity as a singularity in the Borel plane. And one can even extract the effects of superconductivity from the integral Bethe ansatz equations.

We start this chapter by reviewing some fundamentals about the superconductivity and the BCS gap, followed by a conjecture on their relation to resurgence. We review the analysis of the Gaudin--Yang model done in \cite{mr-long}, with some updates from \cite{mmr-antrans}. Then we extend the analysis to the Hubbard model, reviewing the results of \cite{mr-hubbard} for the two-component case at weak coupling and small filling. 

\section{Superconductivity and non-perturbative physics}
\label{sec_introbcs}

A superconductor can be seen as a system of attractive fermions whose ground state is populated by two fermion bound states, Cooper pairs. Such theories have a gap, related to the binding energy of a Cooper pair. This is the essence of BCS theory \cite{bcs}. As we show in this section, the gap is a non-perturbative effect. This sets the ground for the question of how this non-perturbative effect relates to perturbation theory, which will be the problem explored in the rest of the chapter.

\subsection{BCS approximation}

Let us suppose we have the BCS Hamiltonian in one-dimension,
\begin{equation}
H = \sum_{k,\sigma} \epsilon_k c^\dagger_{k,\sigma}c_{k,\sigma}-\frac{g}{L}\sum_{k,k',q} c^\dagger_{k+q,\uparrow}c^\dagger_{-k,\downarrow}c_{-k'+q,\downarrow}c_{k',\uparrow}.
\label{H-BCS}
\end{equation}
In a superconductor vacuum, the Cooper pair operator is expected to acquire a vev,
\begin{equation}
\Delta = \frac{g}{L} \sum_{k} \langle \Omega_S | c_{-k,\downarrow}c_{k,\uparrow} |\Omega_S\rangle.
\label{Delta-vev}
\end{equation}
We can then approximate the Cooper pairs by a bosonic mean-field,
\begin{equation}
\sum_k c_{-k+q,\downarrow}c_{k,\uparrow} \sim \frac{L\Delta}{g}+\cdots,
\end{equation}
and write the mean field Hamiltonian
\begin{equation}
H-\mu N \sim \sum_{k}\left\{(\epsilon_k-\mu) c^\dagger_{k,\sigma}c_{k,\sigma}-\left(\bar\Delta c_{-k,\downarrow} c_{k,\uparrow} + \Delta c^\dagger_{k,\uparrow} c^\dagger_{-k,\downarrow}\right)\right\}+\frac{L |\Delta|^2}{g}.
\label{mean-field-Ham}
\end{equation}
The vacuum of the mean field Hamiltonian can be found with a Bogliubov transformation parameterized by 
\be
\quad \theta_k = -\frac{1}{2}\cot^{-1}\left(\frac{\epsilon_k-\mu}{\Delta}\right).
\label{thetak-bcs}
\ee
The wave function of the vacuum is then given by
\begin{equation}
|\Omega_S\rangle = \prod_k (\cos\theta_k - \sin\theta_k c^\dagger_{k,\uparrow}c^\dagger_{-k,\downarrow})|\Omega\rangle,
\label{BCS-wavefunction}
\end{equation}
where $|\Omega\rangle$ is the vacuum of the $c_{k,\sigma}$ algebra, see \cite{as} for details. We can think of \eqref{BCS-wavefunction} as a squeezed state of Cooper pairs.

Plugging the vacuum  wavefunction \eqref{BCS-wavefunction} back into the vev \eqref{Delta-vev}, leads to an equation for $\Delta$,
\begin{equation}
\Delta = \frac{g}{L}\sum_{k}\sin\theta_k \cos\theta_k = \frac{g}{2L}\sum_{k}\frac{\Delta}{\sqrt{\Delta^2+(\epsilon_k-\mu)^2}}.
\label{BCS-gap-pure}
\end{equation}
This consistency equation for $\Delta$ is the BCS gap equation, and it allows us to solve for the gap itself. It can be derived in an alternative way which is to plug the mean-field vacuum \eqref{BCS-wavefunction} into the original Hamiltonian and treating it as a variational approximation to the ground state with parameters $\theta_k$. Then \eqref{thetak-bcs} and \eqref{BCS-gap-pure} are requirements for the optimal solution.

The Gaudin--Yang Hamiltonian \eqref{delta-GY} for spin $1/2$ particles ($\kappa=2$) is not of the form \eqref{BCS-wavefunction}. The $\delta$ interaction induces a vertex with all spin $\uparrow$ particles and with all spin $\downarrow$, in addition to the BCS vertex. However we can still use \eqref{BCS-wavefunction} as a variational approximation of the vacuum and optimize with respect to $\theta_k$. 
The additional vertices in the Hamiltonian can then be absorbed as corrections to the one-particle energy $\epsilon_k$ in the variational problem, see
\cite{montse,quick,fw} for details. This is a strictly worse description of the ground state than the Bethe ansatz, which is exact. Nevertheless, it is useful in highlighting the non-perturbative nature of the gap and the superconductor nature of the vacuum. 

Following the procedure outlined, the BCS gap equation for the Gaudin--Yang model is
\begin{equation}
\frac{c}{2\pi}\int_\IR\frac{\rd k}{\sqrt{\xi_k^2+\Delta_{\rm BCS}^2}} =1,
\label{GY-gap}
\end{equation}
with
\begin{equation}
\xi_k = k^2-\mu-cn.
\end{equation}
We can also relate the chemical potential $\mu$ to the density $n$ through $\partial E/\partial \mu=-n$, leading to
\be
n={1\over 2 \pi} \int_\IR  \left( 1- {\xi_k \over {\sqrt{\xi_k^2 + \Delta_{\rm BCS}^2}}} \right)\rd k,
\label{n-mu-gy}
\ee
and the energy of the optimal BCS wave-function is
\be
{E_{\rm BCS} \over n}= {1\over 2 \pi n} \int_\IR  \, \xi_k \left( 1 - {\xi_k \over {\sqrt{\xi_k^2 + \Delta_{\rm BCS}^2}}} \right)\rd k + {c n \over 2} +\mu-{\Delta_{\rm BCS}^2 \over 2c n}. 
\label{e-bcs}
\ee

To find and approximate expression for the gap in the weak coupling limit, we start by introducing
\be
b= \mu+cn,\quad \alpha^4=\Delta_{\rm BCS}^2+b^2, 
\ee
as well as
\be
\label{modulus}
m= {1\over 2} \left( 1+ {1\over {\sqrt{1+ \Delta_{\rm BCS}^2/b^2}}} \right). 
\ee
In terms of these variable, the gap equation \eqref{GY-gap} cap be compactly written with  an elliptic integral of the first kind,
\be
\label{K-gap}
K(m)= {\pi  \alpha \over c}.
\ee
Meanwhile, equation \eqref{n-mu-gy} is an elliptic integral of the second kind,
\be
n= {b \over c} -{ \alpha^2 \over c}+{2 \alpha \over \pi} E(m).
\ee
And finally, the energy per particle is simply  
\be
{E_{\rm BCS} \over n}= \frac{2 \alpha^3}{3 \pi n} \left((2m-1)E(m)+(1-m)\frac{\pi\alpha}{c}\right) -\frac{cn}{2}-\frac{\Delta_{\rm BCS}^2}{2cn}. 
\label{bcs-em}
\ee
This formulation, which is exact within the context of the BCS approximation, facilitates taking the limit $c\rightarrow 0$, where
\begin{equation}
m \sim 1-\frac{\Delta_{\rm BCS}^2}{4b^2}+\cdots.
\end{equation}
Using the known properties of exponential integrals, we eventually arrive at
\be
\label{bcs-gap}
\Delta_{\rm BCS} \approx 8  k_F^2 \, \re^{-{\pi^2 \over2 \gamma}}, 
\ee
where $k_F^2$ is the Fermi energy. The normalised ground state energy $e_{\rm BCS}=E_{\rm BCS}/n^3$, when computed with \eqref{bcs-gap} and \eqref{bcs-em} yields
\be
e_{\rm BCS}(\gamma) \approx {\pi^2\over 12}-{\gamma \over 2} - 2 \pi^2 \re^{- \pi^2 / \gamma}. 
\label{egs-bcs}
\ee
Where we have a non-perturbative correction coming from $\Delta_{\rm BCS}^2$.

The gap \eqref{bcs-gap} comes from the variational optimization of the BCS wave-function, which is merely approximate. Meanwhile, the Bethe ansatz wave-function is exact, and it is possible to extract the gap at weak coupling from it, as done in \cite{ko,zhou-exact,frz}. We review an extension of that calculation in chapter \ref{cha_spin}. 
The leading expression for the gap calculated from the Bethe ansatz is
\be
\label{spin-gap}
{\Delta} \approx {16 k_F^2\over \pi} {\sqrt{\gamma \over \pi}} \re^{-{\pi^2 \over 2 \gamma}}. 
\ee
The exponential weight is identical to the BCS case but the overall power of $\gamma$, which is a NLO correction, is not.

There are two main lessons to take away from this calculation. The first is that the gap is indeed an exponentially suppressed effect, as seen in \eqref{bcs-gap} and \eqref{spin-gap}. The second is that the effect of the gap in the ground state energy is proportional to $\Delta_{\rm BCS}^2$. 

\subsection{Non-perturbative gaps and renormalization}

The non-perturbative nature of the superconductor gap can be connected to our discussion on non-perturbative effects in relativistic asymptotically free field theories in chapter \ref{sec_renormalons}. While renormalization is not necessary in these non-relativistic theories, it is a valid procedure in the Wilsonian sense. In particular, we might consider integrating out high energy modes and focus on modes closer and closer to the Fermi surface, which in one dimension are simply the Fermi points $\pm k_F$. Because the dispersion relation is linearised in the effective theory close to the Fermi surface, it approximately becomes  a relativistic theory, in the appropriate units. This effective theory has a $\beta$-function
\begin{equation}
\beta(g) = \mu \frac{\rd g(\mu)}{\rd\mu} = \beta_0 g^2 + \beta_1 g^3+\cdots.
\label{beta-qmb}
\end{equation}
Then, there is an RG-invariant non-perturbative scale in the theory,
\begin{equation}
\mathcal{I}(g) =  \mu  \re^{\frac{1}{\beta_0 g^2}}\left( g^2\right)^{\frac{\beta_1}{\beta_0^2}}\exp\left\{-\int_0^{g(\mu)} \left(\frac{1}{\beta(g')}-\frac{1}{\beta_0 g'^2}+\frac{\beta_1}{\beta_0^2g'}\right)\rd g'\right\}.
\label{I-gap}
\end{equation}

Since the gap is a non-perturbative phenomenon, it is natural to guess that $\Delta \propto \mathcal{I}(g)$. In \cite{ls}, the correct gap for the two-component Gaudin Yang model was found using this strategy. Conceptually, we see the similarities between the superconductor gap in non-relativistic field theories and the dynamically generated scale $\Lambda$, as well as the conventional IR renormalon, in asymptotically free theories. We will revisit this argument when we discuss the multi-component Gaudin-Yang model in chapter \ref{cha_spin}.

\section{Resurgence and the superconductor gap}
\label{sec_superconductivity_conjecture}

In \cite{mr-long}, we conjectured that in a superconductor system, perturbation theory should be a non Borel summable asymptotic series.
This lack of summability should be seen as a manifestation of the fact that the perturbative expansion is ignorant of the non-perturbative superconductor gap. Focusing on the ground state energy, we conjectured that for $A$ such that
\begin{equation}
\Delta^2 \propto \re^{- \frac{A}{\gamma}},\quad \gamma\ll 1,
\end{equation}
then the Borel transform of the perturbative series for the energy density should have a pole at
\begin{equation}
\zeta = A.
\end{equation}
We can write a slightly more refined version of this conjecture by also accounting for the subleading correction. I.e., if 
\begin{equation}
\Delta \propto \gamma^{\frac{b}{2}}\re^{- \frac{A}{2\gamma}},\quad \gamma\ll 1,
\end{equation}
then the asymptotic behavior of the perturbative coefficients of the  energy density is given by
\begin{equation}
c_m \sim C A^{b-m} \Gamma(m-b),\quad m\gg 1,
\label{asym_gap}
\end{equation}
where $C$ is some constant.

In a way, this is a resurgent perspective of the Cooper instability. When one expands at weak coupling, the expansion is around the free gas rather than the gapped phase. This can be seen as expanding around a ``wrong vacuum'' and thus unstable. The resulting perturbative series is still the correct asymptotically approximation, but it cannot capture the non-perturbative physics. This results in the lack of Borel summability. 
Such summability issues can be fixed with a trans-series which requires non-perturbative information. We will see an example where the integral equation, which is exact, can provide the necessary non-perturbative information to complete the perturbative series and reveal further non-perturbative terms.
In addition to the one dimensional models studied in this chapter, we found in \cite{mr-long} suggestive but inconclusive evidence that this relation also holds for three dimensional fermion gases with an attractive $\delta$-interaction.

Lastly, it should be noted that we observe the scale of superconductivity in the perturbative series as a renormalon effect specifically. We will show this in detail in chapter \ref{cha_spin}, while extending the discussion to more general energy gaps. 

\section{Exact series for the Gaudin--Yang model}

Let us focus on the Gaudin--Yang model, introduced in section \ref{sec_iqmb}.

\subsection{Perturbation theory and beyond}

\begin{figure}[t]
\begin{tikzpicture}[scale=1,rotate=+0,very thick]
\draw[] (-1,0.5) -- (-1,-0.5);
\draw[] (1,0.5) -- (1,-0.5);
\begin{feynman}
\draw[photon, cyan] (-1,0) -- (1,0);
\end{feynman}
\node at (2,0) {$\quad \propto\quad\gamma$};
\end{tikzpicture}
\centering
\caption{The four-vertex in the Gaudin--Yang model.}
\label{fig-vertex-GY}
\end{figure}
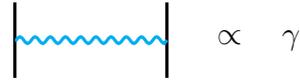

The ground state energy of a quantum many-body system can be calculated using standard Feynman diagram techniques. For a deep pedagogical introduction to field theory techniques in condensed matter, see \cite{as}. As for the Gaudin--Yang model, the interaction term in \eqref{delta-GY} can be written as a four vertex in momentum space, which is often drawn with a ``phonon'' as in figure \ref{fig-vertex-GY}. Combined with the fermion propagator, we can build the appropriate Feynman diagrams.

\begin{figure}[b]
\centering
\includegraphics[width=0.9\textwidth]{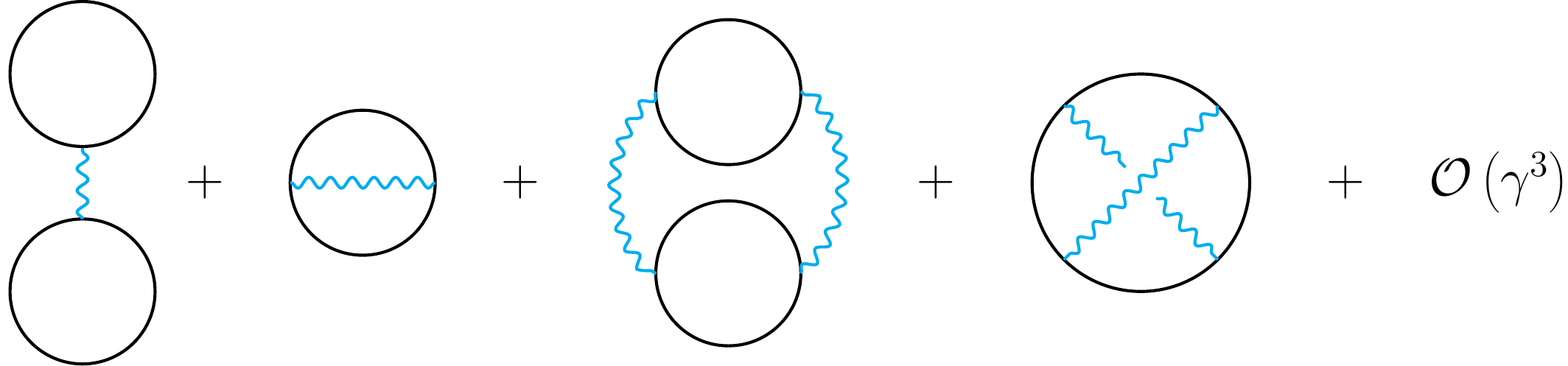}
\caption{Feynman diagrams for the calculation of the ground state energy up to order $\gamma^2$.}
\label{GY-diags}
\end{figure}

Our observable of choice is the ground state energy, which can be expanded as a power series in the dimensionless coupling $\gamma$, defined in \eqref{gammadef},
\begin{equation}
e(\gamma) = \sum_{n\geq 0} c_n \gamma^n.
\label{eGY_powerseries}
\end{equation}
The zeroth-order term of the weak coupling approximation of the ground state energy is naturally the energy of the one dimensional Fermi gas,
\begin{equation}
c_0= \frac{\pi^2}{12}.
\end{equation}
At first order we start considering Feynman diagrams. The first two diagrams in figure \ref{GY-diags} correspond to the Hartree-Fock terms. This is a known textbook calculation. At second order, one must calculate two more complicated diagrams to obtain
\begin{equation}
e(\gamma) \sim \frac{\pi^2}{12} - \frac{\gamma}{2} - \frac{\gamma^2}{12}.
\label{gy_o2}
\end{equation}

Third order is already substantially harder. For this model, it was calculated in \cite{magyar,magyar2} using density function techniques, 
\begin{equation}
c_3=-\frac{\zeta(3)}{\pi^4}.
\label{eGY_to_order3}
\end{equation}
Using Feynman diagram techniques, it was computed for the three dimensional $\delta$-interaction Fermi gas with $\kappa$ components in \cite{bishop}, a computation which was then extended to the one dimensional case in the appendix A of \cite{mr-long}.
 This coefficient is the first to suggest the curious number-theoretical properties of the series, introducing the $\zeta$-function.

Higher orders in $\gamma$ are impractical to obtain with Feynman diagrams. For certain models or limits, some particular families of diagrams become good approximations of the series. Ladder diagrams for example are one such a family (see \cite{steele} for example), as are ring diagrams, which we will discuss in chapter \ref{cha_spin}. Fortunately, thanks to the power of integrability, we can instead extract the exact coefficients without ever computing a single Feynman diagram.

\subsection{Exact series from Volin's method}

Since the Bethe ansatz solution is an exact solution of the problem, it should encode the series \eqref{eGY_powerseries} to all orders. However, even extracting the first few orders is difficult, having been done only to second order analytically in \cite{tw1,tw2} and numerically to tenth order in \cite{prolhac}.
Fortunately, Volin's powerful method, which we review in chapter \ref{cha_volin}, can be applied to the integral equation \eqref{volin_eq_TBA_GY}.

This method obtains a perturbative solution of $f(x)$ in two different limits, where it is approximated by the ansatze \eqref{volin_eq_bulkGY} and \eqref{volin_Rhatansatz}.
The key ingredient for \eqref{volin_Rhatansatz} comes from Fourier transform of the kernel,
\begin{equation}
1-\tilde{K}(\omega) = 1+\re^{-|\omega|},
\end{equation}
whose Wiener--Hopf decomposition is given by 
\begin{equation}
G_+(\omega) = \frac{\re^{ \frac{1}{2\pi}\ri\omega [ \log( -\frac{\ri\omega}{2\pi} ) - 1 ] }}{\sqrt{2\pi}}\Gamma\bigg( \frac{1}{2} - \frac{\ri\omega}{2\pi}  \bigg), 
\label{volin_Gplus_GY}
\end{equation}
as was used in \cite{tw1,tw2,mr-long}. 
 These ansatze have unfixed coefficients $c_{n,m,k}$ and $Q_{n,m}$ which are fixed by recursive algebraic equations. In the appendix section \ref{app-volin-gy}, we work out the details of the specific application to equation  \eqref{volin_eq_TBA_GY} to the first few orders.
Assuming that one has solved for the unknown coefficients to sufficiently high order,
we need only to specify how to obtain the desired observables. 

First the dimensionless coupling \eqref{gammadef} can be related to $B$ to the solution of \eqref{volin_eq_TBA_GY} through
\begin{equation}
\frac{1}{\gamma} = \frac{1}{\pi}\int_{-B}^B f(x)\rd x.
\end{equation}
This integral can be obtained from the bulk solution \eqref{volin_eq_bulkGY} through the  relation \eqref{volin_mom_from_res}, which entails
\begin{equation}
\frac{1}{\gamma} = \frac{2B}{\pi}+\frac{1}{\pi}\sum_{m\geq 0}\frac{c_{1,m,0}}{B^m}.
\label{1overgamma_coefs}
\end{equation}
For the energy itself, the ground state solution also provides the necessary information through
\begin{equation}
e(\gamma)=-\frac{\gamma^2}{4}+\frac{\frac{1}{\pi}\int_{-B}^B x^2f(x)\rd x}{\left(\frac{1}{\pi}\int_{-B}^B f(x)\rd x\right)^3}.
\label{egy_int}
\end{equation}
Combining \eqref{volin_eq_bulkGY} and \eqref{volin_mom_from_res} again, we have that
\begin{equation}
\int_{-B}^B x^2f(x)\rd x = \frac{2B^3}{3}+ B^2 \sum_{m \ge 0} \frac{c_{1,m,0}- 2 c_{1,m,1} +4 c_{1,m,2} }{ B^m} + B \sum_{m \ge 0} \frac{ c_{2,m,0} }{ B^m}. 
\label{x2_coefs}
\end{equation}
Using the coefficients computed in section \ref{app-volin-gy}, \eqref{1overgamma_coefs} and \eqref{x2_coefs} return
\be
\ba
\label{rhox-exp}
\frac{1}{ \gamma}&=\frac{2 B }{ \pi}+ \frac{1}{  \pi^2} \left( \log(\pi B) +1 \right) +\frac{1}{ 2 \pi^3 B} \left( \log(\pi B) + \frac{1 }{ 2} \right) +\CO(B^{-2}), \\
\langle x^2 \rangle &= \frac{2 B^3 }{ 3}+\frac{B^2}{ \pi} \left( \log(\pi B)-1\right)\\
&\qquad+ \frac{B }{ 2 \pi^2} \left( \log^2(\pi B)- \log(\pi B)-\frac{5}{ 2} +\frac{2 \pi^2}{ 3}\right) +
\CO(B^0). 
\ea
\ee
which agree with the results of \cite{tw1, tw2}. The solution of the equation itself $f(x)$ can also be inspected in the bulk regime $x\ll B$, thanks to \eqref{fitselfGY},
\be
\ba
f(x)&= 1 +\frac{1}{ \pi B (1-(x/B)^2)}\\
& +\frac{1 }{ \pi^2 B^2} \left( \frac{1 +  \log(\pi B)  }{ 2  (1-(x/B)^2)} -\frac{\log(B \pi) }{  (1-(x/B)^2)^2}- \frac{x \log\left(\frac{B-x }{ B+x} \right) }{ B(1-(x/B)^2)^2}  \right)+ \CO\left(B^{-3} \right). \\
\ea
\ee
The first term agrees with the results of \cite{iw-gy}. 

Using \eqref{rhox-exp} one can then re-express $B$ as a power series in $\gamma$. To the first few orders we have
\be
B= \frac{\pi }{2\gamma}
+\frac{1}{\pi}  \log (2 )-\frac{1 }{ 2 \pi} \left(\log \left(\frac{2 \pi ^2}{\gamma  }\right)+1\right)
+\CO\left(\gamma \right). 
\label{beta-gamma2}
\end{equation}
When we express $e$ in terms of $\gamma$, there is the fortunate cancellation of all terms in $\log \gamma$. This is precisely what we expect from Feynman diagrams (a similar phenomenon happened in relativistic field theories). 

The ground state energy can then be expanded,
\be
\label{eten}
\ba
& e(\gamma)=\frac{\pi ^2}{12}-\frac{\gamma }{2}-\frac{\gamma ^2}{12}-\frac{ \zeta (3)}{\pi ^4}\gamma ^3-\frac{3  \zeta (3)}{2 \pi ^6}\gamma ^4-\frac{3 \zeta (3)}{\pi ^8}\gamma ^5 -\frac{5  (5 \zeta (3)+3 \zeta (5))}{4 \pi ^{10}}\gamma ^6\\&-\frac{3\left(12 \zeta (3)^2+35 \zeta (3)+75 \zeta (5)\right)}{8 \pi ^{12}} \gamma ^7 -\frac{63  \left(12 \zeta (3)^2+7 \zeta (3)+35 \zeta (5)+12 \zeta (7)\right)}{16 \pi ^{14}}\gamma ^8\\&-\frac{3  \left(404 \zeta (3)^2+240 \zeta (5) \zeta (3)+77 \zeta (3)+735 \zeta (5)+882 \zeta (7)\right)}{4 \pi ^{16}}\gamma ^9
+{\mathcal O}\left(\gamma ^{10}\right)   \ea
   \ee
The series in (\ref{eten}) agrees with the numerical results of \cite{prolhac} as well as with the perturbative calculations. Using Volin's method, we obtained 60 coefficients in their exact form and 168 numerically.

\subsection{Large order behavior}

With the 60 exact coefficients one can obtain important information about the large order behavior of the coefficients. In contrast with relativistic theories, we observe no ``UV phenomena'', i.e. there are no alternating terms in the asymptotic behavior. This significantly simplifies the analysis since convergence is faster. A heuristic way of understanding this advantage is that the alternating terms force us to take the auxiliary series \eqref{auxseries}. So in their absence ``we have twice as many coefficients''. 

\begin{figure}
\centering
\includegraphics[width=\figsize]{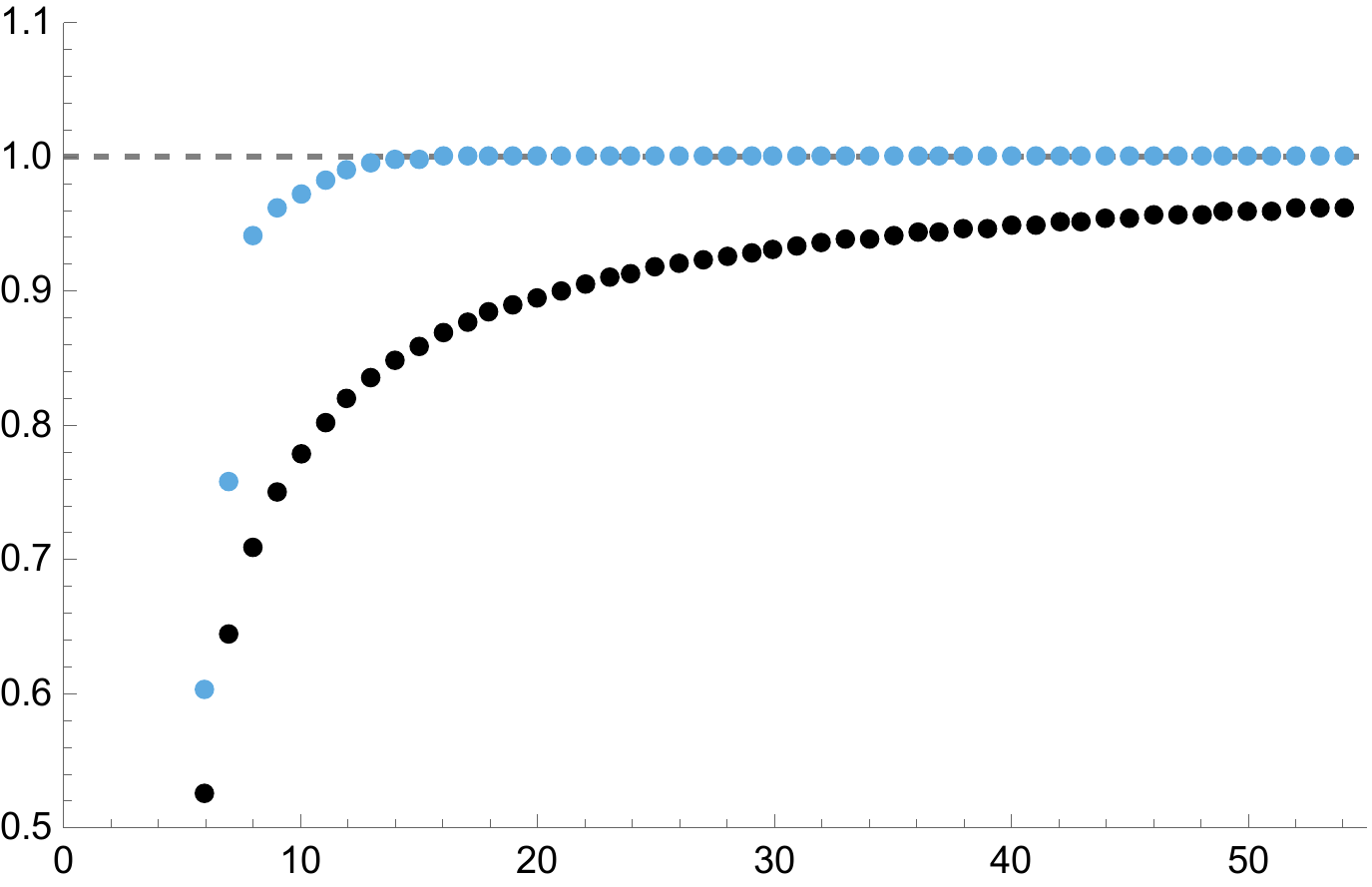}
\caption{Plot of sequence $a_m$ (black) in \eqref{am-seq} and its second Richardson transform (blue).}
\label{gy-plot}
\end{figure}

The leading asymptotic behavior we find is
\begin{equation}
c_m \sim - \frac{1}{\pi}\left(\pi^2\right)^{-(m-1)}\Gamma(m-1), \quad m \gg 1.
\label{GY_LOasym}
\end{equation}
If we define the auxiliary series
\begin{equation}
a_m = -\frac{\pi ^{2 m-1}}{\Gamma (m-1)}  c_m\,,
\label{GY_auxa_series}
\end{equation}
we can test \eqref{GY_LOasym} by verifying that
\begin{equation}
a_m \sim 1, \quad m \gg 1,
\label{am-seq}
\end{equation}
which we plot in figure \ref{gy-plot}. With only 50 coefficients, one can check the $\pi^2$ factor with 6 digits of precision. This asymptotic behavior confirms that the series is asymptotic and  divergent.

From the exact form of the coefficients, we can observe some number theoretical properties. First we observe that the coefficients are structured as
\begin{equation}
c_k = \frac{1}{\pi^{2k-2}}\left(r_k^{(3)} \zeta(3) + r_k^{(5)} \zeta(5) + r_k^{(3,3)} \zeta(3)^2+ r_k^{(3,5)} \zeta(3)\zeta(5) + \cdots\right),\quad k>2,
\end{equation}
where the $r_k^{(j_1,\dots,j_n)}$ are rational numbers and $j_i$ are odd numbers never larger than $k$. The coefficient of order $k=2$ is the only not to follow the pattern with the powers of $\pi$ and this is due to a factor of $\zeta(2)$. For all other coefficients, terms with $\gamma_E$ or $\zeta(2n)$ cancel at the end. This latter pattern was observed for the $O(N)$ sigma model in \cite{volin}, and in all models studied in part \ref{part-iqft}. 
After checking that the cancellation of terms like $\gamma_E$, $\zeta(2n)$, $\log \gamma$, etc. happens at least to order 15, we set such terms to zero in order to push calculations to order 60, as recommended in \cite{volin-thesis}. The success of further numerical tests at higher order validates this assumption.

In \cite{glang}, Lang used the results from \cite{mr-long} to conjecture the exact form of the rational factor of the terms with only $\zeta(3)$ and $\zeta(5)$ in each coefficients $c_k$. The conjectures II and III therein are that
\begin{align}
r_k^{(3)}&= -\frac{(k+2) }{2^{k+1}}\binom{2 k}{k},\quad k\geq 3,\\
r_k^{(5)}&=-\frac{(k+2)}{2^{k+1} (2 k-1) }\binom{2 k}{k} \binom{\binom{k}{2}}{2},\quad k\geq 6.
\end{align}
These conjectures hold for the known first 60 exact coefficients. A curious aspect of these numbers  pointed out in \cite{glang} is that they constitute convergent series. It suggests that the asymptotic nature of the series comes from having to sum more and more $\zeta$ terms at each order.

Since the publication of \cite{mr-long}, we have obtained the numerical expression of the first 168 coefficients in \eqref{eGY_powerseries}. Using these, we can improve upon the published results. As is standard in numerical analysis of large order behavior, see for example \cite{mmbook},
we define the constants $\mu_i$ such that
\begin{equation}
\ba
a_m &= 1 + \sum_{i\geq 1} \frac{\mu_i}{\prod_{k=1}^i (m-1-k)}\\
&=1+\frac{\mu _1}{m-2}+\frac{\mu _2}{(m-3) (m-2)}+\frac{\mu _3}{(m-4) (m-3) (m-2)}+\cdots\,.
\ea
\label{mu_asym}
\end{equation}
We can calculate the $\mu_i$ recursively by inspecting the large order behavior of $a_k$ using Richardson transforms. Optimal convergence was found using the 20-th Richardson transform of the type \eqref{int_RTN}, and we can obtain a substantial amount of digits for the first six $\mu_i$, as is shown in table \ref{GYtab}. The last stable digit is in brackets. These digits are sufficient to write a guess for the exact form of $\mu_i$ in some cases. The guesses we present are correct to all stable digits.

\begin{table}
\centering
\renewcommand{\arraystretch}{1.25}
\begin{tabular}{ c|c|c } 
$i$ & $\mu_i$ numeric & exact guess\\
 \hline
$1$ & $1.499999999999999999(9)$ & $3/2$\\
$2$ & $0.874999999999999(9)$ & $7/8$\\
$3$ & $-1.38955690315959(4)$ & $-\zeta (3)-3/16$\\
$4$ & $-7.8091965963272(7)$ & $-27\zeta (3)/4+39/128$\\
$5$ & $-36.839834641592(2)$ & \\
$6$ & $-217.47093(7)$ & \\
\hline
\end{tabular}
\caption{Numeric estimation of the coefficients in \eqref{mu_asym}}
\label{GYtab}
\renewcommand{\arraystretch}{1}
\end{table}

\begin{figure}
\centering
\includegraphics[width=\figsize]{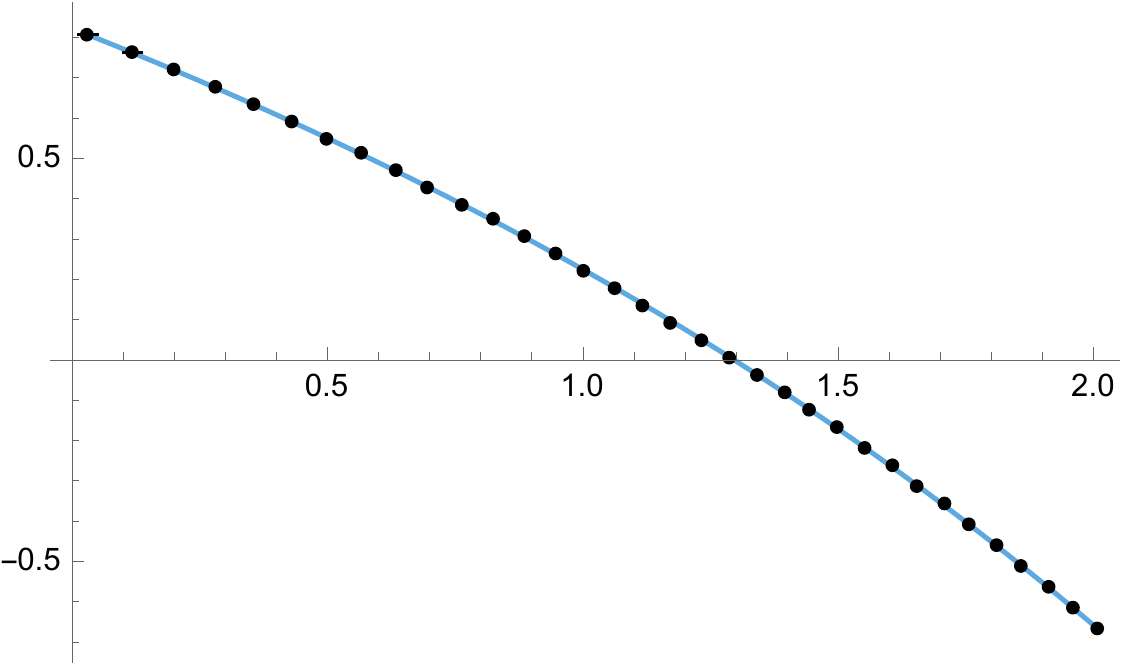}
\\
\includegraphics[width=\figsize]{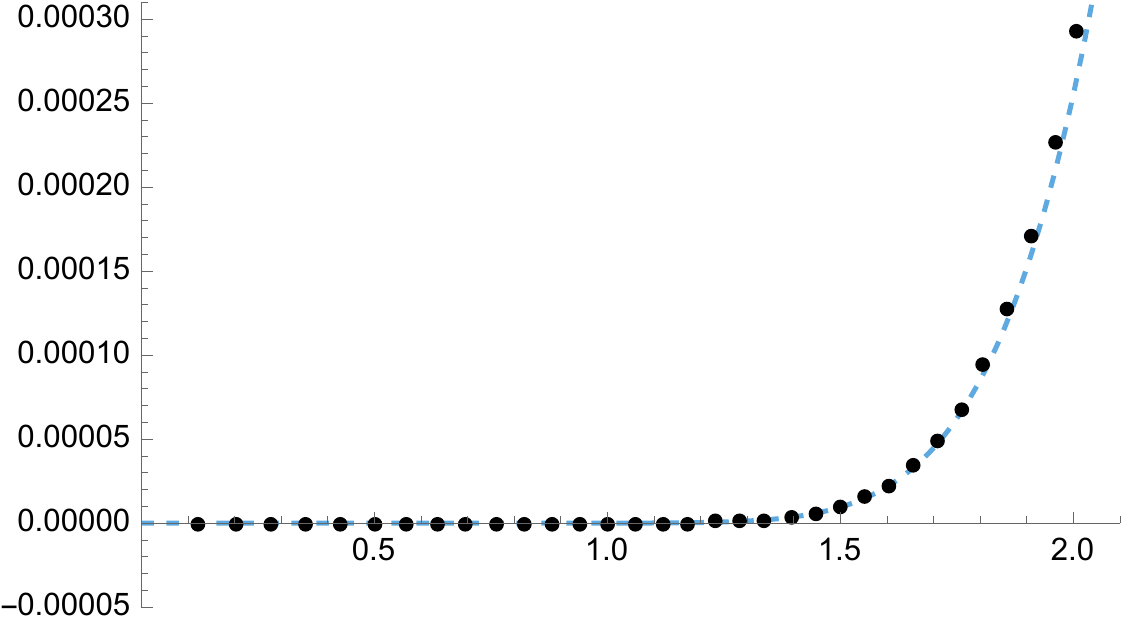}
\caption{
At the top, comparison of $e(\gamma)$ calculated numerically from the integral equation \eqref{volin_eq_TBA_GY} (points, black) and from the Borel summation of the perturbative series (line, blue). At the bottom, the points (black) are the difference between the two calculations while the line (blue, dashed) is the function $\frac{\pi^2}{2} \exp\left({-2 \pi ^2/\gamma }\right)$. These results were found using 80 coefficients in the perturbative series and a quadrature of 100 points in the integral equation. The points correspond to $B=50/(3k+1)$ with $k=0,1,\dots,33$. The $k=0$ point is not included in the lower plot because the numeric error is larger than the precision plotted.
}
\label{gy-real}
\end{figure}

Since the series behaves asymptotically as \eqref{GY_LOasym}, it follows that the series is not Borel resummable. In fact, the Borel summation has a discontinuous imaginary part
\begin{equation}
\disc s(e) \sim -2\ri \re^{-\frac{\pi^2}{\gamma}} \gamma \left(1+\frac{3 }{2 \pi ^2}\gamma+\frac{7}{8 \pi ^4}  \gamma ^2 - \frac{\zeta (3)+\frac{3}{16} }{\pi ^6}\gamma ^3+\cdots\right)+\cdots
\label{imBorel_GY}
\end{equation}
We can still take the real part of the Borel resummation and compare it with the numeric solution of the Bethe ansatz integral equation \eqref{volin_eq_TBA_GY}. We obtain a close match for small $\gamma$, see figure \ref{gy-real}, but eventually these two diverge. We estimate\footnote{The function in figure \ref{gy-real} is guessed from \cite{mmr-antrans}, although in principle one would have to account for discontinuity of the Borel summation of the $\exp\left({- \pi ^2/\gamma }\right)$ terms.} that the difference between the two is of order $\sim \re^{-\frac{2\pi^2}{\gamma}}$. 

We posit that the ground state energy is given by a trans-series of the form
\begin{equation}
e(\gamma) = \sum_{n\geq 0} c_n \gamma^n + \sum_{\ell\geq 1} C_\ell \re^{-\frac{\ell \pi^2}{\gamma}} \gamma^{b_\ell}\sum_{n\geq 0} c_n^{(\ell)} \gamma^n\,.
\label{gy_trans-series}
\end{equation}
It follows from our numerical results and regular resurgence relations that
\begin{equation}
C_1 = \pm \ri,\quad b_1 =1,\quad c_n^{(\ell)} = \frac{\mu_n}{\pi^{2n}}.
\label{numeric_Stokes}
\end{equation}
We will derive analytically the value of $C_1$ and $b_1$ in section \ref{sec-gy-antrans}.

\subsection{Repulsive Gaudin--Yang model}

The Gaudin--Yang model with a repulsive interaction is also within the range of applicability of the Bethe ansatz. This corresponds to $c<0$ in \eqref{delta-GY}. The integral equation describing the ground state is different from the attractive case,
\begin{equation}
f(x) - \frac{1}{2\pi}\int_{-B}^B\int_\IR \frac{\re^{-\ri p x'}f(x-x')}{1+\re^{|p|}}\rd p \rd x' = 2.
\label{repGY_inteq}
\end{equation}
In this case, we define the dimensionless coupling to be
\begin{equation}
\gamma = \frac{|c|}{n},\quad \frac{1}{\gamma} = \frac{1}{4\pi}\int_{-B}^B f(x)\rd x
\end{equation}
and the energy is obtained as
\begin{equation}
e(\gamma) = \frac{E}{n^3}= \frac{\frac{1}{4\pi}\int_{-B}^B x^2 f(x)\rd x}{\left(\frac{1}{4\pi}\int_{-B}^B f(x)\rd x\right)^3}.
\end{equation}
Much like the attractive case, this equation is tractable in the strong coupling limit but naively ill-behaved in the weak coupling limit. It is also of the form \eqref{volin_mostgeneral}, with
\begin{equation}
m=2,\quad p=0,\quad \tilde K(\omega) = \frac{1}{1+\re^{|\omega|}}\,.
\end{equation}
Thus, using
\begin{equation}
G_+(\omega) =\sqrt{2\pi} \frac{\re^{-\frac{1}{2\pi}\ri\omega \left[ \log\left( -\frac{\ri\omega}{2\pi} \right) - 1 \right]}}{\Gamma\left( \frac{1}{2} - \frac{ \ri\omega}{2\pi} \right)}, 
\label{volin_Gplus_repGY}
\end{equation}
in \eqref{volin_Rhatansatz}, and in the bulk limit using the ansatz \eqref{volin_eq_bulkGY}, we can apply Volin's method. The resulting power series is simply \eqref{eten} with $\gamma\rightarrow-\gamma$, as expected.
 
\begin{table}
\centering
\renewcommand{\arraystretch}{1.75}
\begin{tabular}{ r|c|c|l } 
\centering
& $B$ & $\gamma$  & $e_{\text{numeric}}-e_{\text{Borel}}$\\
 \hline
(attractive) GY & $\frac{5}{9}$ & $1.84185\dots$ & $0.000116(2 \pm 8)$\\
\hline
repulsive GY & $\frac{20}{17}$ & $1.85414\dots$ & $(0 \pm 4)\times 10^{-20}$\\
\hline
\end{tabular}
\caption{Comparison of the accuracy of Borel summation in the Gaudin--Yang model for attractive and repulsive interactions.}
\label{tab-gy}
\end{table} 
 
\begin{figure}
\centering
\includegraphics[width=\figsize]{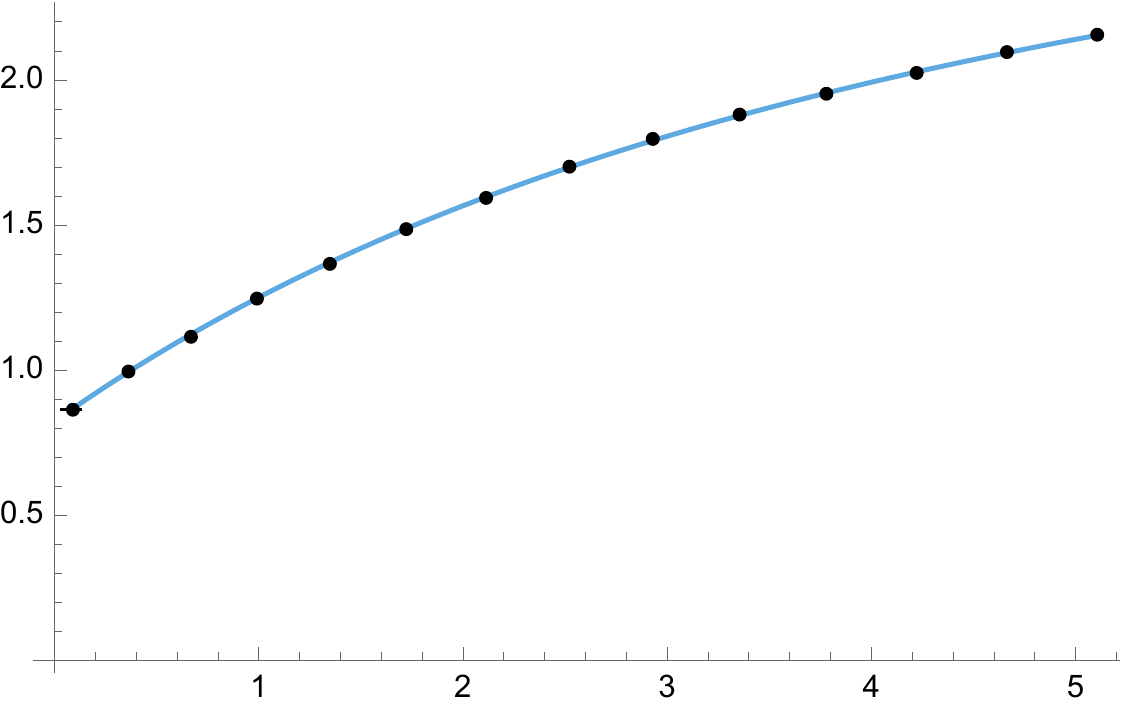}
\caption{
Comparison of $e^{\text{repulsive}}(\gamma)$ calculated numerically from the integral equation \eqref{repGY_inteq} (points, black) and from the Borel summation of the perturbative series (line, blue). The $x$-axis corresponds to $\gamma$. These results used 80 coefficients in the perturbative series and a quadrature of 50 points in the integral equation. The points correspond to $B=20/(3k-2)$ with $k=1,\dots,14$.
}
\label{gy-rep}
\end{figure}

Because of the sign inversion, the resulting coefficients are sign alternating, i.e. 
\begin{equation}
c^{\text{repulsive}}_m \sim - \frac{1}{\pi}\left(-\pi^2\right)^{-(m-1)}\Gamma(m-1), \quad m \gg 1.
\label{GY_LOrep}
\end{equation}
This implies that the series is Borel summable. Indeed, to available precision, the Borel-Padé resummation of the perturbative series matches the energy extracted from the numerical solution of \eqref{repGY_inteq}, see figure \ref{gy-rep}. To check that not even an exponentially suppressed effect could be hiding, we can compare the difference between the numeric solution of the integral equation in the regular case and the repulsive case for similar values of $\gamma$, as we do in table \ref{tab-gy}.

In fact, for repulsive Gaudin-Yang, we find an agreement of 12 digits between the Borel summation and the numeric solution even for $\gamma\sim 5$, as can be seen in figure  \ref{gy-rep}. This allows us to be confident that, in the repulsive case, Borel summation of the perturbative series retrieves the full exact result.

\subsection{Analytic trans-series}
\label{sec-gy-antrans}
In chapter \ref{cha_antrans}, we saw that we can extract the trans-series directly from the Bethe ansatz integral equation. It is thus natural to wonder whether one can do the same for \eqref{volin_eq_TBA_GY} and find \eqref{gy_trans-series} directly. In \cite{tw1,tw2}, Wiener--Hopf techniques were used to extract the first few coefficients of \eqref{eten}. While this approach is inefficient to get higher order of perturbation theory, it provided the ideal template to import the method of chapter \eqref{cha_antrans} to the Gaudin--Yang case as was done in \cite{mmr-antrans}. As we will review here, the first few trans-monomials of \eqref{gy_trans-series} can be obtained analytically in this way.

The first step is to extended \eqref{volin_eq_TBA_GY} into the full real line
\begin{equation}
\frac{f(x)}{2}+\frac{1}{2 \pi}\int_B^B\frac{f(y)}{1+(x-y)^2}\rd y = \Theta(B^2-x^2) + Y(\theta-B) + Y(-\theta-B)\,,
\label{gy_real_line}
\end{equation}
where we extend $f$ to be zero outside the interval $[-B,B]$ and $Y$ is an unknown function. As usual, let $\tilde f$ be the Fourier transform of $f$, decompose the integral kernel as in \eqref{WH-kernel} with \eqref{volin_Gplus_GY}. Furthermore, define
 \be
 g_+(\omega) = \frac{\re^{2\ri B \omega}-1}{\ri\omega},
 \ee
and introduce $Q=G_+ Y_+$ where $Y_+$ is the Fourier transform of $Y$.
We can then cast \eqref{volin_eq_TBA_GY} into the form of \eqref{fourier_WH} 
, as 
\be
\label{int-eq-wh}
\frac{\tilde f (\omega)}{ G_+(\omega) G_-(\omega)}   = 
\tilde \re^{-\ri B \omega}g_+(\omega)+ \re^{\ri B \omega} G_+^{-1}(\omega) Q (\omega) + 
 \re^{-\ri B \omega} G_-^{-1}(\omega) Q (-\omega).
 \ee
with
\begin{equation}
G_+(\omega) = \frac{\re^{ \frac{1}{\pi}\ri\omega [ \log( -\frac{\ri\omega}{2\pi} ) - 1 ] }}{\sqrt{\pi}}\Gamma\bigg( \frac{1}{2} - \frac{\ri\omega}{2\pi}  \bigg)
\label{volin_Gplus_GY-v2}.
\end{equation}
This differs from \eqref{volin_Gplus_GY} by a factor of $\sqrt{2}$ due to the overall normalization of the l.h.s. in \eqref{gy_real_line}.

In terms of the Fourier transform, \eqref{1overgamma_coefs} and \eqref{egy_int} become
 \be
 \frac{ \pi }{ \gamma}= \tilde f(0), \qquad e(\gamma)= -\frac{\gamma^2 }{ 4}-\pi^2\frac{\tilde f ''(0) }{ (\tilde f(0))^3 }. 
 \ee
Turns out it is inconvenient to calculate $\tilde f''(0)$ directly. Instead, as in \cite{tw1} , we can use
 (\ref{int-eq-wh}) to extract
\begin{equation}
\ba
\tilde f(0)&= 2\big(B+ Q(0)\big)\\
-\tilde f''(0) &= \frac{\tilde f(0)}{2} - \tilde g''(0) - 2  \frac{\rd^2}{\rd\omega^2}\left[ \re^{\ri B \omega} \frac{Q(\omega)}{G_+(\omega)} \right]_{\omega=0}.
\ea
\label{eq_f''hat}
\end{equation}

In the same way we had to solve for $Q$ to obtain the expansion of $\epsilon$ in chapter \ref{cha_antrans}, here we must solve for $Q$ to calculate the appropriate expansions of $f$. In fact, the equation for $Q$ is precisely \eqref{Q-eq} using the kernel decomposition \eqref{volin_Gplus_GY} and the appropriate $g_+$. The position of the singularities is, in this case, given by
\be
\label{poles-GY}
\xi_n = \pi(2n-1), \qquad n \in \IN,
\end{equation}
with residues
\be
\sigma^\pm_n =  {\rm Res}_{\omega = \ri\xi_n \pm 0} \,  \sigma(\omega)= \pm \frac{2\pi}{(n-1)!^2}\left( \frac{2n-1}{2\re} \right)^{2n-1}.
\end{equation}

One has to be careful when expanding the integral with $G_-(\omega)g_+(\omega)$ due to a singularity at the origin. This can be regularised by appropriately deforming the contour, as we did in \eqref{h-hankel}. In the end, we have
\be
\begin{aligned}
{1\over 2 \pi \ri} \int_\IR {G_-(\omega') g_+(\omega' ) \over \omega+ \omega'+ \ri 0}\rd\omega' = \frac{1}{\xi}\left( G_+(\ri\xi) - 1 \right) &+ \sum_{n\ge 1} \frac{\re^{-2B\xi_n}\ri\sigma_n^\pm G_-(\ri\xi_n)}{\xi_n(\xi_n + \xi)}\\
&- \frac{1}{2\pi\ri}\int_{\CC^\pm} \frac{\re^{-2B\xi'}\delta G_-(\ri\xi')}{\xi'(\xi'+\xi)}  \rd\xi'.
\end{aligned}
\label{r-source}
\end{equation}

In order to find the exponential effects discussed in this chapter, one needs to solve the equation to order $\re^{-2B}$.
The perturbative calculation was already carried out in \cite{tw1,tw2}.
%
Then one needs to calculate the first exponential corrections to $Q(\omega)$ itself. The technical details follow an identical recipe to that described in chapter \ref{cha_antrans} and can be found in \cite{mmr-antrans}. 
Plugging them in \eqref{eq_f''hat}, we find
%
\begin{equation}
\ba
\tilde f(0) &= 2B + \frac{\log(B\pi) + 1}{\pi} + \mathcal{O}\big(B^{-1}\big) 
\\
&\qquad+ \frac{2\ri\sigma^\pm_1 \re^{-2B\pi}}{\pi^2}\left[ 1 + \frac{1}{2\pi B} + \CO\big(B^{-2}\big) 
\right]+ \cdots,\\
- \tilde f''(0) &= \frac{2}{3}B^3 + \frac{\log(B\pi) - 1}{\pi} B^2 + \mathcal{O}(B) 
\\
&\qquad+ \frac{2\ri\sigma^\pm_1  \re^{-2B\pi}B^2 }{\pi^2} \Biggl[ 1 + \frac{\log(B\pi) + 3/2}{B\pi} +\CO\big(B^{-2}\big) 
\Biggr]+ \cdots.
\ea
\label{eq_f''(0)_result}
\end{equation}

These expansions allow us to encode the dimensionless coupling $\gamma$ as a function of the parameter $B$, a relation which has exponential corrections itself,
\begin{equation}
\ba
\gamma &= \frac{\pi }{2 B} - \frac{\log (B\pi)+1}{4B^2} + \CO\big(B^{-3}\big) - \frac{\ri\sigma^\pm_1 \re^{-2B\pi}}{2 \pi  B^2} \Big(1+ \CO\big(B^{-1}\big) \Big) 
+ \CO\big(\re^{- 4 B \pi}\big).
\ea
\end{equation}
Finally, we can put together the trans-series for the ground state energy
\begin{equation}
e(\gamma) =  \frac{\pi^2}{12} - \frac{\pi}{4B}  + \CO\big(B^{-2}\big) + \frac{3\ri\sigma^\pm_1 \re^{-2B\pi}}{4 \pi  B^2}  \Big( 1+ \CO\big(B^{-1}\big)\Big) + \CO\big(\re^{-4B \pi}\big)
\label{eq_energy_density_exp2}
\end{equation}
which can be re-expressed in terms of the dimensionless coupling only
\begin{equation}
\label{egy}
e(\gamma) = \frac{\pi^2}{12} - \frac{\gamma}{2} +\CO(\gamma^2) \pm \ri \re^{-\pi^2/\gamma}\gamma \big(1+ \CO(\gamma)\big)+ \CO\big(\re^{-2\pi^2/\gamma}\big)
\end{equation}
The leading term of the first exponential sector matches precisely what was conjectured numerically in \eqref{numeric_Stokes}. In principle, if one could push the perturbative corrections to this sector, one could find the $\mu_i$ in table \ref{GYtab}. However, this is very hard to do in practice. It would amount to at least the same complexity of calculating regular perturbative corrections which, before Volin's method, was never done beyond third order. As we observed in chapter \ref{cha_antrans}, it would be interesting to try to mix this method and Volin's to push calculations of higher perturbative correction in exponential sectors.

As for the structure of the exponential sectors themselves, these are dictated by integer combinations of the poles of $\sigma(\omega)$, which in this case amounts to $\exp(-k\pi^2/\gamma)$ for, a priori, all positive integer $k$. This matches the trans-series \eqref{gy_trans-series} which was originally conjectured in \cite{mr-long}.

\subsection{Verifying the conjecture}

Now that we have shown that, at weak coupling, the ground state energy is given by a trans-series expansion and explored methods of obtaining, a natural question is to compare the physical scales of the system with the non-perturbative scale 
\begin{equation}
\gamma \re^{- \frac{\pi^2}{\gamma}}.
\end{equation}
As we saw in \eqref{sec_introbcs}, this is precisely the scale of the superconductor gap
\begin{equation}
\Delta^2 \propto \re^{- \frac{\pi^2}{\gamma}},
\label{expDelta}
\end{equation}
and we verify the conjecture we outlined at the beginning of the chapter.

It should be noted that the BCS estimation of the gap \eqref{bcs-gap} does not get the correct power of the leading term in $\gamma$. However the BCS approximation is not exact and the power of $\gamma$ is a subleading effect when compared to the exponential factor. The important insights of the BCS analysis are the exponential factor and that the contribution of the gap to the ground state energy should be $\sim \Delta^2$ as in \eqref{e-bcs}, which is in accordance with our resurgent result. The more precise gap from integrability \eqref{spin-gap} does obtain both the correct exponential factor and the correct power of $\gamma$. In chapter \ref{cha_spin}, we will extend this match to the multi-component case.

The repulsive case adds a further insight. Due to the repulsive interaction, fermions do not pair up as Cooper pairs at low temperature. Consequently, the absence of superconductivity is matched by a Borel summable perturbative series. This is consistent with our proposal that the lack of Borel summability is a manifestation of superconductivity specifically.

An interesting open question is whether the trans-series derived in section \ref{sec-gy-antrans} can be obtained from the physics of superconductor vacua, in some procedure perhaps akin to a saddle point approximation or, alternatively, an OPE expansion. This would validate the conceptual insight of this conjecture and provide a first-principles procedure which could be extended to non-integrable models, since the approach of section \ref{sec-gy-antrans} relied on the Bethe ansatz. 

\section{Resurgence in the Hubbard model}
\label{sec_hub}
As was introduced in chapter \ref{cha_intro}, the Hubbard model in one dimension is a very rich model. In this section, we will focus on the case with $\kappa=2$ components, which can be both approached as an integrable model or as a superconductor. It is an interesting, more complicated, stage to test the conjecture of section \ref{sec_superconductivity_conjecture}. This section condenses a part of our results from \cite{mr-hubbard}. The analysis of the $\kappa$-component Hubbard model is left to chapter \ref{cha_spin}.

\subsection{BCS gap for the Hubbard model}
Since we have a model of two-component fermions with an attractive interaction, we can use BCS analysis in the spirit of section \ref{sec_introbcs}. This was originally done in \cite{marsiglio} for the particular case of $n=1$, but here, like in \cite{mr-hubbard}, we extend it to any $1\geq n>0$.

We start from the two key BCS relations, which were derived by \cite{nsr} for the Hubbard model in general dimensions. These equations fix  the gap and the density. Respectively, they are given by
\begin{equation}
{1 \over u}= {1\over L} \sum_\bk {1\over E_\bk}, \quad n= {1\over L} \sum_\bk \left( 1- {\epsilon_\bk - \hat \mu \over E_\bk} \right). 
\label{hub_BCS}
\end{equation}
Here the energy $E_\bk$ incorporates the chemical potential shifted by the Hartree correction,
\be
E_\bk = {\sqrt{\left( \epsilon_\bk -\hat \mu\right)^2 + \Delta_{\rm BCS}^2}}, \quad \hat \mu= \mu+ u n,
\ee
 $\epsilon_k$ is given by (\ref{eps-k}) and $\Delta_{\rm BCS}$ is the BCS gap. Our goal is to re-express $\Delta_{\rm BCS}$ as a function of $u$ and $n$, which requires also specifying $\hat\mu$ in terms of the same parameters.

For the case of the one-dimensional Hubbard model, \eqref{hub_BCS} can be expressed as integrals over the Brillouin zone,
\be
\label{gap-hubbard}
\ba
{2 \pi \over u}&=\int_{-\pi}^\pi {\rd k \over {\sqrt{(2 \cos k + \hat \mu)^2+ \Delta_{\rm BCS}^2}}}, \\
n&=1+{1\over 2 \pi}  \int_{-\pi}^\pi {2 \cos k +\hat \mu \over  {\sqrt{(2 \cos k + \hat \mu)^2+ \Delta_{\rm BCS}^2}}} \rd k. 
\ea
\ee
As their integrands might suggest, these equations can be written in the form of elliptic integrals. Let
\be
\rho_\pm = \sqrt{\frac{(\Delta_{\rm BCS}\pm 2\ri)^2+\hat \mu^2}{\Delta_{\rm BCS}^2+(\hat \mu-2)^2}}, \qquad m = - \frac{\rho_+-\rho_-}{4\rho_+ \rho_-},  \qquad \alpha= \Delta_{\rm BCS}^2 + (\hat \mu-2)^2,
\ee
then the gap equation is compactly turned into
\be
\frac{2\pi}{u}=\frac{4 K(m)}{\sqrt{\alpha\rho_+\rho_-}}, 
\ee
where $K(m)$ is the elliptic integral of the first kind. As for the equation that specifies the density, we need $\Pi(\tilde{m},m)$, the elliptic integral of the third kind, where 
\be
\tilde{m}=\frac{1}{2}\left(1-\frac{4+\Delta_{\rm BCS}^2+4}{\alpha\rho_+ \rho_-}\right).
\label{Kmgap}
\ee
With this function, we write
\be
n=1+\frac{1}{2\pi}\frac{4}{\sqrt{\alpha\rho_+ \rho_-}}\left\{ \left(\frac{\alpha}{2\hat \mu}\left(\rho_+\rho_-+1\right)+2-\hat \mu\right)K(m)-2\frac{\rho_+\rho_-+1}{\rho_+\rho_--1}\Pi(\tilde{m},m)\right\}.
\label{Pimn}
\ee
So far we have not done any approximation beyond those of the BCS variational method. Thus, while the equations \eqref{Kmgap} and \eqref{Pimn} are not the exact solution of the model, they encode the exact solution of the BCS framework of \eqref{gap-hubbard}.

We now expand for arbitrary $n$ but $u\ll 1$. This weak coupling limit implies $m\sim 1$. The gap equation leads to
\begin{equation}
\Delta_{\rm BCS} \sim 2(4-\hat \mu^2)\exp\left( -\frac{\pi}{u}\sqrt{1-\frac{\hat\mu^2}{4}}\right). 
\end{equation}
We can use this in the equation for $n$ to fix $\hat \mu$. If we define the chemical potential at weak coupling as 
\begin{equation}
\hat \mu = \mu_0+\mu_1 \Delta_{\rm BCS}^2 + \mathcal{O}(\Delta_{\rm BCS}^4),
\end{equation}
then we find
\begin{equation}
\mu_0= 2 \cos\left(\frac{\pi n}{2}\right)\,,\quad \mu_1= \frac{3\cos\left(\frac{\pi n}{2}\right)+\frac{\pi}{u}\sin(\pi n)}{4(1-\cos(\pi n))}. 
\end{equation}
Plugging back into the gap equation, we find
\begin{equation}
\label{delta-bcs}
\Delta_{\rm BCS}\sim 8 \sin^2\left( {\pi n \over 2} \right) \,\exp\left(-\frac{\pi}{u}\sin\left(\frac{\pi n}{2}\right)\right). 
\end{equation}
This is a weak coupling approximation of the gap within the approximation of BCS theory, which is indeed of the non-perturbative form \eqref{expDelta}.

Like in the Gaudin--Yang case, one cannot trust this result beyond the exponential factor. Fortunately, for the one-dimensional case and for arbitrary filling, a more precise estimation was computed in \cite{ls, wp} using the Bethe ansatz. Reassuringly, it has the same exponential factor, but it has a different power of $u$ at next-to-leading-order. It is given by
\be
\label{del-wp}
 \Delta \sim {4 \over \pi} {\sqrt{2  \sin^3 \left( {\pi n \over 2} \right)}} \, u^{1/2} \exp \left(-{\pi\over u} \sin \left( {\pi n \over 2} \right) \right), \qquad u \rightarrow 0. 
 \ee
In the particular case of $n=1$, the gap can be found to arbitrary order in the coupling, as was done in \cite{lw, marsiglio}. There it is found that
 \be
 \label{all-gap}
 \Delta = {8 \over \pi} \sum_{r=0}^\infty {1\over 2 r+1} K_1 \left( {\pi \over u} (2r+1) \right), 
 \ee
where $K_1(z)$ is a modified Bessel function. When $u\ll 1$ this reduces to a sum of exponential effects, whose leading term agrees with (\ref{del-wp}) when $n=1$.

\subsection{Perturbative expansion at low filling}

As we saw in section \ref{sec_iqmb}, this model is integrable and the Bethe ansatz description of the ground states yields integral equation \eqref{inteq-hub}. Unfortunately, generalizing the approach of section \ref{sec_volin} to such a r.h.s. is technically challenging due to the branch cut of the square root. However, we know that  when $n\rightarrow 0$ we recover \eqref{volin_eq_TBA_GY}. Thus, we can treat the $n\ll 1$ limit as a deformation of the Gaudin--Yang equation \eqref{volin_eq_TBA_GY}.

In order to correctly parameterize this deformation, we need to be careful with the double limits. We find the Gaudin--Yang model at finite coupling when 
\begin{equation}
n\rightarrow 0,\quad u\rightarrow 0,\quad \frac{n}{u}= \frac{1}{\gamma}\quad\text{finite}.
\label{GYlimit}
\end{equation}
On the other hand, the regular weak coupling limit of the Hubbard model is
\begin{equation}
u\rightarrow 0, \quad n \quad \text{finite}.
\end{equation}
In this limit, we have that the ground state is a power series in $u$ parameterized by $n$,
\begin{equation}
E(u,n) = - \frac{4}{\pi}\sin \frac{\pi n}{2} - \frac{u n^2}{2}  + \sum_{m\geq 2} E_m(n) u^m,
\label{hub-Em}
\end{equation}
where the first terms can be computed from the free case and the Hartree correction.
As a consequence of the limit \eqref{GYlimit}, we have that
\begin{equation}
E(u,n) = -2n + n^3 e_{GY}(u/n) + \CO(n^4),
\end{equation}
and more precisely
\begin{equation}
E_m(n) = n^{3-m} \left(c_m +\sum_{k\geq 1}E_m^{(k)} n^{2k} \right)
\label{hub-Ekm}
\end{equation}
where the $c_m$ are the Gaudin--Yang coefficients of \eqref{eGY_powerseries}. Since the coefficients $E_m(n)$ are currently inaccessible for high $m$, we can try to get the series $E_m^{(k)}$. According to our conjecture, the asymptotic behavior of $E_m(n)$ should be dictated by \eqref{del-wp} and thus dependent on $n$. This should be detectable at the level of $E_m^{(k)}$, even if we restrict ourselves to low $k$.

It should be noted that the limit $n\rightarrow 1$ is also an interesting limit of the model, although we will not explore it in this chapter. In particular, for $n=1$  the coefficients $E_m(1)$ are known exactly, thanks to \cite{Misurkin1972,economou,Takahashi1971}, 
\be
\ba
\label{hk_def}
E_{2k}(1)&=-{ (2k-1) (2^{2k+1}-1) ( (2k-3)!!)^3 \over 2^{5 k-3}(k-1)!} {\zeta (2k+1) \over \pi^{2k+1}}, \qquad k \ge 1, \\
 E_{2k+1}(1)&=0. 
 \ea
\ee

We are only left with encoding our two parameters $n$ and $\gamma$ in the integral equation \eqref{inteq-hub}. In the Gaudin--Yang case, we saw that the weak coupling expansion appears in the equation through $\gamma\sim 1/B$, where further corrections must be calculated using the solution to the equation. We want to encode the small $n$ expansion as well, but the integral equation depends only on $B$ and $u$. Thus, we define
\begin{equation}
\beta = B u = \frac{\pi n B}{\frac{1}{\pi}\int_{_B}^B f(x)\rd x}\sim n + \CO(n^2).
\end{equation}
In terms of these parameters, the equation for the ground state can be easily expanded for small $\beta$, which is proportional to the $n$ expansion, as
\begin{equation}
\ba
\frac{f(x)}{2}+\frac{1}{2\pi}\int_{-B}^B\frac{1}{1+(x-x')^2}f(x')\rd x'&= {\mathrm{ Re}}\frac{1}{\sqrt{1-\frac{\beta^2}{B^2}(x-\frac{\ri}{2})^2}}\\
&= \sum_{i=0}^\infty (-1)^i \binom{-1/2}{i} \frac{\beta^{2i}}{B^{2i}}\mathrm{Re} (x-\ri/2)^{2i}.
\label{hub_inteq_pert}
\ea
\end{equation}

By taking only terms in $\beta$ up to some fixed order $\beta^{k}$, we can calculate $f(x)$, and ultimately the terms $E_m(n)$, up to the same order $n^k$ (i.e. $E_m^{(\ell\leq k)}$ in \eqref{hub-Ekm}). With such a truncation, the r.h.s. of \eqref{hub_inteq_pert} is polynomial and is not of the form \eqref{volin_mostgeneral}, but this can be seen as a  extension of the $p=0$ case. The minutae of extending Volin's method to such cases are exposed in detail in appendix A of \cite{mr-hubbard}. In the end, we take the expressions for $\gamma$ and $n$ as series in $1/B$ and $\beta$ and invert them, leading to the desired double expansion.

The series in $n$ should be convergent and we can use our results to interpolate between the Gaudin--Yang model and the $n=1$ case. For example, keeping the driving terms up to $\beta^{10}$ in \eqref{hub_inteq_pert} we can derive
\be
\label{e2-pert}
\ba
&\frac{E_2(n)}{n}=-\frac{1}{12}+\frac{\left(12-\pi ^2\right) n^2}{288}+\frac{\left(12 \pi ^2-\pi ^4\right) n^4}{4608}+ \left(\frac{\pi ^4}{5760}-\frac{61 \pi ^6}{3870720}\right) n^6\\
&+\left(\frac{121 \pi ^6}{9289728}-\frac{277 \pi ^8}{222953472}\right) n^8+ \left(\frac{3917 \pi ^8}{3715891200}-\frac{50521 \pi ^{10}}{490497638400}\right) n^{10} + \CO(n^{12}). 
\ea
\ee
For $n\rightarrow 0$, we see that we recover the term in \eqref{gy_o2}. For $n\rightarrow 1$, we find that it converges\footnotemark $\,\!$  
 towards the value from \eqref{hk_def},
\begin{equation}
E_2(1) = -\frac{7 \zeta (3)}{4 \pi ^3}.
\end{equation}
We can also test the series in $n$ for this coefficient by comparing it with a diagrammatic calculation at finite $n$, checking that the series converges for any $n\in(0,1]$.

\footnotetext{This has a curious number theoretical implication. It suggests a representation of the Apery constant $\zeta(3)$ such that $\zeta(3)= \pi^3\sum_{k\geq 0} a_k \zeta(2k)$, with rational coefficients $a_k$. To our knowledge, such a representation is not currently known, see e.g. \cite{Lupu2019,orr}.}

\subsection{Large order behavior in the double limit}

\begin{figure}
\begin{center}
\begin{subfigure}{\figsize}
\centering
\includegraphics[width=\textwidth]{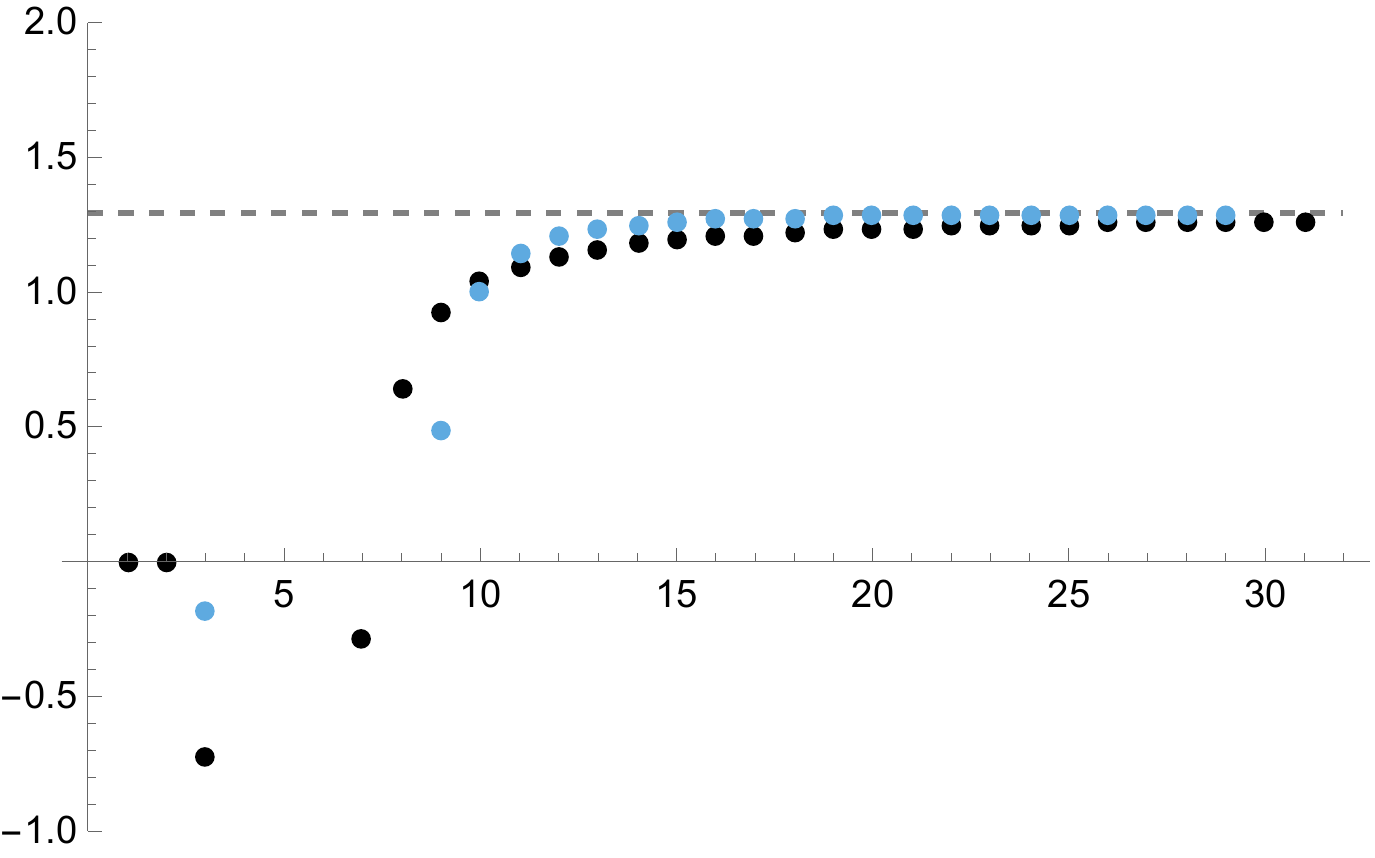}
\caption{Sequence $a^{(1)}_m$}
\end{subfigure}
\\
\begin{subfigure}{\figsize}
\centering
\includegraphics[width=\textwidth]{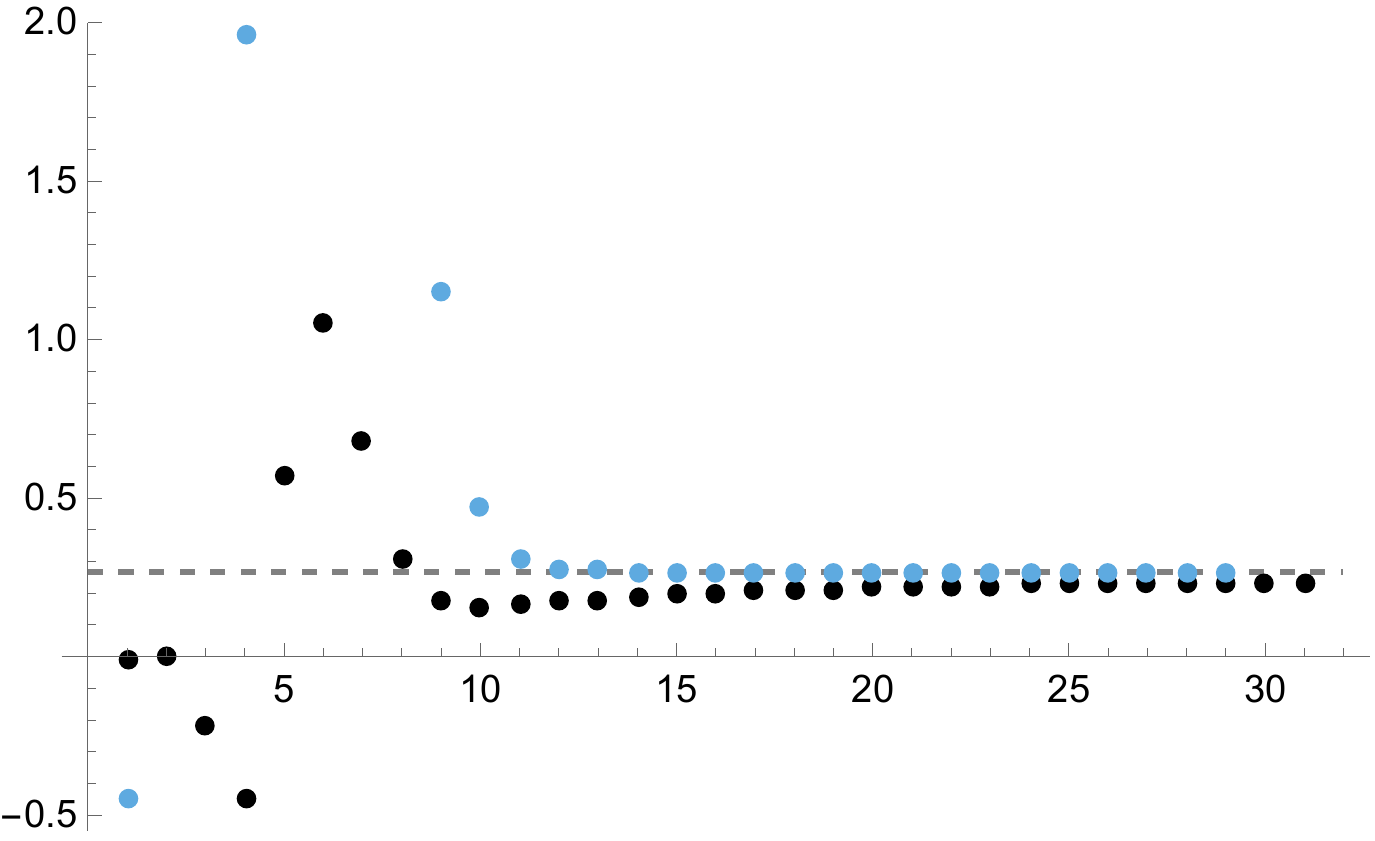}
\caption{Sequence $a^{(2)}_m$}
\end{subfigure}
\end{center}
\caption{
Plot of sequences $a^{(1,2)}_m$ (black), their respective second Richardson transform (blue) and the theoretical prediction from \eqref{a12-theory} (dashed grey line).
}
\label{hub-asym}
\end{figure}

Our conjecture \eqref{asym_gap} combined with gap \eqref{del-wp} predicts that the asymptotic behavior of the $E_m(n)$ in \eqref{hub-Em} should be
\begin{equation}
E_m \sim c(n) \left(2\pi \sin\left(\frac{\pi n}{2}\right)\right)^{1-m} \Gamma(m-1) ,\quad m\gg 1.
\label{asym_HUB_alln}
\end{equation}
The overall constant $c(n)$ is a priori any function of $n$, however we know that in the Gaudin--Yang limit \eqref{GYlimit} we recover the Gaudin--Yang asymptotic behavior \eqref{GY_LOasym} through \eqref{hub-Ekm}. Thus
\begin{equation}
c(n) = - \frac{n^2}{\pi}\left(1+c_1 n^2 + \CO(n^4)\right),
\end{equation}
where $c_1$ is some unknown coefficient. Fortunately, the dependency on $n$ of the geometric factor in \eqref{asym_HUB_alln} is such that we can test it without knowing $c_1$. At each order in $n$ this factor contributes as a polynomial in $m$ of increasingly higher order,
\begin{multline}
\left(\frac{2\pi \sin\left(\frac{\pi n}{2}\right)}{\pi^2 n}\right)^{1-m} =
 1+\left(\frac{\pi ^2 m}{24}-\frac{\pi ^2}{24}\right) n^2
 +\left(\frac{\pi ^4 m^2}{1152}-\frac{\pi ^4 m}{720}+\frac{\pi ^4}{1920}\right) n^4
 \\+\left(\frac{\pi ^6 m^3}{82944}-\frac{\pi ^6 m^2}{46080}+\frac{37 \pi ^6 m}{2903040}-\frac{\pi ^6}{322560}\right) n^6
 +\CO\left(n^8\right)
\end{multline}
And thus the product organizes as
\begin{multline}
c(n) \left(2\pi \sin\left(\frac{\pi n}{2}\right)\right)^{1-m} =
- n^{3-m} \pi^{1-2m} 
\bigg(
1+n^2 \left\{\frac{\pi ^2 m}{24}+\left(c_1-\frac{\pi ^2}{24}\right)\right\}
\\
+n^4 \left\{\frac{\pi ^4 m^2}{1152}+\left(\frac{c_1 \pi ^2}{24}-\frac{\pi ^4}{720}\right) m+\CO\left(m^0\right) \right\}
+\CO\left(n^6\right)
\bigg).
\label{cnsin}
\end{multline}
Taking the leading term in $m\gg 1$ at each order in $n$ we can find the leading asymptotic behavior of the coefficients defined in \eqref{hub-Ekm}. 
With only 30 terms of the series $E_m^{(1)}$ and $E_m^{(2)}$, we can check that they converge towards the predicted behavior \eqref{cnsin}. We can do so by analysing the asymptotics of the sequences,
\begin{equation}
a^{(1)}_m = - \frac{\pi^{2m}E_m^{(1)}}{\Gamma(m)}\sim {\pi^{3} \over  24},\quad 
a^{(2)}_m = - \frac{\pi^{2m}E_m^{(2)}}{\Gamma(m+1)}\sim {\pi^{5} \over  1152}.
\label{a12-theory}
\end{equation}
as we do in figure \ref{hub-asym}.

We then see that the perturbative expansion of the Hubbard model for small $n$ is in agreement with the conjecture from section \ref{sec_superconductivity_conjecture}.
The conjecture can also be tested in the Hubbard model by looking at different limits of $n$. For example, by inspecting the $n\rightarrow 1$ limit and the exact $n=1$ case. Those tests were sucessfully done in \cite{mr-hubbard}.


\chapter[Energy gaps as renormalons][Energy gaps as renormalons]{Energy gaps as renormalons}\label{cha_spin}

In chapter \ref{cha_GY}, we explored the interplay between superconductivity and Borel summation. This analysis could clearly be extended in many directions. Prominently, superconductivity and the BCS gap are found from the binding energy of Cooper pairs. However, we only used the fact that the binding energy is non-perturbative, and this is not exclusive to Cooper pairs. One can extend the connection to the presence of more general bound states. Another aspect left to explore is how the divergence of the series manifests at the level of Feynman diagrams. As we introduced in \ref{sec_renormalons}, this is an important distinction between instantons and renormalons. 

The key tool to address both of these questions is to have a model with arbitrary number of species. Particularly, the multi-component Gaudin--Yang model, which was introduced in section \ref{sec_iqmb}, is a convenient example. On one hand, it lets us study a ground state which is populated by bound states of multiple particles. On the other hand, it introduces a new parameter, the number of components, which allows us to reorganize the perturbative series. This permits a large $N$ analysis which isolates a particular family of Feynman diagrams, ring diagrams, in analogy with section \ref{cha_largeN}. Using this, we show that the divergence found in the previous chapter is indeed due to a renormalon effect. Lastly, we also address the multi-component extension of the Hubbard model.

The contents of this chapter combine the main results from \cite{mr-three}, the analysis in \cite{mr-long} of the multi-component Gaudin Yang model and ring diagrams, as well as the results in \cite{mr-hubbard} for the multi-component Hubbard model and its own ring diagrams.

\section{Gap for the \texorpdfstring{$\kappa$}{k}-component Gaudin--Yang model}
\label{sec-gygap}
As we saw from the Bethe ansatz in section \ref{sec_iqmb}, in the Gaudin--Yang model with $\kappa$ components, a state is populated by bound states of $1 \geq n \geq \kappa$ elementary fermions. In the ground state of the model, all fermions are part of $\kappa$-element bound states, which can be seen as a generalization of Cooper pairs when $\kappa=2$ and ``trions'' when $\kappa=3$. The energy required to break such a state is the energy gap, which is a non-perturbative scale of the theory. In this section, we shall calculate the weak coupling expression of the gap in two ways, assuming the thermodynamic limit \eqref{thl}. The first is directly from the Bethe  ansatz. This is a generalization of the $\kappa=2$ calculation of \cite{ko}, which led to \eqref{spin-gap}. The second one uses only renormalization considerations and extends the analysis of \cite{ls}.

\subsection{From the Bethe ansatz}

In order to calculate the energy gap, we need to calculate how much energy it costs to break a bound state of $\kappa$ particles in the ground state. First we need to inquire what does the bound state break into. Looking at the approximate binding energy of a single bound state in \eqref{energynba}
\begin{equation}
E_m \sim - \left(m\frac{\left(m^2-1\right) c^2}{12}\right),
\end{equation}
we deduce that the first excited state should have a 1 fermion ``bound state'', i.e. a free fermion, and a bound state of $\kappa-1$ fermions. Since the solutions of the Bethe ansatz \eqref{leeNBA} are parameterized simply by the number of bound states of each size $N_i$, this suffices to specify the state. We can then define the gap as the difference in energy between the first excited state and the ground state,
\begin{equation}
\Delta_\kappa = E(1,0,\cdots,0,1,N/\kappa-1)- E(0,0,\cdots,0,0,N/\kappa).
\label{gap_def0}
\end{equation}
 $N$ is the total number of particles which we will take to be large in accordance with \eqref{thl}. 

Recall that in the ground state we have a Bethe roots $\lambda^\kappa_j$ for each of the $N_\kappa=N/\kappa$ bound states, such that the ensemble satisfy 
\eqref{NBAgs}. We can think of the first excited state as a perturbation around this state, since we only affect $\kappa$ particles among $N$ when $N\rightarrow \infty$. We thus define the first excited solution to \eqref{leeNBA} as a deformation of ground state solution, 
\begin{equation}
\lambda^1_1 \equiv k,\quad \lambda^{\kappa-1}_1\equiv \Lambda,\quad \bar{\lambda}^\kappa_j \equiv \lambda_j^\kappa+\frac{\xi_j}{L},\quad \bar{K}^{\kappa}_j = K^\kappa_j +\frac{1}{2}. 
\label{deformed_solution}
\end{equation}
Concretely, $k$ is the momentum of the free fermion, $\Lambda$ characterizes the $\kappa-1$ bound state and $K_j^\kappa$ is shifted by $1/2$ because there is one less $\kappa$ bound state. We can write the gap  \eqref{gap_def0} in terms of these roots using \eqref{energynba},
 \begin{multline}
\Delta_\kappa = k^2 + (\kappa-1)\left(\Lambda^2-\frac{(\kappa-1)^2-1}{12}c^2\right)
\\
+\sum_{j=1}^{N_\kappa-1}\kappa\left((\bar{\lambda}^\kappa_j)^2-\frac{\kappa^2-1}{12}c^2\right)
-\sum_{j=1}^{N_\kappa}\kappa\left(\lambda^2_j-\frac{\kappa^2-1}{12}c^2\right).
\label{Delta_roots}
\end{multline}

Plugging \eqref{deformed_solution} into (\ref{leeNBA}), we find equations for the new roots. For the roots of the new bound states, we find
\begin{align}
Lk &= 2\tan^{-1}\left(\frac{k-\Lambda}{(\kappa-2)c'}\right)+\sum_{l=1}^{N_\kappa-1} 2\tan^{-1}\left(\frac{k-\bar{\lambda}^\kappa_l}{(\kappa-1)c'}\right),\\
(\kappa-1)L \Lambda &= 2\tan^{-1}\left(\frac{\Lambda-k}{(\kappa-2)c'}\right)+\sum_{l=1}^{N_\kappa-1} \sum_{p=1}^{\kappa-1} 2\tan^{-1}\left(\frac{\Lambda-\bar{\lambda}^\kappa_l}{(2\kappa-2p-1)c'}\right).
\label{kLambda}
\end{align}
Meanwhile, the deformed roots of the $\kappa$ particle states are solutions of
\begin{align}
\kappa (L \lambda^\kappa_j+  \xi_j )&
\begin{multlined}[t] =
2\pi K^\kappa_j+\pi + 2\tan^{-1}\left(\frac{\bar{\lambda}^\kappa_j-k}{(\kappa-1)c'}\right)\\
+ \sum_{p=1}^{\kappa-1} 2\tan^{-1}\left(\frac{\bar{\lambda}^\kappa_j-\Lambda}{(2p-1)c'}\right)
+ \sum_{p=1}^{\kappa-1} \sum_{l=1}^{N_\kappa-1}2\tan^{-1}\left(\frac{\bar{\lambda}^\kappa_j-\bar{\lambda}^\kappa_l}{2p c'}\right).
\end{multlined}
\label{NBA0xi}
\end{align}

In the continuum limit, equations \eqref{kLambda} become
\begin{align}
k &= 2\int_{-Q}^Q \rd \lambda f(\lambda) \tan^{-1}\left(\frac{k-\lambda}{(\kappa-1)c'}\right)+\CO\left(\frac{1}{L}\right),\\
\Lambda &= \frac{2}{\kappa-1} \sum_{p=1}^{\kappa-1}\int_{-Q}^Q \rd \lambda f(\lambda) \tan^{-1}\left(\frac{\Lambda-\lambda}{(2\kappa-2p-1)c'}\right)+\CO\left(\frac{1}{L}\right).
\end{align}
These have the solution
\begin{equation}
k=\Lambda=0,
\end{equation}
which can be checked by observing that the integrand on the r.h.s. becomes odd, since $f(\lambda)$ is even.

As for equation (\ref{NBA0xi}), we can remove the $\lambda_j^\kappa$ on the l.h.s. by subtracting \eqref{NBAgs}. Furthermore, we can drop terms of order $L^{-1}$ originating in the perturbation \eqref{deformed_solution}. This reduces the equation to
\begin{multline}
\kappa \xi_j = \pi + 2\tan^{-1}\left(\frac{\lambda^\kappa_j-k}{(\kappa-1)c'}\right)+\sum_{p=1}^{\kappa-1}2\tan^{-1}\left(\frac{\lambda^\kappa_j-\Lambda}{(2p-1)c'}\right)
\\
-\sum_{p=1}^{\kappa-1}2\tan^{-1}\left(\frac{\lambda^\kappa_j-\lambda^\kappa_{N_\kappa}}{2pc'}\right)+\sum_{l=1}^{N_\kappa-1}\sum_{p=1}^{\kappa-1}\frac{4pc'}{L}\frac{\xi_j-\xi_l}{(2pc')^2+(\lambda^\kappa_j-\lambda_l^\kappa)^2}.
\label{NBAxi_precont}
\end{multline}
The last term is the most subtle, because one needs to keep the second term when expanding the arctangent
\begin{equation}
2\tan^{-1}\left(\frac{\bar{\lambda}^\kappa_j-\bar{\lambda}^\kappa_l}{2 pc'}\right) \sim 2\tan^{-1}\left(\frac{\lambda^\kappa_j-\lambda^\kappa_l}{2 pc'}\right)+\frac{4pc'}{L}\frac{\xi_j-\xi_l}{(2 pc')^2+(\lambda^\kappa_j-\lambda^\kappa_l)^2} + \CO\left(\frac{1}{L^2}\right),
\label{xiexpand}
\end{equation}
since the sum over $l$ is proportional to $N_\kappa$ which itself is proportional to $L$ in the limit \eqref{thl}. 

We can now take the continuum limit proper of \eqref{NBAxi_precont},
\begin{multline}
\kappa \xi(\lambda) = \pi +2\tan^{-1}\left(\frac{\lambda}{(\kappa-1)c'}\right)+ \sum_{p=1}^{\kappa-1}2\tan^{-1}\left(\frac{\lambda}{(2p-1)c'}\right) 
\\
- \sum_{p=1}^{\kappa-1} 2\tan^{-1}\left(\frac{\lambda-Q}{2pc'}\right)+ \sum_{p=1}^{\kappa-1}\int_{-Q}^Q \rd\lambda' f(\lambda') \frac{2pc(\xi(\lambda)-\xi(\lambda'))}{(pc)^2+(\lambda-\lambda')^2}.
\end{multline}
The distributions $f(\lambda)$ and $\xi(\lambda)$ appear independently in this equation, which is inconvenient. However, one can use \eqref{pregsc} to reorganize the l.h.s. and the last term of the r.h.s. into
\begin{multline}
2\pi f(\lambda)\xi(\lambda) =\pi +2\tan^{-1}\left(\frac{\lambda}{(\kappa-1)c'}\right)+ \sum_{p=1}^{\kappa-1}2\tan^{-1}\left(\frac{\lambda}{(2p-1)c'}\right)\\
 - \sum_{p=1}^{\kappa-1} 2\tan^{-1}\left(\frac{\lambda-Q}{2pc'}\right)
- \sum_{p=1}^{\kappa-1}\int_{-Q}^Q \rd\lambda' f(\lambda')\xi(\lambda') \frac{2pc}{(pc)^2+(\lambda-\lambda')^2}.
\end{multline}
With the following redefinitions
\begin{equation}
\theta = \frac{\lambda}{c},\quad B = \frac{Q}{c},\quad \Psi(\theta)=f(\lambda)\xi(\lambda),
\label{newdefs}
\end{equation}
the above equation becomes a more transparent equation for $\Psi(\theta)$,
\begin{multline}
2\pi \Psi(\theta) + \int_{-B}^B \rd\theta' K(\theta-\theta')\Psi(\theta')=
\\ \pi +2\tan^{-1}\left(\frac{2\theta}{\kappa-1}\right)+ \sum_{p=1}^{\kappa-1}2\tan^{-1}\left(\frac{2\theta}{2p-1}\right) - \sum_{p=1}^{\kappa-1} 2\tan^{-1}\left(\frac{\theta-B}{p}\right)
\label{eq_Psi}
\end{multline}
where the kernel $K(\theta)$ is given in (\ref{conteq}).
In terms of $\Psi(\theta)$, \eqref{Delta_roots} can be written as
\begin{equation}
\Delta_\kappa= -\kappa c^2 B^2 + \frac{\kappa(\kappa-1)}{4}c^2 + 2 \kappa c^2 \int_{-B}^B \theta \Psi(\theta)\rd \theta.
\label{DeltaPsi}
\end{equation}

In order to calculate the gap at weak coupling, we then need to calculate a weak coupling approximation of $\Psi(\theta)$. As the strategy of \cite{ko} suggests, it is simpler to focus on its anti-symmetric part. Let
\begin{equation}
h(\theta)= \frac{\Psi(\theta)-\Psi(-\theta)}{2}-\frac{\text{sgn}(\theta)}{2}.
\end{equation}
Then $h(\theta)$ solves the integral equation
\begin{equation}
\ba
h(\theta)&+\frac{1}{2\pi} \int_{-B}^B \rd \theta' K(\theta-\theta')h(\theta')=\tau_0(\theta)\\
\ea
\label{eq_h}
\end{equation}
where the driving term is
\begin{multline}
\tau_0(\theta)= -\frac{\text{sgn}(\theta)}{2}+\frac{1}{\pi}\tan^{-1}\left(\frac{2\theta}{\kappa-1}\right)
\\
+\frac{1}{\pi}\sum_{p=1}^{\kappa-1}\tan^{-1}\left(\frac{2\theta}{2p-1}\right) -\frac{1}{\pi}\sum_{p=1}^{\kappa-1}\tan^{-1}\left(\frac{\theta}{p}\right).
\end{multline}
This follows from anti-symmetrizing \eqref{eq_Psi}, making use of the integral
\begin{equation}
\int_{-B}^B\rd \theta'\frac{p\,\text{sgn}(\theta')}{p^2+(\theta-\theta')^2}=2\tan^{-1}\left(\frac{\theta}{p}\right)-\tan^{-1}\left(\frac{\theta+B}{p}\right)-\tan^{-1}\left(\frac{\theta-B}{p}\right).
\end{equation}%
We can see that $h(\theta)$ suffices to calculate the energy gap since \eqref{DeltaPsi} reduces to
\be
\frac{\Delta_\kappa}{c^2}= - 4 \int_{B}^\infty \theta h(\theta)\rd \theta. 
\ee

As in previous chapters of this thesis, the weak coupling limit ($c\ll 1$) implies large $B$ ($B\gg 1$, see \eqref{newdefs}). Expanding \eqref{eq_h} at large $B$ is non-trivial, but it will not surprise the reader of this thesis that Wiener--Hopf techniques can be of great use in these cases. This was done already in  \cite{griffiths, yang-yang} and we follow a similar strategy.

Let $h_0$ be the solution in the $B\rightarrow \infty$ limit, 
\begin{equation}
h_0(\theta) + \frac{1}{2\pi}\int_{-\infty}^\infty \rd \theta' K(\theta-\theta') h_0(\theta')=\tau_0(\theta).
\end{equation}
The Fourier transform of this equation can be solved algebraically
\begin{equation}
(1+\tilde K(\omega)) \tilde h_0(\omega) = \tilde\tau_0(\omega),
\label{tautoh0}
\end{equation}
where in this section we always denote the Fourier transform of a function $g(\theta)$ by $\tilde{g}(\omega)$, and in particular
\begin{equation}
\tilde{K}(\omega) = \int_\IR \rd \theta \re^{\ri \omega \theta} K(\theta) = \frac{\re^{-\left| \omega \right| }-\re^{-\kappa  \left| \omega \right| }}{1-\re^{-\left| \omega \right| }}.
\end{equation}
Explicitly, \eqref{tautoh0} leads to
\begin{equation}
\ba
\tilde{h}_0(\omega)&= -\frac{\ri}{\omega}\frac{1-\re^{-|\omega|/2}}{1+\re^{-\kappa|\omega|/2}}\left(1-\re^{-(\kappa-1)|\omega|/2}\right) \\
&= -\frac{\ri}{\pi}\sum_{n=-\infty}^\infty\frac{\sin\left(\frac{2\pi}{\kappa}\left(n-\frac{1}{2}\right)\right)}{\left(n-\frac{1}{2}\right)\left(\omega-\frac{4\pi\ri}{\kappa}\left(n-\frac{1}{2}\right)\right)}.
\ea
\end{equation}
And conversely,
\begin{equation}
h_0(\theta)=-\frac{1}{\pi}\tan^{-1}\left(\frac{\sin(\pi/\kappa)}{\sinh(2\pi\theta/\kappa)}\right).
\label{h0_theta}
\end{equation}
Unfortunately, $h_0$ is not a good approximation of the solution at large but finite $B$. We can however use it to better approximate $h$.

Let us define $h(\theta)$ outside of the interval $[-B,B]$ such that \eqref{eq_h} still holds. We can take the Fourier transform of \eqref{eq_h} and replace the r.h.s. by \eqref{tautoh0}, which results in
\begin{equation}
\tilde{h}(\omega) + \tilde{K}(\omega) [\tilde{H}_B\star \tilde{h}](\omega) = (1+\tilde{K}(\omega)) \tilde h_0(\omega),
\end{equation}
where we define the Heaviside function supported on the interval $H_B(\theta) = \Theta(B^2-\theta^2)$ and $[ \tilde f \star \tilde g]$ defines the usual convolution between functions $\tilde f$ and $\tilde g$. We can reorganize the above equation into
\begin{equation}
\tilde{h}(\omega) + \frac{\tilde{K}(\omega)}{1+\tilde{K}(\omega)} [(\tilde{H}_B-2\pi \delta )\star \tilde{h}](\omega) = \tilde{h}_0(\omega),
\label{Fourier_h}
\end{equation}
where $\delta$ is the Dirac $\delta$-function.
If we introduce the kernel
\begin{equation}
R(\theta) =\frac{1}{2\pi}\int_\IR \rd \omega  \frac{\re^{\ri \omega \theta}\tilde{K}(\omega)}{1+\tilde{K}(\omega)},
\end{equation}
then the inverse Fourier transform of \eqref{Fourier_h} is
\begin{equation}
h(\theta) - \int_{|\theta'|\geq B}\rd\theta' R(\theta-\theta') h(\theta') = h_0(\theta).
\end{equation}
To leading exponential order, we have  from \eqref{h0_theta} that the r.h.s. grows as
\begin{equation}
\label{hap}
h_0(\theta+B) \sim -\frac{2}{\pi} \sin\left(\frac{\pi}{\kappa}\right)\re^{-\frac{2\pi}{\kappa}(\theta+B)}+\CO(\re^{-\frac{4 \pi}{\kappa}B}).
\end{equation}
Thus, we can introduce $h(\theta+B)\sim r(\theta)$ such that $r(\theta)$ solves
\begin{equation}
\label{req}
r(\theta) = -\frac{2}{\pi} \sin\left(\frac{\pi}{\kappa}\right)\re^{-\frac{2\pi}{\kappa}(\theta+B)}+\int_0^\infty
 \rd\theta' R(\theta-\theta')r(\theta'). 
\end{equation}

Equation \eqref{req} can be solved by conventional Wiener--Hopf techniques.
Following a familiar script, we define the one-sided Fourier transforms of $r(\theta)$
\begin{equation}
\tilde{r}_+(\omega)=\int_0^\infty \rd\theta \re^{\ri \omega \theta} r(\theta),\quad \tilde{r}_-(\omega)=\int_{-\infty}^0 \rd\theta \re^{\ri \omega \theta} r(\theta),
\end{equation}
which are (respectively) analytic in the upper (lower) half plane, and we decompose
\begin{equation}
1-\frac{\tilde{K}(\omega)}{1+\tilde{K}(\omega)} = \frac{1}{R_+(\omega)R_-(\omega)},
\end{equation}
with
\begin{equation}
R_+(\omega)= R_-(-\omega)= \sqrt{\kappa } \frac{ \Gamma \left(1-\frac{\ri \omega }{2 \pi }\right) }{\Gamma \left(1-\frac{\ri \kappa  \omega }{2 \pi }\right)}\re^{ \frac{\ri \omega \left(\log \left(-\frac{\ri \omega }{2 \pi }\right)-1\right)}{2 \pi }-\frac{\ri \kappa  \omega  \left(\log \left(-\frac{\ri \kappa  \omega }{2 \pi }\right)-1\right)}{2 \pi }},
\end{equation}
which is also upper (lower) half plane analytic. The Fourier transform of \eqref{req} can be arranged into
\begin{multline}
\frac{\tilde{r}_+(\omega)}{R_+(\omega)} + \left(\frac{2\ri\sin\left(\frac{\pi}{\kappa}\right)\re^{-\frac{2\pi}{\kappa}B}}{\pi}\right)\frac{R_+(2\pi\ri/\kappa)}{\omega+\frac{2\pi\ri}{\kappa}} =
\\
 R_-(\omega)\tilde{r}_-(\omega) - \left(\frac{2\ri\sin\left(\frac{\pi}{\kappa}\right)\re^{-\frac{2\pi}{\kappa}B}}{\pi}\right)\frac{R_-(\omega)-R_+(2\pi\ri/\kappa)}{\omega+\frac{2\pi\ri}{\kappa}}
\end{multline}
The l.h.s. is analytic in the upper half plane, while the r.h.s. is analytic in the lower half plane. Thus, both sides are equal to an entire function. By inspecting the limit $|\omega|\rightarrow\infty$ in both half planes, we conclude that this entire function must be the constant zero. We then obtain
\begin{equation}
\tilde{r}_+(\omega)=\int_0^\infty \rd\theta \re^{\ri \omega \theta} r(\theta) = -\frac{2}{\pi}\sin\left(\frac{\pi}{\kappa}\right)\re^{-\frac{2\pi}{\kappa}B}\frac{R_+(\omega)R_+(2\pi\ri/\kappa)}{\frac{2\pi}{\kappa}-\ri\omega}.
\label{rplus}
\end{equation}

At last, we can obtain the energy gap from our approximate solution
\begin{equation}
\label{first-gap}
\ba
\frac{\Delta_\kappa}{c^2}& \sim - 4 B \int_0^\infty r(\theta)\rd\theta=-4B\tilde{r}_+(0) \\
&=  \frac{8B}{\pi}\sin\left(\frac{\pi}{\kappa}\right)\frac{\re^{\frac{1}{\kappa }-1} \kappa ^{\frac{1}{\kappa }+1} \Gamma \left(\frac{1}{\kappa }\right)}{2 \pi }\re^{-\frac{2\pi}{\kappa}B}.
\ea
\end{equation}
We can test this result by comparing it with a numerical solution of the integral equation \eqref{eq_Psi}. We did so for $\kappa=2,3,4,7,8$, validating both the exponential coefficient and overall constant factor.

As is by now familiar, $B$ only encodes the weak coupling limit indirectly. In order to express the gap as a function of the dimensionless coupling, we need to use
\be
B= \frac{\pi }{\gamma  \kappa }
+{\kappa  \over 2 \pi} \log (\kappa )-{\kappa -1 \over 2 \pi} \left(\log \left(\frac{4 \pi ^2}{\gamma  \kappa }\right)+1\right)
+\CO\left(\gamma \right). 
\label{beta-gamma}
\end{equation}
This expression was originally found in \cite{mr-long} using Volin's method, which we will revisit one final time in section \ref{ring_gy}. The subleading term is otherwise difficult to find. In analogous computations for $\kappa=2$, it had to be specified numerically, as in \cite{ko}, or indirectly, as in \cite{ls,wp,frz}. Normalizing \eqref{beta-gamma} with the Fermi energy
\be
E_F= {\pi n^2 \over 4},
\ee
we finally find
\begin{equation}
\label{gap-final}
\frac{\Delta_\kappa}{E_F}\sim \left( {\kappa \over 2 \pi} \right) ^{2/\kappa }{64 \over \kappa^2 \Gamma \left(1-\frac{1}{\kappa }\right)}
 \gamma ^{1/\kappa } \re^{- \frac{2\pi^2}{\kappa^2}\frac{1}{\gamma}},\quad \gamma\ll 1.
\end{equation}
Naturally, at $\kappa=2$ this reduces to the aforementioned literature.

\subsection{From renormalization}

As was discussed in section \ref{sec_introbcs}, we can find the weak coupling form of the gap from renormalization alone. From the $\beta$-function coefficients alone, we find
\begin{equation}
\Delta \propto \exp\left(\frac{1}{\beta_0 g(\mu)}\right) g(\mu)^{\frac{\beta_1}{\beta_0^2}}.
\label{gap-beta}
\end{equation}
We can then predict this non-perturbative scale simply by calculating the $\beta$-function.

In most renormalization procedures, like multiplicative renormalization or Wilsonian renormalization, we neglect high energy modes beyond some cutoff $E_0$ and focus on the low energy physics. For quantum gases, these are the excitations around the Fermi sphere (in 1D, the Fermi points). Following the multiplicative RG picture of \cite{m-solyom,solyom}, we construct an effective Hamiltonian with a linearized  dispersion relation near the Fermi surface. We thus start with the free Hamiltonian
\begin{equation}
H_0 = \sum_{k,\alpha} v_F(k-k_F) a^\dagger_{k,\alpha}a_{k,\alpha}+\sum_{k,\alpha} v_F(-k-k_F) b^\dagger_{k,\alpha}b_{k,\alpha}
\label{Hfree}
\end{equation}
where $a$ and $b$ are annihilation operators for right and left moving particles, respectively, and $v_F$ is the Fermi velocity. 
For the Gaudin--Yang model, one needs to consider the only two interacting four-vertices with both incoming and outgoing left and right moving particles, see \cite{solyom}, 
\begin{multline}
\label{H-interacting}
H_I = \sum_{k_1,k_2,k_3,k_4}\sum_{\alpha,\beta}^\kappa\delta(k_1+k_2-k_3-k_4)\\
\times\biggl\{ g_1 b_{k_1,\alpha}^\dagger a_{k_2,\beta}^\dagger a_{k_3,\alpha} b_{k_4,\beta} +  g_2 b_{k_1,\alpha}^\dagger a_{k_2,\beta}^\dagger b_{k_3,\alpha} a_{k_4,\beta}
\biggr\}.
\end{multline}
 
In principle, one can proceed systematically with multiplicative RG, following the template of \cite{m-solyom} (see \cite{solyom} for a review). This a feasible but involved calculation, which we do in 
\cite{mr-three}. Alternatively, one can take a shortcut which is to organize the effective model at the Fermi surface as a relativistic field theory. 

Let us rescale the space dimension by $v_F$, so that the free terms can be written in field theory language as
\begin{equation}
 - \psi^*_+(\partial_0-\partial_1)\psi_+ - \psi^*_-(\partial_0+\partial_1)\psi_-.
\end{equation}
We introduce the Dirac spinor conventions in 1+1 dimensions with signature $(-,+)$,
\begin{equation}
\Psi = \begin{pmatrix}
\psi_+ \\ \psi_-
\end{pmatrix},\quad
\bar{\Psi} = \ri \Psi^\dagger \gamma_0,
\end{equation}
where the $\gamma_\mu$ matrices and the chiral matrix $\gamma_3$ are
\begin{equation}
\quad 
\gamma_0 = \begin{pmatrix}
0&1\\ 1&0
\end{pmatrix},\quad
\gamma_1 = \begin{pmatrix}
0&-1\\ 1&0
\end{pmatrix},\quad
\gamma_3 = \begin{pmatrix}
1&0\\ 0&-1
\end{pmatrix},\quad
\gamma_3 = \gamma_0\gamma_1.
\end{equation}
Then we can wirte a Lorentz-invariant kinetic term equivalent to \eqref{Hfree},
\begin{equation}
\ri \bar{\Psi}\slashed{\partial}\Psi = - \psi^*_+(\partial_0-\partial_1)\psi_+ - \psi^*_-(\partial_0+\partial_1)\psi_-.
\label{free_lag}
\end{equation}

As had been similarly observed in \cite{woy-1, woy-2, Melzer1995}, we can get the interacting vertices \eqref{H-interacting}, by considering the Lagrangian of a specific instance of the chiral Gross--Neveu model or of the Thirring model. We will start by showing the latter. The Thirring Lagrangian with Lie Group $SU(\kappa)$ is  
\begin{equation}
\CL = \ri \bar{\Psi}\slashed{\partial}\Psi- \frac{1}{2}g J_\mu^\alpha J^{\mu\alpha},\quad J_\mu^\alpha = \bar{\Psi} \gamma_\mu T^\alpha \Psi,
\label{ThirringLag}
\end{equation}
where $\Psi_{\alpha}$ is a Dirac spinor in the fundamental representation of $\mathfrak{su}(\kappa)$, and $T^\alpha,\, \alpha= 1,\cdots,  \text{dim}(\mathfrak{su}(\kappa))$ are a basis for the algebra elements such that $\text{Tr}[T^\alpha T^\beta] = \frac{1}{2} \delta^{\alpha \beta}$. This is also the field content of the chiral Gross--Neveu model, see e.g. \cite{forgacs-chiral}.

To reduce the Thirring Lagrangian to \eqref{H-interacting} we need only to use two Fierz identities. One is that of the internal symmetry group $SU(\kappa)$, and is 
 \begin{equation}
(T^\alpha)_{ab}(T^\alpha)_{cd} = \frac{1}{2}\left(\delta_{ad}\delta_{cb}-\frac{1}{N}\delta_{ab}\delta_{cd}\right).
\label{fierzcolour}
\end{equation}
The other is related to spacetime symmetries, concretely to the Clifford algebra, which is
\begin{equation}
(\gamma_\mu)_\alpha^\beta (\gamma^\mu)_\gamma^\delta  = (\mathbb{I})_\alpha^\delta(\mathbb{I})_\gamma^\beta- (\gamma_3)_\alpha^\delta(\gamma_3)_\gamma^\beta,
\label{fierzspin}
\end{equation}
They suffice to expand the currents as
\begin{equation}
\label{int-JJ}
\ba
-\frac{1}{2} g J_\mu^\alpha J^{\mu\alpha} &= \frac{g}{4}\left((\bar{\Psi}\cdot\Psi)^2-(\bar{\Psi}\cdot\gamma_3\Psi)^2 +\frac{1}{N}(\bar{\Psi}\cdot\gamma^\mu\Psi)(\bar{\Psi}\cdot\gamma_\mu\Psi)\right)
\\&= -g(\psi^*_+ \cdot \psi_-)( \psi^*_- \cdot \psi_+) - \frac{g}{N}(\psi^*_+ \cdot \psi_+)(\psi^*_- \cdot \psi_-).
\ea
\end{equation}
The first line is the interaction term for the chiral 
Gross--Neveu model in \cite{forgacs-chiral} with $g'=0$. From the second line, we can see $g_1 \propto -g$ and $g_2 \propto -g/\kappa$. Recall that to match the free Lagrangian with \eqref{Hfree} we had to choose units such that $k_F=1$. Restoring units, we can compare \eqref{int-JJ} with \eqref{delta-GY} to fix
\begin{equation}
g = \frac{(-2c)}{v_F} = - \frac{\kappa\gamma}{\pi},
\label{gtogamma}
\end{equation}
where we used $v_F = 2 k_F$.

The $\beta$-function is known for both the generic Thirring model \cite{destri} and the chiral Gross--Neveu model \cite{forgacs-chiral,gross-neveu}. The first two coefficients of the $\beta$-function for $g$, in the convention \eqref{beta-qmb},  are
\begin{equation}
\beta_0 = \frac{\kappa}{2\pi},\quad \beta_1 = \frac{\kappa}{4\pi^2}.
\label{beta-pi}
\end{equation}
This agrees with the direct multiplicative renormalization calculation done in \cite{mr-three} and a similar calculation for the $SU(\kappa)$ Hubbard model in \cite{solyom-kappa}. Neither references the relativistic field theory model and both treat $g_1$ and $g_2$ as a priori independent.

Through \eqref{gap-beta} this predicts
\begin{equation}
\Delta \propto \re^{\frac{2\pi}{\kappa g}}g^{\frac{1}{\kappa}} \sim \re^{-\frac{2\pi^2}{\kappa^2 \gamma}}\gamma^{\frac{1}{\kappa}},
\end{equation}
where we use the normalization \eqref{gtogamma} to rewrite the gap in terms of $\gamma$. This agrees perfectly with the gap from the Bethe ansatz \eqref{gap-final}.  Of course, the latter has the advantage of calculating the overall constant. However, as we saw in the previous chapter and will see again in the next section, for the purposes of resurgent analysis the most important feature is the dependency on $\gamma$ of the non-perturbative scale.

\section{Resurgence with ring diagrams}
\label{ring_gy}

The generalization of the Gaudin--Yang model to $\kappa$ components is interesting for multiple reasons. First, since it is still integrable we can again apply the generalized Volin's method to obtain the perturbative series to high order. In this section, we then test the conjecture of section \ref{sec_superconductivity_conjecture} for more general energy gaps. Second, we compare the large $\kappa$ limit of the series with ring diagrams and use them to extract the asymptotic behavior of the series, similar to section \ref{cha_largeN}.

\subsection{Perturbative series for the multi-component Gaudin--Yang model}

As we saw in section \ref{sec_iqmb}, the Gaudin--Yang model is integrable for fermions with $\kappa$ components and the ground state energy in such cases is described by the integral equation \eqref{gscont}, which is of of the form \eqref{volin_mostgeneral} with $p=0$, $m=\kappa$. In this section, we write $\mathsf{f}$ and $\mathsf{B}$, according to \eqref{mr-long-convention}, to avoid confusion with the $f$ and $B$ of the previous section. 
The Wiener--Hopf decompostion of the kernel \eqref{K-multigy} gives
\begin{equation}
G_+(\omega) = 
\frac{
\re^{
-\frac{\ri \omega}{2\pi} \left(2 \left(1-\frac{1}{\kappa }\right) \log (\pi )-\frac{2 \log \left(\kappa\right)}{\kappa }-2 \left(1-\frac{1}{\kappa }\right) \log \left(-\frac{\ri \omega}{\re}\right)\right)
}
}{\sqrt{\kappa }}
\frac{\Gamma \left(1-\frac{\ri \omega}{\pi }\right)}{\Gamma \left(1-\frac{\ri \omega}{\pi  \kappa }\right)}.
\end{equation}
And we can carry out Volin's method using  \eqref{volin_Rhatansatz} and \eqref{volin_eq_bulkGY}, as we did in chapters \ref{cha_volin} and \ref{cha_GY}.

In this model, the dimensionless coupling is given by 
\eqref{observables-notebook}.
It is natural to introduce a 't Hooft coupling
\begin{equation}
\lambda = \left(\frac{\kappa}{2}\right)^2 \gamma = \left(\frac{1}{\pi} \int_{-\mathsf{B}}^\mathsf{B} \mathsf{f}(x)\rd x\right)^{-1},
\end{equation}
which leads to a convenient normalization for the energy
\begin{equation}
e(\lambda,\kappa) = \frac{\kappa^2 E}{4 n^3} = - \frac{\kappa^2-1}{3\kappa^2}\lambda^2 +\frac{\frac{1}{\pi}\int_{-\mathsf{B}}^\mathsf{B} x^2\mathsf{f}(x)\rd x}{\left(\frac{1}{\pi}\int_{-\mathsf{B}}^\mathsf{B} \mathsf{f}(x)\rd x\right)^3}.
\end{equation}
Applying Volin's method we find
\begin{equation}
\ba
e(\lambda,\kappa) &= \frac{\pi ^2}{12}
-\frac{(\kappa-1)}{\kappa}
\lambda +\frac{(\kappa -1) }{3 \kappa }\lambda ^2
-\frac{4  \zeta (3)}{\pi ^4 \kappa ^2}(\kappa -1)\lambda ^3
-\frac{12 \zeta (3)}{\pi ^6 \kappa ^3}(\kappa -1)^2 \lambda ^4\\
&-\frac{48  \zeta (3)}{\pi ^8 \kappa ^4}(\kappa -1)^3\lambda ^5
-\frac{40 (\kappa -1)^2}{\pi ^{10} \kappa ^5} \left(5 (\kappa -1)^2 \zeta (3)+  (\kappa^2 -\kappa+1) \zeta (5)\right)\lambda^6\\
&+\CO\left(\lambda ^7\right).&
\ea
\label{eg_kappa}
\end{equation}
We have computed 45 exact coefficients of this expansion.

\begin{figure}
\centering
\begin{subfigure}{\figsize}
\includegraphics[width=\textwidth]{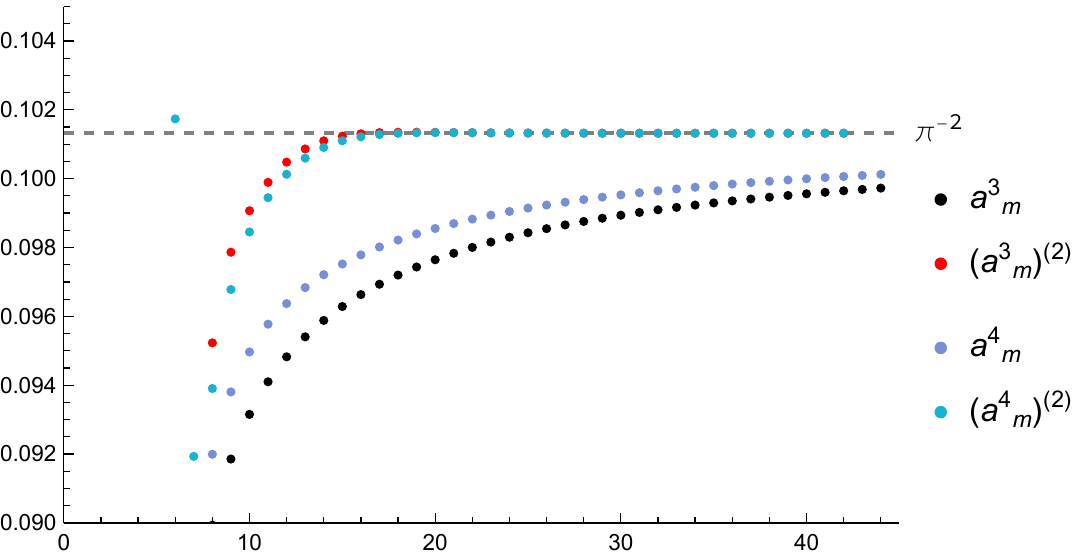}
\caption{Sequences $a^\kappa_m$}
\end{subfigure}
\\
\begin{subfigure}{\figsize}
\includegraphics[width=\textwidth]{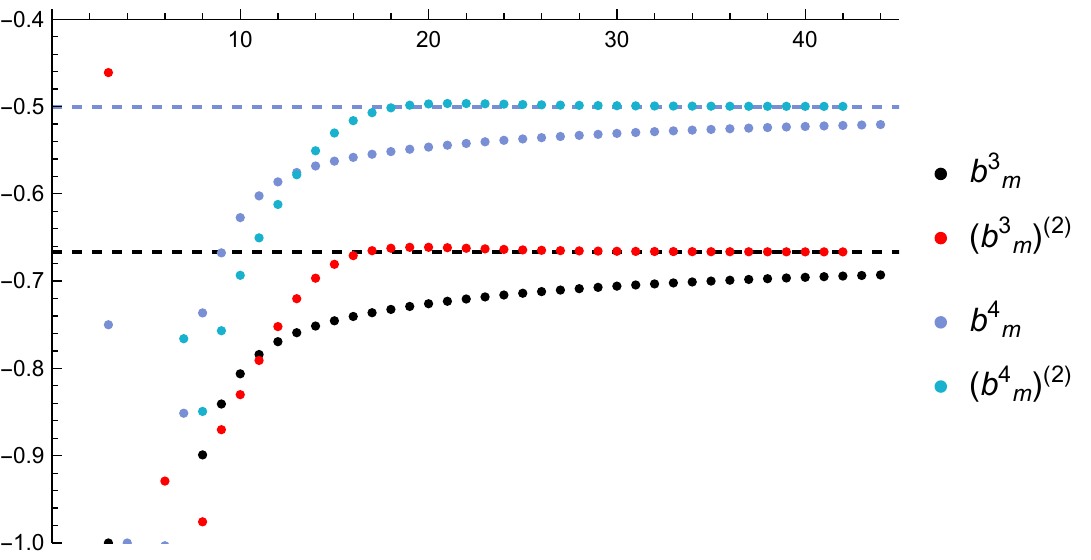}
\caption{Sequences $b^\kappa_m$}
\end{subfigure}
\caption{Plot of sequences $a^\kappa_m$ and $b^\kappa_m$ in \eqref{test-1-2} for $\kappa=3$ and $\kappa=4$ along their respective second Richardson transforms and theoretical predictions. }
\label{gy-kappa-asym}
\end{figure}

At finite $\kappa$, we can verify numerically that the large order behavior of these coefficients is
\begin{equation}
c_m \propto - \left(\pi^2\right)^{-m+\frac{2}{\kappa}}\Gamma\left(m-\frac{2}{\kappa}\right),\quad m \gg 1.
\label{GYkappa_asym}
\end{equation}
We can test this by analysing limits such as
\begin{equation}
a^\kappa_m = \frac{c_{m+1}}{m c_m} \sim \frac{1}{\pi^2},\quad b^\kappa_m = \frac{\pi^2 c_{m+1}}{c_m} - m \sim - \frac{2}{\kappa}.
\label{test-1-2}
\end{equation}
We plot some of these tests in figure \ref{gy-kappa-asym}. The associated ambiguity of the Borel summation is of the form
\begin{equation}
\disc s \left(e(\lambda,\kappa)\right) \propto  \lambda^{\frac{2}{\kappa}} \re^{-\frac{\pi^2}{\lambda}} \sim  \gamma^{\frac{2}{\kappa}} \re^{-\frac{4\pi^2}{\kappa^2\gamma}},
\label{BorelPartKappa}
\end{equation}
which matches $\Delta^2$ in \eqref{gap-final} or \eqref{gap-beta}. Thus, the conjecture of section \ref{sec_superconductivity_conjecture} seems to apply to energy gaps in general.

\subsection{Ring diagrams in the Gaudin--Yang model}
One of the main interests of the $\kappa$ component case is that it allows us to take the $\kappa\rightarrow \infty$ limit. We organize the perturbative series as
\begin{equation}
e(\lambda,\kappa) = \sum_{n\geq 0} e^{(n)}(\lambda)\kappa^{-n}.
\label{en_GY}
\end{equation}
The first term, of order $\kappa^0$, is trivial,
\begin{equation}
e^{(0)}(\lambda) = \frac{\pi^2}{12}-\lambda,
\end{equation}
being nothing more than the free gas energy and the Hartree correction.

\begin{figure}
\centering
\includegraphics[width=0.9\textwidth]{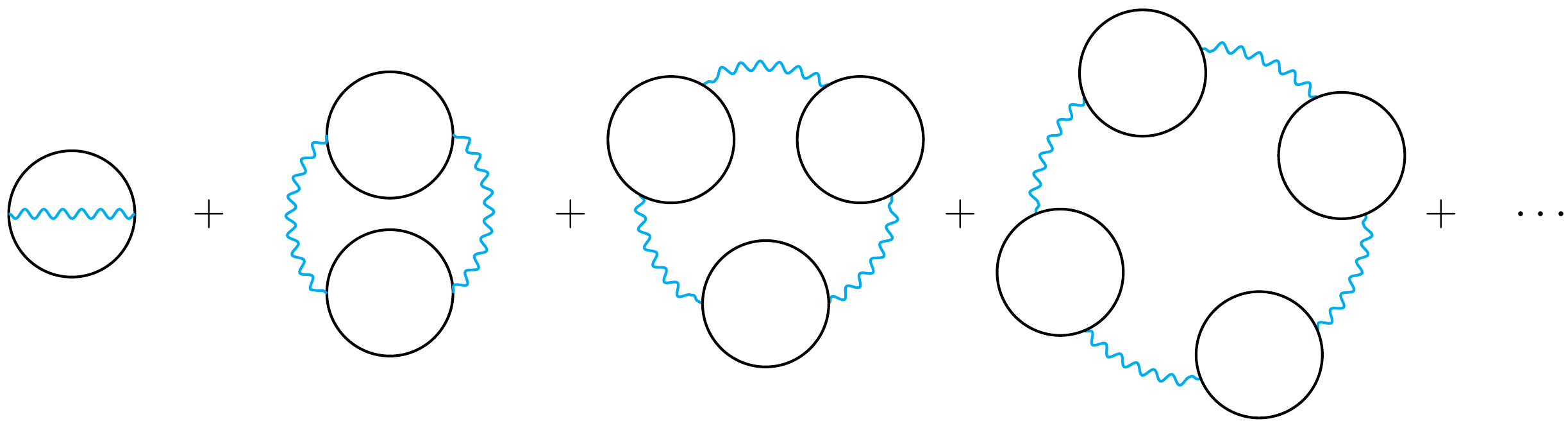}
\caption{Ring diagrams in the Gaudin--Yang model. They contribute to the ground state energy at leading order in $\kappa\gg 1$.}
\label{GY-rings}
\end{figure}

The sub-leading correction in $1/\kappa$ is much more interesting. To the first few orders we have
\begin{equation}
\ba
e^{(1)}(\lambda)&=\lambda-\frac{\lambda^2}{3}-\frac{4  \zeta (3)}{\pi ^4}\lambda^3-\frac{12  \zeta (3)}{\pi ^6}\lambda^4-\frac{48
   \zeta (3)}{\pi ^8} \lambda^5-\frac{40  (5 \zeta (3)+\zeta (5))}{\pi ^{10}}\lambda^6\\
   &-\frac{120
   (7 \zeta (3)+5 \zeta (5))}{\pi ^{12}} \lambda^7-\frac{168  (21 \zeta (3)+35 \zeta (5)+4
   \zeta (7))}{\pi ^{14}}\lambda^8\\
   &-\frac{1344  (11 \zeta (3)+35 \zeta (5)+14 \zeta (7))}{\pi
   ^{16}}\lambda^9+\CO\left(\lambda^{10}\right).
\ea
\label{e1_GY}
\end{equation}
This expression should have a simple interpretation at the level of Feynman diagrams. Beyond first order in $\lambda$, the dominant diagrams in the large $\kappa$ limit are ring diagrams, like those in figure \ref{GY-rings}. 
In condensed matter, these diagrams are often used in the context of RPA calculations. Rederiving these coefficients from ring diagrams lets us resum them, like in section \ref{cha_largeN}, to find the asymptotic behavior from the discontinuity.

\begin{figure}[h]
\centering
\begin{tikzpicture}[scale=0.75,rotate=+0,very thick]
\pgfmathsetmacro\r{1}
\pgfmathsetmacro\R{1}
\begin{feynman}
\draw[] (0,0) arc(0:360:\R);
\draw[photon, cyan] (-3*\R,0) -- (-2*\R,0);
\draw[photon, cyan] (0,0) -- (\R,0);
\end{feynman}
\end{tikzpicture}
\caption{The polarization loop in the Gaudin--Yang model.}
\label{fig-loop-GY}
\end{figure}
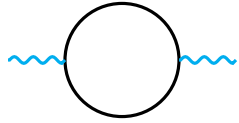

In order to construct ring diagrams, one starts with the propagator for one dimensional free fermions,
\begin{equation}
G(k,\omega) = \frac{1}{\omega-k^2 + \ri \epsilon\,\mathrm{sign}(|k|-k_F)}.
\label{Gfree}
\end{equation}
The building block for the ring diagrams is the polarization loop, also known as the Lindhard function. If we think of the fermion four vertex as the exchange of a phonon, the polarization loop can be thought of as a phonon with momentum $p$ decaying into a fermion/anti-fermion pair that collapses back into a phonon, see figure \ref{fig-loop-GY}. It is given by
\begin{equation}
\Pi(p,\omega)= - \ri \int_\IR \frac{\rd k}{2\pi} \int_\IR \frac{\rd \omega'}{2\pi} G(p+k,\omega+\omega')G(k,\omega').
\end{equation}
For each fermion loop, we also need to multiply by $\kappa$ since we sum over all possible species of fermions. 

The polarisation loop can be calculated explicitly, see for example \cite{electron-liquid}. To do so, we write the loop integral as
\begin{multline}
\Pi(p,\omega)= - \ri \int_\IR \frac{\rd k}{2\pi} \int_\IR \frac{\rd \omega'}{2\pi} \frac{1}{\omega'-k^2+ \ri \epsilon\,\mathrm{sign}(|k|-k_F)}\\ \times\frac{1}{\omega+\omega'-(k+p)^2+ \ri \epsilon\,\mathrm{sign}(|k+p|-k_F)},
\end{multline}
and integrate $\omega'$ by closing the integral in the upper half plane. The resulting integral is non-zero only if a single pole of the integrand has a positive imaginary part. This results in
\begin{equation}
\Pi(p,\omega)=  \int_\IR \frac{\rd k}{2\pi}\left(\frac{\Theta(k_F-|k|)\Theta(|k+p|-k_F)}{\omega+ k^2-(k+p)^2+2\ri\epsilon}+\frac{\Theta(|k|-k_F)\Theta(k_F-|k+p|)}{-\omega-k^2+(k+p)^2+2\ri\epsilon}\right).
\end{equation}
We will evaluate the polarisation loop at imaginary frequency, so we can drop the regulators in the integrals. After some changes of variables in both integrals, we find
\begin{equation}
\Pi(p,\ri\omega)=  - \int_\IR \frac{\rd k}{2\pi} \Theta(k_F-|k-p/2|)\Theta(|k+p/2|-k_F) \frac{k p}{(kp)^2+(\omega/2)^2}.
\end{equation}
This integral is simple to carry out by considering the cases $|p|>2k_F$ and $|p|<2k_F$ separately, which lead to same final result
\begin{equation}
\Pi(p,\ri\omega) = - \frac{1}{4\pi p}\log\left(\frac{\left(\frac{p}{2}+k_F\right)^2+\left(\frac{\omega}{2p}\right)^2}{\left(\frac{p}{2}-k_F\right)^2+\left(\frac{\omega}{2p}\right)^2}\right).
\label{lindhard}
\end{equation}

To put together the ring diagram with $n$ bubbles we need only to string polarization loops with the same momentum $p$ intercalated with the vertex $V(p)$.  In the end, we must divide by the appropriate symmetry factor, which corresponds to $n$ translations and one reflection. For a general potential in a one dimensional system, the ring diagram contribution is then 
\begin{equation}
\frac{\ri\kappa^n}{2n}\int_\IR \frac{\rd p}{2\pi}\int_\IR \frac{\rd \omega}{2\pi} \left(V(p)\Pi(p,\omega)\right)^n.
\end{equation}
Since the interaction is a $\delta$-potential, the potential term is simply a constant $V(p)=-2c$. After a Wick rotation and symmetrization of the $\omega$ integral we find
\begin{equation}
-\frac{\kappa^n}{n}\int_\IR \frac{\rd p}{2\pi}\int_0^\infty \frac{\rd \omega}{2\pi} \left(-2c\Pi(p,\ri\omega)\right)^n.
\label{ringdiag}
\end{equation}
Plugging \eqref{lindhard} in \eqref{ringdiag} and recalling that $k_F = \pi n/\kappa$ and $ \lambda = (\kappa/2)^2c/n$, we find
\begin{equation}
e^{(1)}(\lambda) = \lambda- \frac{\pi}{4}\sum_{n\geq 0}^\infty \frac{1}{n}\left(\frac{4\lambda}{\pi^2}\right)^n\CI_n,
\label{ringe1}
\end{equation}
where
\be
\CI_n = \displaystyle \int_0^\infty \rd y\, y \int_0^\infty \rd \nu \left(\frac{1}{2y}\log\left(\frac{\left(y/2+1\right)^2+\left(\nu\right)^2}{\left(y/2-1\right)^2+\left(\nu\right)^2}\right)\right)^n.
\label{ring_diagram_In}
\ee
One can calculate the integrals $\CI_n$ analytically for lower and numerically for further values, although for too high $n$ it becomes very unstable. Indeed the diagrammatic \eqref{ringe1} agrees with the exact series \eqref{e1_GY}. 

\subsection{Asymptotic behavior from ring diagrams}

Finally, we will show that the asymptotic behavior of the ring diagrams is responsible for the $\pi^2$ factor in \eqref{GY_LOasym} and \eqref{GYkappa_asym}, and so we can claim that the divergence of the perturbative series \eqref{eten} and \eqref{eg_kappa} is due to a renormalon effect. In fact, other known families of diagrams, like ladder diagrams, are sub-leading in $1/\kappa$ and do not grow sufficiently fast. Ring diagrams are thus isolated at large $\kappa$, which also protects them from cancellation with other families.

We start by ressumming the energy at order $1/\kappa$, \eqref{ringe1}, using the integral representation of the ring diagrams \eqref{ring_diagram_In}, which leads to
\be
\label{e1-int}
e_1(\lambda)= \lambda - {\pi \over 4} \int_0^\infty \rd y \, y \int_0^\infty \rd \nu \, \left[ \log \left( 1-{\kappa^2 \gamma  \over \pi^2} F( y, \nu)\right)+ {\kappa^2 \gamma  \over \pi^2} F( y,\nu) \right], 
\ee
where
\be
F( y,\nu)={1\over 2 y} \log \left( {(y/2+1)^2 + \nu^2 \over (y/2-1)^2 + \nu^2 }  \right). 
\ee
As we saw in section \ref{cha_largeN}, this type of integrals has a non-perturbative imaginary piece which must be cancelled by effects beyond the reach of Feynman diagrams. This piece originates from the region of integration where the argument of logarithm is negative, that is
\be
1-{\kappa^2 \gamma  \over \pi^2} F(y,\nu, \gamma)<0.
\label{subregion}
\ee
This condition is satisfied in a bounded subregion of the quadrant $y>0 \wedge \nu>0$. The upper boundary of this subregion is defined through the equality case of \eqref{subregion}, which can be rewritten as the curve
\be
\nu^2 = {-(y-2)^2+ \re^{-2 \pi^2 y/\kappa^2 \gamma} (y+2)^2 \over 4 \left(1-\re^{-2\pi^2 y/\kappa^2 \gamma} \right)}, 
\label{eqnu}
\ee
while the lower boundary is defined by $\nu=0$. The two boundaries intersect at points $y=y_\pm$, which will be used as limits of integration.

In order to calculate this non-perturbative effect, it is convenient to re-parameterize the problem in terms of 
\be
\alpha=\re^{-2\pi^2 /\kappa^2 \gamma},
\ee
as well as changing the variable $y$ to 
\be
y = 2 + 4 \alpha u.
\ee
The endpoints $u_\pm$ now solve
\begin{equation}
\re^{-2\alpha\log\alpha u_\pm}\mp \left(\frac{1}{u_\pm}+\alpha\right)=0, 
\label{eq_upm}
\end{equation}
whose solution can be found iteratively as a simple trans-series
\begin{equation}
\ba
u_\pm &=\pm 1+\alpha  (2 \log (\alpha )+1)\pm \alpha ^2 \left(6 \log ^2(\alpha )+6 \log (\alpha )+1\right)+\cdots\,.\\
\ea
\label{equpm}
\end{equation}
In terms of the new variables, the imaginary part of $e_1(\lambda)$ is given by
\begin{equation}
\int_{u_-}^{u_+} 4\alpha(2+4\alpha u)\nu(u)\rd u ,
\label{nuint}
\end{equation}
where $\nu(u)$ is defined by the curve \eqref{eqnu}.

In order for the integration to be well defined in the small $\alpha$ expansion, it is convenient to rexpress the integrand as a function of $(u-u_-)$ and $(u-u_+)$. This can be done by plugging equations \eqref{eq_upm} into \eqref{eqnu}. We find an expansion
\begin{multline}
\nu(u) = 2\alpha\sqrt{(u_+-u)(u-u_-)}\\
\times\left(1-\alpha ^2 \left(u_--u_++6\right) \log (\alpha ) (\log (\alpha )+1)+ \CO(\alpha^3)\right)\,.
\end{multline}
This expansion can be done systematically to arbitrary order. Terms of the form $u^n(u-u_+)^{m+1/2}(u-u_-)^{k+1/2}$ are simple to integrate into simple expressions of $u_+$ and $u_-$ which can then be easily expanded using \eqref{equpm}. Had we written $u_\pm$ as an expansion in $\alpha$ before integrating, this would cause a singular behavior of the square root once we integrate. While this issue can be fixed by carefully expanding the integration limits themselves, we found that approach to be computationally much more cumbersome.

Carrying out this scheme to sufficiently high order, we find the trans-series expression of the discontinuous part of $e_1(\lambda)$:
\begin{equation}
\ba 
\disc e_1(\lambda)&= -2\pi\ri\bigg(2 \pi ^2 \re^{- \pi^2 /\lambda }
+\frac{8 \pi^2\re^{-2 \pi^2 /\lambda} }{ \lambda^2}\left(\lambda ^2- \frac{3\pi^2}{2}\lambda +\frac{\pi^4}{2}\right)\\
&
+\frac{6 \pi ^{2} \re^{-3 \pi^2 /\lambda}}{ \lambda ^4} \left(3 \lambda ^4-14\pi^2\lambda^3+ 21\pi^4 \lambda^2- 12\pi^6 \lambda + 
\frac{9\pi^8}{4}\right) +\cdots\bigg)\,.
\ea
\label{trans}
\end{equation}
An important test of the above calculation is that it should give a precise formula for the large order behavior of the coefficients 
$c_\ell^{(1)}$ defined as
\be
e_1(\lambda)= \sum_{ m \ge 0} c_\ell^{(1)} \lambda^\ell. 
\ee
These are the coefficients in \eqref{e1_GY}.
With the tools of chapter \ref{cha_resurgence} in mind, define the coefficients $a_{j,i}$ of \eqref{trans} such that
\be
\disc e_1(\lambda)= -2\pi\ri\sum_{j\ge 1} \sum_{i=0}^{2j-2} a_{j,i} \lambda^{-i} \re^{-j \pi^2/\lambda} .
\label{ime1-aji}
\ee
Then from \eqref{resurgence-relation} we have
\be
\label{lo-csub}
 c_m^{(1)} \sim -\sum_{j\ge 1} \sum_{i=0}^{2j-2} {\Gamma(m+i) \over (\pi^2 j)^{m + i}} a_{j,i}. 
 \ee
By taking terms up to some upper bound in $j$, we can approximate the exact value of $c_m^{(1)}$ with extremely high accuracy. In particular, this means that the dominant asymptotic behavior is
\begin{equation}
c_m^{(1)} \sim - 2\pi^2 \left(\pi^2\right)^{-m}\Gamma(m),
\end{equation}
which agrees with the $\kappa\rightarrow\infty$ limit of \eqref{GYkappa_asym}.

We can see that, even at $\kappa=2$, the leading exponential factor is coherent with \eqref{BorelPartKappa} and the multiple expressions for the gap. This proves that the leading asymptotic behavior discussed in chapter \ref{cha_GY} is indeed a renormalon effect originating in ring diagrams. However, neither at $\kappa=2$ nor at larger $\kappa$ does this calculation get correctly the factor of $\gamma^{2/\kappa}$ in the gap, or equivalently the shift in the $\Gamma$-function for the asymptotic expressions. This is expected since this factor is subleading in the $1/\kappa$ expansion, and thus not captured by ring diagrams which account for the leading contribution only. 

\subsection{Trans-series in multi-component case}

One can extend the analysis of section \ref{sec-gy-antrans} to the multi-component case to find the analytic trans-series. Using convention \eqref{multi-gy-convention} for the integral equation,
the function $\sigma(\omega)=G_-(\omega)/G_+(\omega)$ has poles at
\begin{equation}
\omega =  \frac{2\pi\ri\ell}{\kappa}.
\end{equation}
These poles lead to non-perturbative terms proportional to $\re^{- \frac{4\pi\ell}{\kappa}B}$, which contribute to the energy with an overall factor of $1/B^2$ and further perturbative corrections. 
Using \eqref{beta-gamma}, this means that the full trans-series for the ground state energy is of the form
\begin{equation}
e(\lambda,\kappa) = \sum_{m\geq 0} c_m \lambda^m + \sum_{\ell\geq 1}  \CC_\ell^\pm \re^{- \frac{\ell\pi^2}{\lambda}} \lambda^{\frac{2\ell}{\kappa}-2(\ell-1)} \sum_{m\geq 0} c^{(\ell)}_{m} \lambda^m.
\end{equation}
This is consistent with \eqref{egy} for $\kappa=2$, with \eqref{BorelPartKappa} for $\ell=1$ and with \eqref{trans} for $\kappa\rightarrow \infty$. The first $\CC^\pm_\ell$ could be calculated  explicitly using the methods described in section \ref{sec-gy-antrans}.

\section{Renormalons in the Hubbard model}

The one-dimensional Hubbard model can also be extended to $\kappa$ components. However, for $\kappa>2$ the model is not integrable, and as such we cannot obtain its exact perturbative series. Nevertheless, the ring diagram calculation can still be done at large $\kappa$. As we saw in the previous section, this analysis alone can be instructive about the structure of the trans-series even at $\kappa=2$. Furthermore, one can use the non-perturbative scale detected this way to conjecture a form for the energy gap in the multi-component Hubbard model. 

For the $\kappa$-component Hubbard model, it will be useful to define
\begin{equation}
\widetilde{n} = \frac{2}{\kappa} n,
\end{equation}
such that
\begin{equation}
k_F = \frac{\pi \widetilde{n}}{2},\quad \widetilde{n}_{\kappa=2}=n.
\end{equation}
In the weak coupling limit $u\rightarrow 0$, the ground state energy is approximated by the perturbative series 
\be
\label{pert-ser}
E(u, n ;\kappa)=  -{2 \kappa \over \pi} \sin \left( {\pi \widetilde n \over 2} \right)  - {1\over 4} \kappa (\kappa-1) u \widetilde n^2 + \sum_{\ell\ge 2} E_\ell(n; \kappa) u^\ell. 
\ee
By taking the limit analogous to \eqref{GYlimit}, we can recover the perturbative series for the Gaudin--Yang model with $\kappa$-components at first non-trivial order,
\be
\label{e-scal}
E(u, n;\kappa)=-\kappa \widetilde n + \widetilde n^3 e_{\rm GY}(\lambda; \kappa)+ \CO(\widetilde n^4). 
\ee

We can also reorganize the ground state energy in the $\kappa\rightarrow \infty$ limit. Let us introduce the 't Hooft coupling
\begin{equation}
\upsilon = \kappa u.
\end{equation}
Then we define the limit
\begin{equation}
u\rightarrow 0,\quad \kappa, n \rightarrow \infty,\quad \widetilde{n},\upsilon\quad \text{finite}.
\end{equation}
In this 't Hooft-like limit, the ground state energy can be written as
\be
\label{1nexp}
E(u,n;\kappa)= \sum_{r=0}^\infty e_r (\upsilon, \widetilde n) \kappa^{1-r}. 
\ee
The leading order term is encapsulated by the free theory term and the Hartree diagram,
\be
e_0(\upsilon, \widetilde n)=  -{2  \over \pi} \sin \left( {\pi \widetilde n \over 2} \right)  - {1\over 4} \upsilon \widetilde n^2,
\ee
while the next to leading order include the Fock diagram and the dominant diagrams in the large $\kappa$ limit. These are, once again, the ring diagrams of figure \ref{GY-rings}.

As in the previous section, ring diagrams are constructed from the polarization loop. For the one-dimensional Hubbard model, this is a standard calculation, see \cite{coleman} for example. It is given by
\be
\Pi(q, \ri \omega)= -{1\over 8 \pi \sin\left({q\over 2} \right)} \left\{ F \left(q+   \pi \widetilde n , {\omega \over 4 \sin\left({q\over 2} \right)}  \right)- F\left(q-   \pi \widetilde n, {\omega \over 4 \sin\left({q\over 2} \right)}  \right) \right\},
\ee
where
\be 
F(q, \omega)= -\frac{1 }{ {\sqrt{1+ \omega^2}}}\log\left[ 
\frac{1+\sqrt{\omega ^2+1}+\frac{\omega ^2}{2 \cos ^2\left(\frac{q}{4}\right)}}
{1-\sqrt{\omega ^2+1}+\frac{\omega ^2}{2 \cos ^2\left(\frac{q}{4}\right)}} 
\right]. 
\ee
Adding the contribution from the Fock diagram, the ground state energy is approximated by
\be
e_1(\upsilon, \widetilde n)= {\upsilon \widetilde n^2 \over 4} - \sum_{\ell \ge 2} {(-2  \upsilon)^\ell \over \ell} \int_{-\pi}^\pi  {\rd q  \over 2 \pi} \int_0^\infty {\rd \omega \over 2 \pi}  \left( \Pi(q, \ri \omega)  \right)^\ell,
\ee
at next-to-leading order in $1/\kappa$.

Let us define the ``ring diagram'' energy, which is responsible for the asymptotic series beyond first order,
\be
\label{ring-ser}
E^{\rm ring}(n; \kappa)= \sum_{\ell\ge 2} E^{\rm ring}_\ell (n; \kappa) u^\ell = - \sum_{\ell \ge 2} {(-2  \upsilon)^\ell \over \ell} \int_{-\pi}^\pi  {\rd q  \over 2 \pi} \int_0^\infty {\rd \omega \over 2 \pi}  \left( \Pi(q, \ri \omega)  \right)^\ell.
\ee
It is also useful to define the integrals
\begin{equation}
\widetilde{E}_\ell(\widetilde{n}) = - \frac{1 }{ \ell} \int_{-\pi}^\pi  {\rd q  \over 2 \pi} \int_0^\infty {\rd \omega \over 2 \pi}  \left(-2 \Pi(q, \ri \omega)  \right)^\ell = \kappa^{-\ell} E^{\rm ring}_\ell (n; \kappa),
\label{Etilde-n}
\end{equation}
which depend on $n$ and $\kappa$ only through $\widetilde{n}$.
The series \eqref{ring-ser} series can be resummed into
\begin{equation}
\label{rpa-energy}
E^{\rm ring}(n; \kappa) = \int_{-\pi}^\pi \frac{\rd q}{2\pi}\int_{\IR} \frac{\rd \omega}{2 \pi} \left\{ - \upsilon\Pi(q, \ri \omega)+\frac{1}{2} \log\left(1+2\upsilon \Pi(q,\ri \omega)\right) \right\}.
\end{equation}
This integral representation has an ambiguous imaginary part when the coupling is small due to the argument of the logarithm becoming negative. Following again the discussion of section \ref{cha_largeN}, this is equivalent to the ambiguity of the Borel summation and encodes the asymptotic behavior of the perturbative series in the form of a trans-series.

\subsection{Trans-series from the Hubbard model ring diagrams}

Let $\widetilde n<1$. Using the symmetries of the polarization loop, 
\be
\Pi(q,\ri \omega)=\Pi(-q,\ri \omega)=\Pi(q,-\ri \omega)=\Pi(-q,-\ri \omega),
\ee
we can write the imaginary part of (\ref{rpa-energy}) as the integral
\begin{equation}
\Sigma(\upsilon,\widetilde{n})=\frac{1}{\pi}\int_0^\pi \rd q \int_0^\infty \rd \omega \, \Theta\left(-(1+2\upsilon\Pi(q,\ri \omega))\right) .
\end{equation}
Let us define the curve $Q_0(q)$ which satisfies
\begin{equation}
 -\Pi(q, \ri Q_0(q))= \frac{1}{2\upsilon},
 \label{curve_Q0}
\end{equation}
with endpoints at $q_\pm$ such that
\begin{equation}
-\Pi(q_\pm,0)= \frac{1}{2\upsilon}.
\label{qpm_def}
\end{equation}
We can then write the area of the subset of the quadrant where the logarithm has a negative argument,
\begin{equation}
\Sigma(\upsilon,\widetilde{n})
= \frac{1}{\pi}\int_{q_-}^{q_+} \rd q \, Q_0(q).
\label{negarea}
\end{equation}

In order to facilitate the expansion as a trans-series, it is useful to redefine the key variables. We introduce the trans-series parameter 
\be
\alpha = \exp\left(-\frac{2 \pi}{\upsilon}\sin\left(\frac{\pi \widetilde{n}}{2}\right)\right), 
\ee
and change variables from $q$ to 
\begin{equation}
w = \frac{1}{2 \alpha}\left(\frac{\sin(q/2)}{\sin(\pi\widetilde{n}/2)}-1\right).
\end{equation}
In this variable, the end points \eqref{qpm_def} satisfy \eqref{eq_upm}, and thus can be expanded as in \eqref{equpm}.

It is also convenient to re-express $Q_0(q)$ as
\begin{equation}
\nu(q)= \alpha^{-2}\left(\sqrt{\frac{1}{16} Q_0(q)^2\csc ^2\left(\frac{q}{2}\right)+1}-1\right). 
\end{equation}
Which satisfies
\begin{equation}
\ba
\frac{\nu}{4}&\left(\left(\alpha ^2 \nu +2\right) \csc ^2\left(\frac{\pi \widetilde{n}}{2}\right) \left(1-\alpha^2 \re^{2\alpha  \log \alpha  \left(2w+\alpha  \nu +2 \alpha ^2 \nu  w\right)}\right)
\right.\\ &\left.
-2 (2 \alpha  w+1) \left(1+\alpha ^2 \re^{2\alpha  \log \alpha  \left(2w+\alpha  \nu +2 \alpha ^2 \nu  w\right)}\right)\right) = g_+(w,\nu)g_-(w,\nu),
\ea
\label{eqwnu2}
\end{equation}
where
\begin{multline}
g_\pm(w,\nu)=(w-w_\pm)\pm\re^{2 \alpha\log\alpha w_\pm} 
\left(
1+\alpha  w_\pm 
\left(
1-\re^{2 \alpha\log\alpha (w-w_\pm) }
\right)
\right.\\ \left.
-\re^{2 \alpha  \log\alpha  \left(w-w_\pm\right)} \left(\alpha  w_{\pm}-(1+\alpha w) \re^{\alpha ^2 \log\alpha\nu  (2 \alpha  w+1)}\right)
\right).
\end{multline}
In terms of $\nu$, the integral \eqref{negarea} is given by
\begin{equation}
\Sigma(\upsilon,\widetilde{n}) =
  \frac{16\alpha}{\pi}\sin^2 \left(\frac{\pi  \widetilde{n}}{2}\right)
 \int_{w_-}^{w_+} 
(1+2\alpha w)\sqrt\frac{ (1+\alpha^2 \nu)^2-1
}{1-\sin ^2\left(\frac{\pi  \widetilde {n}}{2}\right) (2 \alpha  w+1)^2}
\rd w.
\label{nu_int}
\end{equation}

The form of equation \eqref{eqwnu2} is chosen such that it naturally expands in powers of $\alpha$ and $\log\alpha$, allowing to recursively solve for $\nu(w)$ at each order. To next to leading order we obtain
\begin{multline}
\nu(w)= 2  \tan ^2\left(\frac{\pi  \widetilde{n}}{2}\right)\left(w-w_-\right) \left(w_+-w\right)\\
+4 \alpha  \left(w \left(w-w_-\right) \left(w_+-w\right) \tan ^4\left(\frac{\pi  \widetilde{n}}{2}\right)\right)+\CO\left(\alpha ^2\right).
\end{multline}
Plugging into \eqref{nu_int}, one finds the trans-series, 
\be
\ba
\Sigma(\upsilon,\widetilde{n})&=16 \, \re^{-\frac{4 \pi}{\upsilon}\sin\left(\frac{\pi \widetilde{n}}{2}\right)}\sin \left(\frac{\pi  \widetilde{n}}{2}\right) \tan ^2\left(\frac{\pi  \widetilde{n}}{2}\right)\\
&+\re^{-\frac{8\pi}{\upsilon}\sin\left(\frac{\pi \widetilde{n}}{2}\right)}\left(\frac{2048 \pi ^2}{\upsilon ^2} \sin ^7\left(\frac{\pi  \widetilde{n}}{2}\right) \csc ^2(\pi  \widetilde{n})
+\CO\left(\frac{1}{\upsilon}\right)
\right)\\
& + \CO\left(\re^{-\frac{12\pi}{\upsilon}\sin\left(\frac{\pi \widetilde{n}}{2}\right)}\right).
\ea
\label{trans_ren_hub}
\ee

\begin{figure}
\begin{center}
\begin{subfigure}{\figsize}
\centering
\includegraphics[width=\textwidth]{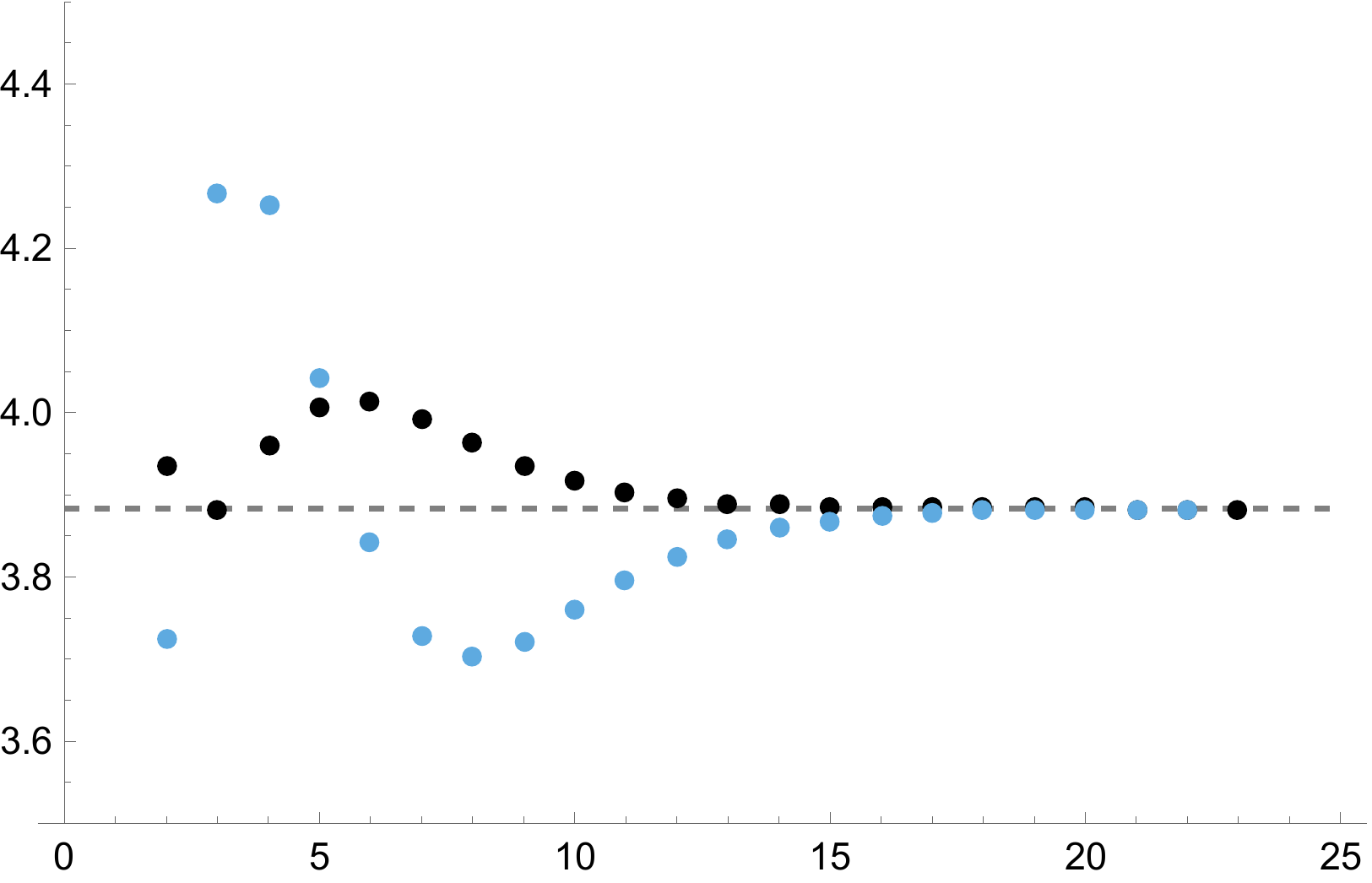}
\caption{$\widetilde{n}=0.2$}
\end{subfigure}
\\
\begin{subfigure}{\figsize}
\centering
\includegraphics[width=\textwidth]{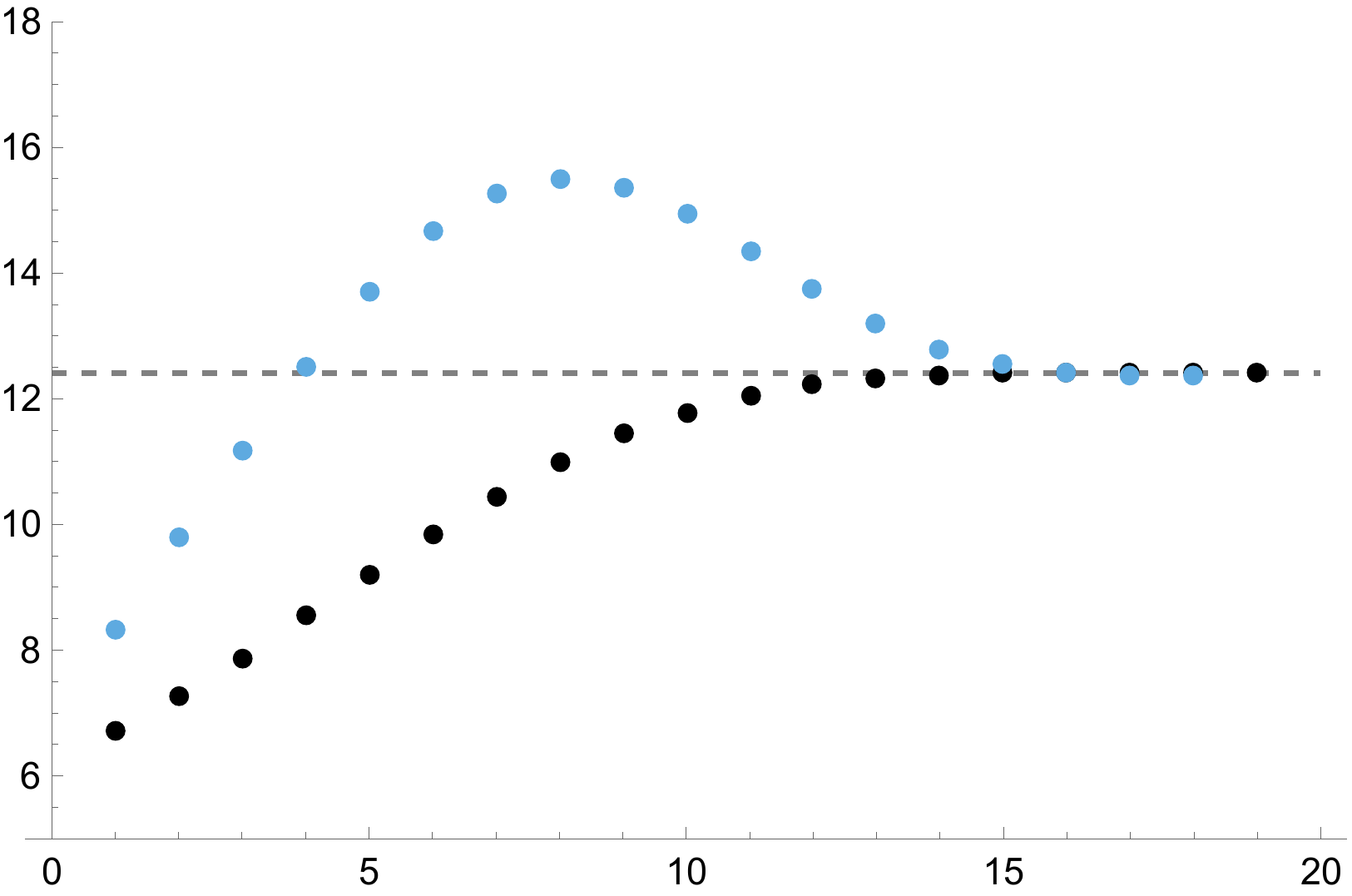}
\caption{$\widetilde{n}=0.9$}
\end{subfigure}
\end{center}
\caption{
Plot of the sequence in \eqref{Etilde-lim} (black), its first Richardson transform (blue) and the theoretical prediction (dashed grey line) for two values of $\widetilde{n}$. The $\widetilde{E}_\ell(\widetilde{n})$ are calculated numerically from their integral representation \eqref{Etilde-n}.
}
\label{hub-rings-n}
\end{figure}

The Borel singularities associated with \eqref{trans_ren_hub}, in the Borel plane dual to $u$, would be at the points 
\be
\zeta= \frac{4 \pi \ell }{\kappa}\sin\left(\frac{\pi n}{\kappa}\right), \qquad \ell \in \IZ_{>0}. 
\ee
In particular, this implies the following factorial growth for ring diagrams
\begin{equation}
E_\ell^\text{ring}(n,\kappa) \propto \left[\frac{4\pi}{\kappa}\sin\left(\frac{\pi n}{\kappa}\right) \right]^{-\ell}\ell!.
\end{equation}
In terms of the integrals defined in \eqref{Etilde-n}, we have
\begin{equation}
\frac{\ell\, \widetilde{E}_{\ell}(\widetilde{n})}{\widetilde{E}_{\ell+1}(\widetilde{n})} \sim 4\pi \sin\left(\frac{\pi\widetilde{n}}{2}\right),
\label{Etilde-lim}
\end{equation}
which is verified numerically, see figure \ref{hub-rings-n}. For $\kappa=2$, this prediction agrees with \eqref{asym_HUB_alln}. This shows that the divergence of the perturbative series in the Hubbard model is also due to a renormalon effect.

As we saw in the case of Gaudin--Yang, ring diagrams can identify the exponential factor but not the leading power of the coupling. In terms of \eqref{asym_gap}, we can guess $A$ but not $b$\footnote{Here we assume that $A$ is $\kappa$-independent when expressed in terms of the 't Hooft coupling. However as we saw in chapter \ref{cha_antrans}, this is sometimes not the case.}. However, we can take the $\kappa=2$ case \eqref{asym_HUB_alln} to conjecture that $b$ is independent of $n$. If that is the case, since the limit $n\rightarrow 0$ recovers the Gaudin--Yang model, we can read $b=2/\kappa$ from \eqref{GYkappa_asym}. Thus, we conjecture that energy gap of this model is 
\begin{equation}
\Delta \propto u^{\frac{1}{\kappa}} \exp\left(-\frac{2\pi}{\kappa u}\sin\left(\frac{\pi n}{\kappa}\right)\right),
\end{equation}
which is manifested as a renormalon effect.

In fact, this formula can be backed up by renormalization arguments. The natural dimensionless coupling in this model is given by
\begin{equation}
g = \frac{2 u}{\pi v_F},\quad v_F = 2 \sin \frac{\pi n}{\kappa}.
\end{equation}
The leading coefficients of the $\beta$-function for this coupling were calculated in \cite{solyom-kappa}, and they are
\begin{equation}
\beta_0 = - \frac{\kappa}{2}, \quad \beta_1 = \frac{\kappa}{4}.
\end{equation}
These are identical to the Gaudin--Yang case since, with appropriate normalization, it is also related to the chiral Gross--Neveu model when close to the Fermi surface, see \cite{Melzer1995,woy-1,woy-2}. Therefore, the above conjecture agrees with \eqref{gap-beta}. This argument for the gap extends the $\kappa=2$ calculation of \cite{ls}.

\subsection{Ring diagrams at half-filling}

In the preceding analysis we assumed $\widetilde{n} < 1$, which implies $n<\kappa/2$. Since $n<1$ and $\kappa\in \IZ_{\geq 2}$, this is automatically satisfied for all $n$ and $\kappa$ except for $\kappa=2$ and $n=1$. However this is a case of special interest, where we know the exact perturbative series in closed form \eqref{hk_def}. Since for $\widetilde n=1$ the trans-series \eqref{trans_ren_hub} is irredeemably singular, we need to redo the entire calculation. 

We redefine the trans-series coupling to be  
\be
 \alpha = \re^{-\frac{2\pi}{\upsilon}},
 \ee
 and in place of $w$ and $\nu$ we use respectively
\begin{equation}
s = \frac{1}{\alpha}\left(1-\sin \left(\frac{q}{2}\right)\right),\quad t(s)= \alpha^{-1}\left(\sqrt{\frac{1}{16} Q_0^2 \csc ^2\left(\frac{q}{2}\right)+1}-1\right).
\end{equation}
For $\tilde n = 1$, the upper limit of the integral in \eqref{negarea} is fixed at $q=\pi$ (or $s=0$), while the lower limit in $q$ (upper limit in $s$) satisfies
\begin{equation}
s_+ =  \re^{-\frac{2\pi}{\upsilon} s_+}(2-\alpha s_+).  
\end{equation}
The equation for the curve \eqref{curve_Q0} can be re-expressed in the new variables,
\begin{equation}
t=\frac{(s_+-s)+\re^{\frac{2 \pi  \alpha  s_+}{\upsilon }}\left(\alpha  s_+-2\right) + \re^{\frac{2 \pi  \alpha  (\alpha  s t+s-t)}{\upsilon }}(2-\alpha  s)}{1-\alpha \,  \re^{\frac{2 \pi  \alpha  (\alpha  s t+s-t)}{\upsilon }}}.
\label{eqts}
\end{equation}
and then expanded in $\alpha$ as
\begin{equation}
t(s)=(s_+-s)\left\{ 1+\alpha  \left(2-\frac{8 \pi }{\upsilon }\right)+
+\CO\left(\alpha ^2\right)\right\}. 
\end{equation}
The integral itself \eqref{negarea} is much simpler in this case,
\be
\Sigma(\upsilon ,1)=\frac{1}{\pi}\int_0^{s_+} 8 \alpha  (1-\alpha  s) \sqrt{\frac{t\, (2+\alpha  t)}{s\, (2-\alpha  s)}}\,  \rd s ,
\ee
and its trans-series expansion is given by
\be
\ba
\Sigma(\upsilon ,1)&=8 \, \re^{-\frac{2 \pi }{\upsilon }}+\re^{-\frac{6 \pi }{\upsilon }}\left(\frac{96 \pi ^2}{\upsilon ^2}-\frac{64 \pi }{\upsilon }+8\right) \\
&+\re^{-\frac{10 \pi }{\upsilon }}\left(\frac{4000 \pi ^4}{\upsilon ^4}-\frac{4800 \pi ^3}{\upsilon ^3}+\frac{1760 \pi ^2}{\upsilon ^2}-\frac{224 \pi }{\upsilon }+8\right)+ \CO\left(\re^{-\frac{14 \pi }{\upsilon }} \right). 
\label{sigma_half}
\ea
\end{equation}

The leading term of \eqref{sigma_half} implies that for $n=1,\kappa=2$ the ring diagrams grow as
\begin{equation}
E_\ell^{\text{ring}}(1;2) \propto \left(\pi\right)^{-\ell} \ell !,
\end{equation}
which agrees with a numerical calculation of the first 20 ring diagrams at $\widetilde{n}=1$, see table \ref{ring-half-tab}. However, this estimate is much larger than the exact coefficients. The latter can be obtained directly from \eqref{hk_def} and is 
\begin{equation}
E_\ell(1;2) \propto \left(2\pi\right)^{-\ell} \ell !.
\end{equation}
In fact, it can be shown that the positions of the Borel singularities associated with the trans-series \eqref{sigma_half} do not agree with those of the exact Borel transform \eqref{hk_def}, see \cite{mr-hubbard} for further details.


\begin{table}
\centering
\begin{tabular}{|c|c|c|| c| c| c|}
\hline
$\ell$& $E^{\text{ring}}_\ell(1;2)$& $\frac{\ell E^{\text{ring}}_\ell}{E^{\text{ring}}_{\ell+1}}$&$\ell$& $E^{\text{ring}}_\ell(1;2)$& $\frac{\ell E^{\text{ring}}_\ell}{E^{\text{ring}}_{\ell+1}}$ \\
\hline
 1 & -0.25 & 3.68491 & 11 & -7.85371 & 3.14194 \\
 2 & -0.0678443 & 3.14159 &  12 & -27.4960 & 3.14173 \\
 3 & -0.043191 & 3.17730 & 13 & -105.022 & 3.14165 \\
 4 & -0.0407809 & 3.18993 & 14 & -434.577 & 3.14161 \\
 5 & -0.0511371 & 3.17774 & 15 & -1936.61 & 3.14160 \\
 6 & -0.0804615 & 3.16240 & 16 & -9246.61 & 3.14160 \\
 7 & -0.152659 & 3.15194 &  17 & -47092.5 & 3.14159 \\
 8 & -0.339033 & 3.14631 &  18 & -254830. & 3.14159 \\
 9 & -0.862045 & 3.14363 & 19 & -1.46007$\times 10^6$ & 3.14159 \\
 10 & -2.46798 & 3.14244 & 20 & -8.83034$\times 10^6$ & 3.14159 \\
\hline
\end{tabular}
\caption{Numerical calculation of the ring diagrams \eqref{ring-ser} with $\kappa=2$ at half filling $n=1$. The ratio $\ell E^{\text{ring}}_\ell/E^{\text{ring}}_{\ell+1}\sim\pi$ confirms that $E_\ell(1;2)\propto \pi^{-\ell}\ell!$. In contrast with the exact coefficients \eqref{hk_def}, the diagrams for odd $\ell$ are non-zero.}
\label{ring-half-tab}
\end{table}

One should perhaps have expected the above disagreement. Ring diagrams with an odd number of bubbles are non-vanishing for any $\widetilde{n}$, but at $n=1$ and $\kappa=2$ we know that the exact odd coefficients are zero. Thus, we know that at least for the odd coefficients there had to be another diagrammatic contribution which grows similarly to the ring diagrams. It is thus conceivable that there is a different family of factorially divergent diagrams, which is subleading at large $\kappa$ but at $\kappa=2$ and $n=1$ cancels the large order behavior of the ring diagrams at both odd and even coefficients. This would explain why the ring diagrams overestimate the asymptotic growth for this case. In any case, this illustrates the limitations of extrapolating from a single sequence of diagrams.

\section{Concluding remark on non-relativistic systems}

The emerging picture of this chapter complements that of chapter \ref{cha_GY}. Not only are superconductor gaps manifest in the large order behavior of perturbation theory, but so are energy gaps more generally. Furthermore, they manifest as renormalon effects specifically, and they appear in precisely the same form as ring diagrams in field theory, see section \ref{cha_largeN}. 

One could wonder what these examples say about the relation between renormalon effects and renormalization. On one hand, these are theories where renormalization is not at all necessary: all diagrams are finite. So in that sense these renormalons are akin to renormalons in super-renormalizable theories, such as the linear sigma model of \cite{mr-2dren}. On the other hand, close to the Fermi surface these theories do resemble renormalizable field theories, and using effective RG revealed itself to be a consistent good predictor of the large order behavior across our different examples. 

The deeper connection, however, is likelier to be the contrast between the trivial vacuum of perturbation theory and the gapped true quantum vacuum. We found this in the $O(N)$ sigma model at large N in  section \ref{cha_largeN}, it is found in the linear sigma model of \cite{mr-2dren}, and it is the case of all these gapped non-relativistic fermion models. We revisit this discussion in chapter \ref{cha-outlook}.

\part{Conclusion and Appendices}
\label{part-final}

\chapter{Outlook}
\label{cha-outlook}
\setlength{\epigraphwidth}{.65\textwidth}
\epigraph{
\raggedleft
Quand on ne réfléchit pas, on se croit le maître de tout;\\
quand on y réfléchit, on voit qu'on n'est maître de rien.\footnotemark}
{Voltaire, \\
Dictionnaire Philosophique, 2\textsuperscript{ème} éd. (1822)}
\footnotetext{``When we don't reflect, we think ourselves masters of everything. // When we do reflect, we realize we are masters on nothing.''}

In this final short chapter, we collect some overall conclusions and sketch possible future directions.

\section{Summary}

In part \ref{part-iqft}, we reviewed the results of \cite{mmr-antrans, mmr-theta}. We analyzed asymptotically free integrable field theories with a chemical potential $h$.
We showed that one can use the Bethe integral equation for the ground state to extract the trans-series expansion of several observables, namely the free energy. The exact value of the observable can then be retrieved from the trans-series representation through Borel summation, as we numerically tested. The non-perturbative scales of these trans-series can be found from the poles of the function $\sigma(\omega)$, which is constructed from the S-matrix.

From these trans-series, non-perturbative physics can be read. In most models studied, the leading non-pertubative effect is tied to an isolated Borel singularity. It matches the leading IR renormalon predicted by Parisi and 't Hooft. This term can be rewritten as a constant proportional to $m^2$ and corresponds to the $-F(0)$ term in the definition of the free energy. 

In the bosonic models we studied, we find a family of singularities proportional to $N$ (or $N-2$), which we associated with instanton effects. In $O(N>4)$ NLSM, for example, they seem to correspond to unstable instantons concretely. Meanwhile, in the $O(3)$ model and some coset models, there are sectors which seem to come from stable instanton effects.

More importantly, in both the Gross--Neveu model and bosonic models, with the exception of the NLSM, we find a sequence of non-perturbative effects proportional to an $N$-dependent rational number. These trans-series sectors survive the large $N$ limit and become the ``canonical'' renormalons predicted by Parisi and 't Hooft. However, at finite $N$, these ``new renormalons'' are notably distinct from the OPE-motivated lore. UV renormalons can also be studied with these techniques, and they seem to match the standard expectations. We can also match the leading new renormalon with ring diagrams in the large $N$ limit, validating their labeling as a renormalon.

At the end of this part, we reviewed the consequences of introducing a topological $\vartheta$ angle with $\vartheta=\pi$, such that integrability is preserved. This is possible in the $O(3)$ non-linear sigma model and in Fendley's coset models. In the former, the trans-series obtained from the Bethe ansatz changes according the topological charge of instanton sectors. We observe that effects related to the cancellation of ambiguities from perturbation theory are unaffected, which is consistent the fact that perturbation theory itself remains the same. In Fendley's coset models, we observe a similar modification, with the leading real non-perturbative contribution changing sign but not the imaginary ambiguity. This happens for both instanton-like and renormalon-like effects, providing evidence that renormalon contributions are affected by the topological $\vartheta$ angle.

In part \ref{part-qmb}, we put together the results of \cite{mr-long,mr-hubbard,mr-three}. We studied non-perturbative effects in one dimensional non-relativistic integrable models, using a combination of large order behavior analysis based on Volin's method, resummation of ring diagrams, and exact trans-series obtained with Wiener--Hopf method. We showed that, in both the Gaudin--Yang model with $\kappa$ components and the one dimensional Hubbard model, the leading Borel singularity for the ground state energy is specified by the square of the energy gap. In the particular case of two-component systems, this gap corresponds to the BCS gap. The fact that the factorial divergence associated with this singularity comes from ring diagrams shows that the gap manifests as a renormalon effect. Furthermore, in the Gaudin--Yang model, we showed that all non-perturbative correction are proportional to even powers of the gap. We expect that the connection between the leading Borel singularity and the square of the energy gap holds for other gapped fermion models, as conjectured in section \ref{sec_superconductivity_conjecture}.



\section{What makes a renormalon}

We had set out to learn about the physics of renormalons, but one of our major results is rather what we had to unlearn. Namely, the conventional IR renormalon positions \eqref{IR-ren} do not hold in general. If we go back to the motivation of these integer positions, they come from the classical dimension of operators in the OPE.
Since the free energy is not amenable to the OPE, the mismatch is unexpected but not dramatic. Nonetheless, it leads to further questions. 

First, the $N$-dependent rational positions of the new renormalons cannot correspond to classical dimensions at all. These renormalon effects then do not come from condensates, at least not directly. We might assume that \eqref{IR-ren} does hold for the observables which can be expanded with an OPE, which has mostly been tested only at large $N$. However, the contrast between the new renormalon of the free energy and the traditional renormalons of the OPE weakens the proposal of ``renormalon universality''. This universality is the idea that the same condensates would contribute to different observables \cite{beneke}. But the structure of the trans-series for the free energy is radically different to that of the two point function, for example. 

Second, assuming that there is no such ``renormalon universality'', this means that if there is a semi-classical interpretation of renormalons it would have to be very subtle. Saddle-points of the action are observable independent. While it is not unthinkable that some saddle points might only contribute to some observables and not others, there is now the extra question of explaining why the free energy and the two point functions have such distinct trans-series.

On the topic of a hypothetical semi-classical interpretation of renormalons, it is not clear what our results concerning the $\vartheta=\pi$ angle imply. One could imagine the change of sign in the real contributions is a consequence of the topological charge of ``renormalon configurations''. This is what happens with instantons. 
However, that the minus sign does not fully factor out as a phase for the leading renormalon effect, in contrast with instantons, suggests that the renormalon is \textit{not} a saddle point effect. The topological charge of a saddle configuration would affect both real and imaginary parts equally. This puts a difficult test to any candidate semi-classical construction. 
Physically, there seems to be a subtle relation between renormalons and the structure of the quantum vacuum, and the latter changes radically under a $\vartheta=\pi$ angle, with the theory becoming gapless. It remains possible, if not likely, that there is no straightforward semi-classical realization of renormalons.

The appearance of renormalons in the non-relativistic models cemented the picture we developed at the end of section \ref{cha_largeN}. In a model which has a trivial vacuum and a gapped phase realized by the true vacuum, perturbation theory will be captured by Feynman diagrams in the trivial vacuum and the gap will manifest as a renormalon effect. In the Fermi quantum gases, the trivial vacuum is the uninteracting free gas and the gapped phase is the gas of Cooper pairs of bound states. But this is no different from the $\chi=0$ and $\chi\neq 0$ phases we discussed for the large $N$ limit of the $O(N)$ non-linear sigma model. 

It should be stressed that perturbation theory around the trivial vacuum is by no means wrong. When the exact result is available and expanded at weak coupling, it reproduces the same series. The Bethe ansatz solutions, from which we obtained most perturbative series in this thesis, are exact descriptions of the gapped phase. And yet the perturbative coefficients are matched by Feynman diagrams calculated in the trivial vacuum.
While the simple idea remains that a renormalon signals a non-perturbative phase ignored by the perturbative series, it is more of a slogan than a calculation-oriented principle.

\section{Open questions}

Starting from the results presented in this thesis, there are many avenues to be pursued in order to better understand renormalons, instantons and applications of resurgence to asymptotically free QFT. 

While the method of chapter \ref{cha_antrans} is in principle capable of extracting the entire trans-series, in practice it is not. It would be interesting to develop a method which allows the calculation of more terms in the higher sectors of the trans-series. Ideally, if one could combine the algebraic recursion of Volin's method with non-perturbative contributions, the trans-series structure could be studied in more detail. Particularly, the resurgence relations between the higher sectors, about which we speculated from the pole structure and the $\vartheta=\pi$ case, could be better understood.

The results of chapter \ref{cha_antrans} should in principle be matched by first principle calculations which do not rely on integrability. In particular, the $O(N)$ sigma model, with emphasis on the $O(3)$ case, only receives corrections from the IR pole and instanton corrections. The would-be stable instanton sector of the $O(3)$ sigma model should be reproducible from the path integral, but this is a thorny calculation yet to be done. One aspect that such a calculation could illuminate is the role of the chemical potential $h$ from the point of view of the path integral. It seems to do more than just set an external scale. For example, from the definition of $\CF(h)$ it might regulate some divergences, which would be important for instanton calculations. 

It would also be interesting to know whether or how the $h$ field changes the instanton configurations. The effect of the coupling to $h$ on renormalons might also be instructive. We know from the PCF model that the choice of charge at least ``selects'' which non-perturbative terms are manifest, being able to turn off the new renormalon contribution in the FKW choice of $Q$. It would be interesting to understand microscopically why this selection happens.

The ultimate goal of this research would be to develop some form of ``renormalon calculus''. A first principles microscopic prescription that would reproduce the renormalon terms, both new and conventional, and that would be applicable to any QFT. It could come from the path integral, the OPE or some other avenue. At the moment, it remains unclear what such procedure would be.

Lastly, there are still many unexplored questions concerning the non-relativistic models.
The methods of chapter \ref{cha_antrans} have not been applied to the Hubbard model, and it might be possible to carry them out at finite $n$. It would also be interesting to explore the phenomenon of ``gaps as renormalons'' in higher dimensional systems. Perhaps the relationship of renormalons to the path integral or to the OPE is more clear in non-relativistic systems, so some foundational exploration might also pay off. Finally, there are other non-relativistic integrable systems where our techniques might be applicable.

The Lieb--Liniger model, whose study we leave to appendix \ref{Lieb--Liniger}, is perhaps the simplest of all models studied in this thesis. Curiously, the non-perturbative scale remains somewhat unclear. At the moment, it has only been identified as a renormalon through analogy with the relativistic linear sigma model.

The path towards non-perturbative QFT will have to cross the foggy question of renormalons. We have made advances towards charting the physics of renormalons in the reduced world of integrable theories. However, this is only the beginning of a map. The understanding of QFT might have advanced immensely since the days of throwing infinities under the rug, but there is still progress to be made.

\appendix

\chapter{Details on Volin's method}
\label{app_volin}

In this appendix, we summarize some minor but useful technical aspects of the generalized Volin's method detailed in chapter \ref{cha_volin}. The aim is to provide an explanation for some of the strategies used in implementing the method, as well as provide some general guidelines for the reader interested in implementing it and generalizing it.

\section{Volin's method in the Gaudin--Yang model}
\label{app-volin-gy}

We work out the ``matching procedure'' for a concrete example in detail. 
For Gaudin--Yang the integral equation is \eqref{volin_eq_TBA_GY}. 
Using \eqref{volin_Gplus_GY}, the appropriate edge limit ansatz is found by setting $m=2$ and $p=0$ in \eqref{volin_Rhatansatz},
\begin{equation}
 \label{eq_edgeGY}
  \hat R(s)=\frac{\re^{ -\frac{s }{  \pi} \left(\log \frac{ s }{ \pi}-1\right)  }}{ {\sqrt{\pi}}}  \Gamma\left( \frac{s}{  \pi}+\frac{1}{ 2} \right)\times\left(\frac{1}{s}+\frac 1{Bs} \sum_{n,m=0}^\infty \frac{ Q_{n,m}(\log B) }{ B^{m+n} s^{n}}\right).
\end{equation}
We can expand this for small $s$ as
\begin{multline}
\hat R(s) = 
\bigg[
\frac{1}{s}
+\frac{1}{\pi }\left((1-\gamma_E) -\log \frac{4s}{\pi}\right)
+\\
\frac{s}{4 \pi ^2} \left(2 (1-\gamma_E)^2+\pi ^2-4(1-\gamma_E)  \log \frac{4s}{\pi}+2 \log ^2\frac{4s}{\pi}\right)+ \CO\left(s^2\right)
\bigg]
\\
+\frac{1}{B}\bigg[
\frac{Q_{0,0}}{s}
+\frac{Q_{0,0}}{\pi }\left((1-\gamma_E) -\log \frac{4s}{\pi}\right)
+ \CO\left(s\right)
\bigg]
+\CO\left(\frac{1}{B^2}\right).
\label{volin_edge_expansion}
\end{multline}
Meanwhile, we can take the bulk ansatz we found for this problem, \eqref{volin_eq_bulkGY}, with $x=B+z/2$ and expand it for $B\rightarrow\infty$ and $z/B\rightarrow 0$. We find
\begin{equation}
\ba 
R(B&+z/2) = 
-\log \left(\frac{z}{4 B}\right)+\frac{z}{4 B}+\CO \left(\frac{z^2}{B^2}\right)\\
+\frac{1}{B}&\bigg[
\begin{multlined}[t]
\frac{B}{z} \left(c_{1,0,1} \log \left(\frac{z}{4 B}\right)+c_{1,0,0}\right)\\
+\frac{1}{4} c_{1,0,0}-\frac{1}{4} c_{1,0,1}
-\frac{1}{4} c_{1,0,1} \log \left(\frac{z}{4 B}\right)
+\CO \left(\frac{z}{B}\right)
\bigg]\end{multlined}\\
+\frac{1}{B^2}&\bigg[
\begin{multlined}[t]
\frac{B^2 \left(c_{2,0,2} \log ^2\left(\frac{z}{4 B}\right)+c_{2,0,1} \log \left(\frac{z}{4 B}\right)+c_{2,0,0}\right)}{z^2}
+\frac{B}{z} \bigg(c_{1,1,2} \log ^2\left(\frac{z}{4 B}\right)
\\+\left(c_{1,1,1}-\frac{1}{2} c_{2,0,1}-\frac{1}{2} c_{2,0,2}\bigg) \log \left(\frac{z}{4 B}\right)+c_{1,1,0}-\frac{1}{4} c_{2,0,1}\right)
+\CO \left(\frac{z^0}{B^0}\right)
\bigg]\\
+\cdots.\end{multlined}
\ea
\label{volin_bulkexpand}
\end{equation}
We take the inverse Laplace transform term by term of \eqref{volin_bulkexpand}, see the next section for details.
Organizing the terms in powers of $1/B$ and $s^n\log(s)^m$, we find at order $1/B^0$, 
\begin{equation}
\begin{multlined}
\frac{1}{s}+c_{1,0,0}
-
c_{1,0,1} (\log (4  B) +\gamma_E)-c_{1,0,1} \log (s)
\\
+s \bigg(
c_{2,0,2} \left(-2 ( 1 -\gamma_E )  \log (4 B)+\log ^2(4  B)+(1-\gamma_E)^2-\frac{\pi ^2}{6}+1\right)\\
+c_{2,0,1} ( 1 -\gamma_E  -\log (4  B))+c_{2,0,0}\\
+\log (s) \left(c_{2,0,2} (2 \log (4  B)-2 ( 1 -\gamma_E) )-c_{2,0,1}\right)
+c_{2,0,2} \log ^2(s)
\bigg)
+\CO(s^2).
\end{multlined}
\label{volin_matchzs_11}
\end{equation}
The leading term at order $m=3$ from \eqref{volin_eq_bulkGY} would appear in  \eqref{volin_bulkexpand} as $1/B^3(B/z)^3$, which would transform into a term $s^2/B^0$, i.e. at next order in \eqref{volin_matchzs_11}. Thus, we must take more terms simultaneously in the $1/B$ and $z/B$ expansions in the bulk limit, and in $s$ and $1/B$ in the edge limit, as we go to higher order. Comparing \eqref{volin_matchzs_11} with \eqref{volin_edge_expansion} we can solve for the $c_{n,m,k}$.  

We already set $c_{0,0,1}=1$ but if we had not done it, it would be fixed now due to $1/s$ term on both sides.
The constant term mixes $c_{1,0,0}$ and $c_{1,0,1}$, so we start with the $\log(s)$ term instead fixing
\begin{equation}
c_{1,0,1} = \frac{1}{\pi},
\end{equation}
and then we can fix
\begin{equation}
c_{1,0,0} = \frac{1+\log(\pi B)}{\pi}\,.
\end{equation}
Then at $\CO(s)$ the easiest is to start with $s \log^2(s)$
\begin{equation}
c_{2,0,2} = \frac{1}{2\pi^2},
\end{equation}
and then $s\log(s)$ and $s$ in this order,
\begin{equation}
\ba
c_{2,0,1} &= \frac{\log (\pi  B)}{\pi ^2},\\
c_{2,0,0} &= \frac{3 \log ^2(\pi  B)+2 \pi ^2-3}{6 \pi ^2}.
\ea
\end{equation}
Here we see already some important patterns. A conceptual one is that the $c_{n,m,k}$ can and must depend on $\log B$. A computational one is that at each order in $s$ we fix a family of coefficients by starting at the one with the highest $k$ (corresponding to the highest power of $\log$) and work down, recursively using our results.

At order $1/B$, the transform of the bulk gives
\begin{multline}
\frac{c_{1,0,1}}{4 s}+c_{2,0,1} \left(\frac{1}{2} \log (\pi  B)-\frac{\gamma }{2}+\frac{1}{4}\right)
+c_{2,0,2} \left(\frac{1}{2} \log (\pi  B)-\frac{\gamma }{2}+\frac{1}{2}\right)
+c_{1,1,0}\\
+c_{1,1,1} (-\log (\pi  B)+\gamma -1)\\
+c_{1,1,2} \left(-2 \gamma  \log (\pi  B)+\log ^2(\pi  B)+2 \log (\pi  B)+\gamma ^2-2 \gamma -\frac{\pi ^2}{6}+1\right)\\
+ \log(s)\left(c_{1,1,2} (2 \log (\pi  B)-2 \gamma +2)-c_{1,1,1}+\frac{1}{2} c_{2,0,1}+\frac{1}{2} c_{2,0,2}\right)\\
+c_{1,1,2} \log ^2(s)
+ \CO(s),
\end{multline}
where we write for compactness,
\begin{equation}
\gamma = 1 - \gamma_E + \log\frac{\pi}{4}.
\end{equation}
Comparing the $1/s$ term with \eqref{volin_edge_expansion}, we find 
\begin{equation}
Q_{0,0} = \frac{1}{4\pi},
\end{equation}
and then we fix the bulk as before working up in orders of $s$ and each order of $s$ working down in powers of $\log(s)$. We must use the coefficients $c_{1,0,k}$ and $c_{2,0,k}$ to find the coefficients $c_{1,1,k}$. We solve
\begin{equation}
c_{1,1,2} = 0,\quad 
c_{1,1,1} = \frac{\log (\pi  B)+1}{2 \pi ^2},\quad
c_{1,1,0} = \frac{1+ 2\log  (\pi  B)}{4 \pi ^2}.
\end{equation}
The rest of the procedure is qualitatively the same with an increasing number of coefficients at each order. One alternates between using the leading ansatz of the edge limit \eqref{volin_Rhatansatz} to fix the $c_{n,m,k}$ and then using these coefficients to fix the $Q_{n,m}$. In the end, it can be verified that all unknown coefficients are fixed and that any additional coefficients would be set to zero. 

%
\section{Laplace transforms for series with logarithms}
\label{app_laplace}

When carrying out Volin's method it is necessary to take the Laplace transform of products of $s$ and $\log(s)$. These transformations require some care to define, for example through analytic continuation, and are often omitted in standard  references. 
In this section, we derive some useful formulae.

We begin by observing the following elementary identities
\begin{align}
\log(Y)^k&= \left[\partial^{(k)}_x Y^x\right|_{x=0} 
\label{lap_x_trick1}\,,\\
\log(Y)^k&= \left[\frac{\partial^{(k+1)}_x\left(x Y^x\right)}{k+1}\right|_{x=0}
\label{lap_x_trick2}\,,\\
\mathcal{L}(s^{a})&= \int_0^\infty \rd s\, \re^{- z s} s^a = \frac{\Gamma(a+1)}{z^{a+1}}\label{lap_lap_s}\,,
\end{align}
where $\mathcal{L}(z)$ is the  Laplace transform. We can use \eqref{lap_x_trick1} followed by \eqref{lap_lap_s} to compute the following useful Laplace transform:
\begin{equation}
\ba
\CL(s^n \log(s)^k)&=\int_0^\infty \rd s\, \re^{- z s} s^n \log(s)^k 
=\int_0^\infty \rd s\, \re^{- z s} \partial_x^k(s^{x+n})\bigg|_{x=0}\\
&= \partial_x^k\left(\frac{\Gamma(x+n+1)}{z^{x+n+1}}\right)\bigg|_{x=0}\\
&= \sum_{t=0}^k\binom{k}{t}\frac{\partial_x^{t}\left(z^{-x}\right)^{t}}{z^{n+1}}\partial_x^{k-t}\Gamma(x+n+1)\bigg|_{x=0}\\
&= \frac{1}{z^{n+1}}\sum_{t=0}^k\binom{k}{t}\left(\log\left(\frac{1}{z}\right)\right)^{t} \left[\partial_x^{k-t}\Gamma(x+n+1)\right|_{x=0}.
\label{lap_log_pos}
\ea
\end{equation}
This expression holds for non-negative integer $n$ and, by analytic continuation, to any $n$ which is not a negative integer. For negative integer $n$, which appear in fermionic models, the derivatives of the $\Gamma$-function in \eqref{lap_log_pos} are divergent. Let us use \eqref{lap_x_trick2} instead,
\begin{equation}
\ba
\mathcal{L}(s^{-n}\log(s)^k)&=
\int_0^\infty \rd s\, \re^{- z s}\frac{\log(s)^k}{s^n}=\int_0^\infty \rd s\, \re^{- z s}\frac{1}{k+1}\partial_x^{k+1}(x s^{x-n})\bigg|_{x=0}\\
&=\frac{1}{k+1}\partial_x^{k+1}\left(\frac{x \Gamma(x-n+1)}{z^{x-n+1}}\right)\bigg|_{x=0}\\
&=\frac{z^{n-1}}{k+1}\sum_{t=0}^{k+1}\binom{k+1}{t}\left(\log\left(\frac{1}{z}\right)\right)^{t} \left[\partial_x^{k+1-t}(x \Gamma(x-n+1))\right|_{x=0}\,,
\label{lap_log_neg}
\ea
\end{equation}
where the factor of $x$ in the derivative cancels the pole of the $\Gamma$-function and one obtains a well defined value. 

We note that \eqref{lap_log_neg} seems to implicitly need some regularization, since the original integral in the Laplace transform considered wouldn't converge. 
We can also extend \eqref{lap_log_neg} to $k=0$ to define a consistent choice of $\mathcal{L}(s^{-n})$:
\begin{equation}
\mathcal{L}(s^{-n})=
 \frac{(-1)^n }{(n-1)!}z^{n-1} \left(\log (z)- \psi ^{(0)}(n)\right)
\,.
\label{lap_neg}
\end{equation}
However, for the purposes of Volin's method, the polygamma term does not contribute.
We use the same approach to calculate the inverse Laplace transform, neglecting  $\delta$-function terms, 
\begin{equation}
\CL^{-1}\left(z^n \log^m(z)\right)=\frac{1}{s^{n+1}}\sum _{k=0}^m \binom{m}{k} (-\log (s))^k \left[\partial^{m-k}_x\frac{1}{\Gamma (-n-x)}\right|_{x=0}\,,\quad s\neq 0\,.
\label{invlap}
\end{equation}
In this case, this procedure is well defined and well known for negative integer  and non-integer $n$. 

Due to the subtleties with regularization,
\eqref{invlap} and \eqref{lap_log_neg} are not actually inverses of each other for negative integer $n$. Using one then the other will only differ by derivatives of the $\delta$-function in $s$-space (i.e. $\delta(s)$, $\delta'(s)$, etc.), and conversely by positive integer powers of $z$ in $z$-space. With more care, these terms can be accounted for in \eqref{invlap} but they do not have any impact in the matching procedure so we will neglect them. Thus, since it is computationally advantageous to work in $z$-space, we can simply solve the equations ignoring terms with positive powers of $z$ and no $\log(z)$. For bosonic models, $n$ is always half-integer and \eqref{invlap} is the inverse of \eqref{lap_log_pos}.

\section{Implementation details}

The expansion of the bulk ansatze \eqref{volin_eq_bulkGY}, 
\eqref{pre_eq_bulkGN}, \eqref{volin_bulk} and \eqref{eq_bulkLL}
when $z/B\ll 1$ can be obtained with a combination of the Taylor expansions of $\log(1+x)$ and $(1+x)^n$. We can organize the key elements with following definitions. Let $k$ be a positive integer and $n\in\frac{1}{2}\mathbb{Z}$,
\begin{equation}
(1+x)^n(\log(1+x))^k =\sum_{a=0}^\infty x^a y_{a,n,k}
=\sum_{a=0}^\infty x^a
\left(
\sum_{b=0}^{a}\binom{n}{b} \hat{B}_{a-b,k}\left(\frac{(-1)^{i+1}}{i}\right)
\right)
,
\end{equation}
where $\hat{B}_{j,k}(x_i)=\hat{B}_{j,k}(x_1,x_2,\cdots,x_{j-k+1})$ are the ordinary Bell polynomials.
The bulk ansatz for bosonic relativistic systems \eqref{volin_bulk} can then be expanded as
\begin{align}
R(z)&=A\sum_{m=0}^\infty\sum_{n=-m}^\infty\sum_{t=0}^{n+m}\frac{\left(\log\frac{4B}{z}\right)^t}{B^m z^{n+1/2}}\sum_{k=t}^{n+m}\sum_{r=\max[0,-n]}^m c_{n+r,m-r,k} F_{n,t,k,r}\\
F_{n,t,k,r}&=\binom{k}{t}2^{1-2 r}(-1)^{k}(y_{r,-n-r-1/2,k-t}+2\varepsilon(k)y_{r-1,-n-r-1/2,k-t})\\
\varepsilon(k)&=k\bmod 2.
\end{align}
While for fermionic-type models, \eqref{pre_eq_bulkGN} expands as
\begin{align}
R(z)&=A\sum_{m=0}^\infty\sum_{n=-m+1}^\infty\sum_{t=0}^{n+m-1}\frac{\left(\log\frac{4B}{z}\right)^t}{B^m z^n}\sum_{k=t}^{n+m-1}\\
&\qquad\times\sum_{r=\max[0,1-n]}^m c_{n+r,m-r,k} F'_{n,t,k,r}\\
F'_{n,t,k,r}&=\binom{k}{t}4^{- r}(-1)^{k}(y_{r,-n-r,k-t}+2\varepsilon(k)y_{r-1,-n-r,k-t}),\\
 \varepsilon(k)&=(k-1)\bmod 2.
\end{align}
With minimal modifications one finds similar expansions for \eqref{volin_eq_bulkGY} and \eqref{eq_bulkLL}.

We could take the term-by-term inverse Laplace transforms of the bulk expansions using \eqref{invlap}. However, this would require taking derivatives of $\Gamma$-function. Since in the edge limit we need to take such derivatives anyway, it is more efficient to not transform the bulk and use \eqref{lap_log_pos} and \eqref{lap_log_pos} to Laplace transform the edge expansion.

In relativistic bosonic models, we can write the edge limit ansatz \eqref{volin_Rhatansatz} as
\begin{equation}
\ba
\hat{R}(s) &=A \left(R_0(s) + R_Q(s)\right)\\
R_0(s) &= \re^{- 2a s \log(4 B s)}\frac{\phi(s)}{\sqrt{s}}\left(\frac{1}{s+\frac{1}{2}}\right)\\
R_Q(s) &= \re^{- 2a s \log(4 B s)}\frac{\phi(s)}{\sqrt{s}}\left(\frac{1}{Bs}\sum_{M=0}^\infty \sum_{n=0}^M \frac{Q_{n,M-n}}{B^M s^n}\right)
\ea
\label{edgeB}
\end{equation}
We define
\begin{equation}
\ba
\phi_{j,k}&= \lim_{x\rightarrow 0} \partial_x^{j} \left(\Gamma(k-2a x)\phi(x)\right)\\
\varphi_{j,k}&=\lim_{x\rightarrow 0} \partial_x^{j} \left(\Gamma(k-2a x)\frac{\phi(x)}{x+\frac{1}{2}}\right)&= 2 \phi_{j,k} - 2 j \varphi_{j-1,k}.
\ea
\label{phiij}
\end{equation}
We then expand \eqref{edgeB} in small $s$ and use \eqref{lap_log_pos} to take the Laplace transform term-by-term. After some reorganization of the derivatives, the Laplace transform of $\hat R(s)$ is given by
\begin{equation}
\ba
\CL[R_0](z) &=\sum_{n=0}^\infty\sum_{t=0}^n\frac{\left(\log\frac{4B}{z}\right)^t}{z^{n+1/2}} \frac{(-2a)^t}{t!(n-t)!}\varphi_{n-t,n+1/2}\\
\CL[R_Q](z) &=
\sum_{m=0}^\infty\sum_{n=-m}^\infty\sum_{t=0}^{n+m}\frac{\left(\log\frac{4B}{z}\right)^t}{B^m z^{n+1/2}}\\
&\qquad\times\sum_{c=\max[0,t-n-1]}^{m-1}\frac{(-2a)^t\phi_{n-t+c+1,n+1/2}}{t!(n-t+c+1)!}Q_{c,m-c-1}.
\ea
\end{equation}
For fermionic type models, the expansion is analogous but one must separate the cases which use \eqref{lap_log_pos} from \eqref{lap_log_neg}. For non-relativistic models, one only needs to change the definition of the $\varphi_{j,k}$ to have $\phi(x)/x$ instead.

A significant part of optimizing Volin's method lies with computing \eqref{phiij} in an efficient way. Since $\phi(s)$ is usually a ratio of $\Gamma$-functions, it amounts to computing the Taylor expansion of such ratios. If all the $\Gamma$-functions are evaluated at integers or half integers when $s=0$, the best approach, suggested by \cite{abbh1,abbh2,bbv}, is to exploit the recursion relation
\begin{equation}
\ba
\frac{1}{\Gamma(x)}&=\sum_{n\geq 1} g_k x^k,\\
g_{n\leq 0} &= 0,\quad
g_{1} = 1,\quad
g_{2} = \gamma_E,\\
g_{n> 2} &= \frac{1}{1-n}\left(\sum _{k=2}^n (-1)^k \zeta (k) g _{n-k}-\gamma_E g_{n-1}\right).\\
\ea
\label{Gamma-recursion}
\end{equation}
This recursion relation needs to be combined with the reflection formula and the standard expressions for $\Gamma(n+x)$ and $\Gamma(n+1/2+x)$ in terms of $\Gamma(x)$, see e.g. \cite{nist-gamma}.

When the $\Gamma$-functions are evaluated at rationals or unspecified variables, as it happens with the PCF model using FKW charges, it is better to calculate the straightforward $n$-th derivative of the logarithm of the ratio and then solve recursively for the $n$-th derivative of the ratio itself using the lower $m$-th derivatives, with $m<n$. This idea is detailed in appendix E of \cite{volin-thesis}.

\section{Spurious factors}
%

As we commented in the previous sections, many terms can be ignored in Volin's method without impacting the final result. Here we derive why two important classes of such terms can be safely dropped.

Let
\begin{equation}
\hat{R}(s)= R \re^{\beta s- 2 a s \log s} \phi(s) \left(\frac{1}{s+\frac{1}{2}}+Q(s,B)\right),
\label{usualR}
\end{equation}
where $R$, $\beta$ and $a$ are constants and $\phi(s)$ is usually a product of $\Gamma$-functions with a factor of $1/\sqrt{s}$ depending on whether the system is of fermionic or bosonic type.
The expansion $e/\rho^2 [\alpha]$ does not depend on $a$ and hence one might  solve the problem using
\begin{equation}
{\overline{R}}(s)= R \re^{- 2a s \log s } \phi(s) \left(\frac{1}{s+\frac{1}{2}}+Q(s,\overline B)\right),
\label{barR}
\end{equation}
instead.
This obtains the same coefficients in $\alpha$ with much less computational effort. 

Using the properties of the Laplace transform, it can be shown that using \eqref{barR} instead of \eqref{usualR}, simply amounts to doing the matching not in the limit $\theta = B + z/2$ with $z/B\ll 1$, but instead in the limit $\theta= B + (z+\beta)/2$  with $z/B\ll 1$. To make sense of the results, we consider that the $\overline B$ used in the anstaze is related to $B$ in the integral equation by $\overline B = B+\beta/2$. It can be shown that the calculation of $\rho$ and $\epsilon$ is indifferent to this shift and once we express everything in terms of $\alpha$, this procedural difference disappears.

This argument can be extended to the case where $\beta$ is a function of $B$. We usually organize the exponent as in \cite{volin-thesis},
\begin{equation}
\beta s - 2 a s \log s = - 2 a s \left(\log(4 B s) - \log\frac{B}{B_0}\right),\quad \log B_0= -\frac{\beta }{2a}-\log 4.
\end{equation}
It was observed in \cite{volin,volin-thesis} that the edge ansatz
\begin{equation}
{\overline{R}}(s)= R \re^{- 2a s \log(4 \overline{B} s)} \phi(s) \left(\frac{1}{s+\frac{1}{2}}+Q(s,\overline B)\right),
\label{barR-nolog}
\end{equation}
produces the same final expansion $e/\rho^2[\alpha]$ with the advantage that $\log(B)$ terms do not appear in intermediate calculations. 

This simplification is phrased in \cite{volin,volin-thesis} as saying that the final result is independent of $B_0$ and then it can be chosen as $B_0=B$. If we consider $\beta' = 2 a \log(B/B_0)$, we can justify it with a shift of the matching limit. Now the integral equation $B$ is related to $\overline{B}$ in the bulk and edge ansatze by $B=\overline{B}+a\log(\overline{B}/B_0)$. This can be explicitly tested by solving the problem with the full $\hat{R}(s)$ and with the reduced ${\overline{R}}(s)$ and matching them by transforming between $B$ and $\overline{B}$. While this shows why the pattern happens with Volin's method,
the developments of \cite{mmr-antrans,bbhv} prove this phenomenon to all orders in the trans-series, with the notable exception of the $O(3)$ model. This is achieved by the introduction of the intermediate coupling $v$, which we did for the Gross--Neveu model in \eqref{v-def} and was done for bosonic models by \cite{bbhv}, removing the $\log B$ terms from the onset.

It should be noted that some care must be taken with the definition of $\alpha$. In the case of bosonic models, for example, the full definition of $\alpha$ with $B_0$ factored out is
\begin{multline}
\frac{1}{\alpha }+\left(a-\frac{1}{2}\right) \log \alpha =
B - \left(a-\frac{1}{2}\right) \log B +\log \left(\tilde{\rho }\right)\\
+ a \log \left(\frac{B}{B_0}\right)
+\log \left(\frac{\mathfrak{C} G_+(i) k}{2^{1+2a} \re^\beta \sqrt{2 \pi }}\right),
\end{multline}
where $\tilde{\rho}$ is defined in \eqref{rhot}.

Finally, as noted in \cite{volin,volin-thesis}, one can also observe that terms with $\gamma_E$, $\zeta(2n)$ or $\log 2$ disappear from the final series (the last of which is contingent on a good choice of $\mathfrak{C}$).  While proving this might require some number theoretical analysis of Feynman diagrams, it has been checked to very high order and indirectly by numeric tests, including those of chapter \ref{cha_volin}. When calculating the coefficients exactly, it is then a significant speed up to set such terms to zero directly in the expansion of the $\Gamma$-function, e.g. in \eqref{Gamma-recursion}.

\chapter{Asymptotic series for the Lieb--Liniger model}
\label{Lieb--Liniger}

In section \ref{sec_iqmb}, we introduced the Lieb--Liniger model, a model of bosons with a repulsive $\delta$-interaction. It is one of the oldest cases of integrable quantum many body systems. While its physics is distinct from the Gaudin--Yang model, its described by a similar integral equation. In this appendix, we will explore the application of the generalized Volin's method to the Lieb--Liniger model. This provides additional details to the identical calculation in \cite{mr-ll}.

\section{Volin's method for the Lieb--Liniger model}
The ground state energy is characterized by the integral equation \eqref{ll-inteq},
from which we extract the dimensionless coupling and the energy in a familiar way
\begin{equation}
\frac{1}{\gamma} = \frac{1}{4\pi}\int_{-B}^B f(x) \rd x, \quad e(\gamma) = \frac{E}{n^3} = \frac{\frac{1}{4\pi}\int_{-B}^B x^2 f(x) \rd x}{\left(\frac{1}{4\pi}\int_{-B}^B f(x) \rd x\right)^3}.
\label{ll-obs}
\end{equation}

The application of Volin's method to \eqref{ll-inteq} is more subtle than the examples discussed in the main text. The equation for the resolvent is
\begin{equation}
(1-D) R^+ - (1-D) R^- = - 4\pi \ri.
\label{LL_bulkeq}
\end{equation}
This case is similar to \eqref{volin_eq_gen_bosonic}, but there is a constant on the r.h.s. .
In the bulk limit, we introduce $u = x/B$ and expand
\begin{equation}
\ba
R^+ (x) &\approx \sum_{m\geq 0} \frac{1}{B^{m-1}} R^+_m(u)\\
D &= 1 + \frac{1}{B}\ri \partial_u + \cdots\,,
\ea
\end{equation}
which at leading order results
\begin{equation}
\partial R^+_0(u) + \partial R^-_0(u) = 4\pi.
\end{equation}
With some luck and complex analysis, one can find a solution to this equation that satisfies the analytical properties of the resolvent. A simpler way is to take the known leading order solution of \eqref{ll-inteq}, which can be found in \cite{ll,popov},
\begin{equation}
f(x) \sim 2 \sqrt{B^2-x^2}.
\end{equation}
From here, we can calculate explicitly the resolvent using its definition \eqref{volin_res_def} and we find
\begin{equation}
R(B u) = -2\pi B \left(\sqrt{1-u^2}-u\right) + R_1(u)+\frac{1}{B} R_2(u) +\cdots\,.
\end{equation}
It is easy to see that $\{R(B u)+2\pi B \left(\sqrt{1-u^2}-u\right)\}$ satisfies an equation of the form \eqref{volin_eq_gen_bosonic}, up to an overall power of $B$. Thus, we modify \eqref{volin_bulk} to find the appropriate bulk ansatz
\begin{multline}
R(x)=2\pi B\left( x/B - \sqrt{(x/B)^2-1}\right)\\
+\sum_{m=0}^\infty\frac{1}{B^{m}}\sum_{n=0}^m\sum_{k=0}^{m+1} c_{n,m-n,k}\frac{(x/B)^{k\bmod2}}{(x^2/B^2-1)^{n+1/2}}\log\left(\frac{x-B}{x+B}\right)^k\,.
\label{eq_bulkLL}
\end{multline}

The Fourier transform of the kernel of \eqref{ll-inteq} is
\begin{equation}
\tilde K(\omega)= \re^{-|\omega|},
\end{equation}
which leads to Wiener--Hopf function
\begin{equation}
G_+(\omega) = \frac{ \re^{\frac{\ri \omega}{2 \pi } \log \left(-\frac{\ri \omega}{2\pi\re}\right)}}{\sqrt{-\ri \omega}}\Gamma \left(1-\frac{\ri \omega}{2 \pi }\right).
\end{equation}
This is, up to overall constants, the same kernel as the $O(3)$ sigma model.
Naively, one could try to plug this into \eqref{volin_Rhatansatz} with $m=2$ and $p=0$, but it would diverge due to $G_+(0)=\infty$. Nevertheless, one can keep the form of the edge limit ansatz with an overall unfixed constant, since it will be fixed by the known leading term of \eqref{eq_bulkLL} when we carry out the matching procedure. We then have
\begin{equation}
\hat R(s) = A(B) G_+(2 \ri s)\left(\frac{1}{s}+\frac{1}{B s}\sum_{m=0}^\infty \sum_{n=0}^{m+1} \frac{Q_{n,m+1-n}}{B^{m}s^n}\right).
\label{Rhat_LL}
\end{equation}

An alternative way of motivating this ansatz is to approach the edge limit as is done in \cite{volin,mr-ll}. We take the inverse Laplace transform of \eqref{LL_bulkeq} and we find an equation for the discontinuity of $\hat{R}(s)$ along the negative real axis,
\begin{equation}
\sin(s) \left(\re^{-\ri s} \hat R(s-\ri \epsilon) -\re^{\ri s} \hat R(s+\ri \epsilon) \right) = \frac{1}{s-\ri \epsilon}-\frac{1}{s+\ri \epsilon}\,, \quad s<0.
\end{equation}
By construction, $\hat R(s)$ must be analytic everywhere else. For $s<0$, the right-hand side forces the discontinuity of $\hat R(s)$ to have poles at the same positions $\sin(s)$ has zeroes. Furthermore, the expansion of the solution around the edge forces the limit $s\rightarrow\infty$ to be a sum of powers $1/s^n$ like in \eqref{Rhat_infinity}. This constrains the form of the ansatz to \eqref{Rhat_LL}.

We can quickly fix $A(B)$ by looking into the very first term in the matching procedure. From the bulk ansatz, we have
\begin{equation}
R(B+z/2) = - 2\pi \sqrt{Bz}+\cdots
\end{equation}
which transforms into
\begin{equation}
\hat R(s) = \frac{\sqrt{B\pi}}{s^{3/2}}+\cdots\,.
\end{equation}
Meanwhile expanding \eqref{Rhat_LL} returns
\begin{equation}
\hat R (s) = \frac{A(B)}{\sqrt{2} s^{3/2}}+\cdots,,
\end{equation}
and thus
\begin{equation}
A(B) = \sqrt{2\pi B}.
\end{equation}
Once this is fixed we can carry out Volin's method as usual, and obtain the desired observables from \eqref{volin_mom_from_res},
\begin{equation}
\ba
\frac{1}{4\pi}\int_{-B}^B f(x)\rd x &=\frac{B^2}{4} + \frac{1}{4\pi}\sum _{m=0}^M \frac{c_{0,m,0}-2 c_{0,m,1}}{B^{m-1}}+\CO\left(\frac{1}{B^M}\right),\\
\frac{1}{4\pi}\int_{-B}^B x^2 f(x)\rd x &=
\frac{B^4}{16}
+\frac{1}{4 \pi }\sum _{m=0}^M \frac{\frac{1}{2} c_{0,m,0}-\frac{5}{3} c_{0,m,1}+4 c_{0,m,2}-8 c_{0,m,3}}{B^{m-3}} \\
&\qquad
 +\frac{1}{4 \pi }\sum _{m=0}^{M-1} \frac{c_{1,m,0}-2 c_{1,m,1}}{B^{m-2}} 
+\CO\left(\frac{1}{B^{M-2}}\right).
\ea
\end{equation}

The resulting series for the energy is
\begin{equation}
\ba
&e (\gamma)=\gamma -\frac{4 \gamma ^{3/2}}{3 \pi }+\left(\frac{1}{6}-\frac{1}{\pi ^2}\right) \gamma ^2
+\frac{3 \zeta (3)-4}{8 \pi ^3}\gamma
   ^{5/2} +\frac{3 \zeta (3)-4}{24 \pi ^4}\gamma ^3
   \\
   &+\frac{60
   \zeta (3)-45 \zeta (5)-32}{1024 \pi ^5}\gamma ^{7/2}
   +\frac{3 \left(4 \zeta (3)+6 \zeta (3)^2-15 \zeta
   (5)\right)}{2048 \pi ^6}\gamma ^4 \\
   &+\frac{-4368 \zeta (3)+6048 \zeta (3)^2+2520 \zeta (5)-8505 \zeta
   (7)+1024 }{786432 \pi ^7}\gamma ^{9/2} +\CO(\gamma^{5}).
   \ea
   \label{ll_eg}
\end{equation}
We have calculated 50 coefficients of this series. Defining
\be 
e(\gamma) = \gamma \sum_{m\geq 0} c^{\textrm{LL}}_m \gamma^{m/2},
\ee
the large order behavior of the coefficients is given by
\begin{equation}
c^{\textrm{LL}}_m \sim (8\pi)^{-m}\Gamma(m),
\end{equation}
which is not Borel summable and implies a leading non-perturbative ambiguity of the form
\begin{equation}
\disc s
(e)(\gamma)\propto\re^{-\frac{8\pi}{\sqrt{\gamma}}}.
\label{disc-ll}
\end{equation}
Unfortunately, it is not evident how to extend the analysis of section \eqref{sec-gy-antrans} to the Lieb--Liniger case in order to analytically find the Stokes constants associated with this ambiguity. This is due to technical complications of having $G_+(0) =\infty$ in a quantum many body system (unlike the case of integrable field theories where a divergent $G_+$ was workable, as we saw in section \ref{sec_bosonic}). Notable progress in this direction has recently been done by \cite{bbv}, and while their techniques are not applied to the ground state energy it is likely they could be extended.

Unlike the Gaudin--Yang model, there is no superconductor interpretation available for this non-perturbative scale. Since this is a model of bosons, one could postulate that this non-perturbative scale comes from an instanton effect. However, as is argued in \cite{mr-2dren}, there are no known solutions to the equations of motion with finite positive real action for the field-theoretical description of the model.

Due to the difficulties with identifying an instanton solution, it is natural to conjecture whether this effect is also due to a renormalon effect. The relativistic analogue of the Lieb--Liniger gas is given by the linear $O(N)$ sigma-model which was studied in \cite{mr-2dren}. There it was found that, at large $N$, the perturbative series of the free energy is dictated to by an IR renormalon associated with ring diagrams. This perturbative series must be calculated by expanding around the classical, symmetry broken, vacuum, but quantum mechanically $O(N)$ symmetry cannot be broken, as predicted by the Coleman-Mermin-Wagner theorem. It seems that the renormalon effect is the ``price'' paid by expanding around a ``false vacuum''. This case was discussed in section \ref{cha_largeN}. Similarly, the leading large order behavior in the non-linear sigma model is due to an IR renormalon and it is continuous with the leading ambiguity in the $O(3)$ model. However, in the $O(3)$ model the IR renormalon and the leading instanton effect appear at the same position in the Borel plane.

A similar effect to the linear sigma model might happen for the Lieb--Liniger model. However, in contrast with the sigma models and Gaudin--Yang cases, we cannot use the large $N$ limit to single out families of diagrams. In spite of the existence of an $SU(N)$ extension of the model \cite{andersen-largeN,pastu-N,nogueira}, the dominant and sub-dominant diagrams in the large $N$ limit do not contribute to the ground state energy. In fact, it was argued in \cite{Yang2011} that the large $N$ limit of the ground state energy of any 1D boson gas with repulsive $\delta$-interaction reduces to that of the regular, $N=1$, Lieb--Liniger model.

\section{Capacitance of a circular plate}

As we discussed in chapter \ref{cha-integrability}, the integral equation for the Lieb--Liniger model overlaps with Love's equation for the circular plate capacitor. In the context of that problem, ``weak coupling'' means small plate separation.
As originally observed in \cite{mr-ll}, the application of Volin's method to the Lieb--Liniger model and thus Love's equation provides an algorithm to calculate this expansion to arbitrary order.
In appropriate units, it was found in \cite{loveeq} that the capacitance is given by
\begin{equation}
C = \frac{1}{4\pi}\int_{-1/K}^{1/K} f(x)\rd x,
\end{equation}
where $K$ is the separation between the circular plates and $f$ is the solution of \eqref{ll-inteq}. Using \eqref{eq_bulkLL} and \eqref{volin_mom_from_res} with $B=1/K$, we have that
\begin{equation}
C =  \frac{1}{4 K} + \frac{1}{2\pi}\sum _{m=0}^\infty K^m \left(c_{0,m,0}-2 c_{0,m,1}\right).
\end{equation}
Since this is the same formula that extracts $1/\gamma$ in \eqref{ll-obs}, this series is calculated as an intermediary step when calculating \eqref{ll_eg}. To the first few orders we find
\begin{multline}
C = \frac{1}{4 K}+\frac{\ell _K-1}{4 \pi }+\frac{K}{16 \pi ^2} \left(\ell _K^2-2\right)
+\frac{K^2}{64 \pi ^3} \left(2 \ell _K^2-3 \zeta (3)-1\right)\\
+\frac{K^3}{384 \pi ^4} \left(-2 \ell _K^3+6 \ell _K^2+\ell _K (3+9 \zeta (3))-24 \zeta (3)\right)\\
+\frac{K^4}{12288 \pi ^5} \big(16 \ell _K^4-96 \ell _K^3+\ell _K^2 (48-144 \zeta (3))\\
+\ell _K (48+912 \zeta (3))+(6-720 \zeta (3)-405 \zeta (5))\big)
+\CO\left(K^5\right)
\label{cap-series}
\end{multline}
where $\ell_K = \log (16\pi/K)$. Using our method, as published in \cite{mr-ll}, this series was calculated to order $K^7$ by \cite{Reichert2020}. 

Recently, the analysis of the disk capacitor problem has been expanded in \cite{bbv} to include non-perturbative effects. There, they adapt and extend the integral equation techniques of \cite{mmr-antrans}, reviewed in chapter \ref{cha_antrans}, to be applicable to the Love equation \eqref{ll-inteq}. Using this formalism, they find the non-perturbative corrections to the capacitance, which agree with the non-perturbative scale \eqref{disc-ll} found in \cite{mr-ll}.
Furthermore, they relate the perturbative series \eqref{cap-series} to the series for $e$ and $\rho$ in the $O(3)$ non-linear sigma model, which allows them to generate beyond 300 terms.